# THE EFFECT OF THE GRAVITATIONAL MASS ON THE ELECTROMAGNETIC RADIATION FROM AN OBLIQUE, RELATIVISTICALLY ROTATING DIPOLE

BY

**Anwar Saleh Al-Muhammad**

A Thesis Presented to the
DEANSHIP OF GRADUATE STUDIES

**KING FAHD UNIVERSITY OF PETROLEUM & MINERALS**
DHAHRAN, SAUDI ARABIA

In Partial Fulfillment of the
Requirements for the Degree of

**MASTER OF SCIENCE**

In

**PHYSICS**

**December 2002**

# KING FAHD UNIVERSITY OF PETROLEUM & MINERALS

## DHAHRAN 31261, SAUDI ARABIA

## DEANSHIP OF GRADUATE STUDIES

This thesis, written by Anwar Saleh Al-Muhammad under the direction of his thesis advisor and approved by his thesis committee, has been presented to an accepted by the Dean of Graduate Studies, in partial fulfillment of the requirements for the degree of **MASTER OF SCIENCE IN PHYSICS**.

**Thesis Committee**

1/1/2003

Dr. Thamer Al-Aithan
Chairman

1/1/2003

Professor Asghar Qadir
Co-Chairman

1/1/2003

Professor Osama Al-Dabbousi
Member

1/1/2003

Professor Harry Mavromatis
Member

1/1/2003

Dr. Ali Al-Shukri
Member

Department Chairman
Dr. Ali Mohammad Al-Shukri

Dean of Graduate Studies
Prof. Osama A. Jannadi

14-1-2003

Date

***To my mother and father***

***and to my children***

***Hassan and Fatima***

# Acknowledgements

I am most grateful to Dr. Asghar Qadir, who guided me from A to Z to accomplish this thesis. Also, I would like to thank, other members of the thesis committee, Drs. Thamer Al-Aithan, Osama Al-Dabbousi, Harry Mavromatis and Ali Al-Shukri, for their useful comments and discussions. Drs. F. D. Zaman and Fazal Mohammad are greatly thanked for their useful contribution to solve the differential equation. Finally, I would like to thank the Physics Department of the King Fahd University of Petroleum and Minerals for the facilities, they provided to complete this thesis.

# TABLE OF CONTENTS

# LIST OF TABLES

# LIST OF FIGURES

# خلاصة الرسالة

**الاسم : أنور صالح علي آل محمد**

**عنوان الرسالة : أثر كتلة الجذب على الأشعاعات الكهرومغناطيسية المنبعثة من ثنائي القطب الذي يدور بزاوية مع محور الدوران بسرعة عالية نسبية**

**التخصص: فيزياء**

**التاريخ: شوال– ١٤٢٣هـ**

بعد اكتشاف النجوم الوماضة (millisecond pulsars) التي لا يزيد الزمن الدوري لوميضها عن واحد من ألف من الثانية، فقد أصبح نموذج النجوم النيوترونية التي تدور حول نفسها بسرعة نسبية مقاربة لسرعة الضوء أكثر قبولاً في الأوساط العلمية لتفسير الومضات المرصودة. كذلك فإن اكتشاف نجوم وماضة ذات زمن دوري أقل من ذلك أصبح أمراً ممكناً. تبحث هذه الرسالة أثر كتلة الجذب للنجم الوماض على مقدار الإشعاعات المنبعثة من النجم عند تلك السرعات العالية. وذلك باستخدام مبادئ النظرية النسبية الخاصة والعامة في إطار نموذج ثنائية الأقطاب المتصل بللنجوم الوماضة والتي تدور بشكل منحرف مع محور المجال المغناطيسي. استخدمنا طريقة ديباولس و انجروسو وقادر لتعميم طريقة هاكستون وروفيني التي وضعت لشحنة واحدة، لكي تطبق على ثنائي القطب بإدخال أثر كتلة الجذب.

فعند حل المعادلة التفاضلية للإشعاع بطريقة حسابية باستخدام برنامج ( Mathematica 4.1 ) وبطريقة تحليلية أمكن رسم طيف الطاقة لثنائيات مختلفة الكتلة. هذه الطيوف أظهرت انخفاض ملموس في معامل التضخيم النسبي ($\gamma^4$) بخانتين إلى ثلاث خانات عند تغير الكتلة من ٠,٥ إلى ٣ كتل شمسية بحسب زاوية عند كمية الحركة المنخفضة نسبياً . وهذا يدل على أن معظم كمية الحركة الزاوية للنجم النيتروني (ثنائي الشحنة) يظل كطاقة حركية دورانية بدلاً من تحوله إلى طاقة إشعاعية. إيضاً فإن الإنخفاض في الطاقة نتيجة للزيادة في الكتلة هو تقريباً مستقل عن عدد كمية الحركة الزاوية وسرعة النجم العالية الدورانية. كذلك فإن الطاقة الإشعاعية تتناسب مع مربع جيب زاوية ميل محور الدوران عن محور المجال المغناطيسي. وهذا يشابه العلاقة التقليدية بينهما. بينما يضمحل الإشعاع عند الأرقام العالية لكمية الحركة فيتلاشى أثر الكتلة. ومن ذلك نستدل على أن التضخم في الطاقة الإشعاعية الذي ينشأ عند الأخذ بالإعتبار مبادئ النظرية النسبية الخاصة فقط، يتوقع أن يقل بسبب الزيادة في كمية الحركة الزاوية الناتجة عن مبادئ النسبية العامة نتيجة عدم إهمال أثر الكتلة.

# Abstract

**Name:** **Anwar Saleh Al-Muhammad**

**Title:** **The Effect of the Gravitational Mass on the Electromagnetic Radiation from an Oblique, Relativistically Rotating Dipole**

**Major:** **Physics**

**Date:** **December, 2002**

After the discovery of the millisecond pulsars the model of a highly relativistic neutron star (NS) became the most acceptable to explain the observed pulses. Also, the discovery of sub-millisecond pulsar is possible. In this thesis, we study the effect of the gravitational mass of a pulsar on the amount of the electromagnetic radiation by using both the special and general theories of relativity. Relying on the magnetic dipole model of the pulsar, we use the extension of the work of *Haxton-Ruffini* [31] for single charges by *DePaolis-Ingrosso-Qadir* [32] for an obliquely rotating magnetic dipole, to incorporate the effect of the gravitational mass.

By using the numerical (evaluated by Mathematica 4.1) and analytical solutions of the differential equation for the radiation, we construct the energy spectra for different masses of the dipole. These spectra show that, in relatively low angular momentum $l$, the effect of the gravitational mass is very significant in suppressing the relativistic enhancement factor ($\gamma^4$), which had been found [27, 28, 32], by two to three orders of magnitude, as the mass changes from 0.5 $M_\odot$ to $3M_\odot$. It is an indication that most of the angular momentum of the NS is retained as rotational kinetic energy instead of being radiated as an electromagnetic energy, $E$. Also, the suppressing in radiation energy is more or less independent of the angular momentum, $l$, and the high rotational velocity, $\beta$. We also found that $E \sim \sin^2\chi$, where $\chi$ is the inclination angle of the obliquity, which is similar to the classical behavior. However, in the very high l, the whole radiation suppresses and the effect of mass is neglected. It indicates that the (special) relativistic enhancement expected is lost to the (general) relativistic increase of angular momentum after incorporating the effect of mass.

**Master of Science Degree**

**King Fahd University of Petroleum and Minerals**

**Dhahran, Saudi Arabia**

**December, 2002**

# CHAPTER 1

# INTRODUCTION

## 1.1. DEVELOPMENT OF STELLAR STUDY

The word “Astronomy” comes from the Greek words “Astrs” (stars) and “nomy” (study)> It was originally used only to signify the study of stars, but now covers a wide range of celestial objects such as, planets, asteroids, nebulae, etc. It has been developing since the ancients first started observing the heavens. The first known recorded astronomical observation was made in China (about 3000 B.C). The initial study was mostly qualitative. Hipparchus drew up the very first (known) catalogue of stars showing their brightnesses and positions in about 100 B.C. However, that work was not sufficiently precise. In 150 A.D Ptolemy provided the first precise catalogue of stars, which used numbers to record their positions [1]. Aristotle stated the very first qualitative cosmology in about 350 B.C. In his cosmological system, he regarded the Universe as composed of five elements: earth, water, air, fire and aether, whose “atoms” had the shape of the five Platonic solid. He considered the Earth as the center of the Universe (the geocentric view). The domes of the Moon, Venus, the Sun, Mars, Jupiter, Saturn, and the “fixed stars” surround them. That view remained dominant for 1800 years [2].

The next major developments came with the Muslims. For example in the 10th-century, Al-Sufi presented one of the most comprehensive lists of stars in his book on the subject. There were further developments in improving precision of observation, and on the theoretical side, from the Muslim civilization. Afterwards (1054 A.D) astronomers around the world, observed a "guest star" in the sky. In particular there are references to it, with proper placement, in the Chinese and Muslim civilizations. Oddly enough, there are no records among the European civilizations of that time. The remains of the first recorded stellar explosion, called a *supernova*, formed what came to be known later as the Crab Nebula. Ulugh Beg built one of the most famous of the pre-telescopic observatories in Samarkand. In about 1420 A.D, he constructed a star catalogue, which is found later to be of better accuracy than an angle of 0.05° of arc [2, 3].

After the decline of the Islamic civilization, further developments came from Europe in the Renaissance. At the beginning of that period (1512 A.D), Copernicus reestablished a cosmological view in which the Sun is considered as the center of the Universe (heliocentric) instead of the Earth. In 1572 A.D Tycho Brahe proved that a supernova he observed had been a star. The most important of those developments was the use of the telescope to observe the heavens by Galileo in 1609. The telescope was used for all astronomical observations thereafter. Soon thereafter, 1609 - 1618 Kepler stated his three laws of planetary motion [2]. Rekeory discovered the first reported binary star (Mizar in the Big Dipper constellation) in 1651 by using a telescope.

Sir Isaac Newton published his famous laws of motion and gravitation in his book (*Principia Mathematika*), in 1687 and used infinitesimal calculus to deduce Kepler's laws from his law of Gravitation. They have been used to study the motion of objects in the Universe including stars. In 1718 A.D. Edmond Halley showed that stars move with respect to each other. (This kind of motion is called "proper motion".) He pointed out that they seem to be fixed because they are so far away. In 1783, John Goodricke explained the varying brightness of the first known variable star, Algol. It was the first known Eclipsing star. Then, in 1838, Friedrich Bessel measured the distance of a star by using the "parallax method". It uses the small change of angle of the star due to the change of the position of the Earth over a period of six months. In 1868, Huggins used the *Doppler*

*Effect* (change of wavelength due to speed of the source of the wave) to determine the speed of a star along the line of sight [4].

Two major revolutionarily theories in physics arose in the 20$^{th}$ century [5]. The first is the *quantum theory*, starting with the work of Max Planck in 1900. He used it to explain the spectra of blackbody radiation. The second was the establishment of *special relativity* by Einstein in 1905 A.D. It predicted the equivalence of mass and energy. This led to a better understanding of the energy generation mechanism of stars through nuclear fusion [2]. In fact those two theories drew up the main outlines of the modern development of physics, including stellar study over the 20$^{th}$ century.

By using relativity, quantum mechanics, and radioactivity (discovered by Mme. Curie in 1898) astrophysicists proposed several sequences of fusion reactions (nucleosynthesis), called chains, to describe energy generation by stars. They include chains and cycles, such as the proton-proton (p-p) chain, and the carbon-nitrogen-oxygen (C-N-O) tricycle. Hans Bethe proposed this cycle during the 1930's. [2]

A new method [2] to study stellar spectra was invented in about 1910. Ejnar Hertzsprung in Denmark and Henry Russell in the United States independently plotted a graph, called an *H-R diagram*. In that diagram, they plotted the intrinsic brightness of the stars versus their temperature. They noticed that all points that they drew are located (as seen later) in limited regions of the diagram. Sir Arthur Eddington (1924) explained the relationship between the mass and the brightness of the star, which showed that the most massive stars are the hottest and they are of the highest luminosity [6]. The discovery of such a relationship led to a much deeper understanding of stellar structure [7], and used all the ideas and concepts that had been examined previously for the stars, as can be seen in the next section.

By using quantum statistics, Chandrasekhar and Fowler (1931) showed that a star should collapse under gravity to a very high density ($\sim 10^5$ gm cm$^{-3}$) and small radius (~5000 km). Such a star was later called a white dwarf (WD). Kuiper first observed these stars in 1935 [4]. Soon after the discovery of a neutron by Chadwick in 1932, Landau (1932), Gamow (1934), Baad and Zwicky (1934) and Oppenheimer (1939) showed the possibility of a neutron star (NS), of about the same mass as a white dwarf but with much less volume (~ 10 km diameter). [4]

The year 1967 was critical for astrophysicists because one of the most remarkable discoveries was made on 28 Nov of that year. That discovery was of periodic pulses emitted by an object in the constellation Cepheus. Jocelyn Bell and Anthony Hewish at Cambridge University detected this object and it was the first discovered pulsar. Later on, Hulse and Taylor (1975) discovered the first binary star pulsar, which emitted X-rays [8]. The discovery of the first millisecond pulsar was eight years later at 1982 by a team of the University of California [2]. The last essential event in recent memory is live observation of supernova in 1987 A.D., which expanded the knowledge of astronomers about stellar explosion [1]. Since the beginning of those discoveries and observations, intensive studies have been concentrated to construct suitable models for those stars. Astrophysicists have tried to fit the models with the observed radiation from those stars. A neutron star is the best candidate model for the pulsar because it agrees most closely with observation.

In our research, we will use the multipole expansion of the radiation field from an oblique relativistically rotating magnetic dipole. For this purpose we use the results of Haxton and Ruffini for a circular orbit of a charged particle in a Schwarzschild geometry [31]. Not only the effect of high speed [27], but also the effect of the Schwarzschild mass on the multipole expansion of the radiation field will be worked out. Then the total radiation field and its asymptotic behavior (at infinity) will be worked out. These results are expected to be relevant for pulsars. Thus, it will shed further light on the neutron star model of the pulsar.

Before going on to the actual research, one should clarify some points relative to the subject. In order to do that, we first give a general overview of stars, which includes, visualization of stars, their types, stellar evolution and radiation. Then the structure of normal stars and their modeling will be considered. The second chapter of this thesis will treat degenerate stars, mainly the white dwarfs. Pulsars will be discussed next. Then, fast rotating dipoles will be treated in detail, without incorporating the mass of NS, by using special relativity theory and electrodynamics. In the fifth part, a differential equation incorporating the mass of NS will be solved numerically and analytically for radiation emitted by a relativistically rotating dipole. It will be solved numerically by using a power series around the surface of a NS and analytically by using asymptotic analysis at a sufficiently far point from it. Then we compare this result with that obtained with out incorporating the mass in to show its effects. The results will be discussed in the last part.

# 1.2. GENERAL OVERVIEW

A star can be visualized as a huge ball of glowing gas. The Sun is the closest star to us. There are several billions of billions of stars in the observable Universe (about $10^{22}$ stars). They are very far and the nearest star - other than the Sun - is more than $4 \times 10^{13}$ km away. Stars started forming about 10 billion years ago. However, new stars are still forming. The Sun itself was probably formed from a rotating mass of gas and dust about 4.1-4.2 billion years ago [1, 4].

## 1.2.1. CLASSIFICATION OF STARS

People started to classify stars according to their groups (constellations), the position in the sky, movement (regular for the so-called "fixed stars" and irregular for the "planets", (which is Greek for "wanderers") and apparent brightness of the stars. But the recent classification of the stars is built, using many star characteristics.

### 1.2.1.1. CLASSIFICATION BY SIZE

Stars can be classified [2, 3, 6, 7] according to their size:

**1.** Big stars like giants and super-giants. They have a high luminosity. Hydrogen in their core is almost burned, so they are generating energy by further fusion, the process by which elements combine with each other to form heavier elements by nuclear reactions. Blue giants and super-giants have higher pressures and densities than the smaller normal stars. Red giants have lower pressures and densities than the smaller normal stars.

**2.** Medium main-sequence stars (like the Sun). They make up about 90 per cent of the stars that can be seen from the Earth. Unlike giants, main-sequence stars still burn hydrogen into helium through nuclear fusion deep inside the star.

**3.** Small stars like white dwarfs and neutron stars. White dwarfs are much smaller than main-sequence stars, but with higher temperatures and densities. White dwarf stars no longer have a supply of energy from fusion, so gravity within white dwarfs shrinks them to their small size. If their gravity overcomes the electron or neutron degeneracy, it can convert them to neutron stars or black holes respectively. The former occurs only if the mass of the star is more than the Chandrasekhar limit, 0.7 (1.4) solar masses for a cold (hot) star [1, 2, 7]. For the latter the limit is 3.2 solar masses ($M_\odot$), as was shown by Fang and Ruffini, independently [10].

## 1.2.1.2. CLASSIFICATION BY TEMPERATURE

Stars can be classified according to their spectral type, which depends on their temperature, in 7 groups. Each one is divided into 10 sub-groups. The foregoing kinds of stars (excluding black holes) can be any one of those groups. From hottest to coolest, the spectral types are lettered as follow: O, B, A, F, G, K, M. The sun is a type G (yellow) dwarf, which lies in subdivision 2 [2, 7].

#### 1.2.1.3. CLASSIFICATION BY COLOR

The above classification of stars can also be expressed according to their observing color filter, because the star seems to be brighter if we observe it through the filter of the same color. Those filters are from hottest to coolest as follows: U, B, V, R, I, J, H, K, L, M, N. The standard set of filters, which has been defined to use in photo-electric systems, has three main filters (spectral regions), one in the ultraviolet U, the second is blue B, and the third is yellow or visual V. Hence, observing stars with this UBV filter is called *three-color photometry*. Also, they can be classified according to their filter, for example, our Sun is a type V star [2]. Another way of measuring the temperatures of stars is by using the color indexes which are the difference between the apparent magnitudes measured in two spectral regions. One of the important color indexes is B-V (blue minus visual) index.

### 1.2.2. HERTZSPRUNG-RUSSELL (H-R) DIAGRAM

In order to understand the relationship between the color and luminosity of a normal star,

Hertzsprung and Russell [2, 7] plotted a diagram for the purposes of classification. As mentioned before, the absolute magnitude of the stars is plotted against their temperature i.e. O, B, A, F, G, K, M or their color index i.e. B-V. For completeness, it is essential to distinguish between apparent and absolute magnitude. The former measures the brightness of the stars as they appear from Earth, which depends on their distances. The latter measures the brightness of the stars *as if they were at the same distance*. Traditionally, the unit distance used is 10 parsec (1 parsec = 3.2616 light year ≈ 3.1 × $10^{13}$ km). H-R diagram for nearby and bright stars can be seen in Fig. 1.1.

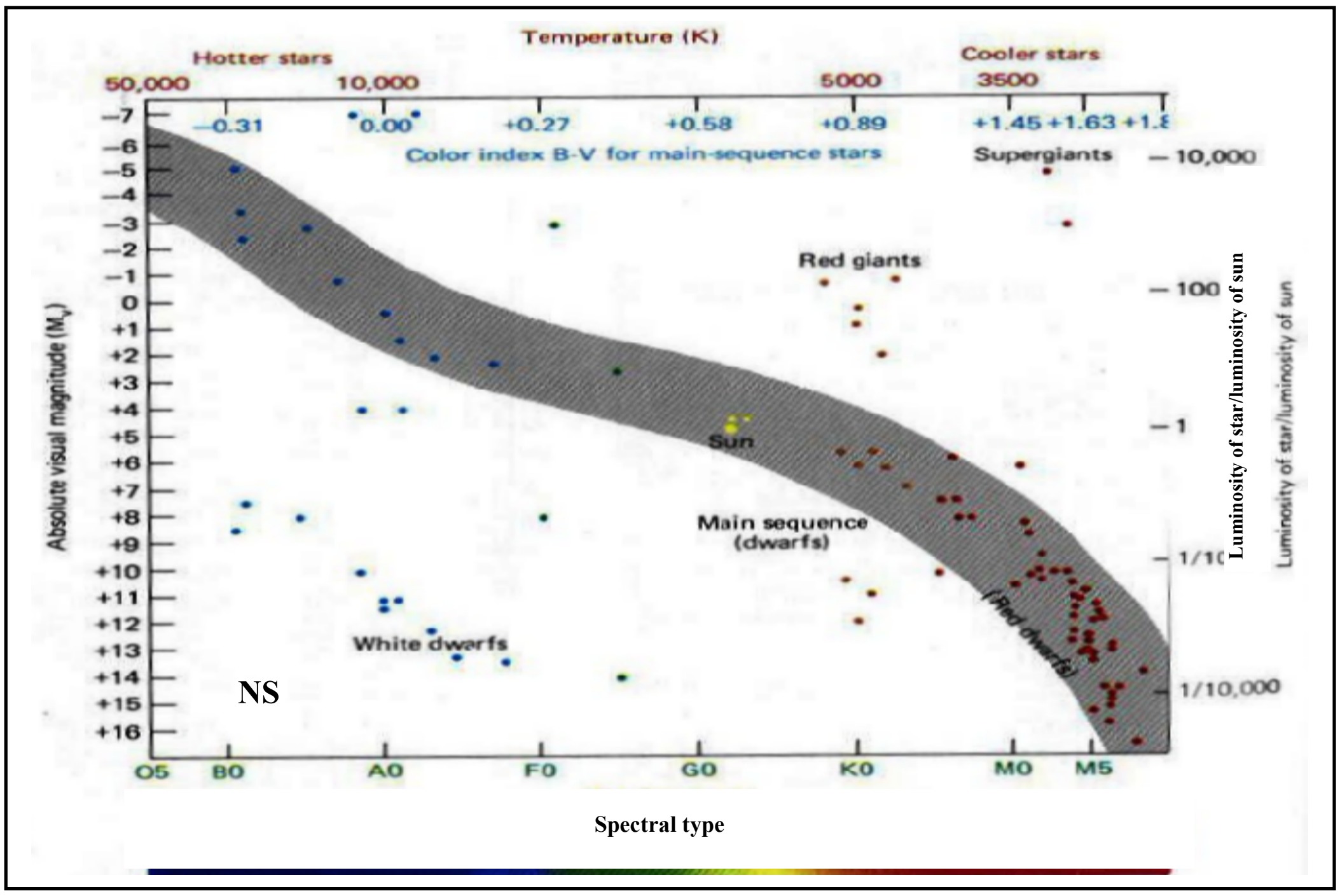

**Fig.1.1.** Hertzsprung-Russell (H-R) diagram, which includes both nearby and bright stars. The spectral-type axis (*bottom*) is equivalent to the temperature axis (*top*).The absolute magnitude axis (*left*) is equivalent to the luminosity axis (right).

The relation [7] between absolute (*M*) and apparent (*m*) magnitude is

$$M = m + 5 - 5\log_{10} d, \tag{1.1}$$

where $d$ is the distance of the star in parsecs. Notice that all stars lie in a limited region of the diagram. Also, it can be considered as one of the proper and easy ways to study most of the stellar characteristics. One can also use the H-R diagram to classify the stars according to their luminosities see Figs. 1.2. It can also be used also, to study the motion, the distances of the stars, and the evolution of the stars.

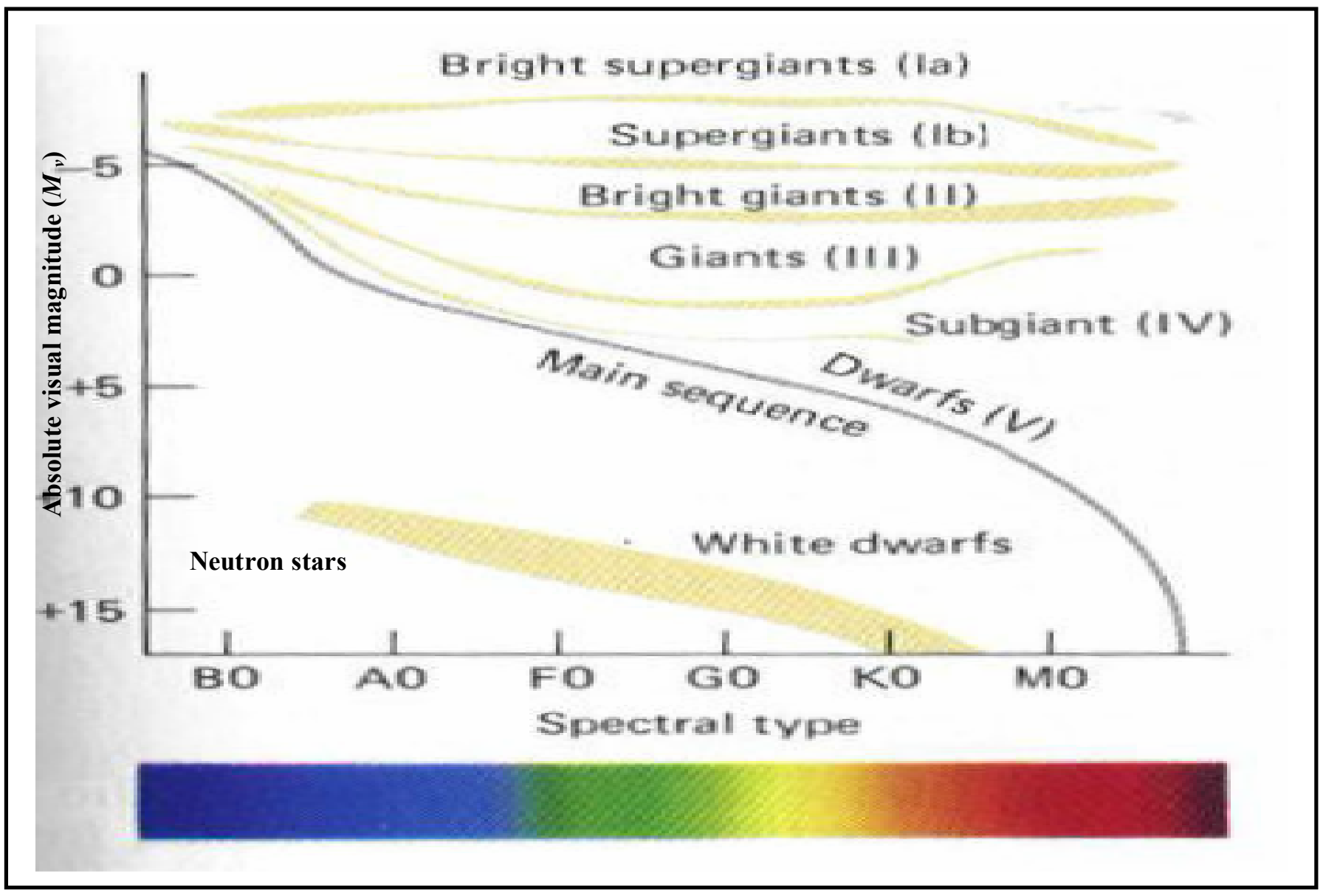

**Fig.1.2** Hertzsprung-Russell (H-R) diagram, which shows the luminosity classification of stars. They are specified by their name or numbers shown in parentheses.

# 1.3. THE SINGLE ORDINARY STAR

In order to understand the main-sequence single star, we have to answer three questions. First, how do the stars form? Second, what keeps a star in equilibrium? Third, how do the stars generate energy?

## 1.3.1. FORMATION OF STARS

New stars are continually forming from condensation in the interstellar gas and dust. The source of gas and dust might be the outer atmosphere of giant stars or the remnants of stellar explosions. If there is any minor density enhancement the dust will contract and keep on contracting because of gravity (self-gravitation) [2, 12]. To work out the extent of density fluctuations required for star formation we appeal to the "*virial theorem*". It states that when a self-gravitating system of particles contracts, half the gravitational potential energy is radiated and half becomes thermal energy. The density required making the inward directed self-gravitation overcome the outward directed forces due to thermal motion and hydrostatic pressure is $\sim 10^{-19}$ g cm$^{-3}$. The interstellar density is $\sim 10^{-23}$g cm$^{-3}$ or lower. Thus there has to be a thousand-fold density enhancement for stars to form. The evolutions of stars mainly depend on the mass of protostars, see the evolution summary for high and low mass protostars is shown in Fig. 1.3.

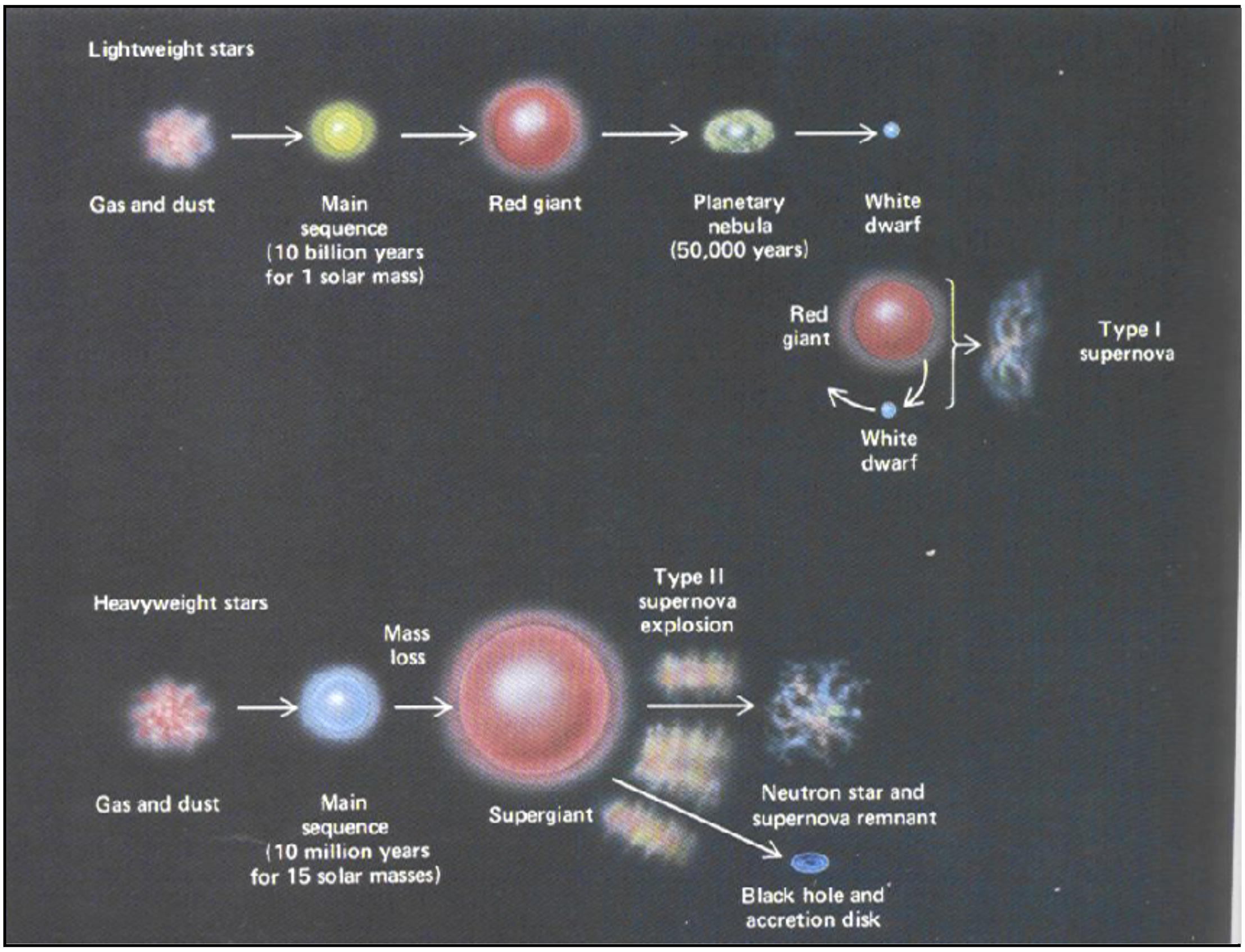

**Fig.1.3.** A summary of the stages of stellar evolution for lightweight and heavyweight stars.

At this stage the dust cloud is called a "protostar". One can draw [2] the lifetime path of stars in H-R diagram to form what is called an "evolutionary track". The mass of the gas and dust is the dominant factor in determining the evolutionary track and the time that the star needs to contract from a protostar to a normal star. The very massive stars need to short period of time to contract by comparison with the less massive stars. For example a star that has a mass of 10 solar mass ($10M_{\odot}$) needs only $2 \times 10^5$ years to reach the hydrostatic equilibrium while that of 1 solar mass ($1M_{\odot}$) needs $5 \times 10^7$ years, see their evolutionary track in Fig. 1.4. In hydrostatic equilibrium of the normal star, the outward force balances the inward force as will be discussed in the next section.

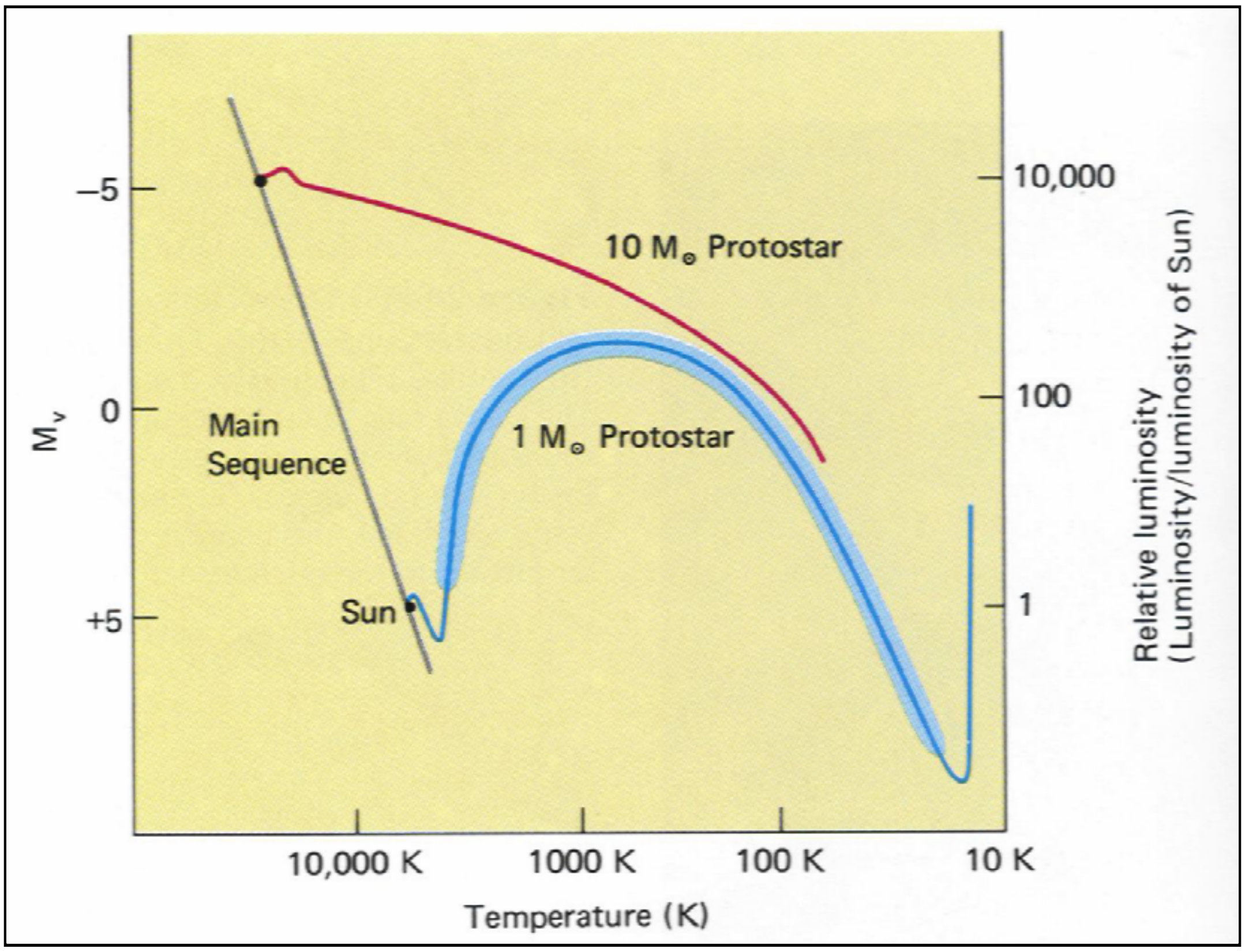

**Fig.1.4.** (H-R) diagram for two protostars, one of 1 solar mass ($1M_{\odot}$) and the other of 10 solar mass ($10M_{\odot}$).

## 1.3.2. HYDROSTATIC EQUILIBRIUM OF STARS

To understand why a star does not collapse upon itself, or break up, one must examine the forces at work in the stellar interior. Consider a small piece of a star with mass m at a distance r (less than the radius) from the center of the star. The forces [9] acting on this mass unit are:

(i) The force of gravity ($\mathbf{F_g}$) acting toward the center;

(ii) The force of buoyancy (**$F_{gas}$**) due to the pressure difference acting on the mass element;

(iii) The force of radiation pressure (**$F_{rad}$**) acting outward on the mass element;

(see the forces acting on the unit mass of the normal star in Fig. 1.5).

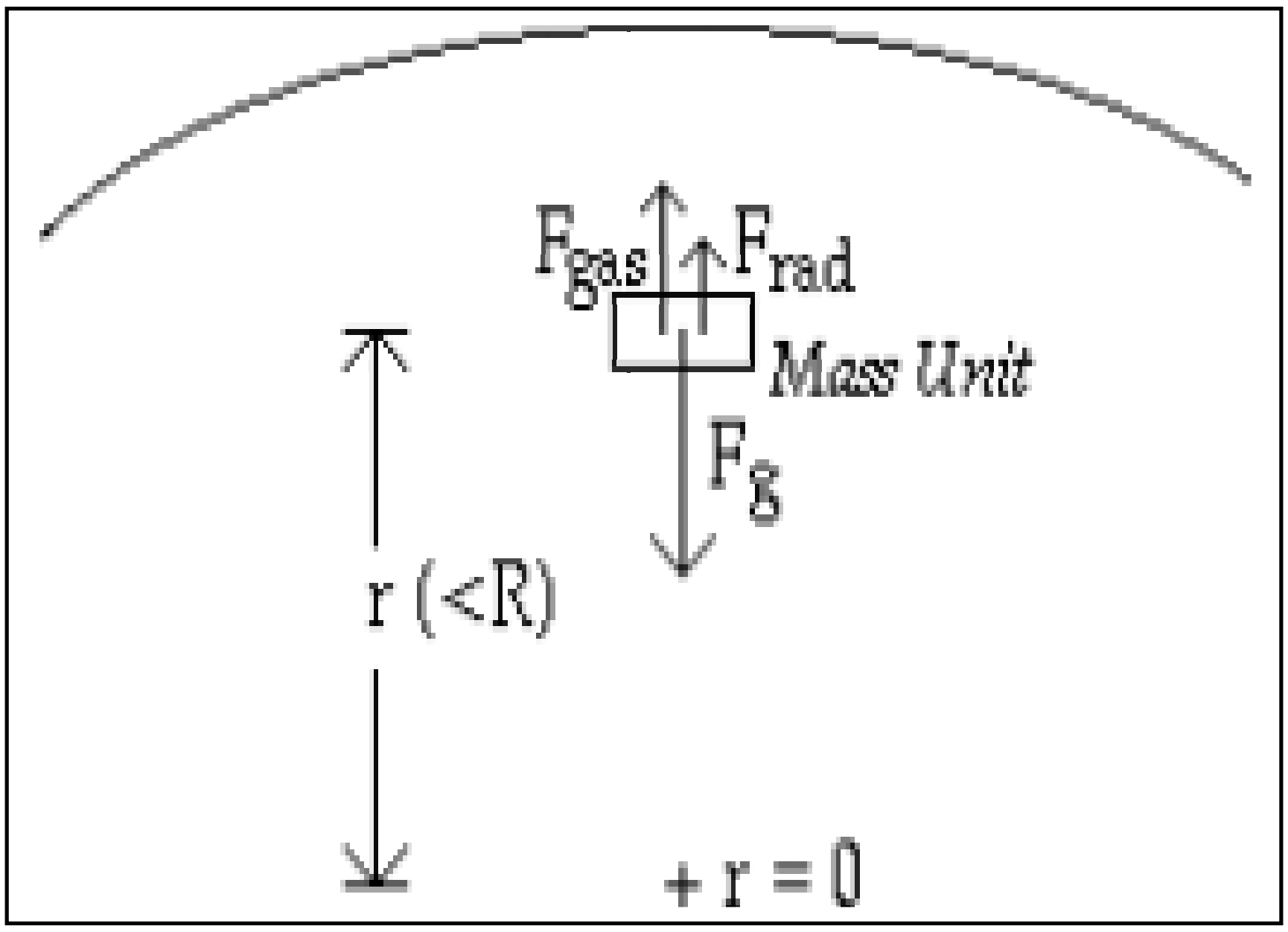

**The center of star**

**Fig.1.5.** Free-body diagram of a unit mass of the normal star.

For the star to be stable, there must be no net contraction or expansion on the part of these mass elements throughout the stellar interior. For each part and by extrapolation the entire star, to be in equilibrium the net force must be zero:

$$\mathbf{F}_{\mathbf{net}} = \mathbf{F}_{\mathbf{g}} + \mathbf{F}_{\mathbf{gas}} + \mathbf{F}_{\mathbf{rad}} = 0. \qquad (1.2)$$

This condition is known as *hydrostatic equilibrium* For a thin shell in a stable star, as shown in Fig.1.6, then (1.2), reduces to

$$\frac{dP}{dr} = -\frac{GM(r)\rho}{r^2}, \qquad (1.3)$$

where $G$ is Newton's gravitational constant, $P$ is the  pressure at the shell, $\rho$ is its matter density and $r$ its radius.

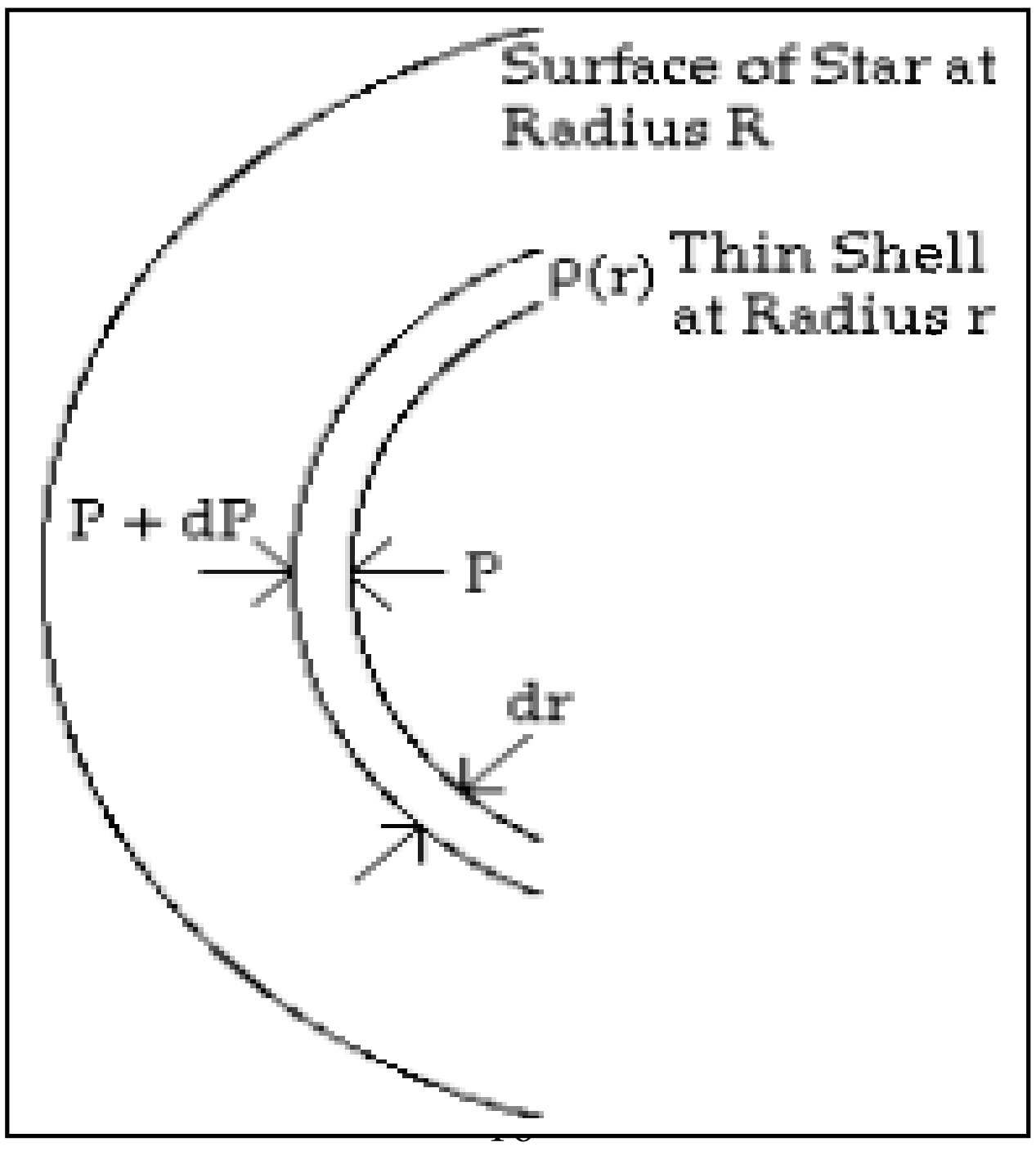

**Fig.1.6.** Thin shell in a normal star under hydrostatic equilibrium.

## 1.3.3. ENERGY GENERATION IN STARS

If gravitational collapse were to be the source of a star's energy generation [2, 12] (as given by the *virial theorem)*, then the ages of stars would have to be much shorter than they are known to be. If there were no extra force that could overcome the gravitational force then the process of collapse would end in a short astronomical period ($\sim 3 \times 10^7$ year). However, some rocks on the Earth have an age of $4 \times 10^9$ years. Hence there has to be some other source of energy generation. *Nuclear fusion* is an adequate source of energy that can keep the stars shining and hold them up against gravitational collapse. Moreover, one can replace $\mathbf{F_{rad}}$ in (1.2) by $\mathbf{F_{fus}}$ to maintain equilibrium.

Nuclear fusion [2] is the process in which light elements are combined to form different heavier elements (*nucleosynthesis*) through nuclear reactions . High pressures and temperatures are needed to start such reactions. These pressures and temperatures can arise from gravitational collapse. As a result of the fusion reactions, there is a slight difference between the masses of the initial elements and the new elements. This difference of mass ($\Delta m$) leads to thermal radiant energy ($E_{rad}$) according to Einstein's special relativity [11],

$$E_{rad} = \Delta mc^2, \tag{1.4}$$

where $c$ is the speed of light ($c$ = 3 x $10^8$ m/s). Nuclear fusion usually occurs in the central part of the stars. Thus, only a small fraction of the original mass is converted to radiated energy. Consequently, the Sun loses only 0.007 of the mass of its central part to keep its present rate of radiation stable for a period of at least $10^{10}$ years [2].

The most probable reaction taking place in main sequence stars is the fusing of hydrogen nuclei to form helium nuclei. There are two ways to obtain such a reaction. One of them is the proton-proton chain (**PP*C***) and the other is the carbon-nitrogen-oxygen (**CNO**) cycle. The probability of each of them depends upon the temperature and composition of the stellar core [9]. In the stars of relatively low temperature ($T < 2 \times 10^7$ K), the former reactions take place, but in stars with higher interior temperature, above $10^8$K, the latter is dominant [2].

The combination of four hydrogen nuclei into helium can be obtained in the Proton-Proton Chain (PPC) reactions [2, 7, 9], which occurs in very massive stars, as follows:

$$H_1^1 + H_1^1 \rightarrow H_1^2 + e^+ + \nu \ ; \tag{1.5}$$

$$H_1^2 + H_1^1 \rightarrow He_2^3 + \gamma \ ; \tag{1.6}$$

$$He_2^3 + He_2^3 \rightarrow He_2^4 + 2H_1^1 . \tag{1.7}$$

The combination of hydrogen into helium through the Carbon-Nitrogen-Oxygen cycle (CNO-cycle) occurs in solar mass stars and goes through the following six steps:

$$C_6^{12} + H_1^1 \rightarrow N_7^{13} + \gamma ; \tag{1.8}$$

$$N_7^{13} \rightarrow C_6^{13} + e^+ + \nu \ ; \tag{1.9}$$

$$C_6^{13} + H_1^1 \rightarrow N_7^{14} + \gamma ; \tag{1.10}$$

$$N_7^{14} + H_1^1 \rightarrow O_8^{15} + \gamma ; \tag{1.11}$$

$$O_8^{15} \rightarrow N_7^{15} + e^+ + \nu ; \tag{1.12}$$

$$N_7^{15} + H_1^1 \rightarrow C_6^{12} + He_2^4 . \tag{1.13}$$

Since we end up with the same amount of carbon, it acts like a catalyst for the reaction. This is why it is called a cycle.

The source of the carbon used in the star for energy generation is the burning of helium through the triple alpha process:

$$He_2^4 + He_2^4 \leftrightarrow Be_4^8 + \gamma \; ; \tag{1.14}$$

$$Be_4^8 + He_2^4 \rightarrow C_6^{12} + \gamma \; ; \tag{1.15}$$

where "alpha" refers to the helium nucleus, which is an alpha particle, and the reverse arrow means that beryllium decays back into helium spontaneously after a very short period of time [7, 9]. If another alpha particle interacts with the beryllium before it decays, carbon is formed.

Through the process of converting hydrogen to helium, 0.007 of the mass of the hydrogen is transformed into energy [2]. It produces high temperatures that keep the stars in equilibrium. This energy is produced in the core of the stars and transported to the outer regions by convection and radiation [2, 9].

# 1.4. BINARY AND VARIABLE STARS

When seen by the naked eye, stars appear to be points, which shine steadily. (The twinkling is due to atmospheric effects.) However, many of them [2] are held together by gravity to form systems of two or more stars (clusters). Systems of two stars are called binary stars. Moreover, the brightness of many stars varies over time, either periodically or non-periodically. Some variable stars are binaries while others are just single stars. A study of such stars helps astrophysicists to determine some important characteristics of stars, and in turn to learn more about their nature.

## 1.4.1. BINARY STARS

Systems of stars are dynamically associated collections of stars. More than half of all stars are to be found in systems. Most of the systems are binaries. Some binary stars can be resolved by visual telescopes and they are called *visual binaries.* Alpha Centauri is a good example of a visual binary. For non-visual binaries, by using their spectra and the periodical change of wavelengths, astronomers can determine whether there are two or more stars revolving around each other. This is a direct application of the Doppler shift. This kind of binary is called a *spectroscopic binary.* The star Cabil is an example [3]. In some cases, the existence of a double star can be detected as a deviation from a straight line in the proper motion of the star across the sky over long periods. This is called an *astrometric binary.* The best example for this type of binary is the brilliant Dog star Sirius, which had a deviation of about 0.38° in 1000 years. [2, 7]

Binary stars can consist of ordinary main-sequence stars, giant stars, dwarf stars or collapsed objects. Many of the X-ray celestial sources turn out to be in binary systems. The source of X-rays comes from hot gases falling from a normal star onto its companion [6], which can be a white dwarf, neutron star, or a black hole. The best examples for this type of binary are HerculesX-1 and Cygnus X-1, in which the pairs consists of a super-giant and a neutron star in the former and a black hole (BH) in the later see Fig. 1.7, [2]. The period of the binary star system varies from a couple of days (1.7 days for HerculesX-1) to a couple of million years for highly separated binary [3].

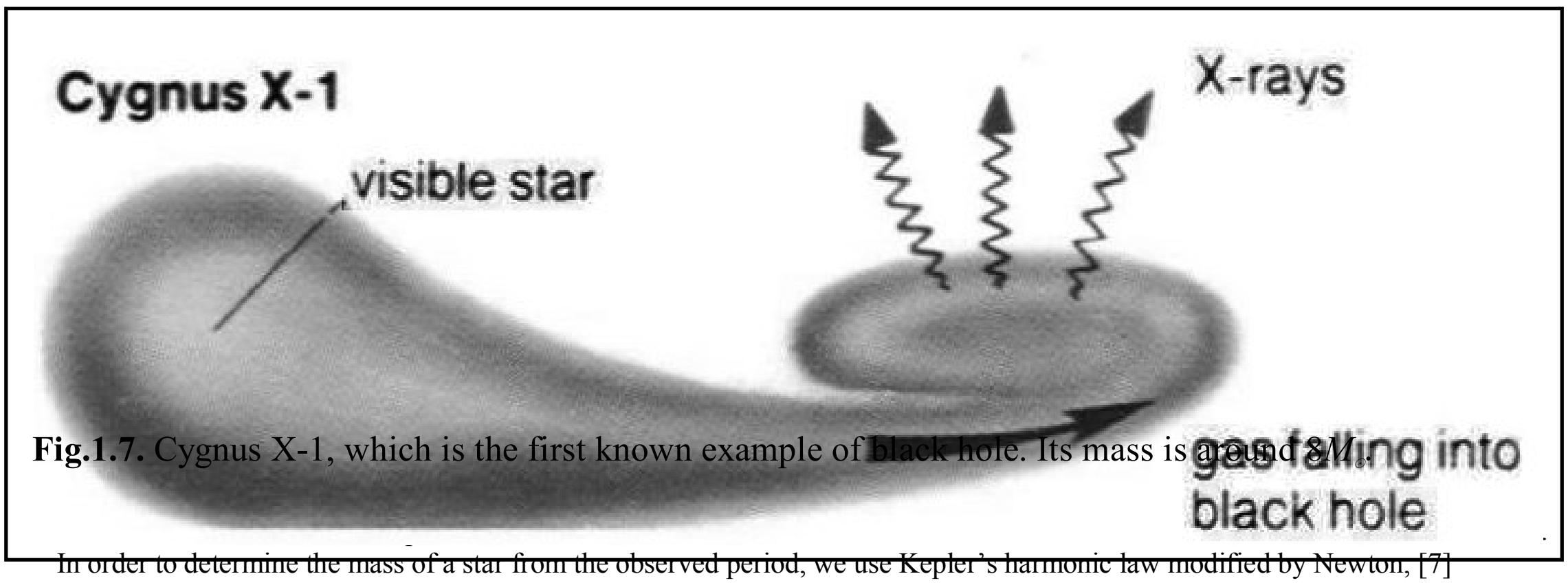

**Fig.1.7.** Cygnus X-1, which is the first known example of black hole. Its mass is around $8M_\odot$.

In order to determine the mass of a star from the observed period, we use Kepler's harmonic law modified by Newton, [7]

$$m_1 + m_2 = \frac{4\pi^2 \times a^3}{G \times p^2}, \tag{1.16}$$

where $m_1$ and $m_2$ are the masses of the two stars, $a$ is their linear distance apart, $p$ is the binary period and $G$ is Newton's constant. The center of their masses lies close to the bigger one. Thus, the masses of the star are inversely proportional to the distances ($d_1$ and $d_2$) between the stars and the center of mass,

$$\frac{m_1}{m_2} = \frac{d_2}{d_1}, \tag{1.17}$$

where $d_1 + d_2 = a$. By observing $p$, one can determine the values of $m_1$ and $m_2$ by solving (1.16) and (1.17). The difficulty of determining the mass of a single star is the same as the difficulty of determining the mass of a single planet, like Mercury or Venus.

## 1.4.2. VARIABLE STARS

Variable stars are stars that change their brightness. Astronomers have a special method to name them [7]. There are three main types of variable stars[2, 6, 7]:
(a) Eclipsing binaries, (b) Pulsating variables, and (c) Exploding stars.

### 1.4.2.1. ECLIPSING BINARIES

Eclipsing binaries are double stars, such as Algol. They move around each other, in such a way that one periodically blocks the other's light. This blocking reduces the total brightness of the two stars as seen from the earth periodically see Fig 1.8. The eclipsing period varies from minutes to years and it can be observed either visually or spectroscopically. [2, 6]

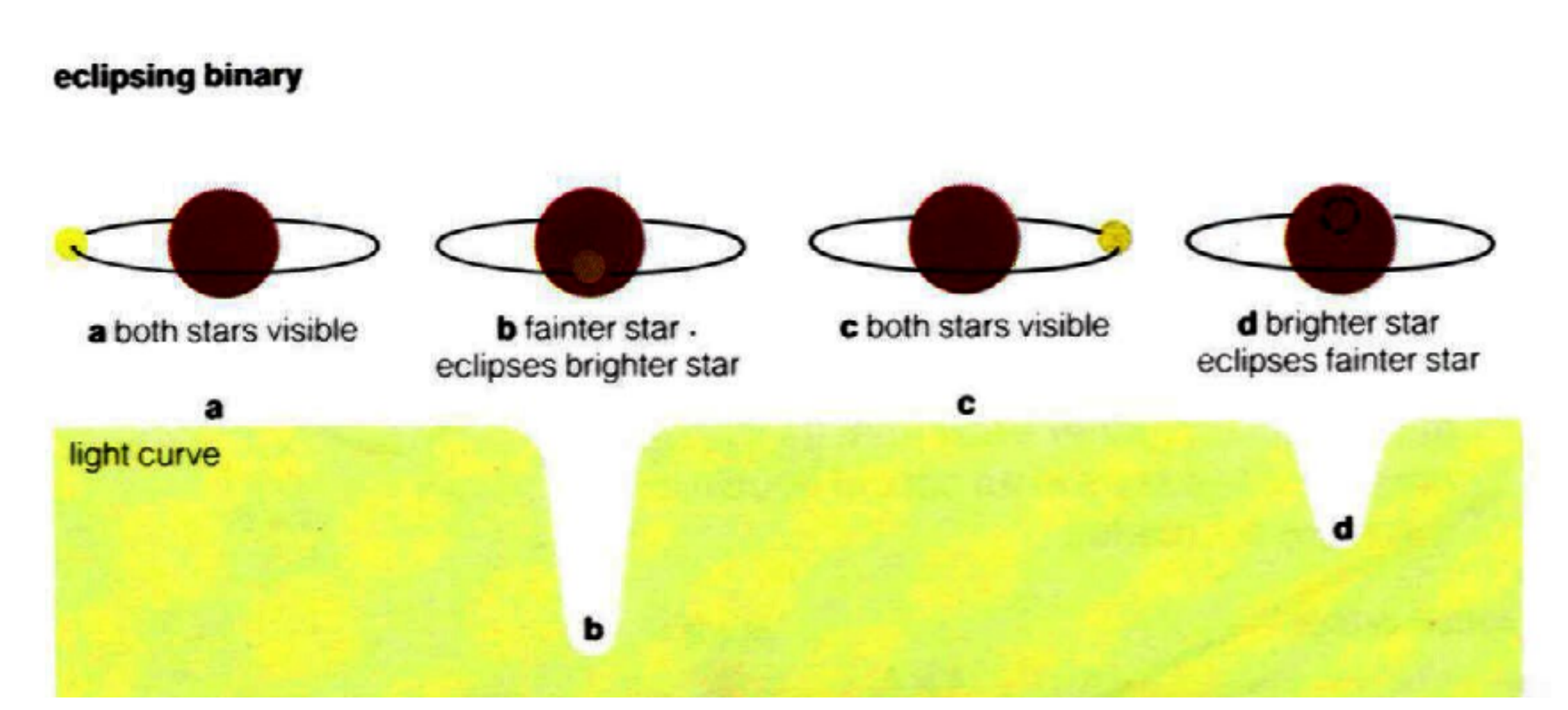

**Fig.1.8.** Eclipsing binary, down curve represents the observed light.

## 1.4.2.2. PULSATING VARIABLES

Pulsations occur when stars have intrinsic fluctuations in their brightness, which can happen in a single star, and are not because of eclipsing. They are due to expansion and contraction of the stars, which in turn makes them hotter and cooler and so emit more and less radiation periodically. There are thousands of stars of this type. The periods of the variation vary from seconds for some stars to years in others. There are three types of these variable stars, classified according to the variation of their brightness, spectrum and the duration of pulsation periods. They are usually studied by using *light curves*, which plot the brightness of the star against time, see Fig. 1.9. [2, 6, 7]

.

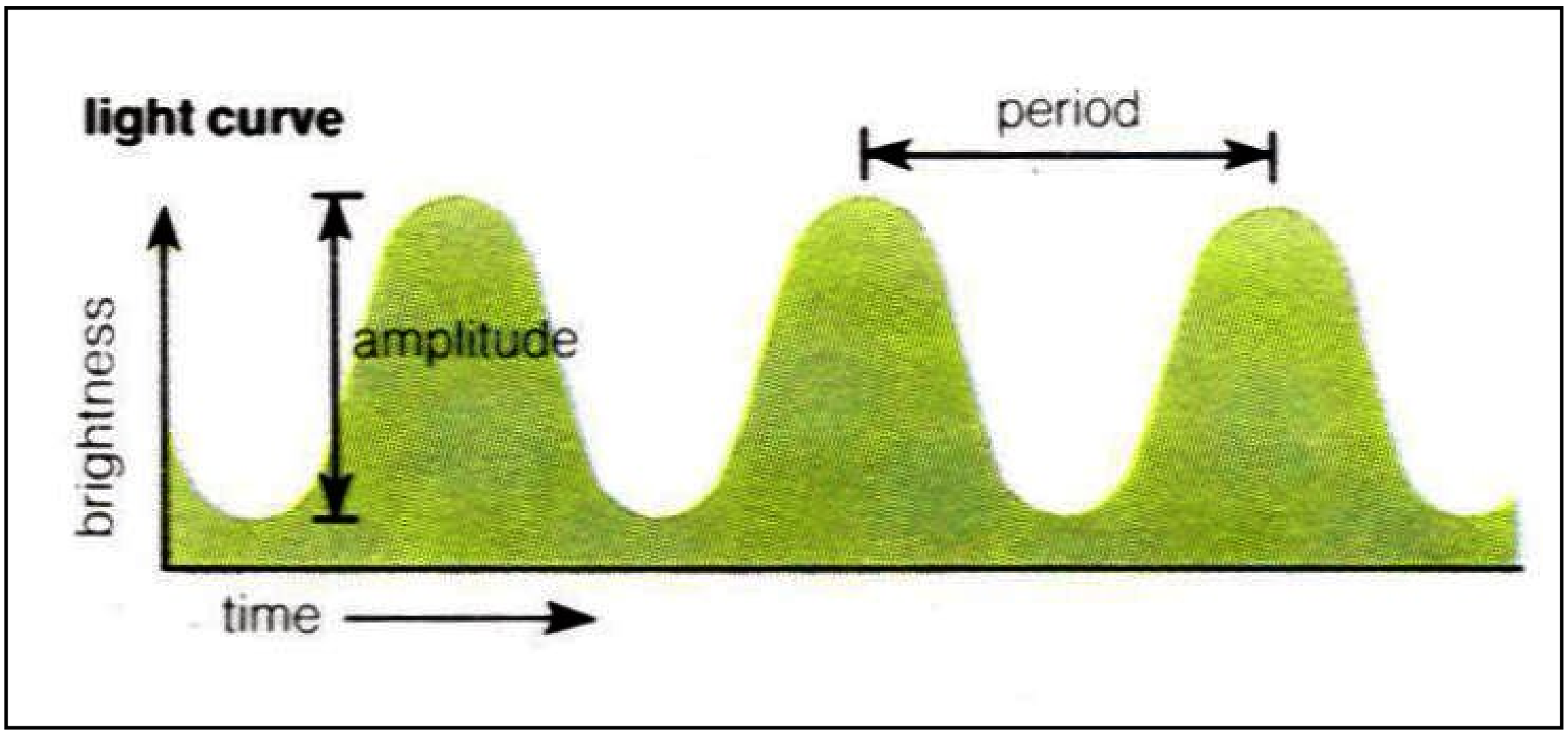

**Fig.1.9.** Light curve showing how the brightness varies in time.

***1.4.2.2.1. Cepheids.*** *They take their name from the star Delta Cephei in the constellation, Cepheus. It was one of the earliest discovered stars of that type, which fluctuates regularly every 5.4 days [7]. Cepheids usually change regularly in brightness by between 0.1 and 2 magnitudes in a period of 1 to 100 days [6]. One can say that the study of Cepheids was the key to understanding the distance scale of the universe, which led to the discovery of other galaxies [2].*

*The importance of Cepheid stars comes from their regular variation in brightness, which is used to plot the relation between the period of the change of their light and their absolute magnitude, see Fig. 1.10. It shows that Cepheids of long period have higher absolute magnitude because they are larger stars, which*

*take more time to contract and expand. Hence, one can determine the distance of a Cepheid from its absolute and apparent magnitude, (1.1), and by means of the inverse square law. [2]*

Cepheids are yellow super-giants stars which are of two types [7]: **type-I cepheids** (classical cepheids); and **type-II cepheids.** In type-I cepheids, the periods of pulsation range from 5 days to many weeks and their visual light often varies by about one magnitude. Polaris is a type-I cepheid with a small range of variation. However, type-II cepheids, which are some times called **W Virginis** stars, have periods of about 12 to 20 days. They have a broader maximum in curve. W Virginis is an example of this type of cepheids.

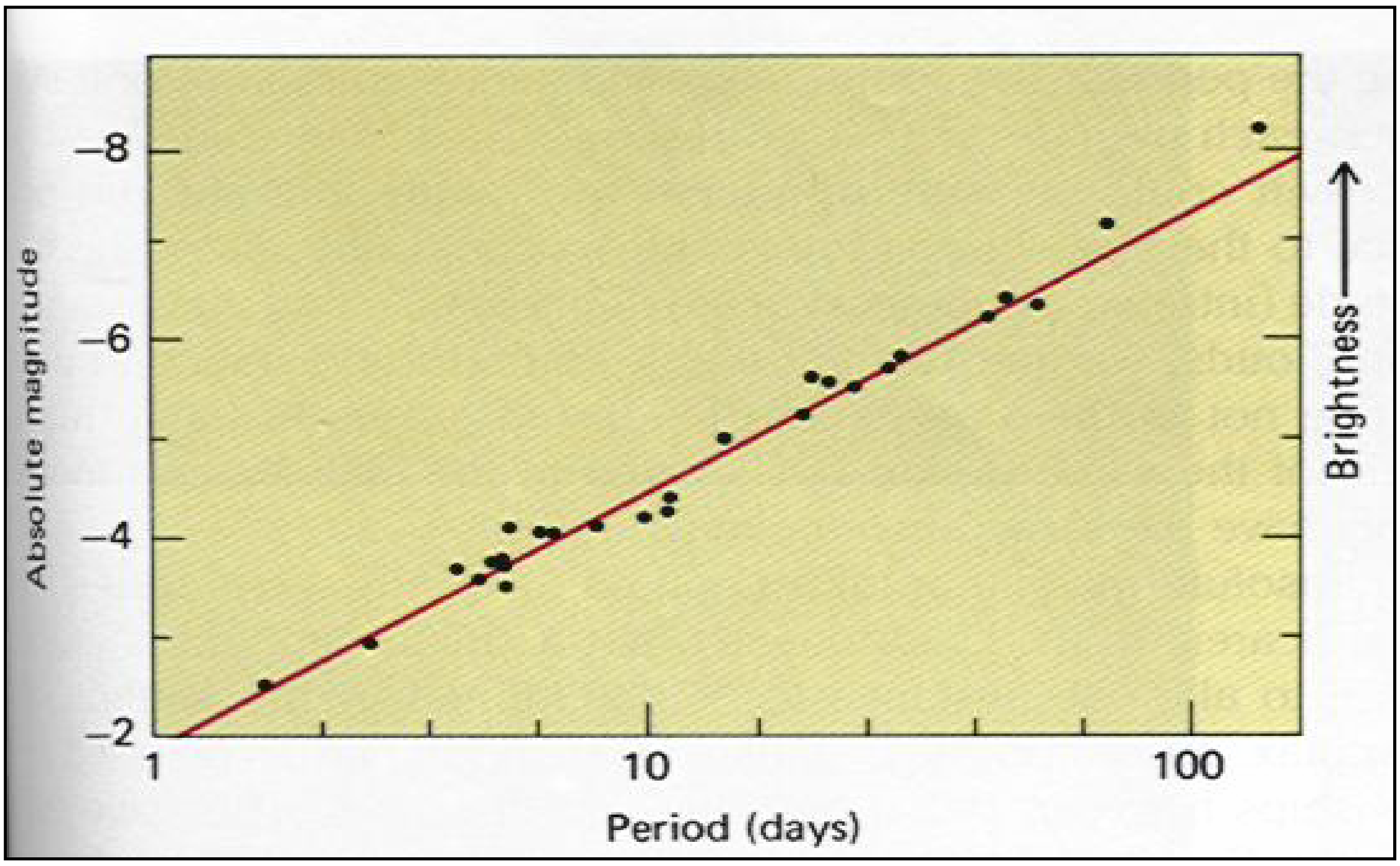

**Fig.1.10.** The period-luminosity relation for Cepheid variable.

**<u>1.4.2.2.2. RR Lyrae Variables</u>**. In some books they are also known as **cluster variables** because they were first detected in the globular clusters around the Lyra constellation. In fact they are blue giants of class A spectrum, whose maximum brightness is bluer. Also, they are known as **short-period** variables because they have a period of less than a day - around half a day. Their variation of brightness is up to 1.5 magnitudes. All stars of this type have an average absolute magnitude of about 0.6 [2, 7].

***<u>1.4.2.2.3. Mira Variables.</u>*** *<u>Mira</u> stars are long-period variables, whose periods can be from 3 months to about two years. They are so named because the star Mira, in the constellation Cetus, fluctuates in brightness over long periods. In fact, Mira stars are giants of spectral type M, and are the most numerous of variable stars. (One should distinguish between pulsating variables and pulsars, which will be discussed later in Ch.3) [2, 7].*

## ***<u>1.4.2.3. EXPLODING STARS</u>***

Exploding can take place in the star through 3 different ways, **nova**, **supernova-I**, and **supernova-II**.

1) **Nova**: is exploding of the matter that accumulates on the surface of a white dwarf in a binary system with an ordinary star, see Fig.1.7. And it may explode if the matter re-accumulates but nova does not destroy the star. [1, 6]
2) **Supernova-I**: In this type, matter that accumulates on the surface of a white dwarf in a binary system makes the white dwarf heavier than the Chandrasekhar limit (Ch.2). So the white dwarf then explodes and throws off solar mass of gas with 12000 km/s, see Fig.1.11 [6].
3) **Supernova-II**: In this type, the mass of white dwarf will be at least 4 times the solar mass, $4M_{\odot}$. And in that several solar mass of gas will be thrown out but with less speed (~ 5000 km/s), see Fig.1.11 [6].

After a supernova takes place, the gas moves off into space to form a shell around the exploded star, that shell is called supernova remnant (SNR). The most famous supernova occurred in the Milky Way in 1054 and produced a huge cloud of rapidly expanding gas (remnant) called the Crab Nebula. The Crab Nebula has a spinning neutron star-a pulsar-at its center. Some time a black hole forms in the center [1, 6].

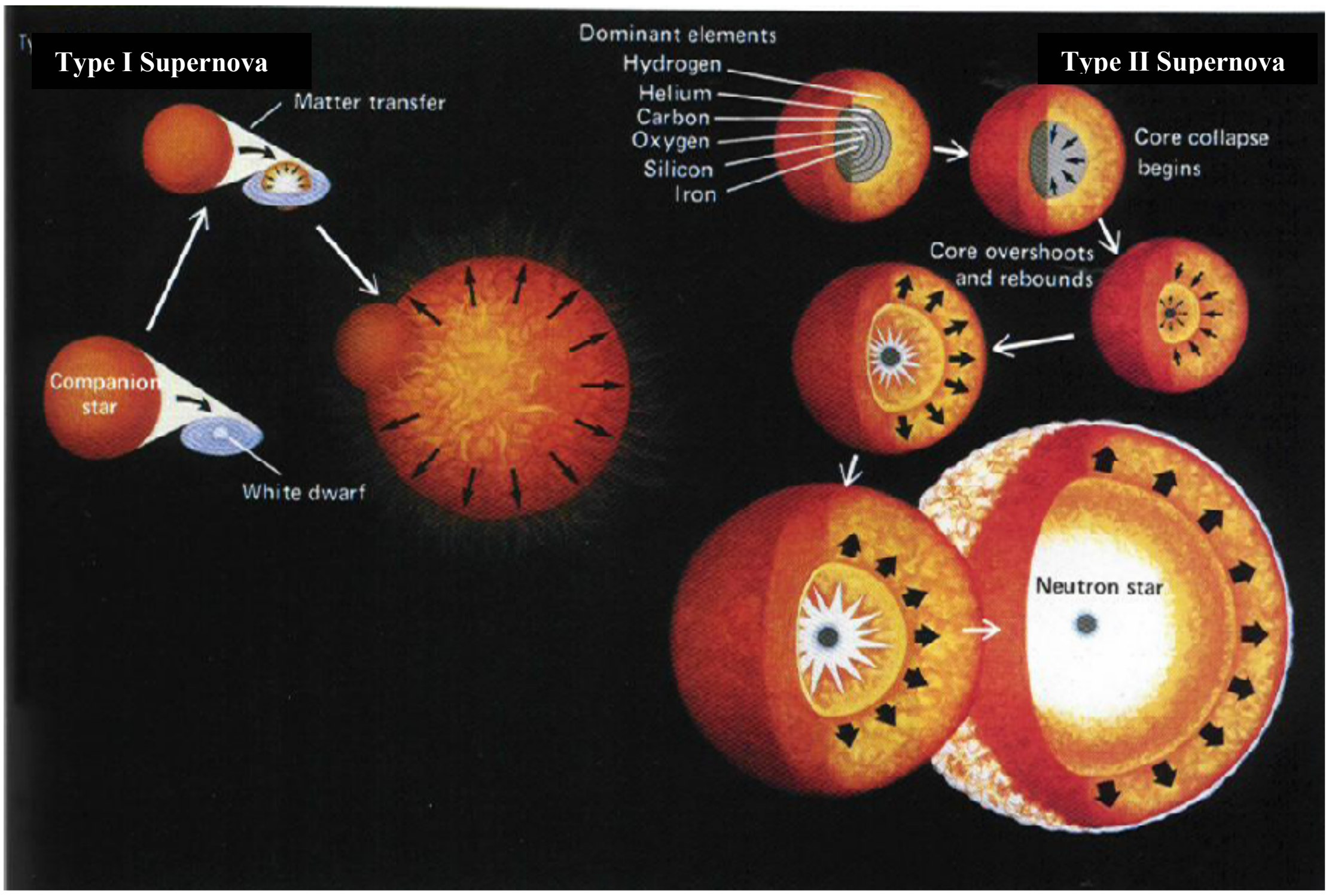

**Fig. 1.11.** The maine sequences of a type I (left) and type II (right) supernova.

## 1.5 THE MASS-LUMINOSITY RELATIONSHIP

The study of binary stars led to the discovery of a relation between the mass and the luminosities of the stars, mainly the main-sequence stars. This relation led to an important result showing that the more massive a star the hotter, and hence more luminous, it is. Not only can the temperature of the star be obtained but also its lifetime and intrinsic magnitude. [2, 7]

In order to show the relation between luminosity ($L_r$) and the mass ($M$) of a star, one can use (1.3), (1.16), and (1.17) along with the equation of state for an ideal gas. That relation [9] can be written in that form,

$$L_r \alpha \frac{R^4}{\kappa M}\left[\frac{M^4}{R^4}\right] \Rightarrow L_r \alpha \frac{M^3}{\kappa}, \tag{1.18}$$

where $\kappa$ is a constant, $L_r$ is the luminosity of the star, $R$ is its radius, and $M$ is its mass. Further, the relation between luminosity and the temperature, $T$, is

$$Lr \ \alpha \ \frac{R^4 T^4}{\kappa M}. \tag{1.19}$$

Finally, the lifetime of the star [9] is proportional to its mass and inversely proportional to its luminosity. So, the relation is

$$\tau = (10 years)\left[\frac{M}{M_{\odot}}\right]^{-2.4}, \tag{1.20}$$

where $\tau$ is the lifetime of the star, $M_{\odot}$ is the mass of the Sun. Thus, the dominant factor for a star is its mass. From the curve showing the relation between mass and luminosity one can estimate the masses [2] of stars of the main sequence, see Fig.1.12. The relations (1.18-20) are applied only on the main-sequence while those for white dwarf and neutron stars will be derived in chapter two and three respectively.

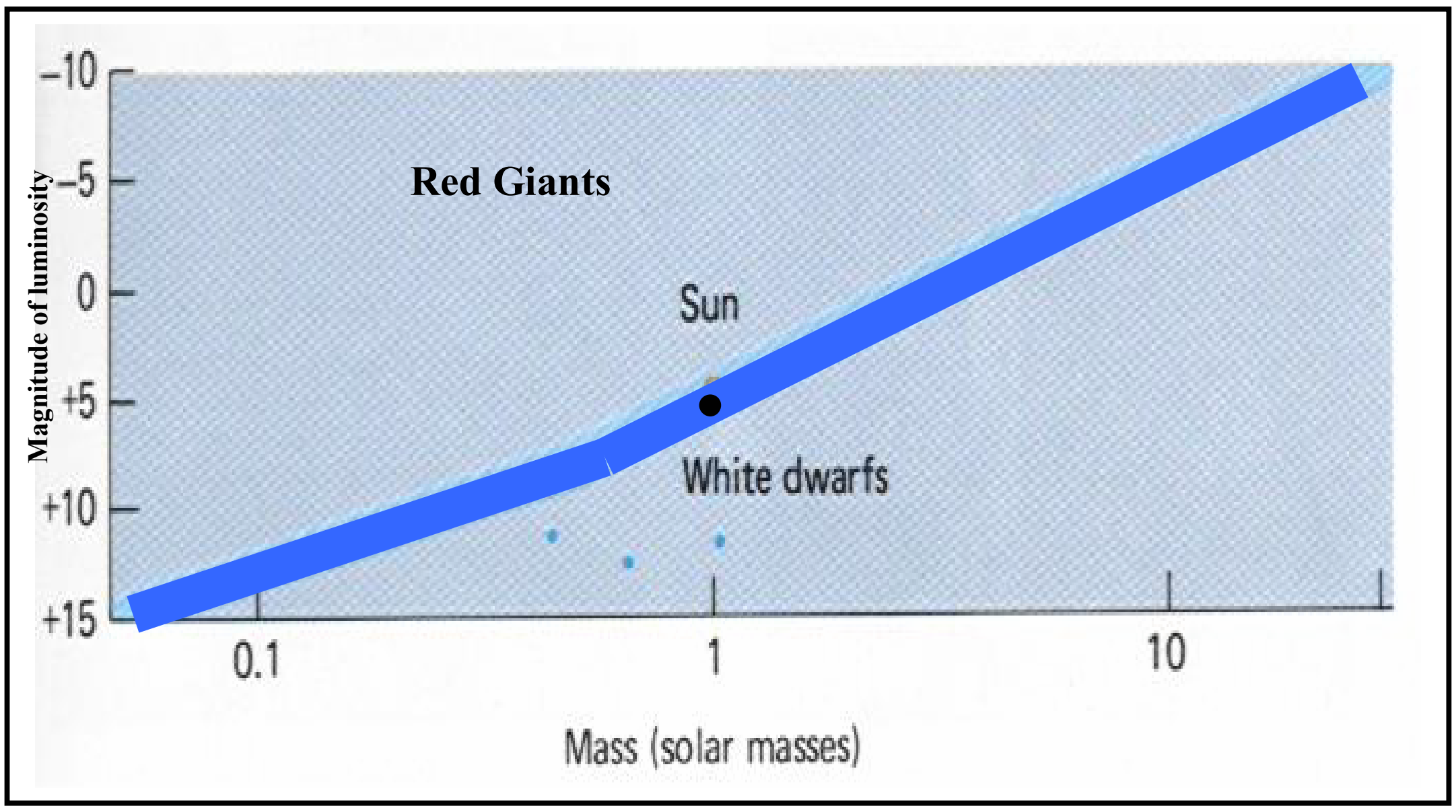

**Fig.1.12.** The mass-luminosity relation, most of the stars lie in the shaded band.

# CHAPTER 2

# COLLAPSED OBJECTS, DEGENERATE STARS (WHITE DWARF)

As we have seen in the last chapter, a star in the first stage transforms hydrogen into helium through the PPC in its core. Helium is then transformed into carbon through the CNC. In that stage, if the star does not have enough mass to fuse carbon into heavier elements, there will be no radiation pressure and so the star will go on contracting. The outer shells are then ejected away as a planetary nebula. The condensed core that remains is called a *collapsed* (or *compact*) *object*, which is characterized by a small radius, high density, and high surface gravity. There are three types of collapsed objects, classified by the degree of compactness: white dwarfs (WD), neutron stars (NS), and black holes (BH) [12]. In this chapter, the first type will be introduced and discussed, while the second will be considered in the next chapters.

White dwarfs are originally red-giant stars of masses in the range of 4 $M_\odot$, where $M_\odot$ is the mass of the Sun. If a star is left with 1.4 $M_\odot$ (*Chandrasekhar limit*) at the stage of collapse, then it will become a WD. Some of them reach that stage either through the phase of being the central star of a planetary nebula or by quiet collapse of a relatively low mass star. The WDs occupy a region that is below and to the left of the main sequence in the H-R diagram, see Fig. 1.1 [2]. The typical values for the WD's radius $R$, density ρ, and the escape velocity $\upsilon_E$ (which is related to the surface gravity) are $R \sim 10^{-2} R_\odot$, $\rho \sim 10^6$ g cm$^{-3}$, and $\upsilon_E \sim 0.02c$ , $R_\odot$ is the solar radius [12]. WD's have low luminosity because of the small surface area, and start out white hot, which gives them the name of "white dwarf". Thus, for example, the companion of Sirius has the solar mass whereas its luminosity is 0.003 times that of the sun. Also, because of its high density it does not show the same relationship between mass and luminosity as the main-sequence stars of § 1.5, see Fig. 1.12. A hot WD has a high temperature in comparison with stars of the same luminosity. In contrast, there are cold WDs, whose the temperature is comparatively low. These are called "black dwarfs" [12].

R. H. Fowler discovered the first theory that led to the understanding of the structure of WDs. He pointed out that the electron gas in the interior of a WD must be highly degenerate. We thus speak of "degenerate stars" [13]. As seen before, in §1.3.2, normal stars are in hydrostatic equilibrium under two opposite forces --- gravity (inward) and radiation pressure resulting from nuclear fusion (outward), see (1.2). However, in case of WDs the outward force generated by electron degeneracy will replace the nuclear force. Hence, we will first introduce the theory of symmetrical and anti-symmetrical states, which is the background of degeneracy. Then, quantum degeneracy will be introduced in detail. Third, it will be applied on WD to construct its mechanical (*Chandrasekhar theory)* and thermal properties.

# 2.1. THE SYMMETRICAL AND ANTI-SYMMETRICAL STATES

In order to understand the principle behind the degeneracy of electrons, it is useful to introduce the symmetric and anti-symmetric wave functions, Ψ, describing the states of the particles. Let us consider a system of *N* similar particles. If the sign of Ψ does not change after interchanging two particles, then such particles are called *bosons*. On the other hand if it changes, they are called *fermions*.

Therefore, if we have particles that can be described by two variables $q_1$ and $q_2$, then

$$\Psi(q_1,q_2) = \Psi(q_2,q_1) \qquad \text{for } bosons \tag{2.1}$$

$$\Psi(q_1,q_2) = -\Psi(q_2,q_1) \qquad \text{for } fermions\ , \tag{2.2}$$

where the first case is called a symmetrical state and the second an anti-symmetrical state. Some particles have symmetrical properties, like photons and some kinds of ions, while others have anti-symmetrical properties, like electrons and nucleons. Degeneracy pressure is generated if all energy states of the fermions are filled. Thus, fermions will overcome any external force that tends to interchange their states. [12, 13]

From a statistical point of view [14], bosons have a *Bose-Einstein*, BE, distribution given by the partition function

$$Z = \prod_s \frac{1}{1 - e^{E_s/kT - \psi}}, \tag{2.3}$$

while fermions have a *Fermi-Dirac* distribution given by the partition function

$$Z = \prod_s \frac{1}{1 + e^{E_s/kT - \psi}}. \tag{2.4}$$

At high *T* both (2.3) and (2.4) will be classical, Boltzmann, distributions

$$Z = \prod_s e^{-E_s/kT} \quad . \tag{2.5}$$

We are interested in fermions because WDs consist mainly of electrons, which generate a degeneracy pressure. Thus

$$\frac{PV}{kT} = \ln Z = \sum_s \ln(1 + e^{-(E_s/kT + \psi)}) \, . \tag{2.6}$$

Also, the total number of fermions is

$$N = \sum_s \frac{1}{1 + e^{(E_s/kT + \psi)}} \, , \tag{2.7}$$

and the total energy is

$$U = \sum_s \frac{E_s}{1 + e^{(E_s/kT + \psi)}} \, . \tag{2.8}$$

In the next section we will discuss further the degeneracy of electrons.

# 2.2. QUANTUM DEGENERACY

In order to understand the basic ideas of a degenerate star, we should introduce the quantum view of the degeneracy. Both the statistical and analytical treatments of the concept of degeneracy will be given in two situations: complete and partial degeneracy.

## 2.2.1 STRONG (COMPLETE) DEGENERACY OF THE ELECTRON GAS

We will start the discussion with Pauli's exclusion principle, which states, *"no two electrons (or, more generally, Fermions) can occupy the same quantum state"*. This principle implies [13] that the number of electrons in configuration *(N(p)dp)* with momentum between $p$ and $p + dp$ in a large volume $V$ can be written as

$$N(p)dp \leq V \frac{8\pi p^2 dp}{h^3}. \tag{2.9}$$

A *completely degenerate electron gas is one in which all the lowest quantum states are occupied*. For Fermions the limit is saturated and so

$$N(p) = V \frac{8\pi p^2}{h^3}. \tag{2.10}$$

The number of electrons with momenta less than $p_0$ ($p_F$) is obtained from

$$N = V \frac{8\pi}{h^3} \int_0^{p_0} p^2 dp, \tag{2.11}$$

or

$$N = V \frac{8\pi}{3h^3} p_0^3. \tag{2.12}$$

Hence, the electron density is

$$n = \frac{N}{V} = \frac{8\pi}{3h^3}\, p_0^3 \,, \tag{2.13}$$

where $p_0$ is just the Fermi momentum, $p_F$, which is given by

$$p_0 = \left(\frac{3n}{8\pi}\right)^{1/3} h \,. \tag{2.14}$$

The density of states, *N(p)* is plotted against *p* in Fig. 2.1

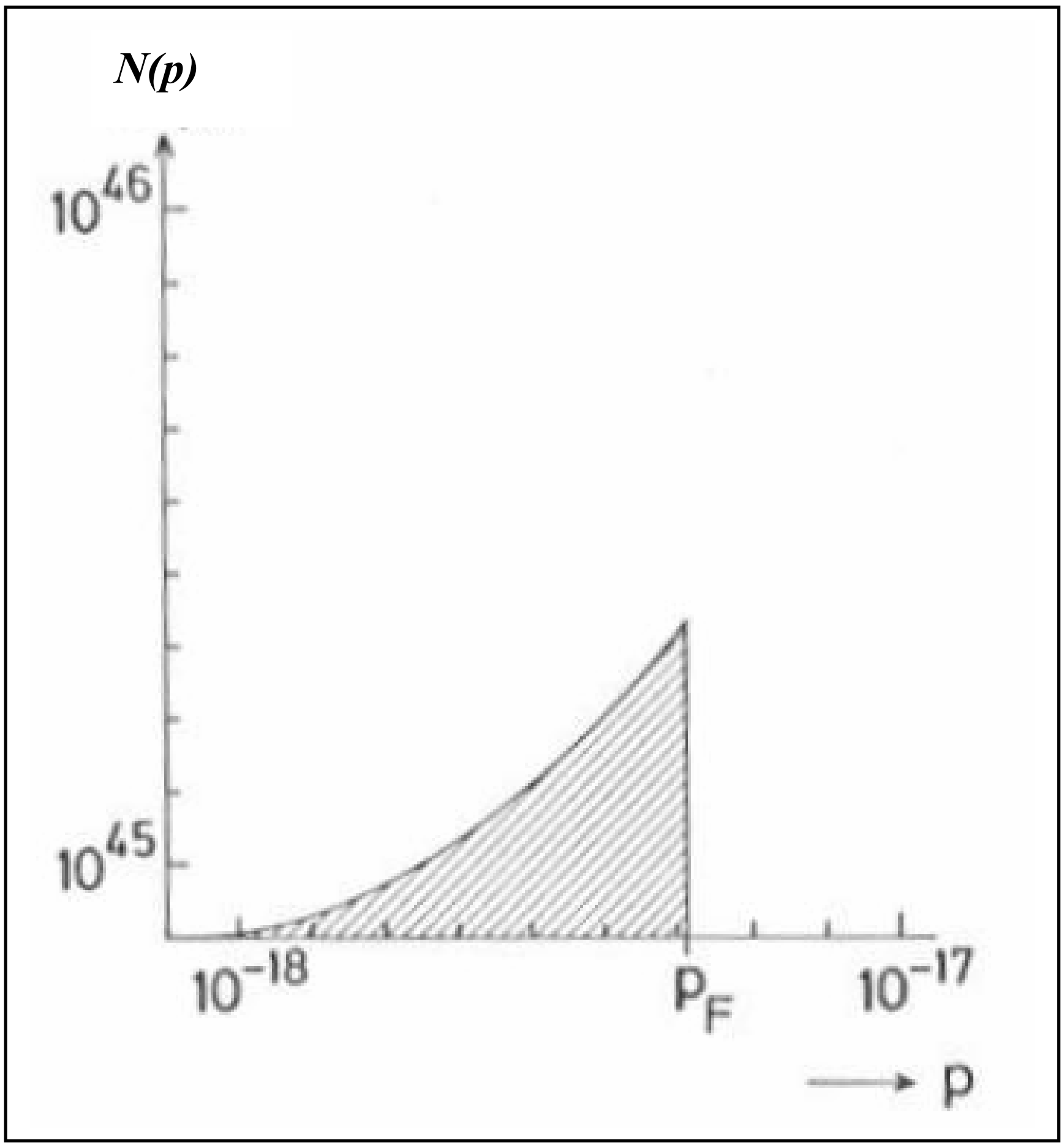

**Fig.2.1.** The density of state *N(p)* against the momentum *p*, in the case of completely degenerate electron gas with $T = 0$ K, and $n = 10^{28}$ cm$^{-3}$.

One of the important quantities for our purposes is the pressure, $P$, which is defined [12] as the mean rate of transfer of momentum across an ideal surface of unit area $A$ in the gas. Thus, from that definition, we find for one particle with velocity $\upsilon_p$,

$$PdV = \frac{\dot{p}}{A}(Adr) = \dot{p}dr = \frac{dp}{dt}dr = \upsilon_p dp\,. \tag{2.15}$$

Also, from (2.2) for all states we find that, for a completely degenerate gas [12, 13, 14]

$$PV = \frac{1}{3}\int_0^{p_0} N(p)p\upsilon_p dp = \frac{1}{3}\int_0^{p_0} V\frac{8\pi p^2}{h^3}p\upsilon_p dp\,, \tag{2.16}$$

and hence

$$P = \frac{8\pi}{3h^3}\int_0^{p_0} p^3\frac{\partial E}{\partial p}dp\,, \tag{2.17}$$

where $E$ is the kinetic energy of the electron. Then, the whole kinetic energy $U$ of the gas is

$$U = \int_0^{\infty} N(p)Edp\,. \tag{2.18}$$

Again, by using (2.10) for a completely degenerate gas,

$$U = V\frac{8\pi}{h^3}\int_0^{p_0} Ep^2dp\,. \tag{2.19}$$

Thus, integrating (2.17) by parts, and using (2.19), we get

$$P_d = \frac{8\pi}{3h^3}E(p_0)p_0^{\;3} - \frac{U}{V}\,. \tag{2.20}$$

This is the general formula for the degeneracy pressure $P_d$ of a particle of momentum $p$.

In the case of classical particles, the relation between $E$ and $p$ is

$$E = \frac{p^2}{2m}\,. \tag{2.21}$$

But the relativistic relation, which is more general, is

$$E = mc^2\left[\left(1 + \frac{p^2}{m^2c^2}\right)^{1/2} - 1\right], \tag{2.22}$$

which gives

$$\frac{\partial E}{\partial p} = \frac{1}{m}\left(1+\frac{p^2}{m^2c^2}\right)^{-1/2} p. \tag{2.23}$$

Hence, (2.23) is just the speed of the particle [14]. Putting it in (2.17), it gives

$$P_d = \frac{8\pi}{3h^3}\int_0^{p_0} \frac{p^4}{\left(1+\frac{p^2}{m^2c^2}\right)^{1/2}} dp. \tag{2.24}$$

Also, the total kinetic energy of the degenerate gas is derived by combining (2.18) and (2.22) to give

$$U = V\frac{8\pi}{h^3}\int_0^{p_0} mc^2\left[\left(1+\frac{p^2}{m^2c^2}\right)^{1/2} - 1\right]p^2 dp. \tag{2.25}$$

In order to integrate (2.24, 25), we may introduce a dimensionless variable, $\theta$, defined by

$$\sinh\theta = \frac{p}{mc}; \qquad \sinh\theta_0 = \frac{p_0}{mc}. \tag{2.26}$$

Then, from (2.22) this equation gives

$$E = mc^2(\cosh\theta - 1) \quad \text{and} \quad \upsilon = c\tanh\theta. \tag{2.27}$$

Thus, (2.12), (2.24), and (2.25) reduce to

$$N = V\frac{8\pi m^3c^3}{3h^3}\sinh^3\theta_0 \tag{2.28}$$

$$P_d = \frac{8\pi m^4c^5}{3h^3}\int_0^{\theta_0}\sinh^4\theta d\theta \tag{2.29}$$

$$U = V\frac{8\pi m^4c^5}{h^3}\int_0^{\theta_0}(\cosh\theta - 1)\sinh^2\theta d\theta. \tag{2.30}$$

Defining

$$x = \sinh\theta_0 = \frac{p_0}{mc} \tag{2.31}$$

$$f(x) = x(2x^2 - 3)(x^2+1)^{1/2} + 3\sinh^{-1}x \tag{2.32}$$

$$g(x) = 8x^3[(x^2+1)^{1/2} - 1] - f(x). \qquad (2.33)$$

By means of this definition, (2.28-30) will reduce to

$$N = V\frac{8\pi m^3c^3}{3h^3}x^3 \Rightarrow n = \frac{N}{V} = \frac{8\pi m^3c^3}{3h^3}x^3 = 5.87\times10^{29}x^3, \qquad (2.34)$$

$$P_d = \frac{\pi m^4c^5}{3h^3}f(x) = 6.01\times10^{22}f(x), \qquad (2.35)$$

$$U = V\frac{\pi m^4c^5}{3h^3}g(x). \qquad (2.36)$$

The numerical values of $f(x)$ and $g(x)$ depend on the value of $x$. They are tabulated by *Chanrasekhar* (1939), [13, p.361] see Table 2.1 and Fig. 2.2.

**Table. 2.1.** The Numerical Values of $f(x)$ and $g(x)$

| $x$ | $f(x)$ | $g(x)$ | $\frac{g(x)}{f(x)}$ | $x$ | $f(x)$ | $g(x)$ | $\frac{g(x)}{f(x)}$ |
|---|---|---|---|---|---|---|---|
| 0. . . . . | 0. | 0. | 1.5 | 2.7. . | 95.17935 | 200.7327 | 2.1090 |
| 0.1. . . | 0.000016 | 0.000024 | 1.5 | 2.8. . | 110.8207 | 235.7072 | 2.1269 |
| 0.2. . . | 0.000505 | 0.000762 | 1.509 | 2.9. . | 128.3012 | 275.1070 | 2.1442 |
| 0.3. . . | 0.003769 | 0.005742 | 1.5233 | 3.0. . | 147.7578 | 319.2942 | 2.1609 |
| 0.4. . . | 0.015527 | 0.023914 | 1.5402 | 3.5. . | 279.8113 | 625.728 | 2.2363 |
| 0.5. . . | 0.046093 | 0.071941 | 1.5608 | 4.0. . | 484.5644 | 1114.466 | 2.2999 |
| 0.6. . . | 0.111126 | 0.17604 | 1.5841 | 4.5. . | 784.5271 | 1846.997 | 2.3543 |
| 0.7. . . | 0.231992 | 0.27348 | 1.6099 | 5.0. . | 1205.2069 | 2893.813 | 2.4011 |
| 0.8. . . | 0.435865 | 0.71358 | 1.6372 | 5.5. . | 1775.1094 | 4334.407 | 2.4418 |
| 0.9. . . | 0.755661 | 1.25849 | 1.6654 | 6.0. . | 2525.7390 | 6257.275 | 2.4774 |
| 1.0. . . | 1.229907 | 2.0838 | 1.6943 | 6.5. . | 3491.599 | 8759.913 | 2.5089 |
| 1.1. . . | 1.902586 | 3.2788 | 1.7233 | 7.0. . | 4710.192 | 11948.818 | 2.5368 |
| 1.2. . . | 2.82298 | 4.9468 | 1.7523 | 7.5. . | 6222.021 | 15939.488 | 2.5618 |
| 1.3. . . | 4.04557 | 7.2052 | 1.7810 | 8.0. . | 8070.587 | 20856.421 | 2.5842 |
| 1.4. . . | 5.62991 | 10.1857 | 1.8092 | 8.5. . | 10302.39 | 26833.12 | 2.6045 |
| 1.5. . . | 7.64053 | 14.0344 | 1.8368 | 9.0. . | $1.296694\times10^4$ | $3.401207\times10^4$ | 2.6230 |
| 1.6. . . | 10.14696 | 18.9115 | 1.8638 | 9.5. . | $1.611672\times10^4$ | $4.254479\times10^4$ | 2.6398 |
| 1.7. . . | 13.22359 | 24.9920 | 1.8900 | 10.0. . | $1.980725\times10^4$ | $5.2591\times10^4$ | 2.6552 |
| 1.8. . . | 16.94969 | 32.4649 | 1.9154 | 20.0. . | $3.192093\times10^5$ | $8.9839\times10^5$ | 2.8144 |
| 1.9. . . | 21.40937 | 41.5338 | 1.9400 | 30.0. . | $1.618212\times10^6$ | $4.6494\times10^6$ | 2.8732 |
| 2.0. . . | 26.69159 | 52.4168 | 1.9638 | 40.0. . | $5.116812\times10^6$ | $1.48596\times10^7$ | 2.9041 |
| 2.1. . . | 32.89010 | 65.3462 | 1.9868 | 50.0. . | $1.249501\times10^7$ | $3.6515\times10^7$ | 2.9224 |
| 2.2. . . | 40.10347 | 80.5689 | 2.0090 | 60.0. . | $2.591280\times10^7$ | $7.6053\times10^7$ | 2.9349 |
| 2.3. . . | 48.43509 | 98.3463 | 2.0305 | 70.0. . | $4.801018\times10^7$ | $1.41346\times10^8$ | 2.9441 |
| 2.4. . . | 57.99311 | 118.9541 | 2.0512 | 80.0. . | $8.190727\times10^7$ | $2.41703\times10^8$ | 2.9509 |
| 2.5. . . | 68.89053 | 142.6823 | 2.0711 | 90.0. . | $13.12039\times10^7$ | $3.87876\times10^8$ | 2.9563 |
| 2.6. . . | 81.24509 | 169.8355 | 2.0904 | 100.0. . | $19.9980\times10^7$ | $5.9206\times10^8$ | 2.9606 |

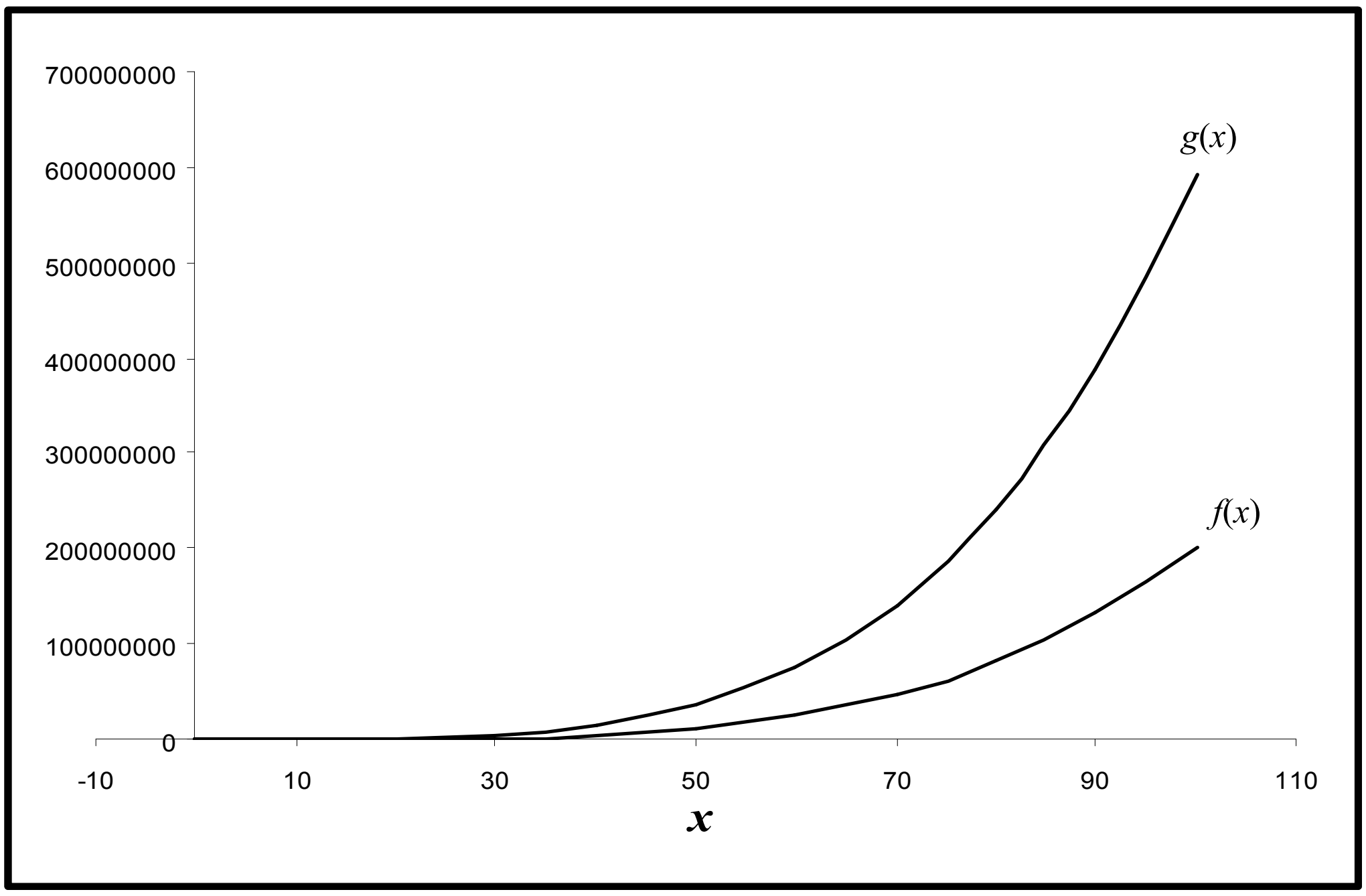

**Fig.2.2.** Plot of $f(x)$ and $g(x)$ versus $x$ as given in table.2.1.

The asymptotic results for $x << 1$ and $x >> 1$ will be particularly useful because the former tends to the non-relativistic case and the latter to the extreme relativistic. In those limits *f(x)* and *g(x)* are given by

$$f(x) \approx \frac{8}{5}x^5 - \frac{4}{7}x^7 + \frac{1}{3}x^9 - \frac{5}{22}x^{11} + ... \qquad \text{for} \qquad x << 1, \tag{2.37}$$

$$f(x) \approx 2x^4 - 2x^2 + 3(\ln 2x - \frac{7}{12}) + (\frac{5}{4}x^{-2}) + ... \qquad \text{for} \qquad x >> 1, \tag{2.38}$$

$$g(x) \approx \frac{12}{5}x^5 - \frac{3}{7}x^7 + \frac{1}{6}x^9 - \frac{15}{176}x^{11} + ... \qquad \text{for} \qquad x << 1, \tag{2.39}$$

$$g(x) \approx 6x^4 - 8x^2 + 7x^2 - 3(\ln 2x - \frac{1}{4}) - (\frac{3}{4}x^{-2}) + ... \qquad \text{for} \qquad x >> 1, \tag{2.40}$$

The physical meaning of the first limit ($x << 1$) is that $p_0 << mc$ see (2.31). Hence, the velocity of the electron is much slower than the speed of light. In that case, if we keep only the first term of (2.37), then (2.35) will be such that

$$P_d = \frac{\pi m^4 c^5}{3h^3} (\frac{8}{5} x^5) . \tag{2.41}$$

But from (2.13, 31)

$$x = \left( \frac{3h^3 n}{8\pi m^3 c^3} \right)^{1/3} = \frac{h}{mc} \left( \frac{3n}{8\pi} \right)^{1/3} , \tag{2.42}$$

substituting (2.42) in (2.41), and using the relation, $\rho = n\mu_e m_u$, we have

$$P_d = \frac{\pi m^4 c^5}{3h^3} \frac{8}{5} \left( \frac{h}{mc} \right)^5 \left( \frac{3n}{8\pi} \right)^{5/3} = \frac{1}{20} \frac{h^2}{m} \left( \frac{3}{\pi} \right)^{2/3} n^{5/3} = \frac{1}{20} \left( \frac{3}{\pi} \right)^{2/3} \frac{h^2}{m m_u^{5/3}} \left( \frac{\rho}{\mu_e} \right)^{5/3} . \tag{2.43}$$

Hence, the relation between $P$ and $\rho$ is a *polytropic relation* of the form $P = K\rho^\gamma$ , where the polytropic exponent (adiabatic index), $\gamma$, is 5/3. Also, in the limit $x << 1$, if we keep only the first term of (2.39), then by means of (2.42), (2.36) will become

$$U = V \frac{\pi m^4 c^5}{3h^3} (\frac{12}{5} x^5) = \frac{3}{2} V \left[ \frac{1}{20} \frac{h^2}{m} \left( \frac{3}{\pi} \right)^{2/3} n^{5/3} \right] = \frac{3}{2} V P_d . \tag{2.44}$$

Similarly for the limit $x >> 1$ (relativistic), keeping the first terms of (2.38) and (2.40), we can write

$$P_d = \frac{1}{8} hc \left( \frac{3}{\pi} \right)^{1/3} n^{4/3} = \frac{1}{8 m_u^{4/3}} hc \left( \frac{3}{\pi} \right)^{1/3} \left( \frac{\rho}{\mu_e} \right)^{4/3} . \tag{2.45}$$

Thus the polytropic exponent, $\gamma$, is 4/3. Also, using the same methods in the non-relativistic case to find the internal energy will give,

$$U = 3 V P_d . \tag{2.46}$$

## 2.2.2. PARTIAL DEGENERACY OF THE ELECTRON GAS

Above the Fermi temperature, not all electrons will have the lowest possible momentum i.e. there are some with $p > p_0$. Moreover, if the temperature is high enough, then *N(p)* will have a Boltzmann distribution. Hence, there must be a smooth transition from the case of strong degeneracy to partial or non-degeneracy. Therefore, *N(p)* will obey the Fermi-Dirac distribution in general, since the electrons are fermions. Thus, (2.1) becomes

$$N(p)dp = V\frac{8\pi p^2 dp}{h^3}\frac{1}{1+e^{E/kT-\psi}} \quad , \tag{2.47}$$

where $\psi$ is called the *degeneracy parameter*, $T$ is the temperature and $k$ is the Boltzmann constant. The distribution function for partial (non-relativistic) degeneracy, (2.47) is plotted against $p$ in Fig.2.3, for $T = 1.9\times10^7$ K and $\psi = 10$.

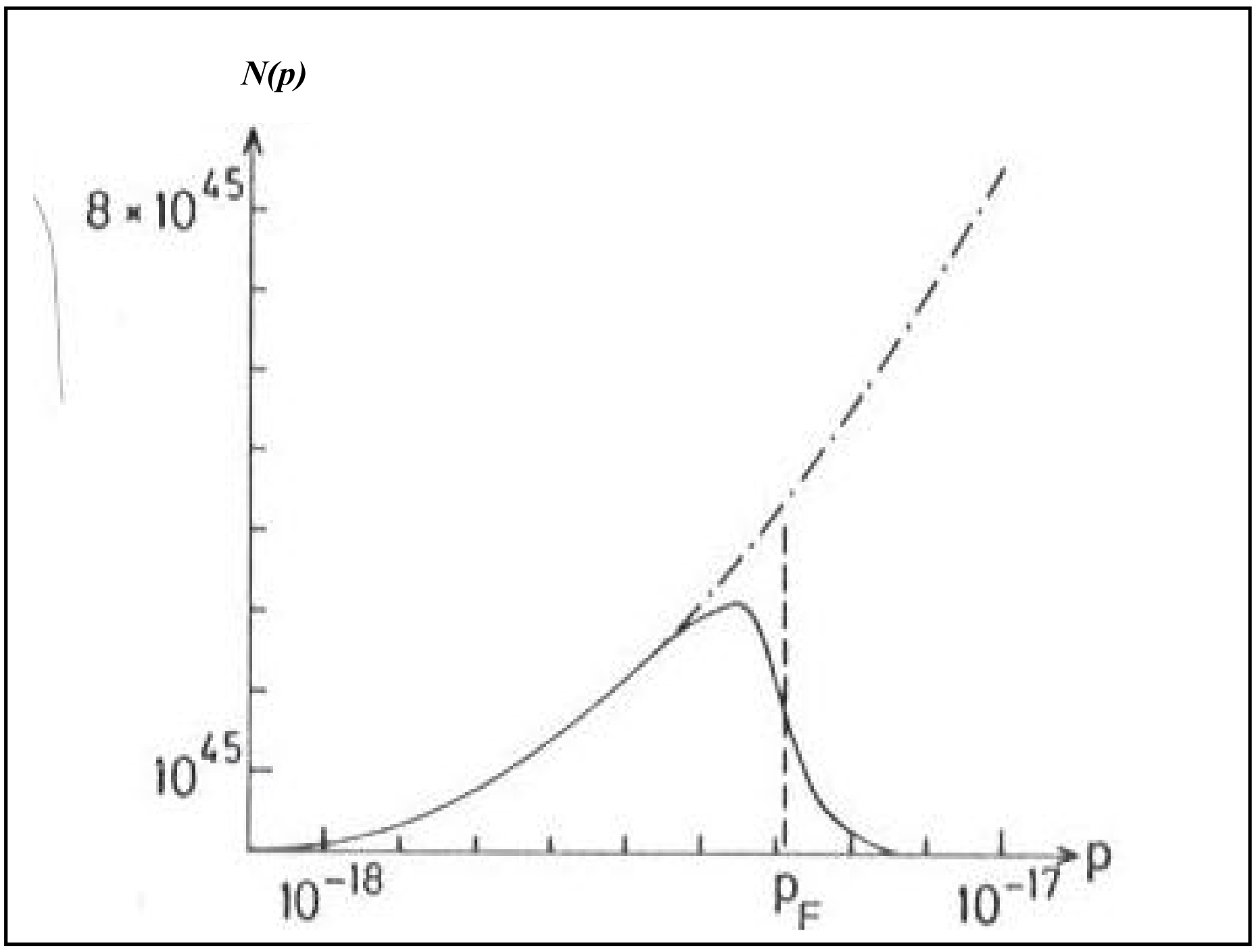

**Fig.2.3**. The density of state *N(p)* against the momentum *p*, in the case of partially degenerate electron gas with $T = 1.9\times10^7$ K, $n = 10^{28}$ cm$^{-3}$, and $\psi = 10$.

Then

$$n(p) = \frac{N}{V} = \frac{8\pi}{h^3}\int_0^\infty \frac{p^2 dp}{1 + e^{E/kT - \psi}}. \tag{2.48}$$

Also, (2.16) and (2.19) become

$$P = \frac{1}{3}\int_0^\infty n(p) p \upsilon_p dp = \frac{1}{3}\frac{8\pi}{h^3}\int_0^{p_0} \frac{p^3}{1 + e^{E/kT - \psi}} \upsilon_p dp, \tag{2.49}$$

$$\frac{U}{V} = \int_0^\infty N(p) E dp = \frac{8\pi}{h^3}\int_0^\infty \frac{Ep^2}{1 + e^{E/kT - \psi}} dp. \tag{2.50}$$

## 2.2.2.1. THE NON-RELATIVISTIC CASE

In the non-relativistic case we substitute the classical relation between $E$ and $p$ (2.21) in (2.48-50), to give

$$n(p) = \frac{8\pi}{h^3}\int_0^\infty \frac{p^2 dp}{1 + e^{p^2/2mkT - \psi}} = \frac{8\pi}{h^3}(2mkT)^{3/2} a(\psi), \tag{2.51}$$

where

$$a(\psi) = \int_0^\infty \frac{\eta^2}{1 + e^{(\eta^2 - \psi)}} d\eta, \tag{2.52}$$

$$\eta = \frac{p}{(2mkT)^{1/2}}. \tag{2.53}$$

From (2.51),

$$a(\psi) = \frac{h^3}{8\pi}\frac{n}{(2mkT)^{3/2}}, \tag{2.54}$$

hence,

$$\psi = \psi\left(\frac{n}{T^{3/2}}\right). \tag{2.55}$$

It is useful to discuss the limiting cases of $\psi$ [12], starting with $\psi \to -\infty$, substituted in (2.54). From (2.52), the value of $a(\psi)$ will be arbitrarily small. Thus, from (2.51), for a finite electron density, $n$, $T \to \infty$. This will be the case of high temperature. In that case $N(p)$ will become a classical, Boltzmann, distribution. In the denominator of (2.47), the 1 can be neglected compared to the second term. Hence,

$$N(p)dp \approx \frac{8\pi p^2 dp}{h^3} e^{-E/kT + \psi}. \tag{2.56}$$

Comparing with the Boltzmann distribution

$$N(p)dp = n\frac{4\pi p^2 dp}{(2\pi mkT)^{3/2}} e^{-E/kT}, \tag{2.57}$$

implying that,

$$e^{\psi} = \frac{h^3 n}{2(2\pi mkT)^{3/2}}. \tag{2.58}$$

In all of the above equations we can replace $E$ by its non-relativistic value (2.21). Thus, in that limit, as in the general case, we have (2.55), which is just the Boltzmann distribution.

On the other hand, if we consider the other limit, $\psi \to \infty$, giving $E_0 = \psi kT$, then

$$\frac{1}{1+e^{E/kT-\psi}} = \frac{1}{1+e^{\psi(E/E_0 - 1)}} \approx \begin{cases} 1 \; for E < E_0 \\ 0 \; for E > E_0 \end{cases}. \tag{2.59}$$

Thus, this discontinuity at $E_0$, where the distribution becomes either 0 or 1, is the same as the case of complete degeneracy. $E_0$ is the same as the Fermi energy $e_f = p^2_f/(2m)$, see (2.11) and Fig.2.1. The numerical values of $\psi$ are moderate in general. Hence, by means of (2.21) and $mdE = pdp$, (2.51) will be

$$n = \frac{4\pi}{h^3}(2m)^{3/2}\int_0^\infty \frac{E^{1/2}dE}{1+e^{E/kT-\psi}}. \tag{2.60}$$

In order to evaluate the above integral, we have to define the Fermi-Dirac integrals

$$F_{\nu}(\psi) = \int_{0}^{\infty} \frac{u^{\nu}}{1+e^{(u-\psi)}} du \ , \qquad (2.61)$$

The numerical value of $F_{\nu}(\psi)$ depends on the values of both $\nu$ and $\psi$, provided in Table. 2.2 [12, p.126]. Also, the plot of $F_{\nu}(\psi)$ against $\psi$ is shown in Fig. 2.4.

**Table.2.2.** The Numerical Values of $(2/3)F_{3/2}(\psi)$, $F_{1/2}(\psi)$, $F_2(\psi)$, $F_3(\psi)$

| $\Psi$ | $\frac{2}{3}F_{3/2}(\Psi)$ | $F_{1/2}(\Psi)$ | $F_2(\Psi)$ | $F_3(\Psi)$ |
|---|---|---|---|---|
| −4.0 | 0.016179 | 0.016128 | 0.036551 | 0.109798 |
| −3.5 | 0.026620 | 0.026480 | 0.060174 | 0.180893 |
| −3.0 | 0.043741 | 0.043366 | 0.098972 | 0.297881 |
| −2.5 | 0.071720 | 0.070724 | 0.162540 | 0.490154 |
| −2.0 | 0.117200 | 0.114588 | 0.266290 | 0.805534 |
| −1.5 | 0.190515 | 0.183802 | 0.434606 | 1.321232 |
| −1.0 | 0.307232 | 0.290501 | 0.705194 | 2.160415 |
| −0.5 | 0.489773 | 0.449793 | 1.134471 | 3.516135 |
| 0.0 | 0.768536 | 0.678094 | 1.803249 | 5.683710 |
| 0.5 | 1.181862 | 0.990209 | 2.821225 | 9.100943 |
| 1.0 | 1.774455 | 1.396375 | 4.328723 | 14.393188 |
| 1.5 | 2.594650 | 1.900833 | 6.494957 | 22.418411 |
| 2.0 | 3.691502 | 2.502458 | 9.513530 | 34.307416 |
| 2.5 | 5.112536 | 3.196598 | 13.596760 | 51.496218 |
| 3.0 | 6.902476 | 3.976985 | 18.970286 | 75.749976 |
| 3.5 | 9.102801 | 4.837066 | 25.868717 | 109.179565 |
| 4.0 | 11.751801 | 5.770726 | 34.532481 | 154.252522 |
| 4.5 | 14.88489 | 6.77257 | 45.20569 | 213.80007 |
| 5.0 | 18.53496 | 7.83797 | 58.13474 | 291.02151 |
| 5.5 | 22.73279 | 8.96299 | 73.56744 | 389.48695 |
| 6.0 | 27.50733 | 10.14428 | 91.75247 | 513.13900 |
| 6.5 | 32.88598 | 11.37898 | 112.93904 | 666.29376 |
| 7.0 | 38.89481 | 12.66464 | 137.37668 | 853.64147 |
| 7.5 | 45.55875 | 13.99910 | 165.31509 | 1080.24689 |
| 8.0 | 52.90173 | 15.38048 | 197.00413 | 1351.54950 |
| 8.5 | 60.94678 | 16.80714 | 232.69369 | 1673.36371 |
| 9.0 | 69.71616 | 18.27756 | 272.63375 | 2051.87884 |
| 9.5 | 79.23141 | 19.79041 | 317.07428 | 2493.65928 |
| 10.0 | 89.51344 | 21.34447 | 366.26528 | 3005.64445 |
| 10.5 | 100.58256 | 22.93862 | 420.45675 | 3595.14883 |
| 11.0 | 112.45857 | 24.57184 | 479.89871 | 4269.86200 |
| 11.5 | 125.16076 | 26.24319 | 544.84118 | 5037.84863 |
| 12.0 | 138.70797 | 27.95178 | 615.53418 | 5907.54847 |
| 12.5 | 153.11861 | 29.69679 | 692.22772 | 6887.77637 |
| 13.0 | 168.41071 | 31.47746 | 775.17183 | 7987.72229 |
| 13.5 | 184.60190 | 33.29308 | 864.61653 | 9216.95127 |
| 14.0 | 201.70950 | 35.14297 | 960.81184 | 10585.40346 |
| 14.5 | 219.75048 | 37.02649 | 1064.00779 | 12103.39411 |
| 15.0 | 238.74150 | 38.94304 | 1174.45439 | 13781.61356 |
| 15.5 | 258.69893 | 40.89206 | 1292.40167 | 15631.12726 |
| 16.0 | 279.63888 | 42.87300 | 1418.09966 | 17663.37576 |
| 16.5 | 301.57717 | 44.88535 | 1551.79837 | 19890.17470 |
| 17.0 | 324.52939 | 46.92862 | 1693.74783 | 22323.71482 |
| 17.5 | 348.51087 | 49.00235 | 1844.19805 | 24976.56198 |
| 18.0 | 373.53674 | 51.10608 | 2003.39907 | 27861.65710 |
| 18.5 | 399.62188 | 53.23939 | 2171.60091 | 30992.31625 |
| 19.0 | 426.78099 | 55.40187 | 2349.05358 | 34382.23057 |
| 19.5 | 455.02855 | 57.59313 | 2536.00711 | 38045.46629 |
| 20.0 | 484.37885 | 59.81279 | 2732.71153 | 41996.46477 |

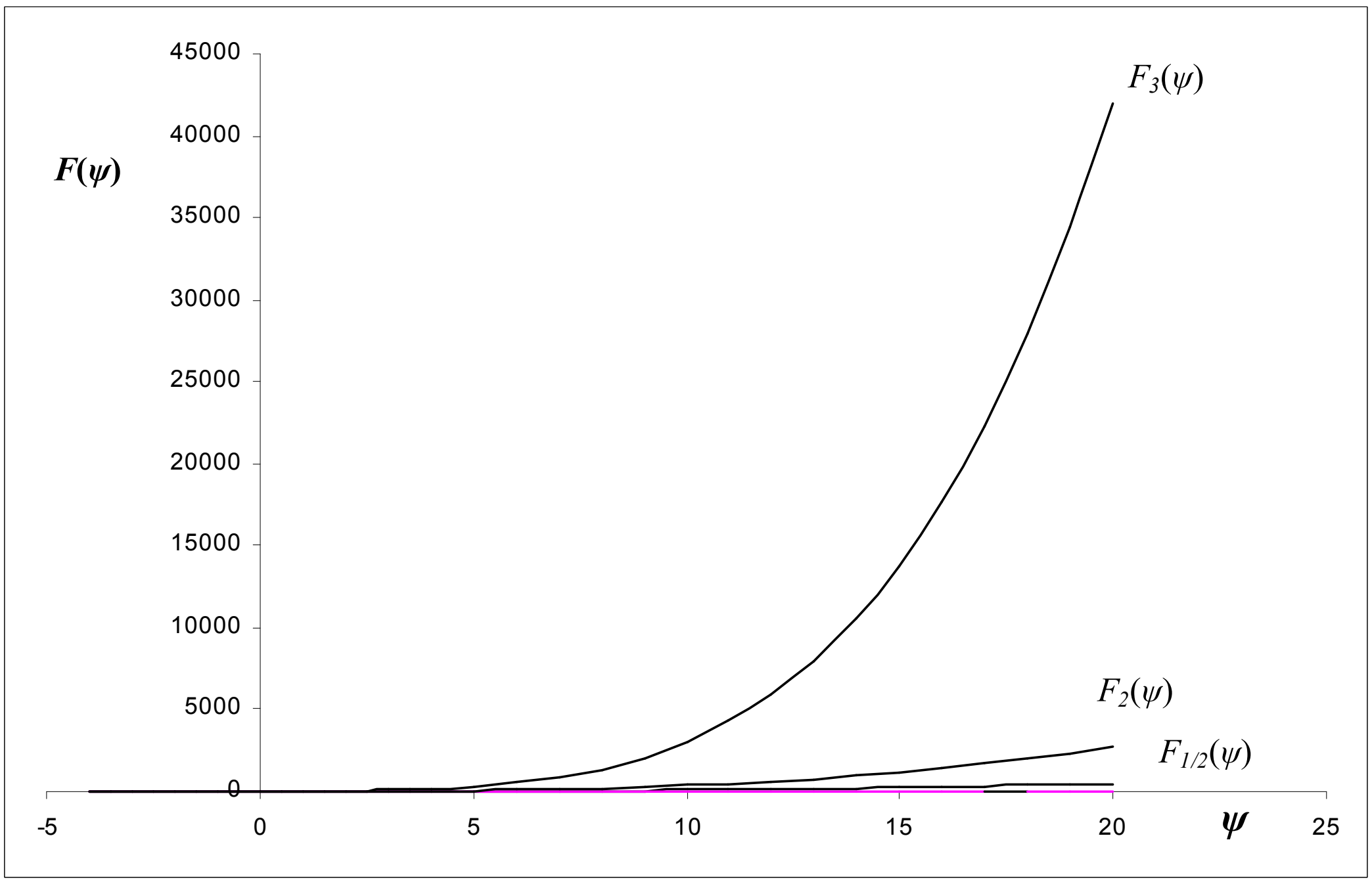

**Fig.2.4.** Plot of $F_{1/2}(\psi)$, $F_2(\psi)$, and $F_3(\psi)$ versus $\psi$ as given in Table.2.2.

Hence

$$n = \frac{4\pi}{h^3}(2mkT)^{3/2} F_{1/2}(\psi)\,. \tag{2.62}$$

This is an example of the general relation, (2.47). The rest of the quantities, i.e. $P$ and $U$, are evaluated in the same way. In the small $T$ limit it behaves like a Pauli parabola, see Fig.2.1, while in the high $T$ limit it behaves like the Boltzmann distribution. The higher the temperature the smoother is the transition from degenerate to non-degenerate around $p_f$, see Fig. 2.3.

The electron pressure $P$ can be written in terms of Fermi-Dirac integrals $F_\nu(\psi)$, by means of $p^3 \upsilon(p)dp = m^4 \upsilon^4 dE = (2mE)^{3/2} dE$. Then (2.49) becomes

$$P = \frac{1}{3}\frac{8\pi}{h^3}(2m)^{3/2}\int_0^\infty \frac{E^{3/2}}{1 + e^{E/kT-\psi}}\,dE\,. \tag{2.63}$$

Writing $E/kT$ as $\eta$, (2.54) can be written as a form of (2.52) as

$$P = \frac{1}{3}\frac{8\pi}{h^3}(2mkT)^{3/2} kTF_{3/2}(\psi), \tag{2.64}$$

and the internal energy $U$, given by (2.50), can be written in the same way as

$$\frac{U}{V} = \frac{4\pi}{h^3}(2mkT)^{3/2} kTF_{3/2}(\psi) = \frac{3}{2}P \ . \tag{2.65}$$

This is in agreement with complete degeneracy in the non-relativistic case (2.44). For the electron gas (2.62-65) define an equation of state. Thus, if $T$ and $n$ are given, then (2.62) yields $\psi$, and hence $P$. Therefore, the final result of partial degeneracy is the same as of complete degeneracy in the non-relativistic limit.

## 2.2.2.2. THE RELATIVISTIC CASE

In the relativistic limit [12], we just replace $p$ by $E/c$ and $\upsilon$ by $c$ in (2.48-49). Then, in the same way, using an approximation and changing variables to derive (2.62, 64) as in the non-relativistic case, we have

$$n = 8\pi(\frac{kT}{hc})^3 F_2(\psi), \tag{2.66}$$

$$P = \frac{8\pi}{3h^3c^3}(kT)^4 F_3(\psi), \tag{2.67}$$

where $F_2$ and $F_3$ are defined by (2.61). In the limit $\psi \to \infty$ (complete degeneracy), $F_\nu(\psi)$ can be expanded as

$$F_\nu(\psi) = \frac{\psi^{\nu+1}}{\nu+1}\left\{1 + 2\left[c_2(\nu+1)\nu\psi^{-2} + c_4(\nu+1)\nu(\nu-1)(\nu-2)\psi^{-4} + ...\right]\right\}, \tag{2.68}$$

where $c_2 = \pi^2/12$, $c_4 = 7\pi^4/720$. Thus, keeping only the first term of (2.68), inserting it in (2.66,67) and eliminating $\psi$, we obtain the same result as for the completely degenerate relativistic case (2.45,46).

In contrast, in the limit of $\psi \rightarrow -\infty$, for high temperatures, again it has a Boltzmann distribution, (2.58). However, in the case of a finite (moderate) value of $\psi$, we have to substitute (2.22) into (2.48-50) without approximation. Therefore, in the case of a moderately relativistic electron gas, we will come up with lengthy expressions that cannot be solved analytically but only numerically, e.g. by using Laguerre polynomials.

In the next section we will apply the previous formulae to the white dwarf WD stars, where the numerical values of the parameters will be inserted to provide a complete picture of WDs. Chandrasekhar [13] derived most of the foregoing theory and its application.

# 2.3. WHITE DWARF STARS

As mentioned in the last section, R. H. Fowler suggested the initial idea of the degeneracy of white dwarfs. He pointed out that the electron gas in the interior of white dwarfs must be highly degenerate. In fact that was the first application of Fermi statistics [14]. Meanwhile, Chandrasekhar [13] wrote a complete theory of WDs. His theory is based upon three main assumptions:

(a) The ideal degenerate electrons produce the pressure of degeneracy, whereas the mass of the spherical WD is provided by the non-degenerate ions of helium;
(b) WDs must be almost entirely degenerate, i.e. the outer fringe of ideal gases of WD is negligible (in fact it is less than one per cent);
(c) Electrons are highly relativistic, i.e. $E \approx pc$.

According to the above assumptions [12, 13, 14] the WD is composed mainly of helium. It has a mass $M \approx 10^{33}$ g packed into a ball of high density ($\rho \approx 10^{7}$ g cm$^{-3}$) with a central temperature $T \approx 10^{7}$ K. Therefore, the energy per particle is of the order of $10^{3}$ eV, which is much more than the energy of helium ionization, hence the whole of the helium in the WD is completely ionized. If the WD consists of $N$ electrons, then there are $N/2$ helium nuclei, each of mass $\approx 4m_p$ ($m_p$ = the mass of proton), hence the total mass of WD is

$$M \approx N(m + 2m_p) \approx 2Nm_p, \tag{2.69}$$

and the electron density is

$$n = \frac{N}{V} \approx \frac{M/2m_p}{M/\rho} = \frac{\rho}{2m_p} \approx 10^{30} \text{ electrons per cm}^3. \tag{2.70}$$

We shall now consider the equilibrium configurations of the WD. If it is spherical, and an adiabatic change in its volume, $V$, produces a change in the energy of gas, from the first law of thermodynamics,

$$dE_0 = -P(n)dV = -P(R)4\pi R^2 dR\,, \tag{2.71}$$

while the change of potential energy is such that

$$dE_g = (\frac{dE_g}{dR})dR = \alpha \frac{GM^2}{R^2} dR\,. \tag{2.72}$$

But if the configuration is in equilibrium, then the net change of energy will be zero, hence

$$dE_o + dE_g = 0\,, \tag{2.73}$$

implying that,

$$P(R) = \frac{\alpha}{4\pi}\frac{GM^2}{R^2}\,. \tag{2.74}$$

From (2.35) we find

$$\frac{\alpha}{4\pi}\frac{GM^2}{R^2} = \frac{\pi m^4 c^5}{3h^3} f(x)\,, \tag{2.75}$$

$$f(x) = \frac{3\alpha h^3}{4\pi^2 m^4 c^5}\frac{GM^2}{R^4} = 6\pi\alpha\left(\frac{\hbar / mc}{R}\right)^3 \frac{GM^2 / R}{mc^2}\,. \tag{2.76}$$

As we assume electrons are highly relativistic, if we keep two terms of *f(x)* in (2.38), then from (2.76) and (2.42)

$$2\left(\frac{3h^3 n}{8\pi m^3 c^3}\right)^{2/3}\left\{\left(\frac{3h^3 n}{8\pi m^3 c^3}\right)^{2/3} - 1\right\} \approx 6\pi\alpha\left(\frac{\hbar / mc}{R}\right)^3 \frac{GM^2 / R}{mc^2}\,. \tag{2.77}$$

Also, from (2.70)

$$n = \frac{N}{V} = \frac{M / 2m_p}{\frac{4}{3}\pi R^3}\,. \tag{2.78}$$

Substituting in (2.77)

$$R \approx \frac{(9\pi)^{1/3}}{2}\frac{\hbar}{mc}\left(\frac{M}{m_p}\right)^{1/3}\left\{1-\left(\frac{M}{M_c}\right)^{2/3}\right\}^{1/2}, \tag{2.79}$$

where

$$M_c = \frac{9}{64}\left(\frac{3\pi}{\alpha^3}\right)^{1/2}\frac{(\hbar c/G)^{3/2}}{m_p^{\ 2}}, \tag{2.80}$$

after inserting the numerical values, and doing some simplification of (2.79), cf. [13, p 421], one can get

$$M_C = \frac{5.75}{\mu_e^{\ 2}}M_{\odot} \approx 1.44M_{\odot} \approx 10^{33}g, \tag{2.81}$$

where, $M_{\odot}$ is the mass of the Sun, and $\mu_e = M/Nm_H$, but $m_H \approx m_p$ , hence from (2.69), $\mu_e = 2$. Also, from (2.79), $R \approx 10^8$cm, that is of the order of an Earth radius.

Therefore, if the mass of a WD increases, then its size decreases as seen from (2.79). Also, there is a limiting mass $M_C$, at which the size vanishes according to (2.79). If $M > M_C$, there is no real solution to (2.79). Thus, all WDs must have a mass less than $M_C$, called the *Chandrasekhar limit*, which fits fully with observation. From (2.76), taking $f(x)$ to be of the order of unity, the radius is of the order of $R \approx 10^8$ cm, which is called the characteristic length, $\ell$ (~ $3.86 \times 10^8$ cm). The mass-radius relationship, (2.79), is shown in Fig. 2.5, where the masses are expressed in terms of $M_C$ and the radii in terms of $\ell$.

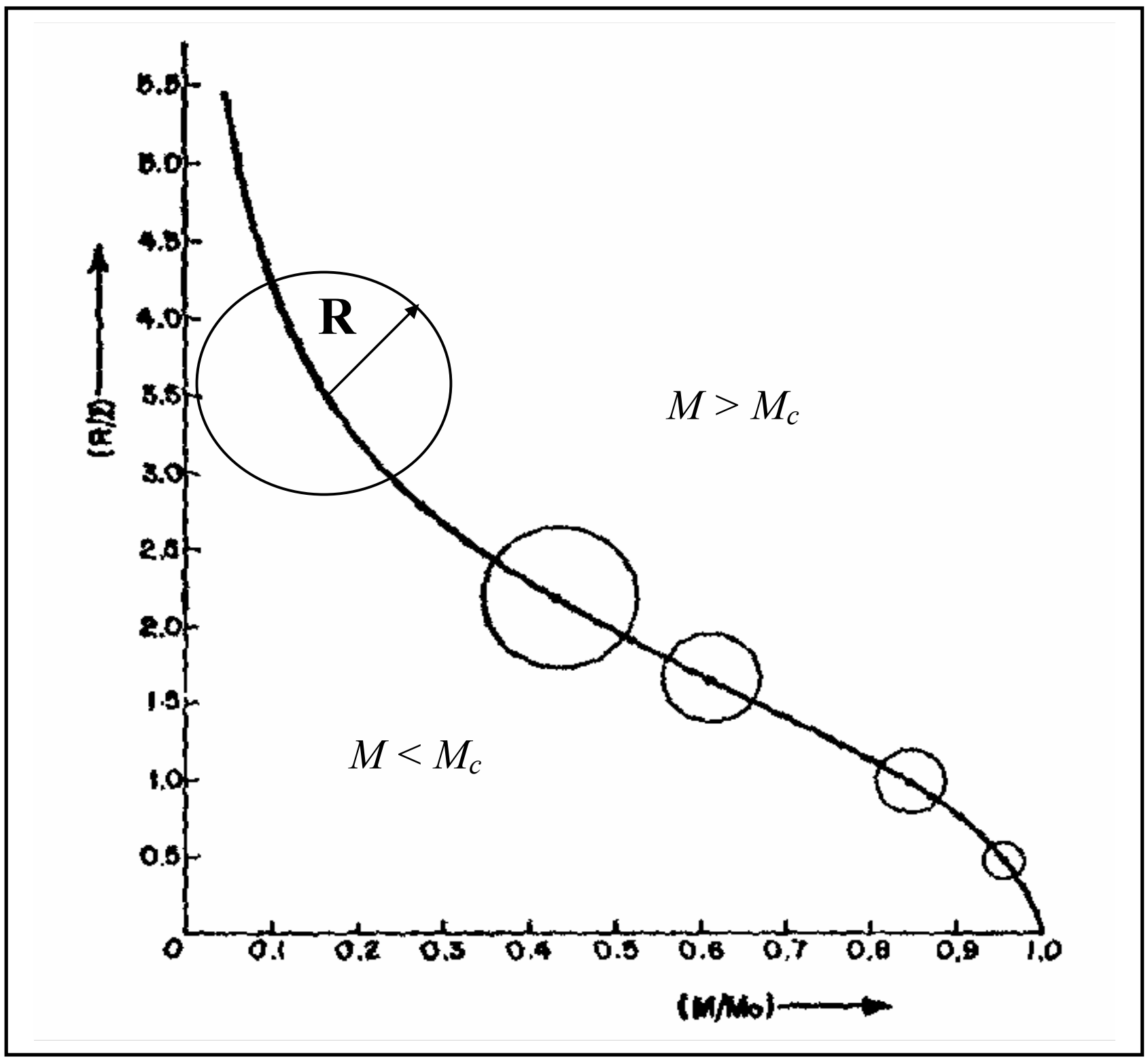

**Fig.2.5.** The radius-mass relationship for WD. The radii are expressed in terms of $\ell$, and masses in terms of $M_c$.

Finally, let us consider what happens physically, when $M < M_c$ and $M > M_c$. In the former case, degeneracy pressure (outward) is greater than the gravitational pressure (inward), hence the star will expand until the electrons become non-relativistic. However, in the second case the opposite occurs. The star collapses so that it is not in equilibrium. In that case we need a different theory, which may be of NSs or BHs. Thus, $M_C$ is the limit of the mass at which the WD is in equilibrium.

# CHAPTER 3

# PULSARS (NEUTRON STARS)

## 3.1. THE DISCOVERY OF PULSARS

One of the most notable astronomical discoveries in the last century was the detection, on 28 Nov. 1967, of clocklike radio pulses emitted by an object in the constellation Cepheus. A young student, Jocelyne Bell, and her supervisor, Anthony Hewish, at the University of Cambridge, made the discovery. It was a direct (but unexpected) result of using a large radio telescope designed to study the interplanetary scintillation of compact radio sources [8].

Initially they thought it was useless signals [2]. In Bell's words: "there was 'a bit of scruff'". After detailed examination of the signal during eight weeks of observation, they determined that it was a set of rapid regular pulses see Fig. 3.1, with one pulse every 1.3373011 s, it was called *pulsar*. When the observers in the same observatory detected three more pulsars (PSRs), it became clear that they had to be natural phenomena.

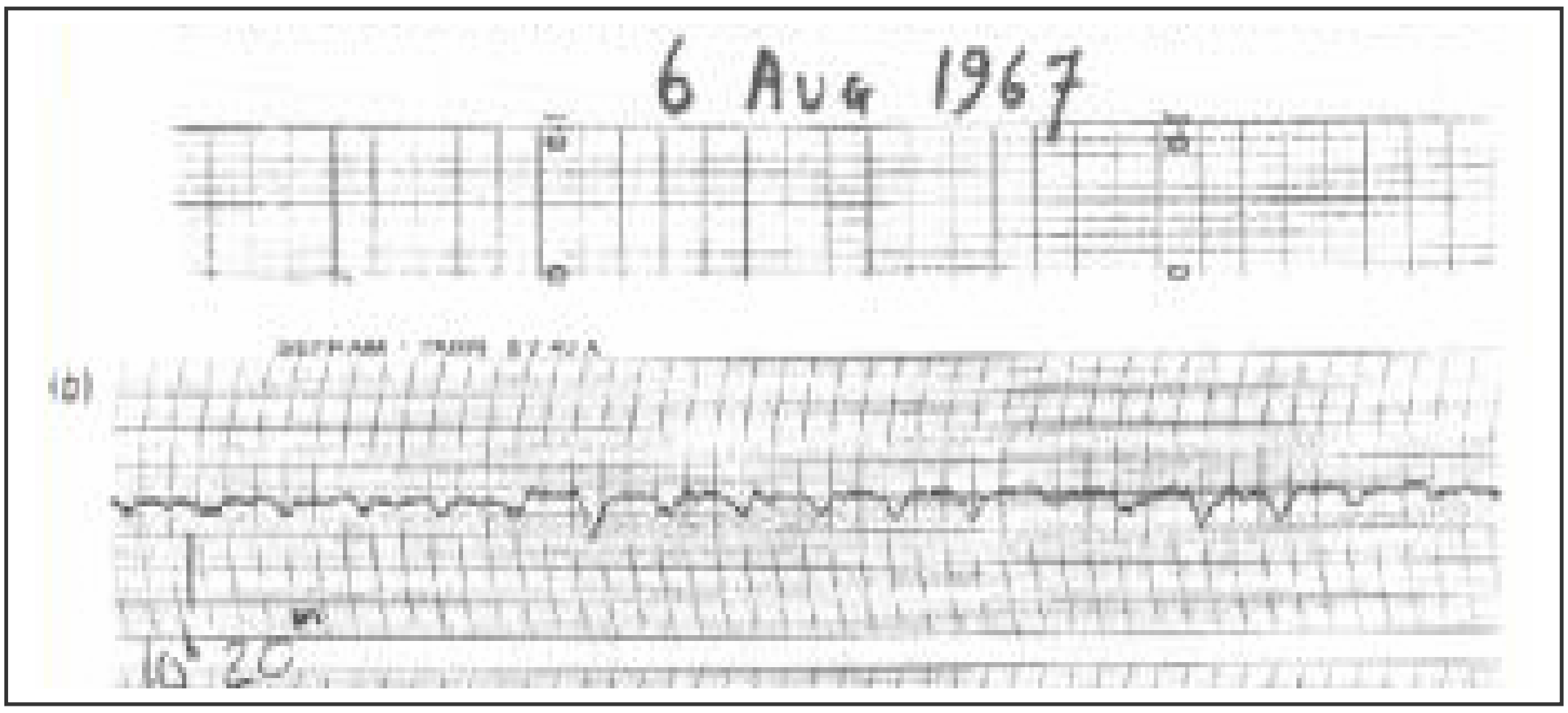

**Fig.3.1.** The first chart record of the periodical pulses of the first discovered pulsar PSR 1919+21.

In late 1968, the group of the Molonglo, Australia, a big radio telescope, announced the discovery of a much faster pulsar, with a period of only 0.089 s. It was in the center of the large SNR, Vela. It was called the *Vela pulsar*. Meanwhile, another group from Harvard University at Green Bank detected two sources of pulses near the Crab Nebula, which is the most widely studied of all supernova remnants (SNR), see Fig. 3.2. Observation later on showed that one of these pulsars is around its center and its period was only 0.033 s. It was the most interesting pulsar because it was an optical pulsar and its period was very short. It was called *Crab pulsar* (PSR 0531+21). Later, Hulse and Taylor (1975) discovered the first *binary pulsar* (SCR B1913+16) [8], which emitted X-rays. As seen from § 1.4, such a system can be used to determine the mass of pulsars and in turn, their evolution. The discovery of the first *millisecond pulsar* (period ~0.0016 s, PSR 1937+21) was eight years later in 1982 [2]. This provides an excellent test for general relativity [2, 8, 22, 37].

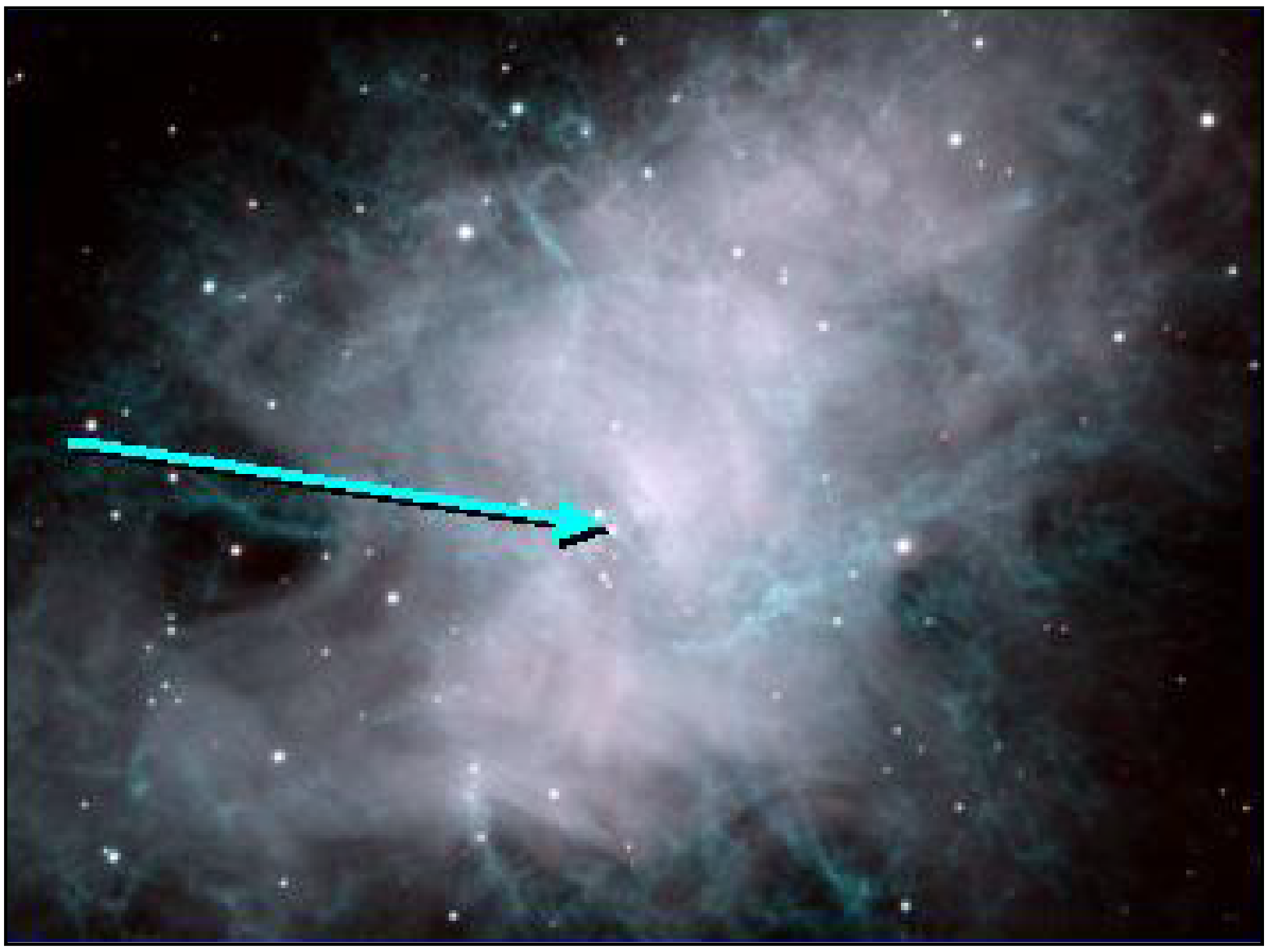

**Fig.3.2.** The Crab Nebula, in the Taurus constellation, it is the remnant of a supernova explosion that became visible on the Earth in 1054 A.D. The arrow points on the position of its pulsar RSR 0531+21.

Consequently, many intensive searches were undertaken: either to find more pulsars, or to model them. As a result, hundreds of pulsars (more than 550 pulsars [28]) have been detected, and about 100,000 active pulsars are estimated to lie in our galaxy [2] (see the distribution of 445 in galactic coordinates in Fig. 3.3). In addition, several models were constructed and discussed. In the rest of this section we will concentrate on the first models and discuss them.

The first models had been narrowed down quickly to coincide with the observed periods of the discovered pulsars. From the start, astrophysicists expected that pulsars are not normal stars but compact objects. The only possible models were the observed WDs and the theoretical neutron stars NSs, which had been proposed in 1934, and will be discussed later in this chapter.

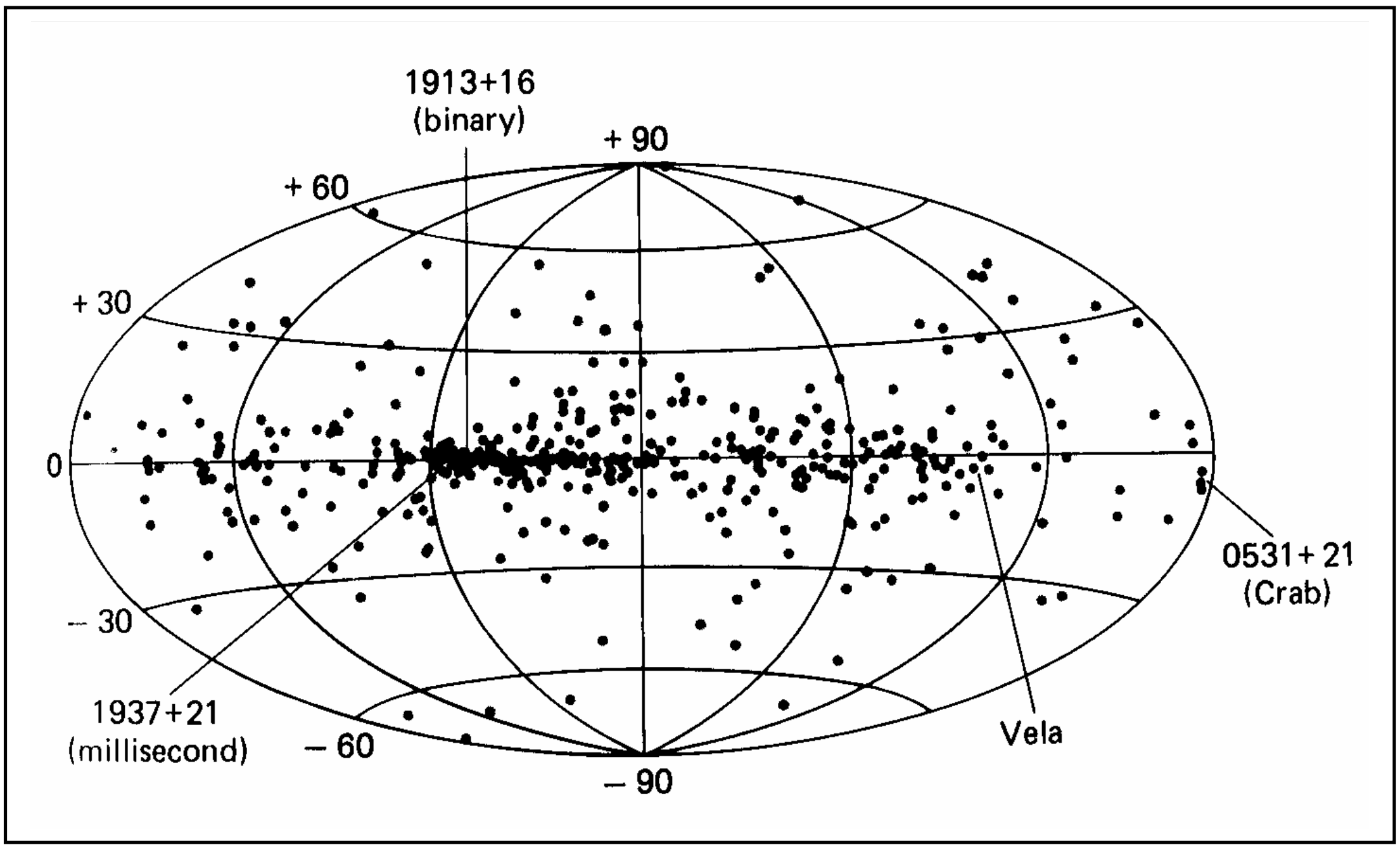

**Fig.3.3.** The distribution of the 445 pulsars on a projection mapping the entire sky, with the plane of Milky Way (our galaxy) along the zero-degree line on the map.

Two main characteristics [2] have to be considered in the pulsars. One is the clock mechanism and the second is the amount of energy radiated. In this part the first models, proposed to fit the first characteristic, are considered. The energy production mechanism will be discussed in detailed in a separate section. That will be the main subject matter of the thesis. Compact objects -WDs or NSs - were the first candidates for pulsars. The reason for this belief was as follows. For Cepheid variable stars, the rough relation between the pulsation period, $p$, and the density of the star, $\rho$, is $p \sim \rho^{-1/2}$. Hence, the densities of the pulsars would be expected to be about $10^3$ – $10^4$ g cm$^{-3}$, which are the typical mean densities for WDs [8]. In addition, the main-sequence stars are too big to turn off and on in such short periods. For example, our Sun would need at least 5 seconds to become dark.

There were three different mechanisms proposed [2, 8] to describe the clock mechanism of the pulsars: radial pulsations (analogous to classical Cepheids), orbital motion, and rotation. From a theoretical point of view, the period of radial pulsation cannot exceed one second in WDs or NSs. But it was less than 0.1 s in the *Crab pulsar*. Thus, radial pulsations were virtually ruled out.

The orbital motion of WDs was also soon ruled out [2, 8] because the orbital period of a pair of WDs in contact, would not be less than 1.7 s. Similarly, if the NSs orbited each other, then as a result of general relativity, there would be a huge loss of energy in the form of gravitational radiation which, in turn, would result in a decrease of their periods. In contrast, observations of various pulsars had shown that in all cases the period was (very) gradually increasing. In order to avoid excessive gravitational radiation, the orbiting object would have to be of mass $<< 10^{-6} M_{\odot}$. Consequently, the orbital motion of NSs was also ruled out. Further, tidal forces would destroy a normal planet of the required low mass.

As long as the first two mechanisms were ruled out, rotation of either WDs or NSs is the only way to explain the first characteristic of pulsars. The rotational WD model was also ruled out because it can only be stable if the period is more than one second. If it were less, then it would be destroyed by centrifugal forces. Therefore, astrophysicists expected that pulsars should be more compact objects than WDs to overcome these forces, even if they rotate at relativistic speeds. As a result, Gold (1968) proposed that rotating NSs provided the best candidate model to obtain periods in the observable range, see Fig. 3.4. The NS would be formed in a supernova explosion [8, 15].

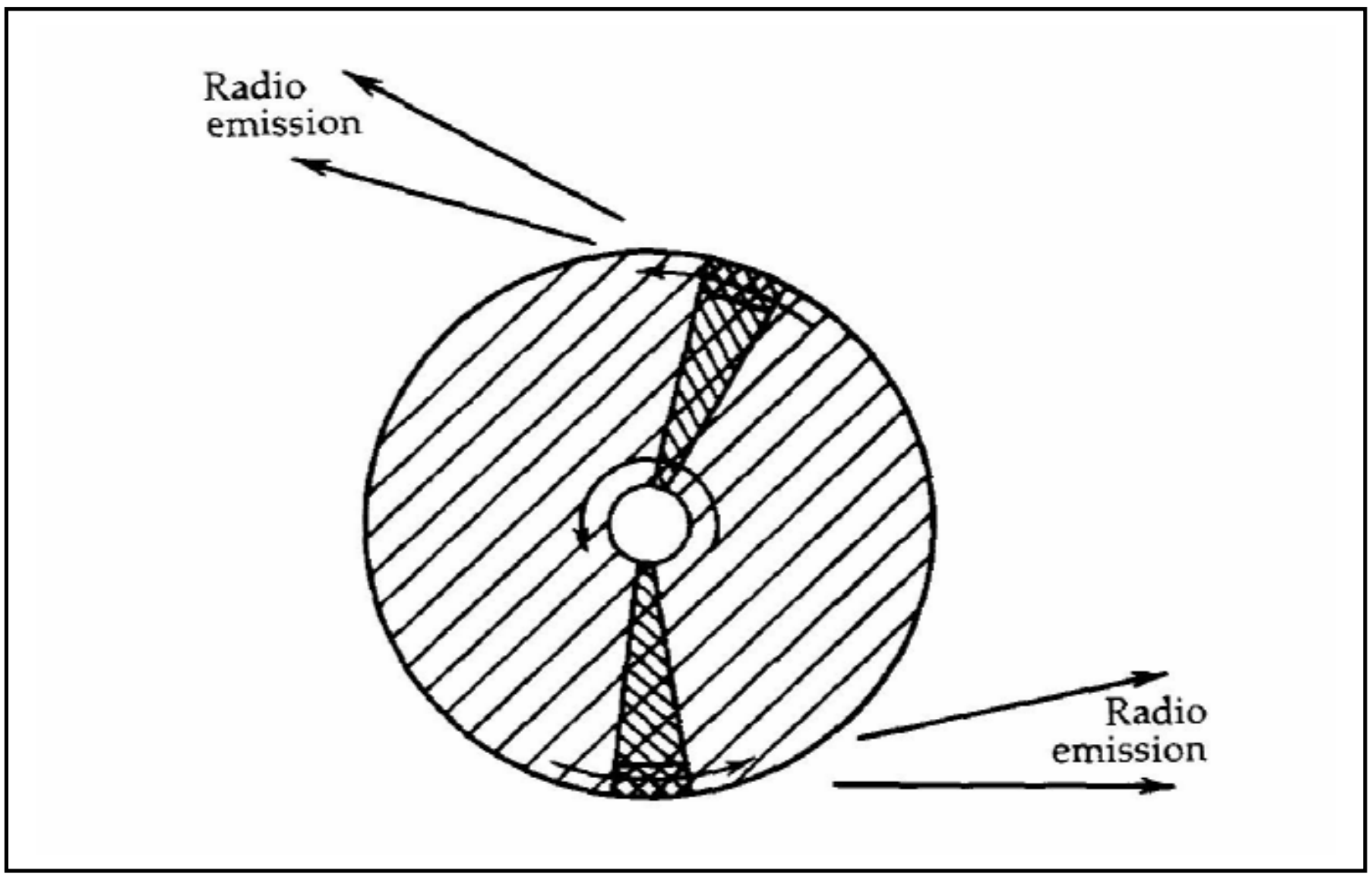

**Fig.3.4.** The Gold's simplest model of a pulsar.

In fact Gold used the idea [15] of Pacini (1967), the first to discuss a rapidly spinning NS, before the discovery of pulsars. There were two main features of Gold's model. The first is that the magnetic axis of the NS does not coincide with its rotation axis. The second is the increase of the pulsation period, which, was confirmed later (1969). For example, the period of the *Crab pulsar* is regularly increasing by 36 nanoseconds a day. The Gold model was the first simple model which led to an acceptable (oblique rotator) model, see Fig. 3.5. The features and characteristics of this model will be considered later.

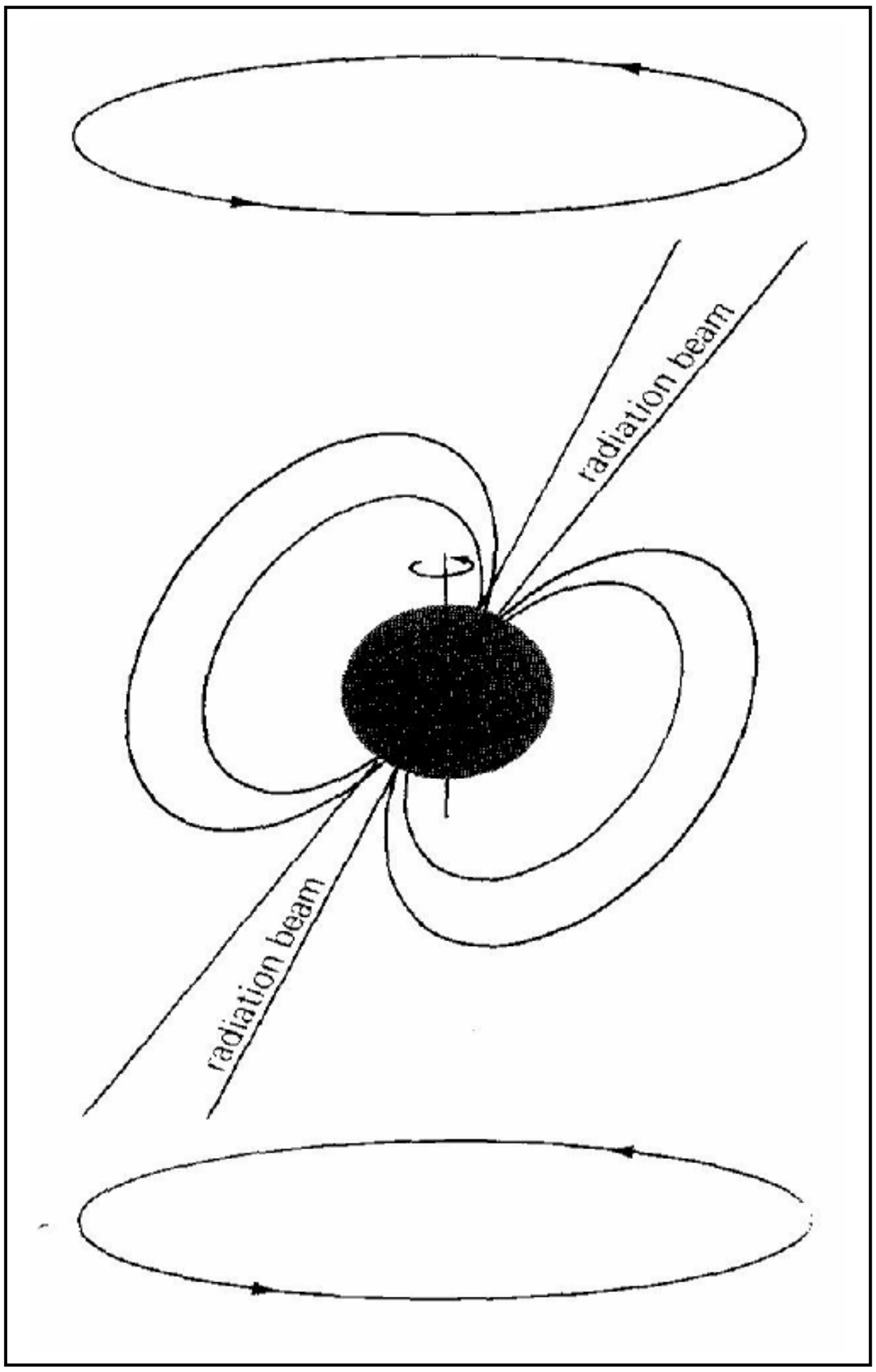

**Fig.3.5.** Oblique rotator model at which the axis of rotation is inclined with the axis of magnetic field.

## 3.2. CHARACTERISTICS OF NEUTRON STARS

Baade and Zwicky [8, 16] introduced the concept of neutron stars (NSs) in 1934, as supernova remnants (SNRs). Later on, Oppenheimer and Volkoff constructed the first models of such high-density objects in 1939 [16]. That idea remained theoretical for 28 years till the discovery of pulsars. That object was the first pulsar discovered.

The NS's characteristics are extreme: density ($\sim 10^{14.5}$ g cm$^{-3}$) like a huge nucleus; rotation velocity (~ c); and magnetic field ($\sim 10^{12}$ gauss). In other words, the mass of the NS is approximately the same as the WD but its radius is around 10 km. Consequently, NSs are of great interest to astrophysicists, especially after the discovery of the millisecond pulsar [2, 8]. So, many attempts were made to construct appropriate models for NSs that fit the observed signals from the various pulsars. In the following chapter we are going to introduce the formation and the structure of NSs and then the hydrostatic models.

### 3.2.1. FORMATION AND INTERIOR STRUCTURE OF NSs

The neutron star is one of the possible terminal stages of a star. It can occur if the core of a massive star (~8 solar masses) collapses, usually after a type-II supernova explosion [2]. As mentioned in § 1.4.2.3, after the supernova explosion the ejecta move off into space to form a shell around the core of the star, the supernova remnant. If the mass of the core lies in a certain range, then that core will form a neutron star [15]. Like WDs, in NSs the (outward) degeneracy pressure (of neutrons in this case) will balance the (inward) gravitational force to give a hydrostatic equilibrium. The lower and upper limit of the mass and the central density, $\rho_c$, as well as the properties of NS mainly depends on the equation of state used. According to most of these equations [22], the lower limits probably lie in the range

$$0.05\, M_{\odot} \lesssim M_{min} \lesssim 0.2\, M_{\odot}, \qquad 10^{13.4}\text{ g cm}^{-3} \lesssim \rho_{c\,min} \lesssim 10^{14}\text{ g cm}^{-3};$$

and the upper limits [12] in the range

$$1.42\, M_{\odot} \lesssim M_{max} \lesssim 3\, M_{\odot}, \qquad 10^{15}\text{ g cm}^{-3} \lesssim \rho_{c\,max} \lesssim 10^{16}\text{ g cm}^{-3}.$$

The reactions, taking place in the formation of NS, will be considered next.

## 3.2.1.1. FORMATION OF NSs

In this section we will answer the question, "under what circumstances and how does the NS reach that equilibrium?" To answer this question let us follow the collapse of the core after the supernova explosion.

Soon after the supernova explosion [2, 7, 12, 15] the core of the supernova starts collapsing at a high temperature (T > $10^{10}$K). As a result of increasing density, the electrons are absorbed by the nuclei, which increases the number of neutrons as

$${}^{0}_{-1}e + {}^{1}_{1}p \rightarrow {}^{1}_{0}n + \nu, \tag{3.1}$$

where $\nu$ is the neutrino, which escapes from the dense core of the NS. Positrons can also be produced if some protons transform to neutrons by

$${}^{1}_{1}p \rightarrow {}^{1}_{0}n + {}^{0}_{1}e + \bar{\nu}. \tag{3.2}$$

Therefore, neutrino-antineutrino pairs can be formed by annihilation of electrons and positrons

$${}^{0}_{-1}e + {}^{0}_{1}e \rightarrow \nu + \bar{\nu}. \tag{3.3}$$

These neutrinos and antineutrinos can carry an enormous amount of energy out of the NS and thereby cool it very rapidly [15]. After one day the temperature can reach $10^9$ K [12]. In addition, neutrinos and antineutrinos can be produced through the reactions

$$ {}^1_0n + {}^1_0n \rightarrow {}^1_0n + {}^1_1p + {}^{\ 0}_{-1}e + \overline{\nu}, \tag{3.4} $$

$$ {}^1_0n + {}^1_1p + {}^{\ 0}_{-1}e \rightarrow {}^1_0n + {}^1_0n + \nu, \tag{3.5} $$

Some calculations seem to show that the temperature decreases to $10^8$ K after one month [15], while some others that it takes 100 years [12] after the NS creation. In addition, the NS may cool even faster because of superfluidity in the core, as we will see in the next section [15].

### 3.2.1.2. INTERIOR STRUCTURE OF NSs

The physical properties of the NS vary with radial distance. The dominant factor in this case is the density, which decreases with radial distance. There are many equations of state that can be derived to describe NSs. Thus, there is no unique theory, defining its interior structure. This issue is still controversial among astrophysicists.

Consider, for example, a NS of $M = (4/3)M_\odot$ [16] and radius 16 km. It is expected that its outermost crust will consist of a mixture of heavy nuclei (${}^{56}Fe$) and degenerate electrons with a relatively low density, ρ ~ $10^4$ g $cm^{-3}$. (This is about the same as the interior density of a WD.) Below the outer crust, the Fermi energy of electrons increases because of increasing density. This leads to the 'neutronization through electron capture' mentioned above, called *'neutron drip'*. Thus, in this layer the density lies between $2.3\times10^{11}$ and $2\times10^{14}$ g $cm^{-3}$, in which there are so many neutron-rich nuclei that form another crystalline lattice, called the *inner crust*. The density $\rho_{dr} \sim 4 \times 10^{11}$g $cm^{-3}$ is called the '*drip density*'. The number of neutrons increases rapidly as the density increases inward. Below the critical temperature ($10^{10}$ K) the Fermi energy of the electrons, $E_F$, increases to about 1000 MeV. In that case the mean energy of the particles ($\sim kT$) is much less than $E_F$. Hence, that case can be regarded as so cold as to allow neutrons to attract each other to form a *superfluid*. At $\rho \sim 2\times10^{14}$ g cm$^{-3}$, which is the same as nuclear density $\rho_{nuc}$, all nuclei will dissolve into neutrons.

The density increases with decreasing radial distance and the material can be thought of as a uniform sea of electrons, protons, and neutrons. The concentrations of these particles can be calculated by considering the reaction

$$n \leftrightarrow p + e, \tag{3.5}$$

in a ratio of $n_n : n_p : n_e = 8 : 1 : 1$. Again at $T < 10^{10}$ K, the neutron-liquid in that region is superfluid and the protons (*Cooper pair*) [5] are in a superconducting state. Both superfluidity and superconductivity conditions can be considered as important factors in understanding the physical and electromagnetic behaviors of NS.

In some models (i.e. for some equations of state), if the density exceeds $10^{15}$ g cm$^{-3}$ there is the possibility of a solid core, which can contain some heavy particles, like $\Sigma^-$ *and* $\Lambda^\circ$ hyperons. A question arises regarding the existence of quark stars as the density increases further. From our knowledge of particle physics [12] we know that at $\rho > 2\,\rho_{nuc}$ the reaction:

$$n \rightarrow p + \pi^-, \tag{3.6}$$

can occur. But $\pi^-$ is a boson, which contributes to form a Bose-Einstein condensate - bosons with zero momentum–under critical temperature. Therefore, there is no contribution to the pressure provided by quarks to balance the increased gravity [12]. In contrast it will contribute to $\rho$, which in turn enhances gravitational collapse.

One can summarize the picture of neutron stars as follows [16]. NSs consists of a core of strange particles (hadrons with non-zero strangeness number), with $\rho \sim 4.7 \times 10^{15}$ gm cm$^{-3}$. The main body consists of protons, neutrons and electrons, with $\rho \sim 5 \times 10^{13}$ gm cm$^{-3}$ and the crust of heavy nuclei, with $\rho \sim 10^{12}$ gm cm$^{-3}$. Their outer region is an atmosphere of a lattice of $\rho \sim 10^{8}$ gm cm$^{-3}$. Finally the outermost crust consists of iron of $\rho \sim 10^{6}$ gm cm$^{-3}$. One can see schematic cross-sections of NSs, based on two different equations of state for total gravitational mass of around $1.33M_{\odot}$ [8] and $1.4M_{\odot}$ [12] in Fig. 3.6 and 3.7 respectively.

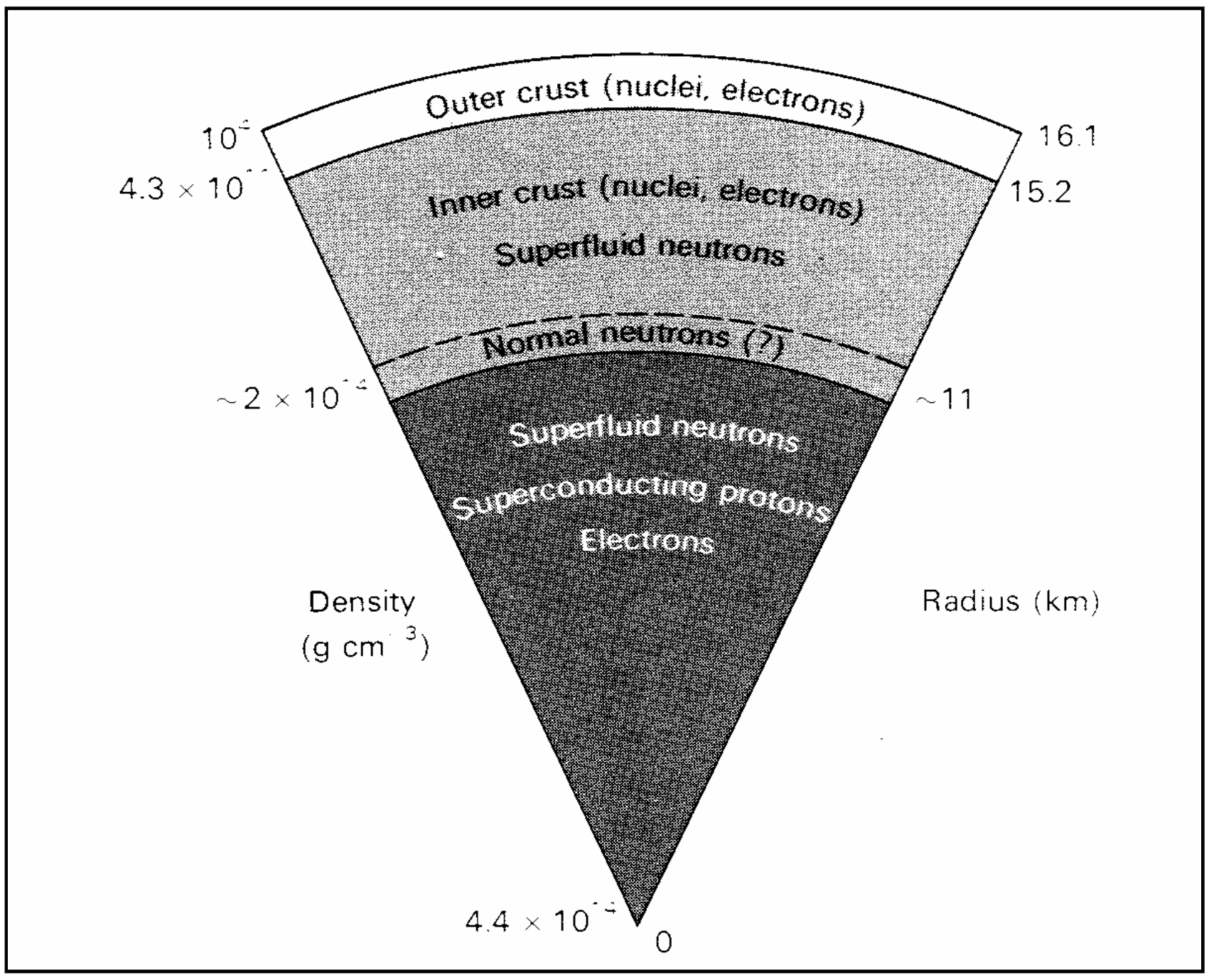

**Fig.3.6.** Cross section of the interior structure of NS model with $M \sim 1.33M_{\odot}$, the values of $\rho$ do not reach $10^{15}$ gm cm$^{-3}$, at which hyperon creation and solidification may occur.

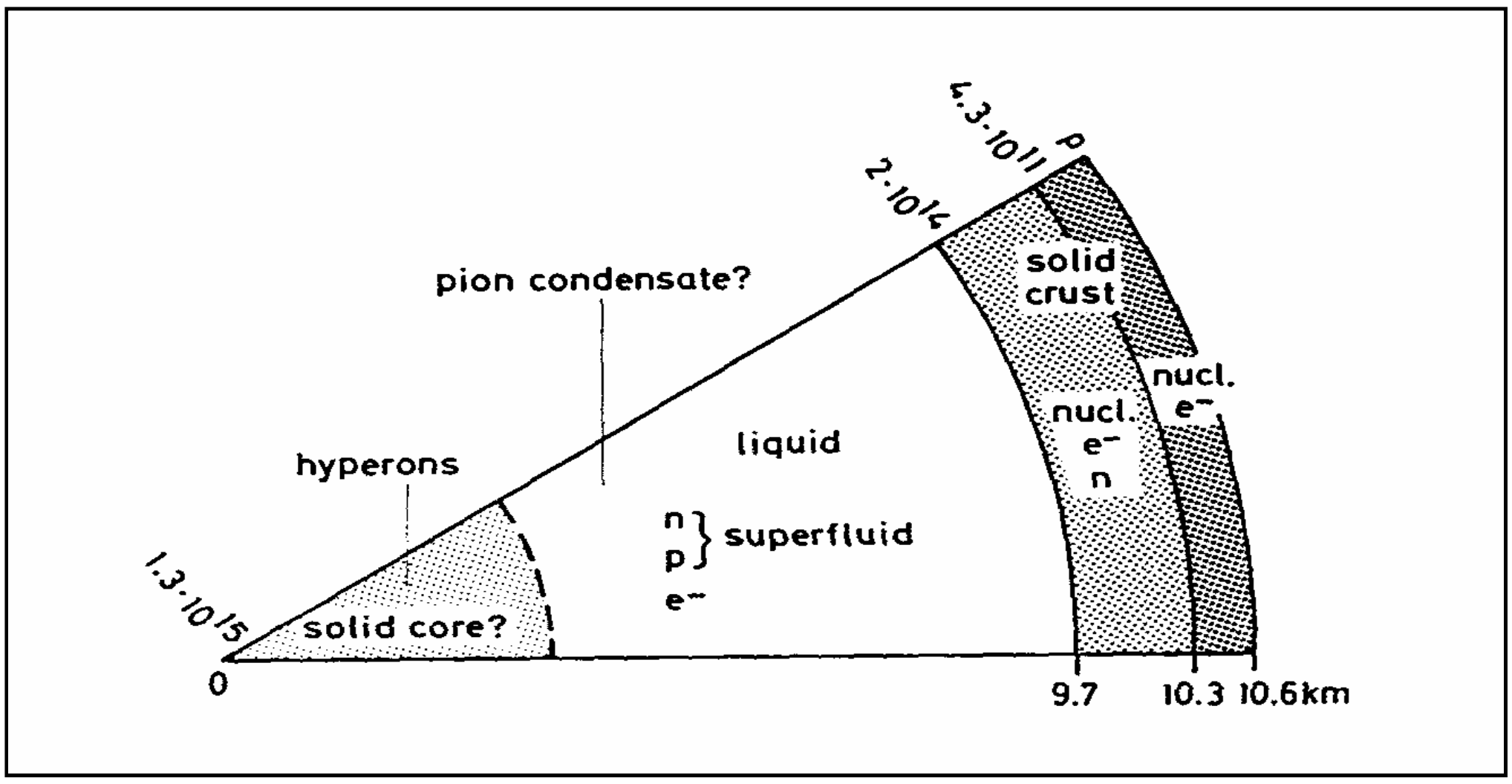

**Fig.3.7.** Cross-section of the interior structure of different NS model with $M \sim 1.4M_\odot$, some of the characteristic values of $\rho$ are indicated along upper radius in gm cm$^{-3}$.

## 3.2.2 THE HYDROSTATIC MODELS OF NSs

As for WDs (Ch.2), in this section we are going to derive the equation of state, $P=P(\rho)$, and then construct hydrostatic models of NSs. Unlike WDs, there are many equations of state and hydrostatic models for NSs. This is due to the extreme characteristics of NSs, which require consideration of other effects such as *general relativity*, *superconductivity*, and *superfluidity*. The equation of state will be derived first without, then including, general relativistic effects.

### 3.2.2.1 NEGLECTING GENERAL RELATIVISTIC EFFECTS

In this section we are going to construct the hydrostatic equilibrium equation for NSs without considering effects of *general relativity*. Nevertheless, we do consider the effects of *special relativity*. From the last section, it is apparent that, for densities up to $\rho_{dr}$, relativistic, degenerate electrons dominate the equation of state, (c.f. also, (2.45)). Increasing the density will lead to increase of the density of neutrons, $n_n$, at the expense of the electrons, $n_e$, which increases the (outward) pressure to balance the gravitational (inward) force, (1.3).

Defining the adiabatic index [12], $\gamma_{ad}$, to be

$$\gamma_{ad} = d(\ln P)/d(\ln \rho)\,. \qquad (3.7)$$

From (2.45), $\gamma_{ad} = 4/3 = \gamma_{cr}$, which is a critical value of the adiabatic index. At $\gamma_{ad} < 4/3$, the gravitational force would be more than the pressure of degeneracy and the star would collapse after initial compression. If $\gamma_{ad} = 4/3$, there is hydrostatic equilibrium, but if $\gamma_{ad} > 4/3$, there is stable dynamic equilibrium. Therefore, around $\rho_{dr}$ , $\gamma_{ad} < 4/3$, which raises $\rho$ up to $7\times10^{12}$g cm$^{-3}$ ,the value at which the pressure of neutrons will contribute more to reach dynamical stability.

If the neutron pressure $P_n$ dominates the outward pressure [12], then we can consider the neutrons as a *fully degenerate gas*, which obeys the same statistics as the electrons. Hence, if $m_e$ is replaced by $m_n$ and $\mu_e$ by 1 (because we have one nucleon per fermion), then (2.43) and (2.45) for non-relativistic and relativistic neutrons can be written respectively as

$$P_d = \frac{1}{20}\left(\frac{3}{\pi}\right)^{2/3} \frac{h^2}{m_n^{8/3}} \rho_0^{5/3}\,, \qquad (3.8)$$

$$P_d = \frac{1}{8m_n^{4/3}} hc\left(\frac{3}{\pi}\right)^{1/3} \rho_0^{4/3}\,, \qquad (3.9)$$

where we assume $m_u = m_n$ and $\rho_0 = n_n m_n$, the rest density.

The mass density, $\rho$, for relativistic particles equals the density of total energy divided by $c^2$. Also, the total energy, $E$, can be written as

$$E = E_0 + T\,, \tag{3.10}$$

where $E_0$ is the rest energy of the particle and $T$ is its kinetic energy. Thus, for relativistic neutrons, $\rho = \rho_0 + u/c^2$, where $u$ is the kinetic energy density of a neutron. This correction was neglected for the WDs, because $\rho_0 >> u/c^2$ for relativistic electrons as well as for non-relativistic neutrons. In contrast, for ultra-relativistic neutrons $\rho_0 << u/c^2$, $\rho \approx u/c^2$. Also, for relativistic particles, see (2.46), $P = u/3$, *i.e.*

$$P_d = \frac{c^2}{3}\rho\,. \tag{3.11}$$

Thus, if $n_n$ is not too high, then the nuclear attraction between neutrons will reduce $P_d$. For very high density, there is degenerate repulsion, which enhances $P_d$.

### 3.2.2.2 INCORPORATING GENERAL RELATIVISTIC EFFECTS

To incorporate the effects of general relativity [12, 22] instead of solving (1.3) one has to solve Einstein's field equations

$$R_{ik} - \frac{1}{2}g_{ik}R = \frac{\kappa}{c^2}T_{ik}\,, \qquad \kappa = \frac{8\pi G}{c^2}\,, \tag{3.12}$$

where $R_{ik}$ is the Ricci tensor, $g_{ik}$ is the metric tensor, $R$ is the Ricci scalar, and $T_{ik}$ is the energy-momentum tensor, which is diagonal for an ideal gas, i.e. $T_{00} = \rho c^2$, $T_{11} = T_{22} = T_{33} = P_d$ . Thus, there are 10 partial differential equations for 10 functions ($g_{ik}$) of 4 independent variables $(t, r, \theta, \varphi)$ in spherical coordinates. The Schwarzschild solution of the Einstein equations represents a static spherically symmetric mass distribution, given by the metric

$$ds^2 = e^{\upsilon}c^2dt^2 - e^{\lambda}dr^2 - r^2(d\theta^2 + \sin^2\theta d\varphi^2)\,, \tag{3.13}$$

where $\upsilon = \upsilon(r), \lambda = \lambda(r)$. Thus, the metric tensor

$$g_{ik} = \begin{bmatrix} e^{\upsilon(r)} & 0 & 0 & 0 \\ 0 & -e^{\lambda(r)} & 0 & 0 \\ 0 & 0 & -r^2 & 0 \\ 0 & 0 & 0 & -r^2 \sin^2\theta \end{bmatrix}. \tag{3.13}$$

Therefore (3.11) can be reduced to the 3 ordinary differential equations:

$$\frac{\kappa P_d}{c^2} = e^{-\lambda}\left(\frac{\upsilon'}{r} + \frac{1}{r^2}\right) - \frac{1}{r^2}, \tag{3.14}$$

$$\frac{\kappa P_d}{c^2} = \frac{1}{2} e^{-\lambda}\left(\upsilon'' + \frac{1}{2}\upsilon'^2 + \frac{\upsilon' - \lambda'}{r} - \frac{\upsilon'\lambda'}{2}\right), \tag{3.15}$$

$$\kappa\rho = e^{-\lambda}\left(\frac{\lambda'}{r} + \frac{1}{r^2}\right) + \frac{1}{r^2}, \tag{3.16}$$

where, $\lambda' = d\lambda / dr$. Multiplying (3.16) with $4\pi r^2$ and integrating it gives

$$\kappa m = 4\pi r(1 - e^{-\lambda}). \tag{3.17}$$

Differentiation of (3.14) and elimination of $\lambda, \lambda', \upsilon', \upsilon''$ by using (3.14-16) gives [12] the so-called *Tolman-Oppenheimer-Volkoff (TOV) equation* for hydrostatic equilibrium in general relativity:

$$\frac{dP}{dr} = -\frac{Gm}{r^2}\rho\left(1 + \frac{P}{\rho c^2}\right)\left(1 + \frac{4\pi r^3 P}{mc^2}\right)\left(1 - \frac{2Gm}{rc^2}\right)^{-1}, \tag{3.18}$$

where *m* is the *"gravitational mass"* inside a radius *r*, defined by

$$m = \int_0^r 4\pi r^2 \rho dr. \tag{3.19}$$

If $r = R$, then $m = M$, where $M$ is not only the rest mass but also the whole energy divided by $c^2$.

Clearly, (3.18) reduces to the usual equation (1.3) as $c \to \infty$. However, for a gravitational field that is not too large, one can use a linear approximation of (3.18), called the *post-Newtonian approximation:*

$$\frac{dP}{dr} = -\frac{Gm}{r^2}\rho\left(1 + \frac{P}{\rho c^2} + \frac{4\pi r^3 P}{mc^2} + \frac{2Gm}{rc^2}\right). \tag{3.20}$$

## 3.2.3 ROTATION OF NSs

As mentioned in §3.2.1, the collapsed core of a star, after a supernova explosion, can become a NS. Usually, the original core has a small spin and a weak magnetic field before collapsing. As a result of the conservation of angular momentum and magnetic flux [5], the equatorial velocity and the magnetic field lines of the NS become large after collapsing. In such a catastrophic collapse, the radius of the star shrinks by about $10^5$ times the original radius. Thus, if the original value of angular momentum is $L$, then [15] the equatorial velocity, $\upsilon$, will be

$$\upsilon = \frac{L}{MR} \alpha \frac{1}{R}, \tag{3.21}$$

where, $R$ is the radius of the star. Therefore, $\upsilon$ will become 100,000 times the original value, which is close to the velocity of light $c$. Also, it is easy to relate the equatorial velocity, $\upsilon$, to the rotational period $p$ by

$$p = \frac{2\pi R}{\upsilon} \alpha R^2, \tag{3.22}$$

which means that its rotational period will become $10^{-10}$ of the original period. For example, if our sun, which has a period of about 27 days, were suddenly to turn into a neutron star, it would rotate around 42,000 times per second.

Similarly, if the original magnetic flux is $\Phi$, then the magnetic field $H$ is related to $R$ by

$$H = \frac{\Phi}{4\pi R^2} \alpha \frac{1}{R^2}, \tag{3.23}$$

whence the magnetic field becomes $10^{10}$ times its original value. It is more than what has ever been achieved in any laboratory in the world. A discussion of the electromagnetic field of the NS will be given in detail in the next section. It is important to note that there is continuous conversion of the rotational kinetic energy of the NS into electromagnetic energy, which decreases $\upsilon$, and in turn increases *p*.

### 3.2.3.1 AGE ESTIMATION OF PULSARS

One of the predictions of the Gold-model is the increase of rotational period for all neutron stars. It was confirmed by observation in 1969 for the Crab pulsar. This established the spinning neutron star model for pulsars. Therefore, the youngest neutron stars should spin faster than older stars and so have less rotational period, *p*. Although the Gold-model directly gives an acceptable explanation and prediction for many aspects of pulsars, there are some other observed quantities, which need further assumptions to fit with the observed data. The *braking index*, defined in (3.25), and its relation to the age of pulsars[17, 18, 19], and the *drift velocity* [19, 20] are some of these quantities.

If the radiation of the pulsar is considered as purely dipole in nature [8, 17, 18], $\omega$ is the pulsar frequency and $\dot{\omega}$ is its derivative then

$$\dot{\omega} = b\omega^3 , \tag{3.24}$$

where *b* is a constant. The dipole radiation is due to the angle between the axis of rotation and that of magnetic field-Gold model.

Defining the braking index, *n* as

$$n = \frac{\omega\ddot{\omega}}{\dot{\omega}^2} , \tag{3.25}$$

the observed $n$ and $\dot{\omega}_p$ for the Crab pulsar (NP 0532) (in 1969, when the age was 915 years) were 2.515 ± 0.005 and 3.9 × $10^{-10}$ Hz/s. Also, its period $p$ = 0.033 s, and rate of change of period $\dot{p}$ = 36.5 ns/day.

If we replace (3.24) with the more general formula [17, 18]

$$\dot{\omega} = b\omega^n, \tag{3.26}$$

the $n$ of the Crab is 16% less than for a pure dipole. Defining the *characteristic time*, $\tau = \omega / \dot{\omega} = p / \dot{p}$, which is a time scale for the loss of rotational energy or the age of pulsar. Thus, integrating [8] (3.26) we have

$$\tau = \frac{\omega_p}{(n-1)\dot{\omega}_p}\left[1-\left(\frac{\omega_p}{\omega_i}\right)\right] \qquad (n \neq 1), \tag{3.27}$$

where $\omega_i$ is the frequency at $t$ = 0. If $\omega_i >> \omega_p$, then (3.27) becomes the characteristic time

$$\tau = \frac{\omega_p}{(n-1)\dot{\omega}_p} \qquad (n \neq 1). \tag{3.28}$$

For dipole radiation $n$ = 3. Then $\tau = \omega_p / 2\dot{\omega}_p \approx 3.9 \times 10^{10}$ s ≈ 1240 years, which is around 30% longer than the historical age (915 years in 1969) of supernova explosion (occurred at 1054 AD). But if $\omega_i$ is not much greater than $\omega_p$, then it is less than the above value. Also, if the braking index has increased with time, then the present value of $\tau$ may underestimate the true age. In fact, $n$ cannot be less than 2, because $\dot{p}$ is not constant, hence it should be within a factor of two as observed. It has been argued that $\omega_i$ cannot exceed 100 Hz as the pulsar would break up due to centrifugal forces. However, a "millisecond pulsar" has a frequency of 1000Hz, so the argument is unreliable. Further, Abramowicz showed that the relativistic centrifugal forces could cause attraction instead of repulsion in the regime of pulsars.

On the other hand, if the radiation is written as a multipole expansion, then (3.24) becomes

$$\dot{\omega} = b\omega^3 + c\omega^5 + d\omega^7 + f\omega^9 + \cdots. \qquad (3.29)$$

With the above equation, the value of the braking index cannot be less than 3. Thus, it must be modified by introducing an additional term, say $a\omega$. To fit with the data for the Crab pulsar we can set $f = g = \cdots = 0$. Then (3.29) becomes

$$\dot{\omega} = a\omega + b\omega^3 + c\omega^5 + d\omega^7. \qquad (3.30)$$

The coefficients in this expansion would vary slowly with time compared with the variation of ω, but they can be considered as constants in a short period. Also, (3.30) has a monopole term as well as a dipole and two other higher multipole radiation terms. The monopole and the higher pole terms are due to very diffuse plasma around the neutron star [17, 18], while the dipole term, as mentioned before, is due to the obliquely rotating magnetic dipole.

Using the assumption of high $\omega_i$ and obtaining the age of the pulsar from the formula [17],

$$\tau = \int_0^T dt = \int_\infty^{\omega_p} d\omega / \omega, \qquad (3.31)$$

using (3.30), one arrives at a long expression [17] for the age of the pulsar. Now the values of the braking index and the age of the pulsar can be used to fit the coefficients of (3.30) to the empirical data. This model, unlike the previous model, which allows wide variation of the initial frequencies of pulsars, takes $\omega_i$ to be much more uniform. Thus, the observed variation in pulsar frequencies is due to the fact they are in different evolutionary stages. The predicted value ($\sim 2 \times 10^3$ Hz) of the initial frequency of pulsars by the second model was believed to have been verified by observations of the claimed pulsar at the center of Supernova 1987A, before the claim was withdrawn.

### 3.2.3.2. PULSAR DRIFT VELOCITY

We take the nascent pulsar to have a very high $\omega_i$, and mass density, $\rho$. As such the gravitational effects in the other aspects of pulsars should not be neglected. An extraordinary property of pulsars is their high *drift velocity*, which is the speed of the pulsar relative to the center of the supernova remnant [19, 20]. It is typically ~ 100 km/s, but in some pulsars can reach 500 km/s [8], see Table.3.1, which lists characteristic ages and drift velocity observed for some pulsars. There are different explanations for this velocity.

One set of proposals of drift velocity [8], appeals only to classical mechanics. There may be perturbations due to other gravitating bodies, some asymmetry in the supernova explosion or some asymmetry in the radiation reaction. However, no detailed mechanism has been worked out for them in a general context. Again, the neutron star could originally have been a member of a close binary system. If one of them exploded as a supernova, the other will fly off with a substantial fraction of its orbital velocity. This would not explain why only the youngest pulsars are inside SNRs and the older ones are not.

**Table. 3.1. The characteristic ages and drift velocity observed for some pulsars.**

*Known proper motions and transverse velocities of pulsars*

| PSR | Characteristic age, $\tau$ ($10^6$ yr) | Proper motion, $\mu$ ($10^{-3}$ arc s yr$^{-1}$) | Distance, $d$ (pc) | Transverse velocity (km s$^{-1}$) | $z$ (pc) |
|---|---|---|---|---|---|
| 0531+21 | 0.0012 | $12 \pm 3$ | 2,000 | 110 | −200 |
| 1237+25 | 2.3 | $102 \pm 18$ | 370 | 190 | 370 |
| 0834+06 | 3.0 | $52 \pm 14$ | 480 | 120 | 210 |
| 1929+10 | 3.1 | $159 \pm 25$ | 110 | 80 | −7 |
| 1133+16 | 5.0 | $365 \pm 36$ | 180 | 310 | 160 |

Another proposal [20, 21] suggests that the drift velocity is due to a general relativistic effect. The collapsed object is considered as formed by a collection of ionized particles falling radially inward in a high magnetic field with a high angular momentum. According to the pseudo-Newtonian (ΨN) force formulation of general relativity [23] there can be an asymmetry between the North and South hemispheres along the axis of rotation [21]. That asymmetry causes a net force between the two hemispheres, which creates an impulse on the pulsar during the time of rapid collapsing. Therefore, the drift velocity has a magnitude equal to the value of the impulse divided by the mass of pulsar along the axis of rotation. If a pulsar of 10 km radius and sufficiently high $\omega_i$, collapses in one second, then this proposal yields a drift velocity of the same order as observed (100 km/s).

The nebulae that surround pulsars are called "*radio nebulae*" because they emit energy as radio waves. The source of such radiation comes from the huge magnetic field of the NS-pulsar. Therefore, the total radiated energy of a radio nebula equals the rotational kinetic energy lost by its pulsar. Kardashov [15], in 1964, was the first to suggest this mechanism for energy radiation from the Crab Nebula. Although there are more than 100 detected radio nebulae, pulsars have been detected in only three. This may be because the radio emission of pulsars is not isotropic (not with the same intensity in all directions), but is confined within a cone of radiation, which may not come to the Earth, see Fig. 3.5.

# 3.3. ELECTROMAGNETIC RADIATION OF PULSARS (NSs)

In the rest of this chapter we will treat the electromagnetic radiations (signals) of the pulsars from a classical point of view. Those detectable signals are the only received information that gives us a chance to understand the physics of pulsars. Models of pulsars should yield a prediction close to the detected radiation. The two main characteristics of the radiation are the regularity of the pulses, "clock behavior", and the huge amount of energy radiated by pulsars. In this chapter, the classical model of NS radiation will be discussed. It is mainly the model proposed [24] by Armin Deutsch, 1955, for a rotating magnetic star in vacuum. The modern models of the pulsars will be discussed in the next chapters. Both special and general relativistic effects will be incorporated there.

## 3.3.1. ELECTROMAGNETIC RADIATION OF SLOWLY ROTATING MAGNETIC STAR (DEUTSCH MODEL)

This was the first model used to understand the physical nature of pulsars. Deutsch [24] constructed this model even before the discovery of pulsars to explain the periodic variation of the magnetic field strength, observed in some stars. It models the star as carrying a magnetic field symmetrical about an axis that is inclined to the rotation axis.

He made certain assumptions regarding the star he proposed: it is spherical, sharply bounded, perfectly conducting (superconductor), isotropic, and rigidly rotating in vacuum. The speed of rotation is so slow, that the surface tangential velocity is much less than the speed of light, i.e. $\upsilon = |\boldsymbol{\omega} \times \mathbf{r}| << c$. Therefore, it is called *the classical model of pulsars*. Also, the static magnetic field **H** is frozen in the star and symmetric about the magnetic axis **M**. On the other hand, Deutsch determined the electromagnetic field by solving Maxwell's equations in three different regions. The first region he considered is the internal region, the second is the externally close region, and the third is externally far region (wave zone) of the star. The last region is the most interesting for us because we lie in it.

## 3.3.1.1. THE INTERNAL ELECTROMAGNETIC FIELD

For an isotropic star one can write the Maxwell equations in SI units as

$$\nabla\cdot \mathbf{E} = \rho/\varepsilon_0, \tag{3.32a}$$

$$\nabla\cdot \mathbf{H} = 0, \tag{3.32b}$$

$$\nabla\times \mathbf{E} = \mu_0\, \partial\mathbf{H}/\partial t, \tag{3.32c}$$

$$\nabla\times \mathbf{H} = \mathbf{J} + \varepsilon_0\, \partial\mathbf{E}/\partial t, \tag{3.32d}$$

where $\varepsilon_0$ and $\mu_0$ are the dielectric constant and magnetic susceptibility of the vacuum and **E**, **H**, **J** and $\rho$ are measured by an observer at rest with respect to the inertial reference system (A) whose origin is at the center of the star. In addition, Deutsch considered the magnetic field at rest with respect to (A) as $\mathbf{H}(r, \theta, \varphi)$ and the solution inside the star rotates with angular velocity $\boldsymbol{\omega}$ along the z-axis is just $\mathbf{H}(r, \theta, \varphi\text{-}\omega t)$. This transformation only gives the classical approximation to the relativistic result, because for high enough velocity, $v = |\boldsymbol{\omega} \times \mathbf{r}| \sim c$, the components of **H** are found from the Maxwell tensor **F**. Consequently, the Lorentz transformation matrix, $\boldsymbol{\Lambda}$, is not simply the identity matrix [25].

In order to solve the Maxwell equations inside the star, we first introduce the necessary condition for **H** to be frozen into the star, which is

$$\mathrm{H}(r, \theta, \varphi, t) = r_0\, \mathrm{H_r}(r, \theta, \lambda) + \theta_0\, \mathrm{H}_\theta(r, \theta, \lambda) + \varphi_0\, \mathrm{H}_\varphi(r, \theta, \lambda), \tag{3.33}$$

where $\lambda$ is an azimuthal coordinate measured from a fixed meridian in the star. Of course this property of **H** occurs because the electric field inside the superconductor star must vanish with respect to the inertial frame (B) momentarily at rest [16, 26]. Therefore, due to rotation the induced drift electric field **E** in the star with respect to (A) can be found as

$$\mathbf{E}_{\mathrm{ind}} = -\mu_0\, \upsilon \times \mathbf{H}. \tag{3.34}$$

Substituting (3.34) into (3.32c) gives

$$\nabla\times (\upsilon \times \mathbf{H}) = -\partial\mathbf{H}/\partial \mathrm{t}. \tag{3.35}$$

The charges and current densities, $\rho$ and **J**, associated with **H** can be determined by (3.32a) and (3.32d). Internal values of **H**, **E**, $\rho$, and **J** are not interesting as they are but to determine the corresponding external fields.

## 3.3.1.2. THE EXTERNAL ELECTROMAGNETIC FIELD

The region external to the star is considered as source-free, i.e. $\rho = 0$ and $\mathbf{J} = \mathbf{0}$. The solution of Maxwell's equations for the electromagnetic fields leads to propagating wave functions. They have spherical components $H_r$, $H_\theta$, and $H_\varphi$ for the magnetic field and $E_r$, $E_\theta$, and $E_\varphi$ for the electric field, with respect to the system (A).

In order to simplify the problem let us consider a unit sphere drawn about the center O. Let (**i**, **j**, **k**) be the unit vector of Cartesian coordinates with an origin O and **k** parallel to **ω** and the unit vector **e** mark the instantaneous direction of the symmetry axis of the internal field **H**. Hence, the angle between **e** and **ω** is just $\chi$, see Fig 3.8. Furthermore, if we consider spherical coordinates ($r$, $\phi$, $\upsilon$) with axis along **e**, then the magnetic field in the star,

$$\mathbf{H} = r_0\, R_1(r)\, S_1(\cos\phi) + \phi_0\, R_2(r)\, S_2(\cos\phi) + \upsilon_0\, R_3(r)\, S_3(\cos\phi), \qquad (3.36)$$

is symmetric about **e**. However, at the surface of the star, $r = a$, the boundary conditions give

$$H^{ext}_{r} = H^{int}_{r} = R_1(a)\, S_1\,, \qquad (3.37a)$$

$$E^{ext}_{\theta} = E^{int}_{\theta} = -\omega\mu_0\, a R_1(a)\, S_1 \sin\theta\,, \qquad (3.37b)$$

$$E^{ext}_{r} = E^{int}_{\phi} = 0. \qquad (3.37c)$$

Therefore, the external field can be determined if we specify $S_1$ function. Using the method developed by W. Hanson for the case of $S_1 = \cos\phi$. Also, by satisfying (3.32b), one can show that

$$S_2 = \sin\psi, \qquad R_2 = -\frac{1}{2r}\frac{d}{dr}(r^2 R_1)\,. \qquad (3.38)$$

The other Maxwell equations (3.32a, 32c, 32d) show that the external fields become a wave propagating outward. Deutsch found the general expression for thesecomponents as a real part of the (3.39-44)

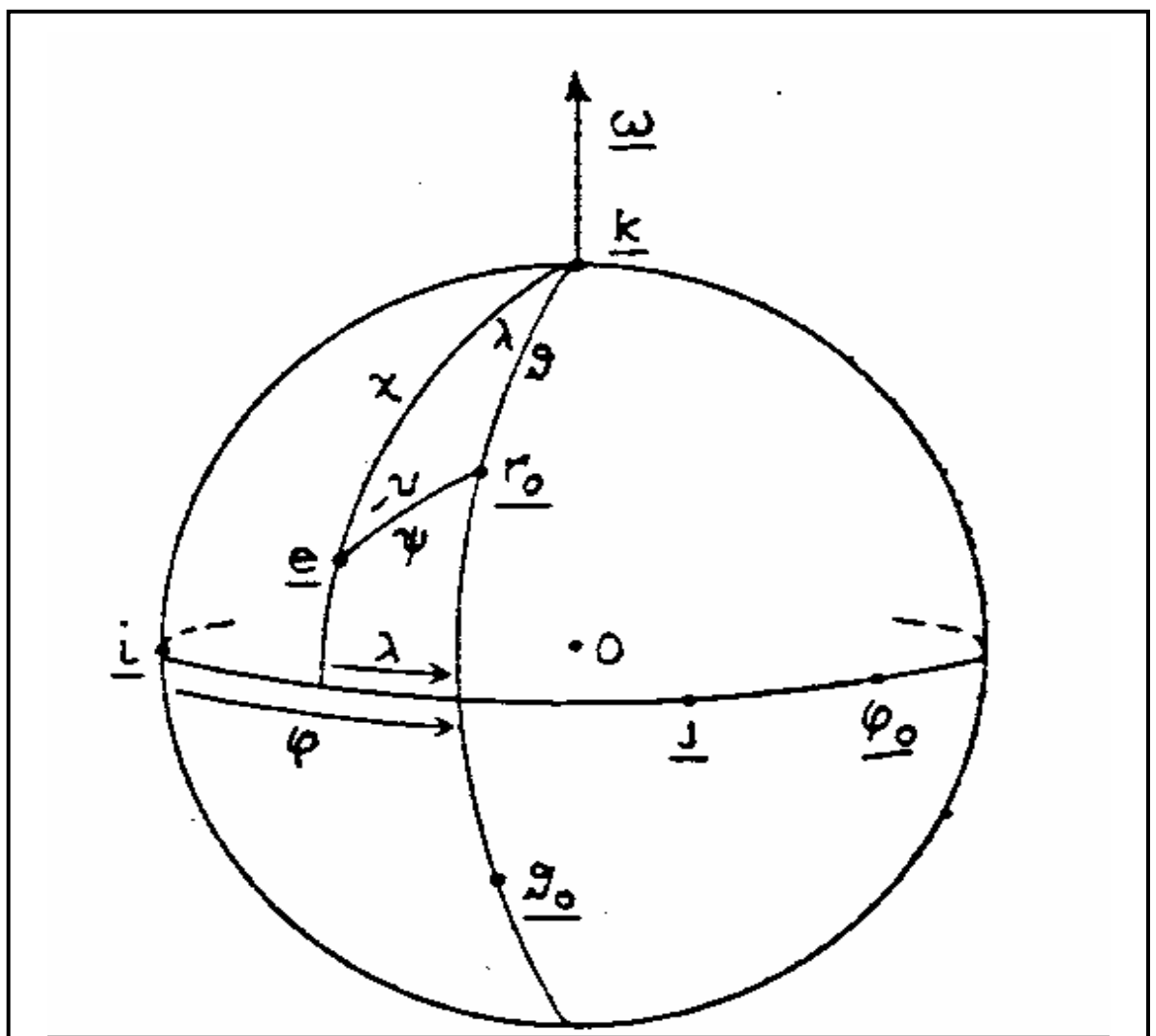

**Fig.3.8.** The unit sphere about the center of magnetic star with angular frequency **ω**.

$$\mathrm{H_r} = R_1(a)\left[\frac{a^3}{r^3}\cos\chi\cos\theta + \frac{h_1/\rho}{(h_1/\rho)\alpha}\sin\chi\sin\theta e^{i\lambda}\right] \tag{3.39}$$

$$\mathrm{H}_\theta = \frac{1}{2}R_1(a)\left\{\frac{a^3}{r^3}\cos\chi\sin\theta + \left[\left(\frac{\rho^2}{\rho h_2' + h_2}\right)_{\rho=\alpha} h_2 + \left(\frac{\rho}{h_1}\right)_{\rho=\alpha}\left(h_1' + \frac{h_1}{\rho}\right)\sin\chi\cos\theta e^{i\lambda}\right]\right\} \tag{3.40}$$

$$\mathrm{H}_\varphi = \frac{1}{2}R_1(a)\left\{\left(\frac{\rho^2}{h_2' + h_2}\right)_{\rho=\alpha} h_2\cos 2\theta + \left(\frac{\rho}{h_1}\right)_{\rho=\alpha}\left(h_1' + \frac{h_1}{\rho}\right)\right\} i\sin\chi e^{i\lambda} \tag{3.41}$$

$$\mathrm{E_r} = \frac{1}{2}\omega\mu_0 a R_1(a)\left\{-\frac{1}{2}\frac{a^4}{r^4}\cos\chi(3\cos 2\theta + 1) + 3\left(\frac{\rho}{\rho h_2' + h_2}\right)_{\rho=\alpha}\frac{h_2}{\rho}\sin\chi\sin 2\theta e^{i\lambda}\right\} \tag{3.42}$$

$$\mathrm{E}_\theta = \frac{1}{2}\omega\mu_0 a R_1(a)\left\{-\frac{a^4}{r^4}\cos\chi\sin 2\theta + \left[\left(\frac{\rho h_2' + h_2}{\rho}\right)_{\rho=\alpha}\frac{\rho}{\rho h_2' + h_2}\cos 2\theta - \frac{h_1}{h_1(\alpha)}\right]\sin\chi e^{i\lambda}\right\} \tag{3.43}$$

$$\mathrm{E}_\varphi = \frac{1}{2}\omega\mu_0 a R_1(a)\left\{\left(\frac{\rho}{\rho h_2' + h_2}\right)_{\rho=\alpha}\frac{\rho h_2' + h_2}{\rho} - \frac{h_1}{h_1(\alpha)}\right\} i\sin\chi\cos\theta e^{i\lambda}. \tag{3.44}$$

In the above relations, $\rho = (\omega/c)r$, $\alpha = (\omega/c)a$, $h_1$ and $h_2$ are the spherical Bessel functions of the third kind, $h_1' = dh_1/d\rho$ and $h_2' = dh_2/d\rho$.

Near the surface of the star, at a distance much less than the wave length i.e. $r \ll c/\omega$, the above expressions reduce to

$$\mathrm{H_r} \approx R_1(a)\frac{a^3}{r^3}[\cos\chi\cos\theta + \sin\chi\sin\theta\cos\lambda] \tag{3.45}$$

$$\mathrm{H}_\theta \approx \frac{1}{2}R_1(a)\frac{a^3}{r^3}[\cos\chi\sin\theta - \sin\chi\cos\theta\cos\lambda] \tag{3.46}$$

$$H_\varphi \approx \frac{1}{2} R_1(a) \frac{a^3}{r^3} \sin\chi \sin\lambda \tag{3.47}$$

$$E_r \approx -\frac{1}{4} \omega\mu_0 a R_1(a) \frac{a^4}{r^4} [\cos\chi(3\cos 2\theta + 1) + 3\sin\chi \sin 2\theta \cos\lambda] \tag{3.48}$$

$$E_\theta \approx -\frac{1}{2} \omega\mu_0 a R_1(a) \frac{a^2}{r^2} \left[ \frac{a^2}{r^2} \cos\chi \sin 2\theta + \sin\chi \left( 1 - \frac{a^2}{r^2} \cos 2\theta \right) \cos\lambda \right] \tag{3.49}$$

$$E_\varphi \approx \frac{1}{2} \omega\mu_0 a R_1(a) \frac{a^2}{r^2} \left( 1 - \frac{a^2}{r^2} \right) \sin\chi \cos\theta \cos\lambda, \tag{3.50}$$

where **H** is perpendicular to **E**. Transforming this to the coordinates ($r, \psi, \upsilon$), we have,

$$H_r = \frac{2M}{r^3} \cos\psi, \qquad H_\psi = \frac{M}{r^3} \sin\psi, \qquad H_\upsilon = 0, \tag{3.51}$$

where

$$M = \frac{1}{2} a^3 R_1(a). \tag{3.52}$$

Therefore, near the surface of the star, the instantaneous behavior of **H** is the same as that of a stationary magnetic dipole with moment $M\mathbf{e}$.

On the other hand, at distances large compared with the wave length, $r >> c/\omega$, the components will be

$$H_r \approx \frac{\omega}{c} a^3 \frac{R_1(a)}{r^2} \sin\chi \sin\theta \sin\left[ \omega\left( \frac{r}{c} - t \right) + \varphi \right], \tag{3.53}$$

$$H_\theta \approx \frac{1}{2} \frac{\omega^2}{c^2} a^3 \frac{R_1(a)}{r} \sin\chi \cos\theta \cos\left[ \omega\left( \frac{r}{c} - t \right) + \varphi \right], \tag{3.54}$$

$$H_\varphi \approx -\frac{1}{2} \frac{\omega^2}{c^2} a^3 \frac{R_1(a)}{r} \sin\chi \sin\left[ \omega\left( \frac{r}{c} - t \right) + \varphi \right], \tag{3.55}$$

$$E_r \approx 0, \tag{3.56}$$

$$E_\theta \approx -\frac{1}{2} \frac{\omega^2 \mu_0}{c} a^3 \frac{R_1(a)}{r} \sin\chi \sin\left[ \omega\left( \frac{r}{c} - t \right) + \varphi \right], \tag{3.57}$$

$$E_\varphi \approx -\frac{1}{2} \frac{\omega^2 \mu_0}{c} a^3 \frac{R_1(a)}{r} \sin\chi \cos\theta \cos\left[ \omega\left( \frac{r}{c} - t \right) + \varphi \right]. \tag{3.58}$$

Fig. 3.9 below shows the field vector distributions in the plane (λ = 0, or equivalently, λ = π), for the case χ = π/4. The magnetic lines of force of **H** are shown as curved lines while arrows represent the instantaneous **E**.

It is easy to determine the surface charge $q$ and current densities $K$ from the boundary conditions, (3.37), and the discontinuities in the surface of the star given by

$$q = \varepsilon_0 (E_r^{ext} - E_r^{\text{int}}), \tag{3.59}$$

$$K_\theta = -(H_\varphi^{ext} - H_\varphi^{\text{int}}), \tag{3.60}$$

$$K_\varphi = H_\theta^{ext} - H_\theta^{\text{int}}. \tag{3.61}$$

The most interesting quantity is the amount of rate of energy radiated by the star and detected on the Earth. In order to determine this theoretically, one has to find the *Poynting vector*, **S**, this represents the energy per unit time, per unit area, transported by the electromagnetic field [25, 26]. It can be written as

$$\mathbf{S} = \frac{1}{\mu_0}(\mathbf{E} \times \mathbf{B}). \tag{3.62}$$

Hence, the rate of energy radiated is

$$-\frac{dW}{dt} = \oint_s \mathbf{S} \cdot d\mathbf{a} = \frac{8\pi}{3}\mu_0 \omega^4 M^2 \sin^2\chi = \frac{8\pi}{3} C. \tag{3.63}$$

Of course, this formula seems valid even if the rotational velocity of the star reaches the speed of light, which is not correct as we see in the next chapter.

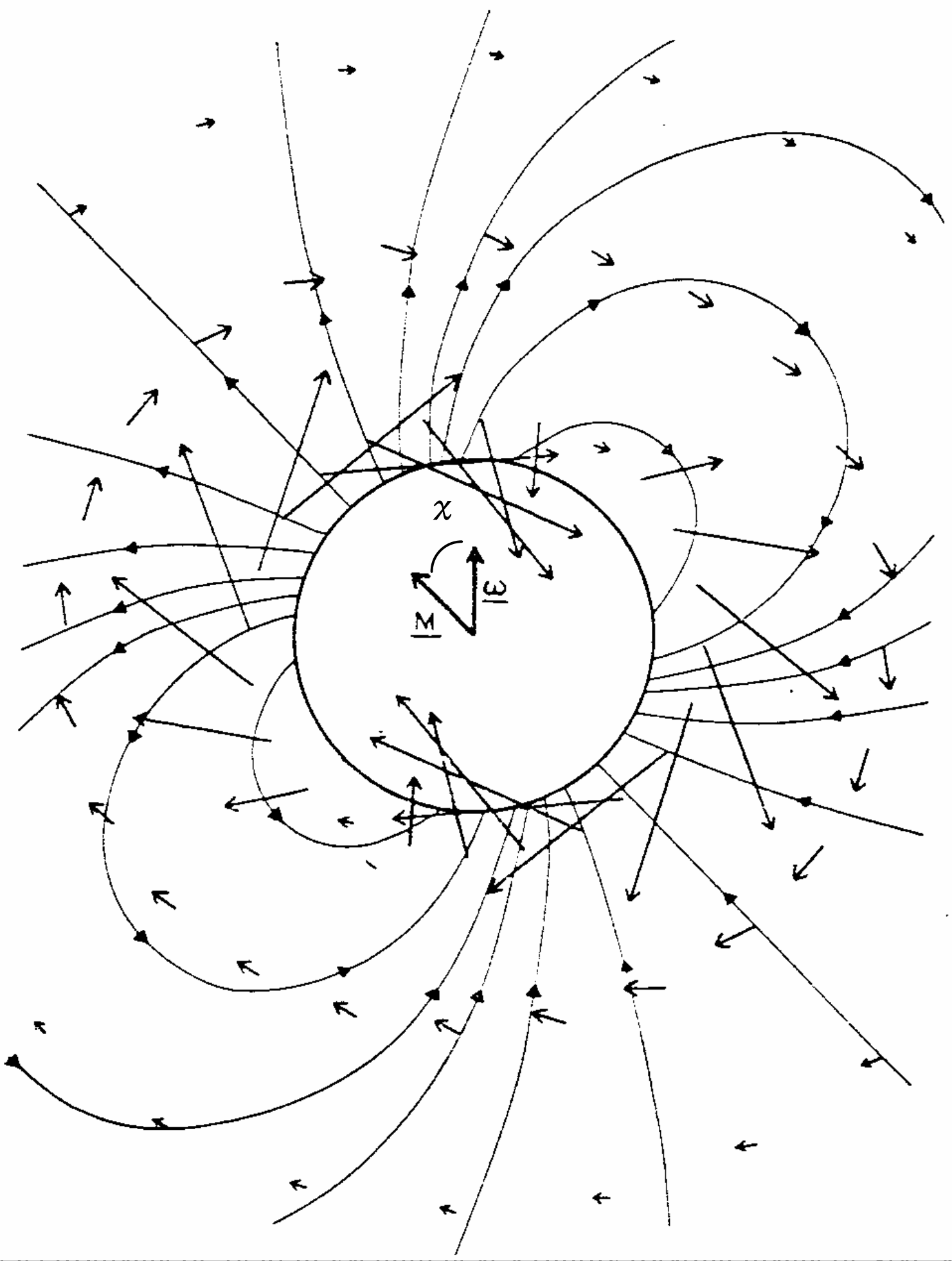

Fig.5.9. The electromagnetic field in vacuum near a rigidly rotating magnetic star, and internal magnetic field is symmetric about **M**, and external has a configuration of magnetic dipole **M**.

# CHAPTER 4

# THE RELATIVISTIC MODEL OF PULSARS (NEUTRON STARS)

The importance of the magnetic field has been very obvious in rotating stars, such as white dwarfs and neutron stars [32]. Similarly, the study of electromagnetic radiation received from pulsars has been very important since the discovery of pulsars (detection of NS). In fact, the only way to understand the nature of pulsars is through the information obtained from their electromagnetic radiation detected on the Earth. In the last chapter we introduced NSs and their characteristics. At the end of the last chapter, we discussed the classical view of electromagnetic radiation from a NS, based on Deustch's work. In that model, a pulsar (NS) is considered as a slowly rotating magnetized star. Also, the rate of change of radiated energy does not diverge when the rotational velocity reaches the speed of light.

On the other hand, using the theory of relativity leads to a different result at high speeds. This chapter introduces the special relativistic models that may fit the observed radiation. The radiation from a fast rotating magnetic dipole [27] will be discussed first. The classical result of Deutsch will be recovered as the low speed limit. The exact solution of the Maxwell equations is derived for the radiation generated by a magnetic dipole rotating with arbitrary high speed in this model. After that, the radiation field of an arbitrarily fast rotating magnetic star is discussed by using the 4-vector formalism [28]. The comparison between these two models is presented next. Then we discuss the electromagnetic radiation spectrum, due to a charge moving in a circular orbit in a Schwarzchild geometry [31]. The extension of this method to treat the radiation field and energy rate of an arbitrarily oblique fast rotating magnetic dipole [32] will be discussed later.

The above pulsar models neglect the effect of gravitational mass for simplicity. Incorporating the effect of the gravitational mass in the electromagnetic radiation will be the subject of the next chapter. That is the main purpose of this thesis.

# 4.1. THE ELECTROMAGNETIC RADIATION FROM A RELATIVISTIC ROTATING MAGNETIC DIPOLE

Since the discovery of a millisecond pulsar in 1983, the idea of a fast rotating NS has been the only candidate model to explain such a pulsar. As mentioned in §3.3.1.1, the main problem with Deutsch's work is the transformation from the static frame to the rotating star. He used the Galilean, instead of the Lorentz, transformation matrix [25, 26]

$$\Lambda = \begin{pmatrix} \gamma & -\gamma\beta & 0 & 0 \\ -\gamma\beta & \gamma & 0 & 0 \\ 0 & 0 & 1 & 0 \\ 0 & 0 & 0 & 1 \end{pmatrix}, \tag{4.1}$$

where $\gamma = 1/\sqrt{1-\beta^2}$ , and $\beta = \upsilon/c$. However, for high enough velocity, $\upsilon = |\boldsymbol{\omega} \times \mathbf{r}| \sim c$, the components of **H** are found from the Maxwell tensor **F** using the Lorentz transformation. Consequently, the result of Deutsch is not valid in the relativistic regime.

On the other hand, Belinsky et al, 1994 [27], treated the electromagnetic radiation of pulsars rotating at relativistic speeds. They simplified the issue by assuming that the radiation is generated by an infinitely thin magnetized rotating rod in the far field (wave zone) where the internal magnetic structure is not important. The exact relativistic formulation coincides with the non-relativistic results obtained by Deutsch in the low speed limit.

Consider an infinitely thin rod of rest length $2l_0$ and rest magnetization density $\mathbf{M}_0$ along z-axis. If the origin of the Cartesian coordinates lies at the center of the rod, then

$$M_{0x} = 0, \tag{4.2a}$$

$$M_{0y} = 0, \tag{4.2b}$$

$$M_{0z} = \frac{\mu_0}{2l_0}\delta(x)\delta(y)[\theta(z+l_0)-\theta(z-l_0)], \qquad (4.2c)$$

where $\delta(x)$ is the Dirac delta function, $\theta$ is the step function, and $\mu_0$ is the absolute value of the total magnetic moment of the rod:

$$\int \mathbf{M}_0 d^3x = \mu_0 \mathbf{k}, \qquad (4.3)$$

which is the only source of magnetization when the rod is at rest. But if it moves, then there is an induced electric field **E** and a polarization **P** inside the rod. Thus, the macroscopic Maxwell equations (3.32) for the average field in the absence of extraneous charge and current ($\rho_e = 0$, $\mathbf{J}_e = 0$) can be written in Gaussian units [25] as:

$$\nabla \cdot \mathbf{D} = 0, \qquad (4.4a)$$

$$\nabla \times \mathbf{H} = c^{-1}\,\partial \mathbf{D}/\partial t, \qquad (4.4c)$$

$$\nabla \cdot \mathbf{B} = 4\pi\, \rho_m, \qquad (4.4b)$$

$$-\nabla \times \mathbf{E} = 4\pi\, c^{-1}\, \mathbf{J}_m + c^{-1}\, \partial \mathbf{B}/\partial t, \qquad (4.4d)$$

where

$$\mathbf{D} = \mathbf{E} + 4\pi\, \mathbf{P}, \qquad (4.5a)$$

$$\mathbf{H} = \mathbf{B} - 4\pi\, \mathbf{M}, \qquad (4.5b)$$

$$\rho_m = -\, \nabla \cdot \mathbf{M}, \qquad (4.5c)$$

$$\mathbf{J}_m = \partial \mathbf{M}/\partial t - c\nabla \times \mathbf{P}. \qquad (4.5d)$$

If the rod is at rest then nothing depends on time, and so $\mathbf{E} = \mathbf{0}$, $\mathbf{P} = \mathbf{0}$, $\mathbf{J}_m = \mathbf{0}$ and $\mathbf{D} = \mathbf{0}$. Hence, from (4.2) and (4.5c) we have

$$\rho_{m0} = \frac{\mu_0}{2l_0}\{\delta(\mathbf{r}-\mathbf{r}_{n0}) - \delta(\mathbf{r}-\mathbf{r}_{s0}\}, \qquad (4.6)$$

where $\mathbf{r}_{n0} = (0, 0, l_0)$ is the position vector of the north pole of the magnet and $\mathbf{r}_{s0} = (0, 0, -l_0)$ of the south pole. The formula (4.6) can be used if the rod has an arbitrary position with a restriction that $|\ \mathbf{r}_{n0} - \mathbf{r}_{s0}\ | = 2l_0$. However, to find $\rho_m$ and $J_m$ for arbitrarily moving rod, we can use $\mu/l$ as a Lorentz invariant then (4.4-6) yield

$$\rho_m = \frac{\mu}{2l}\left\{\delta[\mathbf{r}-\mathbf{r}_{\mathrm{n}}(t)] - \delta[\mathbf{r}-\mathbf{r}_{\mathrm{s}}(t)\right\} , \tag{4.7a}$$

$$\mathbf{J}_m = \frac{\mu}{2l}\left\{\mathbf{v}_n\delta[\mathbf{r}-\mathbf{r}_{\mathrm{n}}(t)] - \mathbf{v}_s\delta[\mathbf{r}-\mathbf{r}_{\mathrm{s}}(t)\right\} , \tag{4.7b}$$

where $\mu$ and $l$ are the values for the moving rod, and $\mathbf{r}_{\mathrm{n}}$, $\mathbf{r}_{\mathrm{s}}$, $\mathbf{v}_{\mathrm{n}}$, and $\mathbf{v}_{\mathrm{s}}$ are the trajectories and the velocities of the north and south poles respectively.

In order to find the electromagnetic fields generated by the moving rod in the wave zone (which is empty space), we have to put **M** = **0** and **P** = **0**. Therefore, from (4.5a-b) the macroscopic fields **D** and **H** coincide with the microscopic values **E** and **B**. Further, the Maxwell equations (4.4) with (4.7) can be solved exactly to give

$$\mathbf{H} = -\frac{1}{c}\frac{\partial \mathbf{a}}{\partial t} - \Delta\psi , \qquad \mathbf{E} = -\nabla\times\mathbf{a}, \tag{4.8}$$

where

$$\mathbf{a} = \frac{1}{c}\int \frac{1}{|\mathbf{r}-\mathbf{r}'|}\mathbf{J}_m d^3x' , \tag{4.9}$$

$$\psi = \frac{1}{c}\int \frac{1}{|\mathbf{r}-\mathbf{r}'|}\rho_m d^3x' . \tag{4.10}$$

One also can obtain this solution from the rigid electrical dipole, which has a length $2l$ and two electrical charges at the ends. Interchanging ($e$, **H**, **E**) → ($\mu/2l$, -**E**, **H**) gives the corresponding solution to the Maxwell equations for the external fields generated by the magnetic dipole. Hence, the final form of the solution for **H** and **E** can be written down immediately by using the famous Lienard-Wiechert potentials for the moving point charge [26]. To do so we have to specify the type of motion, i.e. the trajectories $\mathbf{r}_{\mathrm{n}}(t)$ and $\mathbf{r}_{\mathrm{s}}(t)$. As mentioned in the previous chapter we are interested in oblique rotation in which the rod rotates with angular speed $\omega$ at a certain angle $\chi$ with respect to the z-axis. If the center of the rod is fixed at the origin of the coordinate system, see Fig. 4.1, then $\mathbf{r}_{\mathrm{n}}(t) = -\mathbf{r}_{\mathrm{s}}(t)$ and

$$r_{nx} = l \sin\chi \cos\omega t, \qquad r_{ny} = l \sin\chi \cos\omega t, \qquad r_{nz} = l \cos\chi. \tag{4.11}$$

The components of the magnetic field in Gaussian units, at the wave zone, i.e. $r >> l$, are

$$H_r = \frac{\mu}{2lr^2}\left\{\frac{1-\beta\sin\theta\sin\Omega_1-\beta^2\Upsilon_1-\beta^2\Delta_1}{(1-\beta\sin\theta\sin\Omega_1)^3}-\frac{1+\beta\sin\theta\sin\Omega_2-\beta^2\Upsilon_2-\beta^2\Delta_2}{(1+\beta\sin\theta\sin\Omega_2)^3}\right\}, \tag{4.12a}$$

$$H_\theta = \frac{\mu\omega^2}{2c^2r}\sin\chi\cos\theta\left\{\frac{\cos\Omega_1}{(1-\beta\sin\theta\sin\Omega_1)^3}-\frac{\cos\Omega_2}{(1+\beta\sin\theta\sin\Omega_2)^3}\right\}, \tag{4.12b}$$

$$H_\phi = \frac{\mu\omega^2}{2c^2r}\sin\chi\left\{\frac{\beta\sin\theta-\sin\Omega_1}{(1-\beta\sin\theta\sin\Omega_1)^3}-\frac{\beta\sin\theta+\sin\Omega_2}{(1+\beta\sin\theta\sin\Omega_2)^3}\right\}. \tag{4.12c}$$

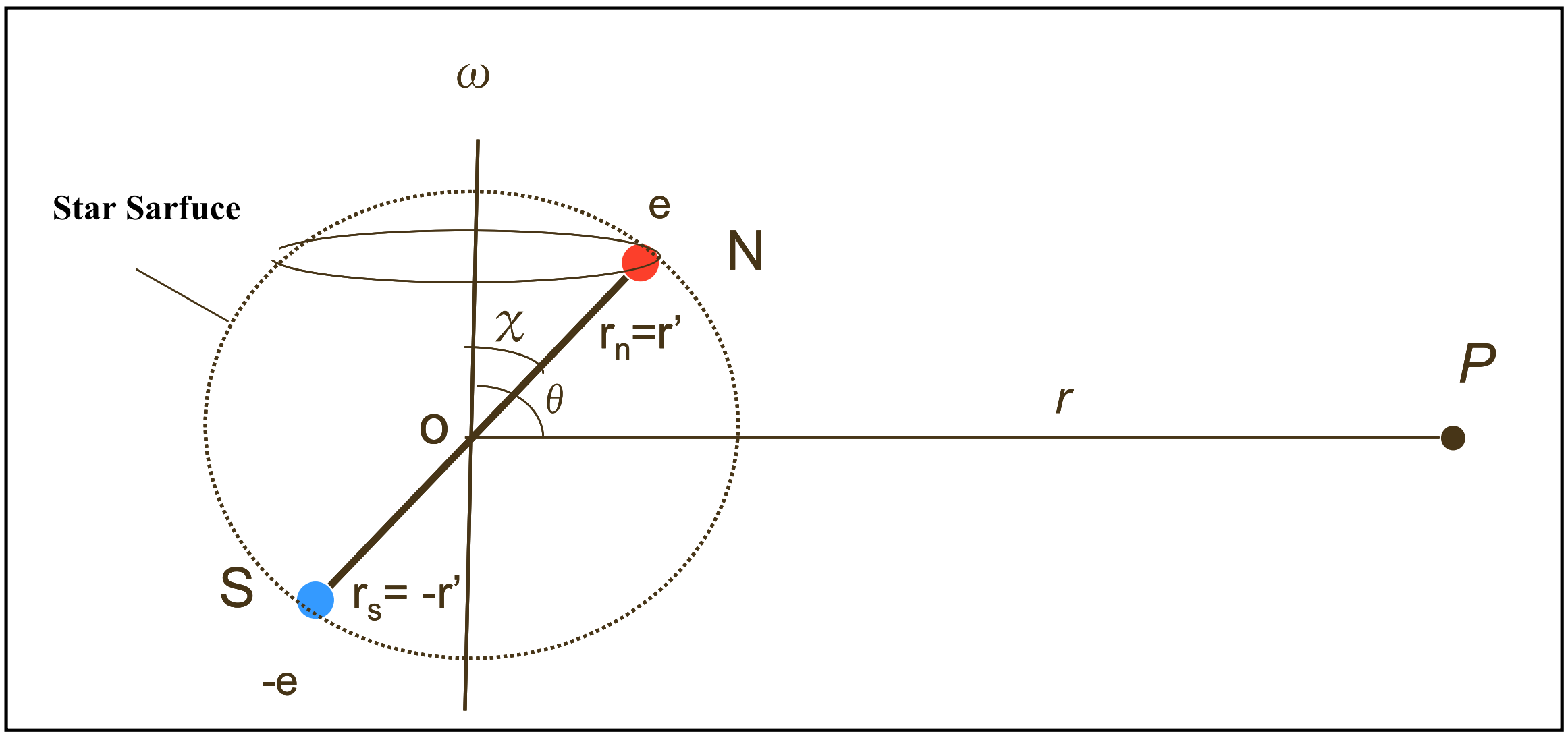

**Fig.4.1.** The oblique dipole rotator, in which the magnetic rod rotates at angular frequency ω, inclined at an angle χ with respect to rotating axis. The dotted sphere represents the neutron star and the point P is the point of observation in the wave zone.

Also, those for the electric field are

$$H_{\phi} = \frac{\mu\omega^2 l}{2c^2 r^2}\left\{\frac{\cos\chi\sin\Omega_1 + \beta\Xi_1 - \beta\cos\chi\sin\theta}{(1-\beta\sin\theta\sin\Omega_1)^3} - \frac{\cos\chi\sin\Omega_2 - \beta\Xi_2 + \beta\cos\chi\sin\theta}{(1+\beta\sin\theta\sin\Omega_2)^3}\right\}, \quad (4.13a)$$

$$E_{\theta} = H_{\phi}, \quad (4.13b)$$

$$E_{\phi} = -H_{\theta}, \quad (4.13c)$$

where

$$\Delta_{1,2} = \cot\chi\sin\theta\cos\theta\cos\Omega_{1,2}, \quad (4.14a)$$

$$\Xi_{1,2} = \sin\chi\cos\theta\cos\Omega_{1,2}, \quad (4.14b)$$

$$\Upsilon_{1,2} = \sin^2\theta\cos^2\Omega_{1,2}, \quad (4.14c)$$

$$\Omega_1 = \phi - \omega t_1; \qquad \Omega_2 = \phi - \omega t_2; \qquad \beta = l\omega c^{-1}\sin\chi, \quad (4.14d)$$

in terms of the retarded times $t_1$ and $t_2$ for the north and south pole. They are related to the time of observation, $t$, at the far point, $P$, by the relations

$$ct_1 - l\sin\chi\sin\theta\cos(\phi - \omega t_1) - l\cos\theta\cos\chi = ct - r, \quad (4.15a)$$

$$ct_2 + l\sin\chi\sin\theta\cos(\phi - \omega t_2) + l\cos\theta\cos\chi = ct - r. \quad (4.15b)$$

In the non-relativistic limit, i.e. $\beta \to 0$, then (4.15) lead to $t_1 = t_2 = t - r/c$ (retarded times coincide) and from (4.12-13) we have

$$H_r = \frac{2\mu\omega}{cr^2}\sin\chi\sin\theta\sin(\phi - \omega t + \omega r/c), \quad (4.16a)$$

$$H_{\theta} = \frac{\mu\omega^2}{c^2 r}\sin\chi\cos\theta\cos(\phi - \omega t + \omega r/c), \quad (4.16b)$$

$$H_{\phi} = -\frac{\mu\omega^2}{c^2 r}\sin\chi\sin(\phi - \omega t + \omega r/c), \quad (4.16c)$$

and for electric field

$$E_r = 0, \quad (4.17a)$$

$$E_{\theta} = -\frac{\mu\omega^2}{c^2 r}\sin\chi\sin(\phi - \omega t + \omega r/c), \quad (4.17b)$$

$$E_\phi = -\frac{\mu\omega^2}{c^2 r}\sin\chi\cos\theta\cos(\phi - \omega t + \omega r/c). \tag{4.17c}$$

Expressions (4.16) and (4.17) coincide with the Deutsch result (3.53-58) using the relation $\mu = a^3\, R_1(a)/2$ and units $\varepsilon_0 = \mu_0 = c^{-1}$ in Deutsch's paper. The important quantity is the total energy radiated by the magnetic dipole in the relativistic regime. It can be averaged over the period of rotation $2\pi/\omega$ through the solid angle $d\Theta$ as follows

$$\overline{I} = \frac{\omega}{2\pi}\int_t^{t+2\pi/\omega} dt \int d\Theta \frac{c}{4\pi} r^2 \mathbf{H}^2 . \tag{4.18}$$

From (4.12), $\mathbf{H}^2$ contains terms for each pole separately and terms arising from the interference between them. The former terms can be integrated easily but the latter can be obtained at some limits. Thus, (4.18) can be written as

$$\overline{I} = \frac{\mu^2\omega^4\sin^2\chi}{3c^3}\gamma^4\left[1 + \frac{1}{\gamma^4}F(\beta,\chi)\right], \tag{4.19}$$

where $F(\beta, \chi)$ represents the interference term

$$F(\beta,\chi) = \frac{3\omega}{16\pi^2}\int_t^{t+2\pi/\omega} dt \int \frac{\left\{\begin{array}{l}\beta\sin\theta(\sin\Omega_1 - \sin\Omega_2) + \cos(\Omega_1 - \Omega_2) - \\ \sin^2\theta\cos\Omega_1\cos\Omega_2 - (\beta\sin\theta)^2\end{array}\right\}}{\left\{(1-\beta\sin\theta\sin\Omega_1)^3(1+\beta\sin\theta\sin\Omega_2)^3\right\}} d\Theta . \tag{4.20}$$

There is no easy solution to (4.20). Rather than trying to find the exact solution, let us evaluate the integral in both non-relativistic and relativistic limits. In the first case $\beta \to 0$, $\gamma = 1$, (4.20) gives $F = 1$ and from (4.19)

$$\overline{I}_{nonrel} = \frac{2\mu^2\omega^4\sin^2\chi}{3c^3} = \frac{2}{3}C , \tag{4.21}$$

which again coincides with Deustch's result (3.63). In the ultra-relativistic limit, $\beta \to 1$, $\gamma$ becomes very large, and the interference term $F \to 0$. Then

$$\overline{I}_{ultrarel} = \gamma^4\frac{\mu^2\omega^4\sin^2\chi}{3c^3} = \frac{1}{2}\gamma^4\overline{I}_{nonrel} . \tag{4.22}$$

Clearly the relativistic factor, $\gamma^4$, is very effective and cannot be neglected at high velocity.

To study the spectrum of radiation emitted by the dipole, one can apply the Fourier series transformation with respect to time to the magnetic field in the wave zone. Thus, the radiation intensity averaged over the period of rotation $2\pi/\omega$ in the solid angle $d\Theta$ with the frequency $n\omega$ is

$$d\bar{I}_n = \frac{c}{2\pi}\left(\left|H_{\theta n}\right|^2 + \left|H_{\phi n}\right|^2\right) r^2 d\Theta, \tag{4.23}$$

where $H_{\theta n}$ and $H_{\phi n}$ are the Fourier coefficients of the $H_\theta$ and $H_\phi$ components of the magnetic field (4.12) corresponding to the frequency $n\omega$. The calculations show that (4.23) can be written for a dipole of length $2l$ as

$$d\bar{I}_n = \frac{\mu^2\omega^2 n^2}{4\pi l^2 c}[1-(-1)^n \cos(2n\beta \cot\chi \cos\theta)]\left[\cot^2\theta \ell_n^2(\xi_n) + \beta^2\left(\frac{d\ell_n(\xi_n)}{d\xi_n}\right)^2\right] d\Theta, \tag{4.24}$$

where $\ell_n$ is the usual Bessel function and $\xi_n = n\beta \sin\theta$. At $\theta = \pi/2$ (the equatorial plan), the radiation intensity vanishes as can be obtained from (4.24). The radiation spectrum in which $\theta = \pi/4$ and $\chi = 5°$ is shown in Fig. 4.2.

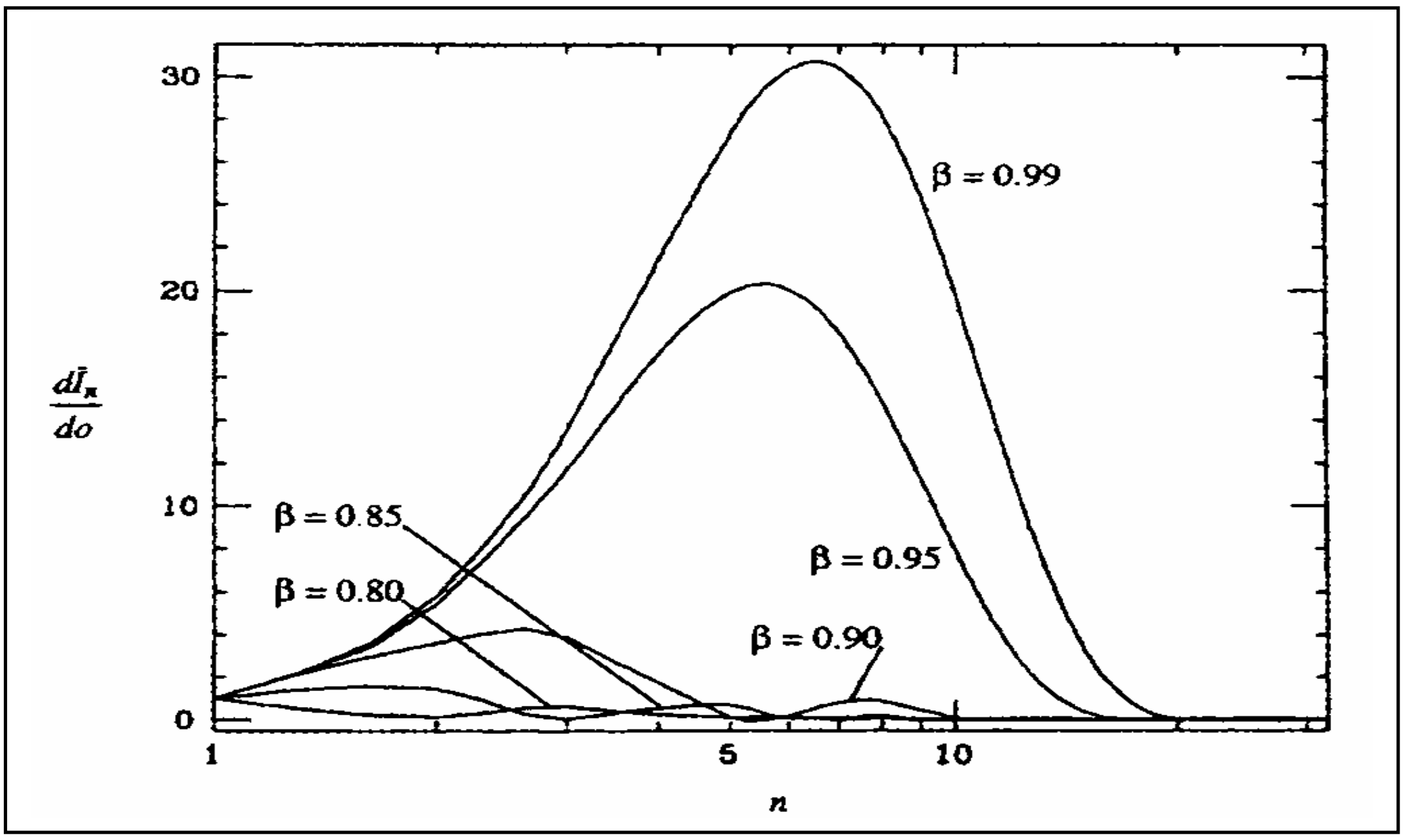

**Fig.4.2**. The spectrum of the radiation in the plane $\theta = \pi/4$ and $\chi = 5°$, for some values of $\beta$ (0.99, 0.95, 0.90, 0.85, 0.80) for the dipole rotating with an angular frequency $\omega$.

# 4.2. ELECTROMAGNETIC RADIATION FROM A RELATIVISTIC ROTATING NEUTRON STAR

In the previous model, the electromagnetic radiation was calculated for an infinitely thin fast obliquely rotating magnetic dipole. It cannot be considered as an extension of Deutsch's work to the relativistic regime, even though it coincides with its results in the non-relativistic limit. Thus, it is a different approach to the problem. One can ask the question whether both results also coincide in the relativistic limit. In order to answer this question we need to reformulate the Deutsch model relativistically. As mentioned before, the main problem of Deutsch's result is the transformation he used between the static frame and the rotating star. It yields a finite result, even if the rotational speed reaches the speed of light. For this reason, the coordinate transformation will be reformulated in this section.

## 4.2.1. THE TRANSFORMATION OF ROTATING COORDINATES

The simple Galilean transformation, used by Deutsch, cannot be valid because it is non-relativistic. Furthermore, the simple Lorentz transformation, which is used in case of high velocity, can only be used between two inertial frames. Several transformations [28] were proposed to solve this problem. The simplest one, proposed by Post [29], uses instantaneous Lorentz transformations.

Using cylindrical coordinates ($t$, $\rho$, $\psi$, $z$), Post assumes that the observer in the non-inertial frame sees the velocity vector orthogonal to the radius vector, $\mathbf{v}'\cdot r'=0$. This gives

$$dt = \gamma dt \qquad d\psi = d\psi' + \gamma\omega dt', \tag{4.25}$$

where, $\gamma^{-2} = 1 - \omega^2 R^2$ (where we use units with $c = 1$), $R$ is the radius of the circle of rotation. Hence, the inverse transformation of (4.25) is

$$dt' = \gamma^{-1}\, dt\,, \qquad d\psi' = d\psi - \omega dt\,. \tag{4.26}$$

Also, the line element is

$$ds^2 = dt'^2 - 2\omega\rho'^2\gamma dt' d\psi' - d\rho'^2 - \rho'^2 d\psi'^2 - dz'^2\,. \tag{4.27}$$

In fact, if the transformation (4.26) is applied to a rotating magnetic dipole, it yields the result of Belinsky et al [27]. For this reason, it is the transformation that will be used to reformulate Deutsch's work.

## 4.2.2. THE APPLICATION OF POST'S TRANSFORMATION TO DEUTSCH'S ANALYSIS

In this subsection we basically follow the work of DePaolis and Qadir [28]. The external fields will only be considered at a distant point. The electromagnetic field splits into the static and dynamic parts. Thus,

$$\mathbf{H} = \mathbf{H_s} + \mathbf{H_d}\,, \qquad \mathbf{E} = \mathbf{E_s} + \mathbf{E_d}\,, \tag{4.28}$$

where the static fields can be obtained from (3.37). The dynamical parts satisfy the full source-free Maxwell equations which yield the vectorial Helmholtz equation [30]

$$\nabla^2\mathbf{H_d} + \omega^2\mathbf{H_d} = 0\,, \qquad \nabla^2\mathbf{E_d} + \omega^2\mathbf{E_d} = 0\,. \tag{4.29}$$

The solutions of (4.29) are plane waves, written in spherical coordinates ($r$, $\theta$, $\phi$) for Deutsch's transformation as

$$\mathbf{H_d} = \mathbf{H_0} e^{-i(\phi-\omega t)}, \qquad \mathbf{E_d} = \mathbf{E_0} e^{-i(\phi-\omega t)}. \tag{4.30}$$

If the Post transformation (4.25) is used instead, the dynamical part will be

$$\mathbf{H_d} = \mathbf{H_0} e^{-i(\phi-\hat{\gamma}\omega t)}, \qquad \mathbf{E_d} = \mathbf{E_0} e^{-i(\phi-\hat{\gamma}\omega t)}, \tag{4.31}$$

where $\mathbf{E_0}$ and $\mathbf{H_0}$ are obtained from boundary conditions, $\hat{\gamma} = (1-\omega^2 a^2 \sin^2\eta)^{-1/2}$ and $\eta$ is the polar angle of the source point. It is not necessary for $\eta$ to be $\chi$, because the magnetic field is confined in the surface of the star, not only in the thin dipole. The average magnetic source in a static star is considered as a simple dipole inclined at an angle $\chi$. Again, solving the Maxwell equations gives [28]

$$H^r = 2\Xi\hat{\gamma}\frac{\alpha\cos K + \sin K}{r\Sigma}\sin\theta\ , \tag{4.32a}$$

$$H^\theta = \omega\Xi\hat{\gamma}^2\left\{\frac{\alpha^2(\Delta\cos K - \Lambda\sin K)}{\Delta^2+\Lambda^2} + \frac{\cos K - \alpha\sin K}{\Sigma}\right\}\cos\theta\ , \tag{4.32b}$$

$$H^\phi = -\omega\hat{\gamma}^2\left\{\frac{\alpha^2(\Delta\cos K + \Lambda\sin K)}{\Delta^2+\Lambda^2}\cos 2\theta + \frac{\alpha\cos K + \sin K}{\Sigma}\right\}, \tag{4.32c}$$

where $\Xi = M\omega\sin\chi/r$, $\alpha = \gamma\omega a$, $\Delta = 6 - 3\alpha^2$ , $\Lambda = 6\alpha - \alpha^3$, and $K = \hat{\gamma}\omega(r-a-t)+\phi$. At ultra-relativistic limit, $\beta = 1$, (4.31) are reduced to

$$H^r = \frac{2M}{ar^2}\sin\chi\sin\theta\cos K\ , \tag{4.33a}$$

$$H^\theta = -\frac{2M}{a^2 r}\sin\chi\cos\theta\cos K\ , \tag{4.33b}$$

$$H^\phi = -\frac{2M\omega^2}{r}\hat{\gamma}^2\sin\chi\sin^2\theta\cos K\ . \tag{4.33c}$$

The electric field components can be obtained by solving Maxwell's equations. The final solutions for the electromagnetic field (**H** and **E**) can be written as

$$\mathbf{H} = \boldsymbol{e}_r H^r(r,\theta,\lambda) + \boldsymbol{e}_\theta H^\theta(r,\theta,\lambda) + \boldsymbol{e}_\phi H^\phi(r,\theta,\lambda)\ , \tag{4.34a}$$

$$\mathbf{E} = \boldsymbol{e}_r E^r(r,\theta,\lambda) + \boldsymbol{e}_\theta E^\theta(r,\theta,\lambda) + \boldsymbol{e}_\phi E^\phi(r,\theta,\lambda)\ , \tag{4.34b}$$

where $\lambda = \phi - \hat{\gamma}\omega t$ (Post transformation).

The average power radiated can be obtained from the Poynting vector, (3.62), by integrating it over one period (4.18). If the source of the magnetic field is a rigid dipole, then $\hat{\gamma}$ becomes γ and the radiated power will be

$$\tilde{I} = \frac{8}{5}\gamma^4 \tilde{I}_{non-rel}, \tag{4.35}$$

where $\tilde{I}_{non-rel}$ is defined by (3.63). But this is inconsistent with the assumption of distribution of the magnetic field over the whole surface of the star. Thus $\gamma^4$ in (4.35) must be replaced by

$$\overline{\gamma^4} = \frac{1}{4\pi}\oiint_{\Theta} \hat{\gamma}^4 d\Theta = \int_0^{\pi/2} \hat{\gamma}^4 \sin\eta d\eta, \tag{4.36}$$

where Θ is the solid angle. Hence, (4.21) would have to be modified by the factor $8\overline{\gamma^4}/5$. The value of $\overline{\gamma^4}$ depends on $\chi$. Therefore, to compare between the result of this model and that obtained by the dipole model, we have to specify $\chi$. This will be done in the next subsection.

## 4.2.3. COMPARISON OF THE RELATIVISTIC DEUTSCH AND THE DIPOLE MODELS

In order to compare the relativistic Deutsch model, obtained in the last subsection, with the dipole model we have to evaluate (4.37) at different values of $\chi$. The physical difference between the two models is due to the distribution of magnetic sources in each model. In the former they have different strengths rotating at different speeds distributed over the surface of the star. In the latter, two sources of equal strength rotate at the same speeds in opposite directions. To simplify (4.37) we assume that the strengths of all sources in the surfaces are the same. Thus, the final result of (4.37) can be reduced to

$$\overline{\gamma^4} = \frac{\oint_\Theta \hat{\gamma}^4 \cos^2(\eta - \chi) d\Theta}{\oint_\Theta \cos^2(\eta - \chi) d\Theta}, \tag{4.37}$$

where, Θ is the solid angle. It is not easy to find the general form of 94.37), hence we will evaluate this integration at three values of $\chi$: π/2, π/4, 0.

For the first case, we have

$$\overline{\gamma^4}(\pi/2) = \frac{3\gamma_1^2}{2a^2\omega^3} + \frac{3}{4a^3\omega^3}\ln\left|\frac{1-a\omega}{1+a\omega}\right|, \tag{4.38}$$

where $\gamma_1 = (1 - a^2\omega^2)^{-1/2}$ which is equal to $\gamma$ in this case. Therefore, the radiation from the dipole (4.35) is much more than that obtained in this model. For $\chi$ = π/4, (4.37) is not easy to find but the leading term is

$$\overline{\gamma^4}(\pi/4) = \frac{3\gamma_1^3}{4\sqrt{2}a\omega}\tan^{-1}(a\omega\gamma_1). \tag{4.39}$$

In the ultra-relativistic limit, $a\omega \rightarrow 1$, we have

$$\overline{\gamma^4}(\pi/4) \approx \frac{3\pi}{8\sqrt{2}}\gamma_1^3 = \frac{\gamma^4}{262}. \tag{4.40}$$

Hence, the radiation factor is more in the relativistic Deutsch model than for the dipole. Finally, we consider $\chi \to 0$. (We cannot take $\chi = 0$, as $\sin\chi = 0$ in that case, and there is no radiation.) In this case we have

$$\overline{\gamma^4}(0) = \frac{3\gamma_1}{2a^3\omega^3}\tan^{-1}(a\omega\gamma_1) - \frac{3}{2a^4\omega^4}\left(\frac{\gamma_1^2}{1+a^2\omega^2\gamma_1^2}\right), \tag{4.41}$$

where the leading term is $3\pi\gamma_1/4$ which is less than that obtained by the dipole model. Thus, the total averaged power radiated in this model is given by

$$\overline{I}_{rel-Deutsch} = \frac{64\pi}{15}\overline{\gamma^4}\mu^2\omega^4\sin^2\chi = \frac{64\pi}{15}\overline{\gamma^4}C. \tag{4.42}$$

Comparing (3.63), (4.22), and (4.42) gives a further numerical factor of 16/5. The difference between these two models in the ultra-relativistic limit is relevant [29] for the periods and the magnetic field proposed by Usov [43]. The magnetic field required may be reduced by an order of magnitude. The maximum radiation of the Deutsch model does not occur at $\chi = \pi/2$, i.e. the axis of magnetic field in the equatorial plane. This happens, because the radiation from rotating magnetic objects is proportional to the number of open magnetic lines. Magnetic open lines are those lines which extent so far out that their speed, $r\omega$, exceeds the speed of light somewhere. Therefore, the polar angles $\theta$ at which they are open satisfy

$$\theta_{open} < \cos^{-1}\left[\frac{\gamma - a^2\omega^2\cos\chi}{1+a^2\omega^2\cos^2\chi}\right], \tag{4.43}$$

see Fig. 4.3 below. We know that the number of open lines and the area of radiation decrease as $\chi$ increases toward $\pi/2$ whereas the intensity of lines increases. Therefore, the maximum radiation power, which is directly proportional to the intensity and inversely to the radiation area, will occur at an angle somewhere between 0 and $\pi/2$.

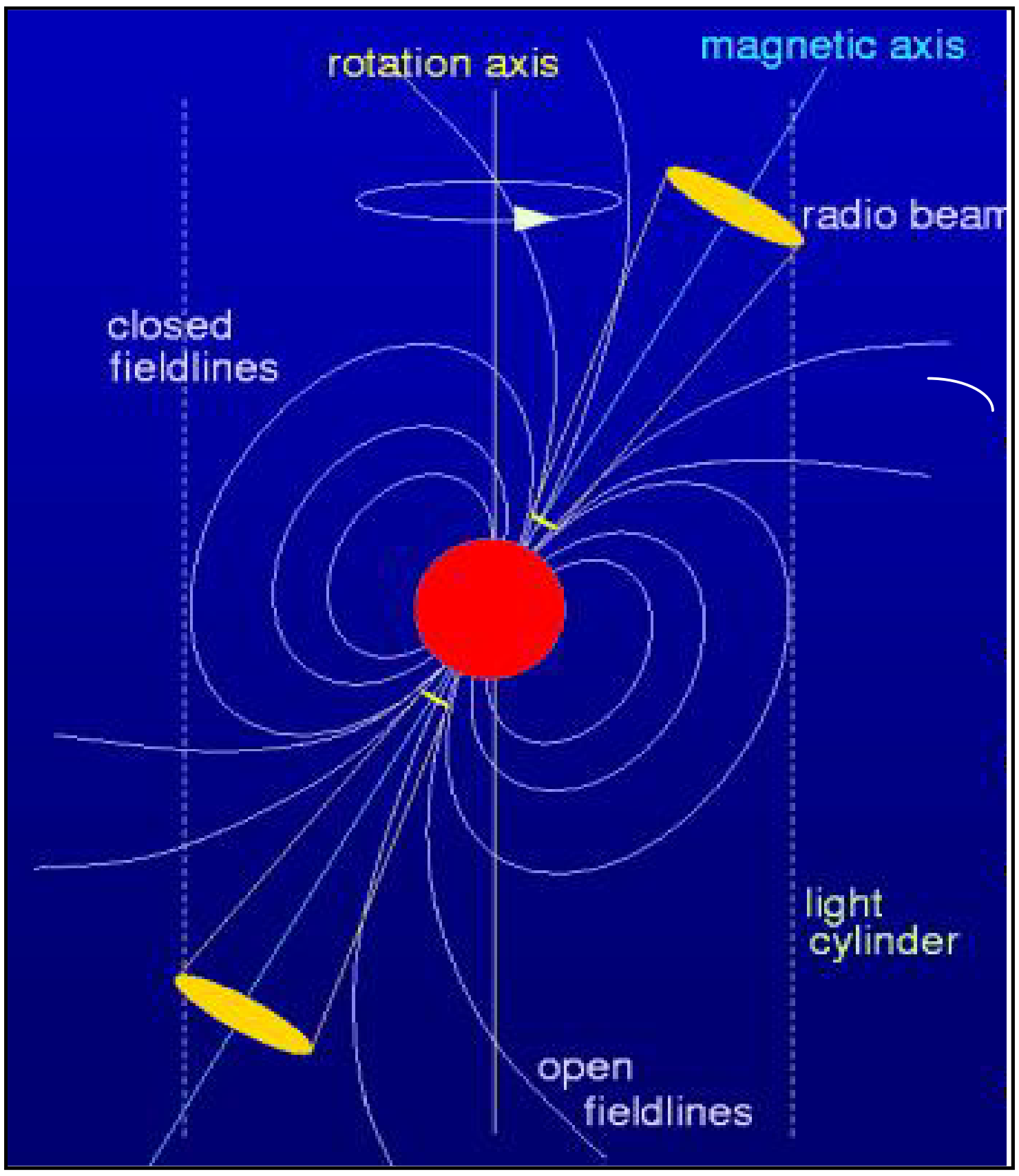

**Fig.4.3.** Oblique rotator. At the surface of the light cylinder, the speed of the field lines exceeds the speed of light. Hence they are open.

# 4.3. THE MULTIPOLE EXPANSION OF THE ELECTROMAGNETIC RADIATION FIELD OF RELATIVISTIC ROTATING CHARGES

In the two previous sections we discussed relativistic models of NSs without incorporating their mass. To incorporate the effect of the mass one has to use general relativity. In other words, one has to take into account the effect of the background geometry on the electromagnetic radiation spectrum of the star. Since: "*space acts on matter, telling it how to move and matter reacts back on space, telling it how to curve*" [23], there should be a modification of the electromagnetic radiation, due to the mass of the star. In this subsection, the radiation spectrum emitted by a single charge moving in a circular orbit in the Schwarzschild geometry [31] will be discussed first. After that, the same method will be followed to discuss the spectrum from an oblique rotating electric (magnetic) dipole [32]. In both cases, the mass of the NS will be neglected. The next step, to incorporate the mass of the NS, will be discussed in the next chapter.

## 4.3.1. THE RADIATION SPECTRUM EMITTED BY A SINGLE CHARGE MOVING IN A CIRCULAR ORBIT

In order to simplify the problem, we take the curvature of spacetime to be caused by the gravitational field of a static, spherical symmetry body. Thus, it can be described by the Schwarzschild metric (3.13), which can be written in spherical coordinates as

$$ds^2 = -\left(1-2GM/rc^2\right)dt^2 + \left(1-2GM/rc^2\right)^{-1}dr^2 + r^2(d\theta^2 + \sin^2\theta d\phi^2). \quad (4.44)$$

Using gravitational units, $G = c = 1$, the metric tensor is

$$g_{ik} = \begin{bmatrix} -\left(1-2M/r\right) & 0 & 0 & 0 \\ 0 & \left(1-2M/r\right)^{-1} & 0 & 0 \\ 0 & 0 & r^2 & 0 \\ 0 & 0 & 0 & r^2\sin^2\theta \end{bmatrix}. \tag{4.45}$$

The Maxwell equations can be written [25, 26] in tensor form as

$$\frac{\partial F^{\mu\nu}}{\partial x_\mu} = 4\pi J^\nu, \qquad \frac{\partial \Re^{\mu\nu}}{\partial x_\mu} = 0, \tag{4.46}$$

where $F^{\mu\nu}$, $\Re^{\mu\nu}$ and $J^\mu$ are respectively the Maxwell tensor, the dual tensor, and the current density 4-vector, $\mu$ and $v$ go from 0 to 3. The first two can be written as

$$F^{\mu\nu} = \begin{bmatrix} 0 & -E_x & -E_y & -E_z \\ E_x & 0 & -B_z & B_y \\ E_y & B_z & 0 & -B_x \\ E_z & -B_y & B_x & 0 \end{bmatrix}, \text{ and } \; F_{\mu\nu} = \begin{bmatrix} 0 & E_x & E_y & E_z \\ -E_x & 0 & -B_z & B_y \\ -E_y & B_z & 0 & -B_x \\ -E_z & -B_y & B_x & 0 \end{bmatrix}, \tag{4.47}$$

$$\Re^{\mu\nu} = \frac{1}{2} \in^{\mu\nu\alpha\beta} F_{\alpha\beta} = \begin{bmatrix} 0 & -B_x & -B_y & -B_z \\ B_x & 0 & E_z & -E_y \\ B_y & -E_z & 0 & E_x \\ B_z & E_y & -E_x & 0 \end{bmatrix}. \tag{4.48}$$

Also, the relation between the 4-vector potential $A^u$ and the field tensor is

$$F^{\mu\nu} = \left( \frac{\partial A^\nu}{\partial x^\mu} - \frac{\partial A^\mu}{\partial x^\nu} \right). \tag{4.49}$$

The 4-vector potential $A^u$ can be related to $J^\mu$ as

$$\Box A^\mu = 4\pi J^\mu, \tag{4.50}$$

where □ is the d'Alembertian operator, which [26] can be written as,

$$\Box = \frac{\partial}{\partial x^\nu} \frac{\partial}{\partial x_\nu} = \nabla^2 - \frac{\partial^2}{\partial t^2}. \tag{4.51}$$

Furthermore, one can derive (see [25] § 5.4, 5.5 and 16.1) the general form for the multipole expansion of the magnetic vector potential, **A**(**x**), in spherical coordinates, by substituting the well known expansion

$$\frac{1}{|\mathbf{x}-\mathbf{x}'|} = 4\pi \sum_{l=0}^{\infty} \sum_{m=-l}^{l} \frac{1}{2l+1} \frac{r_<^l}{r_>^{l+1}} Y_{lm}^*(\theta', \phi') Y_{lm}(\theta, \phi), \tag{4.52}$$

into

$$\mathbf{A}(\mathbf{x}) = \int \frac{\mathbf{J}(\mathbf{x}')}{|\mathbf{x}-\mathbf{x}'|} d^3x', \tag{4.53}$$

where $\mathbf{x}'$ is the position vector of the charge. Thus, the general result at a time $t$ will be

$$\mathbf{A}(\mathbf{x},t) = \sum_{lm} [\mathbf{A}_{lm}^{(1)}(r,t) + \mathbf{A}_{lm}^{(2)}(r,t)] Y_{lm}(\theta,\phi), \tag{4.54}$$

where $\mathbf{A}_{lm}^{(1)}(r,t)$ and $\mathbf{A}_{lm}^{(2)}(r,t)$ are radial and time-dependent functions. Hence, by using (4.54) and (4.50) we can follow the same steps as Haxton and Ruffini [31] to obtain the multipole expansion for the electromagnetic radiation emitted and absorbed by black holes (BHs). The first step is to expand $A_\mu$ and $J_\mu$ in vector harmonics:

$$A_\mu(r,\theta,\phi,t) = \sum_{lm} (-1)^{l+1} \begin{bmatrix} 0 \\ 0 \\ \dfrac{a_{lm}(r,t)}{\sin\theta} \dfrac{\partial Y_{lm}(\theta,\phi)}{\partial \phi} \\ -a_{lm}(r,t)\sin\theta \dfrac{\partial Y_{lm}(\theta,\phi)}{\partial\theta} \end{bmatrix} + (-1)^l \begin{bmatrix} f_{lm}(r,t) Y_{lm}(\theta,\phi) \\ h_{lm}(r,t) Y_{lm}(\theta,\phi) \\ k_{lm}(r,t) \dfrac{\partial Y_{lm}(\theta,\phi)}{\partial\theta} \\ k_{lm}(r,t) \dfrac{\partial Y_{lm}(\theta,\phi)}{\partial\phi} \end{bmatrix}, \tag{4.55}$$

$$4\pi J_\mu(r,\theta,\phi,t) = \sum_{lm} (-1)^{l+1} \begin{bmatrix} 0 \\ 0 \\ \dfrac{\alpha_{lm}(r,t)}{\sin\theta} \dfrac{\partial Y_{lm}(\theta,\phi)}{\partial \phi} \\ -\alpha_{lm}(r,t)\sin\theta \dfrac{\partial Y_{lm}(\theta,\phi)}{\partial\theta} \end{bmatrix} + (-1)^l \begin{bmatrix} \psi_{lm}(r,t) Y_{lm}(\theta,\phi) \\ \eta_{lm}(r,t) Y_{lm}(\theta,\phi) \\ \xi_{lm}(r,t) \dfrac{\partial Y_{lm}(\theta,\phi)}{\partial\theta} \\ \xi_{lm}(r,t) \dfrac{\partial Y_{lm}(\theta,\phi)}{\partial\phi} \end{bmatrix}. \tag{4.56}$$

Then from (4.45) and (4.46) we have

$$\partial(g^{rr}\frac{\partial a_{lm}}{\partial i})/\partial i - g_{rr}\frac{\partial^2 a_{lm}}{\partial t^2} - \frac{l(l+1)a_{lm}}{r^2} = -\alpha_{lm}, \tag{4.57.a}$$

$$\partial(g^{rr}\frac{\partial a_{lm}}{\partial i})/\partial i - g_{rr}\frac{\partial^2 b_{lm}}{\partial t^2} - \frac{l(l+1)b_{lm}}{r^2} = \frac{1}{l(l+1)}(\partial(r^2\psi_{lm})/\partial r - r^2\partial\eta_{lm}/\partial t), \tag{4.57.b}$$

where $b_{lm} = r^2(\partial h_{lm}/\partial 0 - \partial f_{lm}/\partial 0)(l(l+1))^{-1}$. Since the charge, $q$, is confined in an orbit with instantaneous position vector $\mathbf{x}'$, one can use the usual expression for the current density:

$$J^{\mu} = \frac{q}{(-g)^{1/2}}\delta^3(\mathbf{x}-\mathbf{x}')\frac{d\mathbf{x}'}{dt}. \tag{4.59}$$

For a circular orbit of radius $R_0$ we obtain

$$\psi_{lm} = \frac{4\pi q}{r^2 g^{00}}\delta(r-R_0)Y^*_{lm}(\frac{\pi}{2},\omega_0 t), \tag{4.60}$$

$$\eta_{lm} = 0, \tag{4.61}$$

$$\alpha_{lm} = \frac{-4\pi q}{l(l+1)}\omega_0\delta(r-R_0)\left(\frac{\partial Y^*_{lm}(\theta,\omega_0 t)}{\partial\theta}\right)\Bigg|_{\theta=\pi/2}, \tag{4.62}$$

where $\omega_0$ is the orbit frequency. In order to eliminate the time, $t$, we can find the Fourier expansion of (4.57) about the frequency $\omega_0$ to yield

$$\frac{d^2 a_{lm}}{dr^{*2}} + (n^2\omega_0{}^2 - V_1^{\text{eff}})a_{lm} = g^{rr}\frac{4\pi q\omega_0}{l(l+1)}\frac{\partial Y^*_{lm}(\pi/2,\omega_0 t)}{\partial\theta}\delta(n-m)\delta(r-R_0), \tag{4.63a}$$

$$\frac{d^2 b_{lm}}{dr^{*2}} + (n^2\omega_0{}^2 - V_1^{\text{eff}})b_{lm} = g^{rr}\frac{4\pi q}{l(l+1)}Y^*_{lm}(\pi/2,\omega_0 t)\delta(n-m)\frac{d}{dr}\left(\frac{\delta(r-R_0)}{g^{00}}\right), \tag{4.63b}$$

where $dr/dr^* = g^{rr}$, and $V_1^{\text{eff}} = g^{rr}[l(l+1)/r^2]$. From (4.45), $g^{rr} = (1\text{-}2M/r)$.

One way to solve (4.63) is by determining the Green's function $G(r^*, r^{*\prime}, n\omega_0)$, that satisfies

$$\left(\frac{d^2}{dr^{*2}} + (n^2\omega_0{}^2 - V_1^{\text{eff}})\right)G_l(r^*, r^{*\prime}, n\omega_0) = \delta(r^* - r^{*\prime}). \tag{4.64}$$

If we assume that the solutions of (4.64) are purely emitted waves as $r^* \to +\infty$, and purely absorbed waves as $r^* \to -\infty$ then

$$G_l(r^*, r^{*\prime}, \omega) = \frac{\upsilon(r^*, \omega) u(r^{*\prime}, \omega)}{W(\omega)}, \qquad r^* < r^{*\prime} \tag{4.65a}$$

$$G_l(r^*, r^{*\prime}, \omega) = \frac{u(r^*, \omega) \upsilon(r^{*\prime}, \omega)}{W(\omega)}, \qquad r^* > r^{*\prime} \tag{4.65b}$$

where $u(r^*, \omega)$ and $\upsilon(r^*, \omega)$ are the solutions to the sourceless form of (4.63) at infinity and at the surface of the BH respectively. $W(\omega)$ is the Wronskian of $u$ and $\upsilon$, $W(u, \upsilon) = u\upsilon' - \upsilon u'$. Thus

$$a_{lm}(r^*, n, \omega) = \frac{-4\pi q \omega_0}{l(l+1)} \delta(n-m)(l(l+1) - m(m+1))^{1/2} Y^*_{lm+1}(\pi/2, \omega_0 t) G_l(r^*, R_0, \omega), \tag{4.66a}$$

$$b_{lm}(r^*, n, \omega) = \frac{-4\pi q}{l(l+1)} \delta(n-m) Y^*_{lm}(\pi/2, \omega_0 t) \frac{d}{dR_0^*}(G_l(r^*, R_0, \omega)). \tag{4.66b}$$

Therefore from the Maxwell tensor, one can find the flux of radiation through a given spherical surface centered on the star. It will be

$$\left\langle \frac{dE}{d\omega} \right\rangle = \frac{g^{rr}}{4\pi} \sum_{lm} i\omega_0 ml(l+1)\left(\frac{\partial a_{lm}}{\partial i} a^*_{lm} + b_{lm} \frac{\partial b^*_{lm}}{\partial i}\right). \tag{4.67}$$

As $r^* \to \pm\infty$, $a_{lm}$ and $b_{lm}$ will be plane waves functions. Hence, (4.67) can be written as

$$\left\langle \frac{dE}{d\omega} \right\rangle = \sum_{lm} \frac{m^2 \omega_0^2}{4\pi} \left\{ \left| R^e_{lm} \right|^2 + \left| R^o_{lm} \right|^2 \right\}, \tag{4.68}$$

where we have defined:

$$R^e_{lm} = \frac{4\pi q}{(l(l+1))^{1/2}} Y_{lm}\left(\frac{\pi}{2}, 0\right) \frac{d}{dR_0^*} G_l(r^*, R_0^*, m\omega_0), \tag{4.69a}$$

$$R^o_{lm} = \frac{4\pi q \omega_0}{(l(l+1))^{1/2}} (l(l+1) - m(m+1))^{1/2} Y_{lm+1}\left(\frac{\pi}{2}, 0\right) G_l(r^*, R_0^*, m\omega_0), \tag{4.69b}$$

where, $R^e_{lm}$ is zero for odd $l+m$ and $R^o_{lm}$ is zero for even $l+m$ [32].

In the classical limit, in which we neglect the mass of BH ($M \to 0$), from (4.45) we have $g^{rr}=1$. Therefore, (4.63) become

$$\frac{d^2 a_{lm}}{dr^{*2}} + n^2\omega_0{}^2 a_{lm} - \frac{l(l+1)}{r^2} a_{lm} = \frac{4\pi q\omega_0}{l(l+1)} \frac{\partial Y^*_{lm}(\pi/2, \omega_0 t)}{\partial\theta} \delta(n-m)\delta(r-R_0), \quad (4.70a)$$

$$\frac{d^2 b_{lm}}{dr^{*2}} + n^2\omega_0{}^2 b_{lm} - \frac{l(l+1)}{r^2} = \frac{4\pi q}{l(l+1)} Y^*_{lm}(\pi/2, \omega_0 t)\delta(n-m)\frac{d\delta(r-R_0)}{dr}. \quad (4.70b)$$

Thus, solving (4.70) yields a solution, which can be written in terms of the usual spherical Bessel functions $j_l(r)$ and $h_l^1(r)$ at $r > R_0$, as

$$a_{lm} = \frac{-4\pi iq\omega_0}{l(l+1)} \delta(n-m)\omega_0(l(l+1)-m(m+1))^{1/2} Y^*_{lm+1}(\frac{\pi}{2}, \omega_0 t) r(R_0\omega_0 j_l(n\omega_0 R_0) h_l^1(n\omega_0 r)), \quad (4.71a)$$

$$a_{lm} = \frac{4\pi inq}{l(l+1)} \delta(n-m) Y^*_{lm}(\frac{\pi}{2}, \omega_0 t) r\omega_0 h_l^1(n\omega_0 r) \left( \frac{dr' j_l(n\omega_0 r')}{dr'} \right)\Bigg|_{r'=R_0}. \quad (4.71b)$$

With the classical boundary condition, one can derive (4.68) at infinity

$$R^e_{lm} = \frac{4\pi q}{\Delta} Y_{lm}(\frac{\pi}{2}, 0)\frac{d}{dx}\left[x j_l(x)\right]_{x=x_0}, \quad (4.72a)$$

$$R^e_{lm} = \frac{4\pi q R_0\omega_0}{\Delta} \Lambda Y_{lm+1}(\frac{\pi}{2}, 0) j_l(x_0), \quad (4.72b)$$

where $\Delta = [l(l+1)]^{1/2}$, $\Lambda = [l(l+1)-m(m+1)]^{1/2}$, and $x_0 = m\omega_0 R_0$. In the next subsection we will extend this result to the oblique rotating electric dipole.

## 4.3.2. THE MULTIPOLE EXPANSION OF THE ELECTROMAGNETIC RADIATION EMITTED BY A RELATIVISTICALLY ROTATING OBLIQUE MAGNETIC DIPOLE

The results of Haxton and Ruffini (4.63-69) are general forms that have several parameters. Therefore, they can be modified to apply to several cases. For example, they can be extended to determine the electromagnetic radiation from rotating oblique magnetic dipole. To do so, in this subsection we will follow the method used by De Paolis, Ingrosso and Qadir [32].

The previous result was derived to obtain a multipole expansion of the radiation field for a single charge in a circular orbit around a Schwarzschild BH. Thus, it can easily be extended to obtain the radiation for a rotating oblique magnetic dipole, similar to that obtained by Belinsky et al. The first step is to extend (4.72) to the electric dipole rotating around its center in the plane orthogonal to its rotational axis see Fig.4.4. This is done by introducing another opposite charge totally out of phase with the first, at $\theta = \pi/2$. The electric dipole moment is $d = 2qa$, where $a$ is the radius of the orbit (the NS in the Deutsch model, $a = R_0$). From the duality of electromagnetic fields as mentioned in § 4.1, one can replace the electric dipole moment $d$ by a magnetic dipole moment $M$.

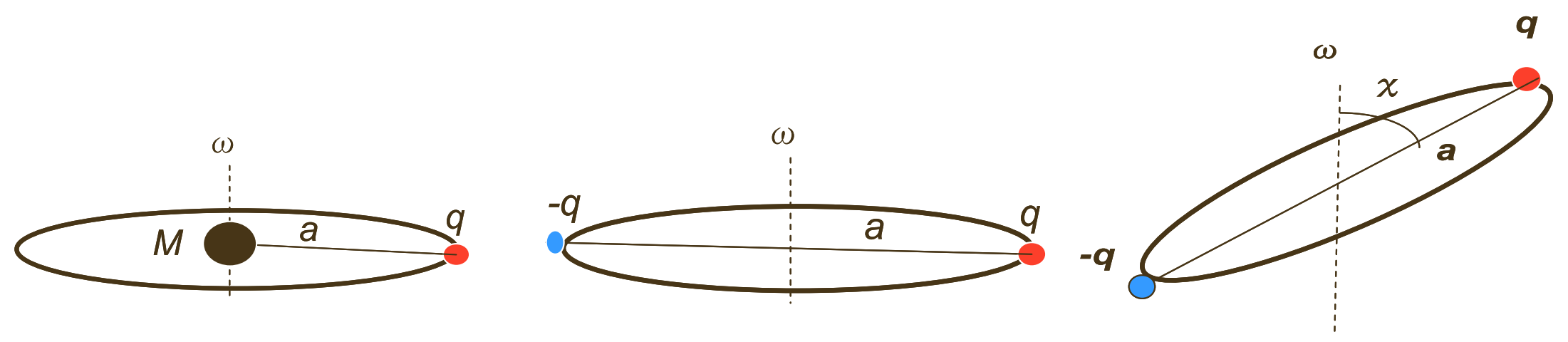

(a)
Single Charge Rotating Around BH ($M \to 0$) (Haxton-Ruffini Model)

(b)
Orthogonal Rotating Dipole

(c)
Oblique Rotating Dipole (Belinsky & De Paolis-Ingrosso-Qadir Model)

**Fig.4.4.** Three different models for a possible source of electromagnetic radiations detected from a BH and NS. Schwarzschild geometry is used in all of these models, in which the gravitational mass, $M$ is neglected.

The second step is to tilt the rotational plane through an angle $\chi$ with respect to the rotational axis. As a result, the magnetic poles will always lie at $\theta = \chi$ and $\theta = \pi - \chi$, instead of $\theta = \pi/2$. Further, $x_0 = m\omega_0 a \sin\chi$ instead of $x_0 = m\omega_0 a$ for the orthogonal rotator. Therefore, (4.72a, b) are replaced by

$$R^e_{lm} = \frac{4\pi M}{a\Delta} K_{lm} \frac{d}{dx}\left[x j_l(x)\right]_{x=x_0}, \tag{4.73a}$$

$$R^o_{lm} = \frac{4\pi M \omega_0 \sin\chi}{\Delta} \Lambda K_{lm+1} j_l(x_0), \tag{4.73b}$$

where, $K_{lm} = [Y_{lm}(\chi, 0) - Y_{lm}(\pi - \chi, \pi)]$.

In order to compare this result with that obtained by Belinsky et al, we have to convert the multipole expansion (4.68) into a generalized Fourier expansion. The summation should be over all $l \geq m$. Hence we have to find

$$\overline{E_m} = \sum_{l \geq m} \overline{E_{lm}}. \tag{4.74}$$

The radiation spectrum is obtained by dropping the summation over $l$ in (4.68). The radiation spectra for a single charge rotating in the plane orthogonal to the rotational axis are shown in Fig. 4.5, and those for a magnetic dipole with $\chi = \pi/2$, rotating with different velocities are shown in Fig. 4.6.

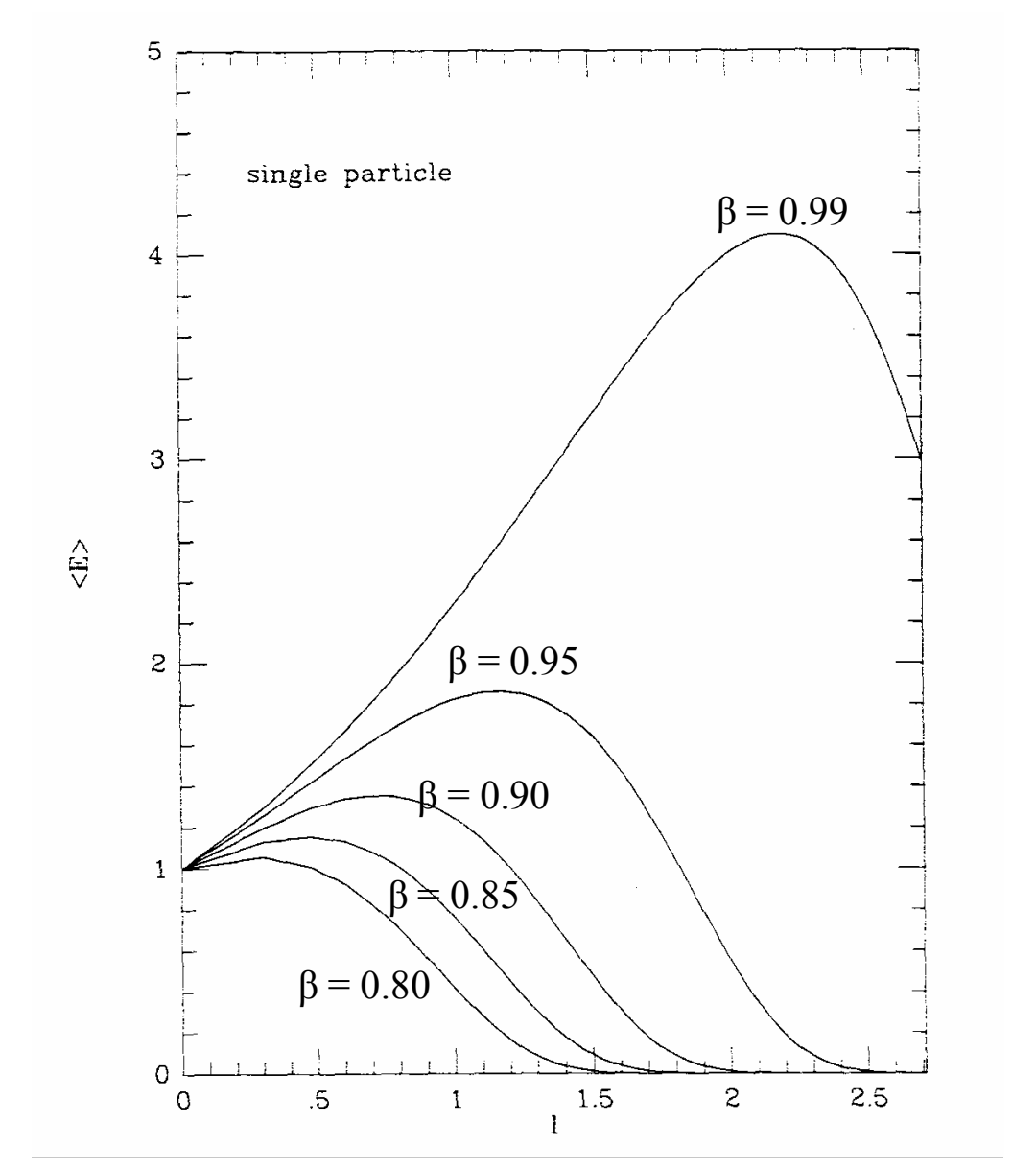

**Fig.4.5.** The radiation spectra for a charged particle rotating on circular orbit with different velocities.

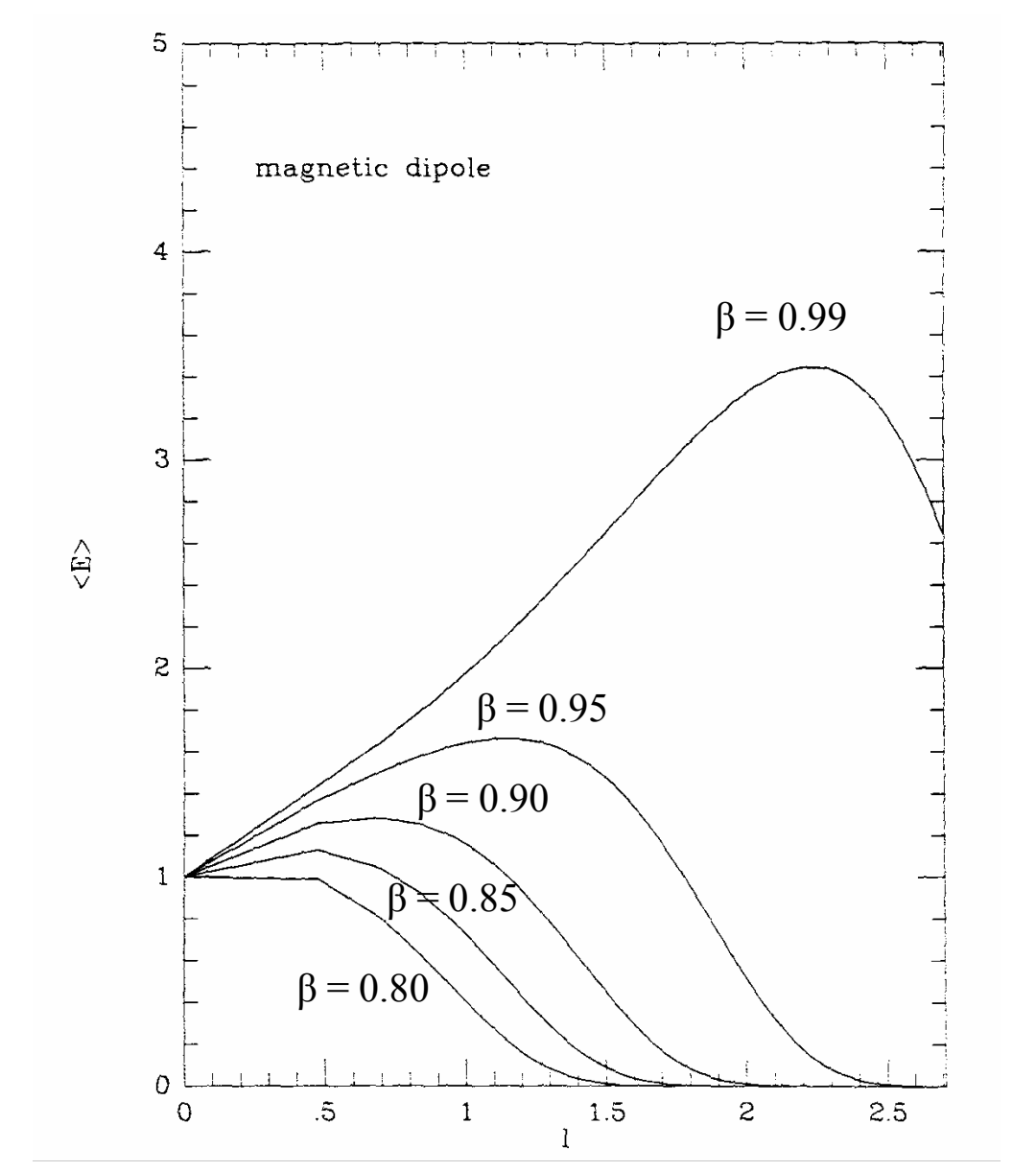

**Fig.4.6.** The radiation spectra for magnetic (electric) dipole with $\chi = \pi/2$ with different velocities.

# CHAPTER 5

# THE EFFECT OF THE GRAVITATIONAL MASS ON THE RADIATION FROM A FAST, OBLIQUE, ROTATING DIPOLE

In this chapter we will incorporate the effect of a gravitational mass for obtaining the multipole expansion of the electromagnetic radiation field of the NS. One of the models, used to do so, is the oblique rotating dipole. After having introduced the method of multipole expansion for a single orbiting charge, §4.3.1, and its extension to the oblique dipole, §4.3.2, we are in a stage to incorporate the effect of a gravitational mass on the multipole expansion of the dipole radiation. To do so, we use the same general relativistic approach in those sections but we do not set the mass of Schwarzschild equal to zero. The main purpose of this chapter is to find the power radiated by the dipole (4.68), which is expected to be relevant for application to neutron star models. This result may fit better with the detected radiation from expected sub-millisecond pulsars.

Constructing the Green's function (4.65) is the cornerstone to solve this problem. To do so, we will first derive the differential equation (4.63), whose solutions, are used to construct the Green's function. This will be done by considering the modifications introduced in §4.3.2, so as to extend the result to an electric (magnetic) dipole. This equation includes the mass term. Some of the possible methods used to solve this equation will be discussed in the second section. In the third section, two methods: numerical method using power series and asymptotic method will be used to obtain the two solutions of the equation: one around the radius of the dipole (the source of the radiation); and at infinity (our location) respectively. These two solutions will be used to construct the Green's function which is in turn used to draw the energy spectra. These spectra may then be compared with those obtained without incorporating the mass in §4.3.2, if possible. In the final section, the results will be discussed, and the conclusions will be stated.

# 5.1. THE GREEN'S FUNCTION WITH GRAVITATIONAL MASS

The general form of radiated energy emitted by a single charge moving in a circular orbit around a Schwarzchild mass [31], §4.3.1, can be obtained from (4.68-69) as

$$\left\langle \frac{dE}{d\omega} \right\rangle = \sum_{lm} \frac{m^2\omega_0^2}{4\pi} \left\{ \left| R^e_{lm} \right|^2 + \left| R^o_{lm} \right|^2 \right\}. \tag{5.1}$$

By extending these equations to a mass-less rotating oblique magnetic dipole [32], §4.3.2, the Green's function is, $G = x\, j_l(x)$, where $j_l(x)$ is the spherical Bessel function. Thus from Eqs (4.69), one can write the components of (5.1)

$$R^e_{lm} = \frac{2\pi M}{a\Delta} K_{lm} \left[ \frac{d}{dx} x j_l(x) \right]_{x=x_0}, \tag{5.2a}$$

$$R^o_{lm} = \frac{2\pi M \omega_0 \sin\chi}{\Delta} \Lambda K_{lm+1} j_l(x_0), \tag{5.2b}$$

where, $x_0 = m\omega_0\, a \sin\chi$, $a$ is the dipole radius, $K_{lm} = [Y_{lm}(\chi,0) - Y_{lm}(\pi-\chi,\pi)]$, $\Delta = [l(l+1)]^{1/2}$, and $\Lambda = [l(l+1) - m(m+1)]^{1/2}$. For even $l$, $Y_{lm}(\chi,0) = Y_{lm}(\pi-\chi,\pi)$. Hence, $K_{lm}$ is zero. Thus, the energy spectra will not include even $l$. Retaining the mass term of (4.45) in (4.63) i.e. $g^{rr} = 1 - (2GM)/rc^2$ and defining a dimensionless variable $x$ as

$$x(r, M) = rc^2/(2GM) - 1 = r/r_s - 1, \tag{5.3}$$

where, $M$ is the gravitational mass of the dipole (NS) and $r_s = 2GM/c^2$ is Schwarzchild radius ($r_s = 2M$, at $G = c = 1$). Thus,

$$dx = \frac{dr}{r_s} \Rightarrow \frac{d}{dr} = \frac{1}{r_s}\frac{d}{dx} \Rightarrow \frac{d^2}{dr^2} = \frac{1}{r_s^2}\frac{d^2}{dx^2}, \tag{5.4}$$

and from (4.45), we have

$$\frac{d}{dr^*} = g^{rr} = \left(1 - \frac{r_s}{r}\right)\frac{d}{dr} = \left(1 - \frac{r_s}{r}\right)\frac{1}{r_s}\frac{d}{dx}. \tag{5.5}$$

But

$$r = r_s(x+1), \qquad \Rightarrow \qquad 1-\frac{r_s}{r} = 1-\frac{1}{x+1} = \frac{x}{x+1}. \tag{5.6}$$

Substituting of (5.4-6) into (4.69), and using the extension to the oblique dipole, §4.3.2, implies that (5.2) can be rewritten as

$$R^e_{lm} = \frac{4\pi q}{\Delta}\frac{1}{r_s}K_{lm}\left[\frac{x}{1+x}\frac{d}{dx}G_l(x)\right]_{x=x_0}, \tag{5.7a}$$

$$R^o_{lm} = \frac{4\pi q\omega_0}{\Delta}\Lambda K_{lm+1}G_l(x_0), \tag{5.7b}$$

where, $x_0$ is given by (5.3) at the position of the pole in the dipole, i.e. $x_0 = a \sin\chi/r_s - 1$. Since the magnetic dipole definition includes the length of the dipole, $2a$, (5.7) becomes.

$$R^e_{lm} = \frac{2\pi M}{a\Delta}\frac{1}{r_s}K_{lm}\left[\frac{x}{1+x}\frac{d}{dx}G_l(x)\right]_{x=x_0}, \tag{5.8a}$$

$$R^o_{lm} = \frac{2\pi M\omega_0}{a\Delta}\Lambda K_{lm+1}G_l(x_0). \tag{5.8b}$$

When tabulating and plotting $dE/d\omega$, (5.1), for a given $l$, we shall denote it by $E_l$ for convenience. However, we will denote the total energy per frequency as $\varepsilon = \sum_{l=0}^{\infty} E_l$.

Therefore, the whole problem can be solved if we have the Green function $G_l(x)$. It is defined by (4.65) as

$$G_l(r^*) = \frac{u(R)\upsilon(r^*)}{W(u,\upsilon)}, \qquad \text{as } R \to \infty \tag{5.9}$$

where, $u$ and $\upsilon$ are the two solutions of (4.64). If $r^* \neq r$, and $G_l(r^*)$ is replaced by $z(r^*)$, then (4.64) can be written as

$$\left(\frac{d^2}{dr^{*2}} + (\frac{n^2\omega_0{}^2}{c^2} - g^{rr}\frac{l(l+1)}{r^2})\right)z(r^*) = 0, \tag{5.10}$$

hence

$$\left(\frac{d^2}{dr^{*2}} + (\frac{n^2\omega_0{}^2}{c^2} - (1-\frac{r_s}{r})\frac{l(l+1)}{r^2})\right)z(r^*) = 0. \tag{5.11}$$

Using (5.5) and

$$\frac{d^2}{d^2r^*}=\left(1-\frac{r_s}{r}\right)\frac{d}{dr}\left(\left(1-\frac{r_s}{r}\right)\frac{d}{dr}\right)=\left(1-\frac{r_s}{r}\right)\left[\left(1-\frac{r_s}{r}\right)\frac{d^2}{d^2r}+\frac{r_s}{r^2}\frac{d}{dr}\right]. \tag{5.12}$$

Substituting (5.12) into (5.11)

$$\left(1-\frac{r_s}{r}\right)^2\frac{d^2z}{d^2r}+\frac{r_s}{r^2}\left(1-\frac{r_s}{r}\right)\frac{dz}{dr}+\left(\frac{m^2\omega_0{}^2}{c^2}-\left(1-\frac{r_s}{r}\right)\frac{l(l+1)}{r^2}\right)z(r)=0\,. \tag{5.13}$$

Using (5.6), we have

$$\left(\frac{x}{x+1}\right)^2\left(\frac{1}{r_s{}^2}\frac{d^2z}{dx^2}\right)+\frac{1}{r_s{}^2(x+1)^2}\left(\frac{x}{x+1}\right)\frac{dz}{dx}+\left(\frac{m^2\omega_0{}^2}{c^2}-\left(\frac{x}{x+1}\right)\frac{l(l+1)}{r_s{}^2(x+1)^2}\right)z(x)=0\,. \tag{5.14}$$

Multiplying (5.14) by $r^2_s(x+1)^3$, gives

$$x^2(x+1)\frac{d^2z}{dx^2}+x\frac{dz}{dx}+\left(r_s^2\,\frac{m^2\omega_0{}^2}{c^2}(x+1)^3-xl(l+1)\right)z(x)=0\,. \tag{5.15}$$

But $r_s = 2GM/c^2$. Substituting it into (5.15)

$$x^2(x+1)\frac{d^2z}{dx^2}+x\frac{dz}{dx}+\left(\lambda^2(x+1)^3-xl(l+1)\right)z(x)=0\,, \tag{5.16}$$

where, $\lambda = (2GMm\omega_0)/c^3 = 2Mm\omega_0 = m\omega_0\, r_s$ $(G = c = 1)$ , which is the term including gravitational mass. Hence, (5.16) can be written in canonical form as

$$\frac{d^2z}{dx^2}+\frac{1}{x(x+1)}\frac{dz}{dx}+\frac{\left[\lambda^2(x+1)^3-xl(l+1)\right]}{x^2(x+1)}z(x)=0\,. \tag{5.17}$$

The rest of this chapter will be devoted to solving this second order linear differential equation, in the desired range. Thus, from (5.5) $u(x)$ is the solution of (5.7) that has an infinite radius of convergence while $v(x)$ is the solution that converges around $x_0$.

# 5.2. THE POSSIBLE METHODS TO SOLVE THE DIFFERENTIAL EQUATION

It is clear that (5.13) for mass-less dipole ($r_S = 0$) becomes

$$r^2 z''(r) + [\left(\frac{m\omega}{c}\right)^2 r^2 - l(l+1)]z(r) = 0\,. \tag{5.18}$$

The solution of (5.18) is $z(r) = r\, j_l(m\omega r)$ [33], where $j_l(m\omega r)$ is the normal spherical Bessel function. It is the Green's function for the massless case [31, 32]. However, (5.16) cannot be solved analytically, but it can be solved numerically. Disregarding the interval of convergence, it can be solved in general by several methods. For example, it can be solved by using the power series method. The power series solutions can be determined for the differential equation as it is [34, 35], or for the transformed differential equation of (5.16) [35, 36]. Also since it has a regular singular point at $x = 0$, it can be solved by using the Frobenius method. Again, the power series solutions, obtained by the Frobenius method, can be determined for the differential equation as it is [34, 35], or for the transformed differential equation [35, 36] which has a regular singular point too. As we are interested in finding one solution of (5.16) at infinity, we may try to change variables [35], or to find the asymptotic limit [36] of the equation as $x \to \infty$. The last method is considered as an analytic, rather than numerical, method. All of these six methods will be presented in this section, for completeness. It will be seen that only some of them are useful for our purposes. Finally, we will approximate (5.16) to match the massive and massless cases, so as to draw the complete spectra. The appropriate ones will be used in the next section.

## 5.2.1. THE POWER SERIES METHOD

The key point in this method is to use the following substitution [34, 35]:

$$z(x) = \sum_n c_n x^n \,. \tag{5.19}$$

To avoid the singular point, $x = 0$, one can shift (6.8) by a finite value, $\alpha$, to obtain

$$z(x) = \sum_n c_n (x-\alpha)^n = \sum_n c_n y^n \,, \qquad y = (x - \alpha). \tag{5.20}$$

Hence, $x = y + \alpha$, $x' = y'$, and $x'' = y''$. The main task is to find the values of $c_n$, or the recurrence relations. To actually compute the series, we have to truncate the summation somewhere. The number of terms we need in (5.18) depends on the desired range. The above substitution (5.18) can be inserted directly in (5.16) or after performing a transformation. The results will be the same but the second method may be longer.

### 5.2.1.1. THE POWER SERIES SOLUTIONS FOR THE EQUATION

Substituting (5.19) in (5.16) gives

$$\begin{gathered}(y+\alpha+1)(y+\alpha)^2\sum_n n(n-1)c_n y^{n-2} + (y+\alpha)\sum_n nc_n y^{n-1} + \\ [\lambda^2(y+\alpha+1)^3 - (y+\alpha)l(l+1)]\sum_n c_n y^n = 0,\end{gathered} \tag{5.21}$$

which can be rewritten as

$$\begin{gathered}\sum_n \lambda^2 c_n y^{n+3} + \sum_n 3\lambda^2(\alpha+1)c_n y^{n+2} + \sum_n [n(n-1) + 3\lambda^2(\alpha+1)^2 - l(l+1)]c_n y^{n+1} + \\ \sum_n [(3\alpha+1)n(n-1) + n + \lambda^2(\alpha+1)^3 - \alpha l(l+1)]c_n y^n + \\ \sum_n [\alpha(3\alpha+2)n(n-1) + \alpha n]c_n y^{n-1} + \sum_n \alpha^2(\alpha+1)n(n-1)c_n y^{n-2} = 0.\end{gathered} \tag{5.22}$$

It is clear that (5.21) has a six term recurrence relation. It can be worked out by taking terms of the same order in $y$. To do so, let $k = n + 3$, $n + 2$, $n + 1$, $n$, $n - 1$, and $n - 2$ in the first, second, third, fourth, fifth, and sixth term of (6.21), respectively. It will become

$$\begin{aligned}
&\left\{[\lambda^2(\alpha+1)^3-\alpha l(l+1)]c_0+\alpha c_1+2\alpha^2(\alpha+1)c_2\right\}y^0+\\
&\left\{\begin{matrix}3\lambda^2(\alpha+1)^2c_0+[1+\lambda^2(\alpha+1)^3-\alpha l(l+1)]c_1+\\ 6\alpha(\alpha+1)c_2+6\alpha^2(\alpha+1)c_3\end{matrix}\right\}y+\\
&\left\{\begin{matrix}3\lambda^2(\alpha+1)c_0+[3\lambda^2(\alpha+1)^2-l(l+1)]c_1+[2(3\alpha+2)+\lambda^2(\alpha+1)^3-\\ \alpha l(l+1)]c_2+3\alpha(6\alpha+5)c_3+12\alpha^2(\alpha+1)c_4\end{matrix}\right\}y^2+ \qquad (5.23)\\
&\sum_{k=3}\left\{\begin{matrix}\lambda^2c_{k-3}+3\lambda^2(\alpha+1)c_{k-2}+[(k-1)(k-2)+3\lambda^2(\alpha+1)^2-l(l+1)]c_{k-1}+\\ [(3\alpha+1)k(k-1)+k+\lambda^2(\alpha+1)^3-\alpha l(l+1)]c_k+\\ [\alpha(3\alpha+2)(k+1)k+\alpha(k+1)]c_{k+1}+\alpha^2(\alpha+1)(k+2)(k+1)c_{k+2}\end{matrix}\right\}y^k=0.
\end{aligned}$$

Since the solutions must be independent ($W(u, v) \neq 0$) , all coefficients of $y$ in (5.22) must be identically equal to zero because $y^k \neq 0$. From the first, second, and third terms of (5.22) we have the following recurrence relations

$$c_2=-\frac{1}{2\alpha^2(\alpha+1)}\left\{[\lambda^2(\alpha+1)^3-\alpha l(l+1)]c_0+\alpha c_1\right\}, \qquad (5.24a)$$

$$c_3=-\frac{1}{6\alpha^2(\alpha+1)}\left\{\begin{matrix}[3\lambda^2(\alpha+1)^2-l(l+1)]c_0+[1+\lambda^2(\alpha+1)^3-\alpha l(l+1)]c_1+\\ 6\alpha(\alpha+1)c_2\end{matrix}\right\}, \qquad (5.24b)$$

$$c_4=-\frac{1}{12\alpha^2(\alpha-1)}\left\{\begin{matrix}3\lambda^2(\alpha+1)c_0+[3\lambda^2(\alpha+1)^2-l(l+1)]c_1+\\ [2(3\alpha+2)+\lambda^2(\alpha+1)^3-\alpha l(l+1)]c_2+\\ [3\alpha(6\alpha+5)]c_3\end{matrix}\right\}. \qquad (5.24c)$$

The general form of the recurrence relation is obtained by equating the coefficients of $y^k$ in (5.22) to zero. It is

$$c_n=-\frac{1}{\alpha^2(\alpha+1)n(n-1)}\left\{\begin{matrix}\alpha(n-1)[(3\alpha+2)(n-2)+1]c_{n-1}+\\ [(n-2)((3\alpha+1)(n-3)+1)+\lambda^2(\alpha+1)^3-\alpha l(l+1)]c_{n-2}+\\ [(n-3)(n-4)+3\lambda^2(\alpha+1)^2-l(l+1)]c_{n-3}+3\lambda^2(\alpha+1)c_{n-4}+\lambda^2c_{n-5}\end{matrix}\right\}. \qquad (5.25)$$

The power series is not meaningful unless it converges. Every convergent series has a radius of convergence $R$. The series (5.19) converges for all $x$ satisfying $|x - \alpha| < R$, where $R > 0$ [34]. The radius of convergence $R$ can be found from

$$R = \lim_{n\to\infty}\left|\frac{c_{n-1}}{c_n}\right| = \left|-\frac{\alpha^2(\alpha+1)}{\alpha(3\alpha+2)} - \frac{(3\alpha+1)}{\alpha(3\alpha+2)}\frac{c_{n-2}}{c_n} - \frac{1}{\alpha(3\alpha+2)}\frac{c_{n-3}}{c_n}\right|, \tag{5.26}$$

hence, $R$ needs to be determined numerically. Roughly speaking, $R$ goes as $\alpha/3$, for large $\alpha$. But this is not a precise value because the $c$'s may depend on $\alpha$ too. Also, it does not help because we want to find the solution at the source of radiation (the surface of NS). Thus, in some cases we may take $\alpha = ac^2/(2GM) - 1 \approx 2.62$, in gravitational units, where $a$ is the radius of the NS, which is assumed to be 8 (~12 km). The number of terms required for convergence, can be evaluated by inserting numerical values. The reasonable number of terms and the rough estimate of the radius of convergence can be obtained from Fig. 5.1. The two independent power series solutions can be obtained by taking $c_0 = 0$ and $c_1 = 1$, or $c_0 = 1$ and $c_1 = 0$.

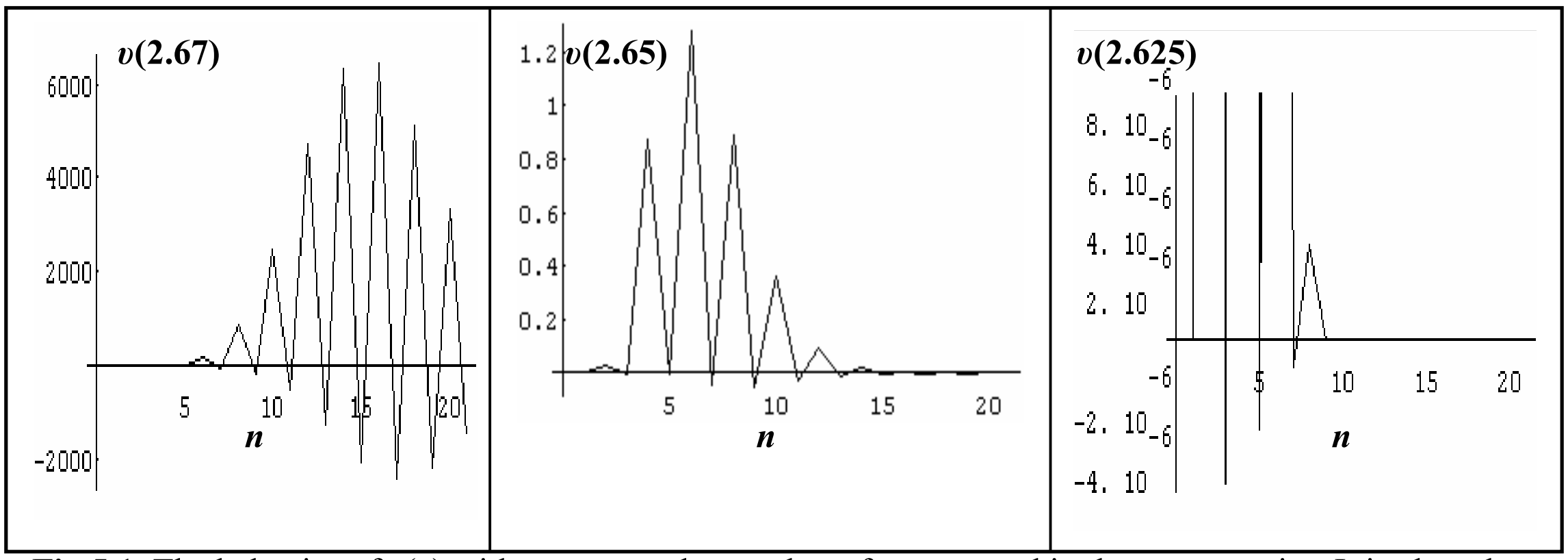

**Fig.5.1.** The behavior of $\upsilon(x)$ with respect to the number of terms, used in the power series. It is clear that, the series converges more as the values of x become closer to the value of α. Also, the number of required terms to achieve convergence becomes less.

### 5.2.1.2. THE POWER SERIES SOLUTION FOR THE TRANSFORMED EQUATION

Equation (5.17) can be written [36] as

$$z''(x) + p(x)z'(x) + q(x)z(x) = 0\ . \tag{5.27}$$

where,

$$p(x)=\frac{1}{x(x+1)}, \qquad q(x)=\frac{\lambda^2(x+1)^3-xl(l+1)}{x^2(x+1)}, \tag{5.28}$$

This can be transformed [36] without loss of generality to an equation of the form

$$Z''+(q-\frac{1}{2}p'-\frac{1}{4}p^2)Z=0, \tag{5.29}$$

where the relation between $Z$ and $z$ is

$$z(x)=Z(x)\mathrm{e}^{-\frac{1}{2}\int p(x)dx}, \tag{5.30}$$

and

$$\mathrm{e}^{-\frac{1}{2}\int p(x)dx}=\sqrt{\frac{x+1}{x}}. \tag{5.31}$$

Thus, instead of solving (5.16), one can solve (5.28) using (5.29). If we use the same power series (5.19) and substitute it in (5.28) we have

$$\begin{aligned}&4[(y+\alpha)^2+2(y+\alpha)^3+(y+\alpha)^4]\sum_{n=1} n(n-1)c_n y^{n-2}+\\&\left\{4\lambda^2(y+\alpha+1)^4-4l(l+1)(y+\alpha)^2-4[l(l+1)-1](y+\alpha)+1\right\}\sum_{n=0} c_n y^n=0.\end{aligned} \tag{5.32}$$

Thus, we have a seven-term recurrence relation of the order ($n$-2, $n$-1, $n$, $n$+1, $n$+2, $n$+3, $n$+4). Therefore, we will come up with more difficult solutions. Hence it will not be helpful. Apparently, we may come up with a similar result, hence no need to repeat it again.

From (5.8-9), we only need to use $\upsilon(x)$, $\upsilon'(x)$, and $\upsilon''(x)$ at the surface of the NS ($x=\alpha$). Thus from (5.20), we only need to the first three coefficients $c_0$, $c_1$, $c_2$, and the rest will no longer be used.

## 5.2.2. THE FROBENIUS METHOD

We tried to obtain the solutions of (5.16) using the Frobenius method [34, 35]. It is a series solution. It can be used to solve second order differential equations having regular singular points. For example, (5.17) has a regular singular point at $x = 0$ but not at $x = -1$, since $x > 0$. Because, $xp(x)$, and $x^2q(x)$ are analytic at $x = 0$, it is a regular singular point,

$$\lim_{x\to 0} x\frac{1}{x(x+1)} \to 1, \quad \text{(finite)}, \tag{5.33a}$$

$$\lim_{x\to 0} x^2 \frac{\lambda^2(x+1)^3 - xl(l+1)}{x^2(x+1)} \to \lambda^2, \quad \text{(finite)}. \tag{5.33b}$$

The key point of this method is to substitute

$$z(x) = \sum_{n=0} c_n x^{n+r}, \tag{5.34}$$

in terms of the dependent variable in the differential equation (5.16) instead of the form (5.19). Here, $r$ is a constant, called the *index* that must be determined from an equation, called the *indicial equation,* obtained by equating the coefficients of the lowest power of $x$ to zero. This method can be used to solve (5.17), or its transformed equation (5.30).

### 5.2.2.1. THE SOLUTIONS FOR THE EQUATION BY USING THE FROBENIUS METHOD

Substituting (5.34) into (5.17) gives

$$x^r \left\{ \begin{array}{l} \sum_{n=0} c_n[(n+r)(n+r-1) + (n+r) + \lambda^2]x^n + \\ \sum_{n=0} c_n[(n+r)(n+r-1) + 3\lambda^2 - l(l+1)]x^{n+1} + 3\lambda^2 \sum_{n=0} c_n x^{n+2} + \lambda^2 \sum_{n=0} c_n x^{n+3} \end{array} \right\} = 0. \tag{5.35}$$

Again putting $k = n$, $n$+1, $n$+2, and $n$+3 in the first, second, third and fourth term of (5.34), respectively, we have

$$x^r \left\{ \begin{array}{l} \sum_{k=0} [(k+r)(k+r-1)+(k+r)+\lambda^2]c_k x^k + \\ \sum_{k=1} [(k+r-1)(k+r-2)+3\lambda^2 - l(l+1)]c_{k-1}x^k + 3\lambda^2 \sum_{k=2} c_{k-2}x^k + \lambda^2 \sum_{k=3} c_{k-3}x^k \end{array} \right\} = 0 . \quad (5.36)$$

From the first term we have

$$c_0[r(r-1)+r+\lambda^2] = 0, \quad (5.37)$$

if $c_0 = 0$, then all the other indices vanish and we will come up with a trivial solution z = 0, which is meaningless. Hence, the indicial equation is

$$r(r-1)+r+\lambda^2 = 0, \quad (5.38)$$

with the two roots

$$r_\pm = \pm i\lambda . \quad (5.39)$$

The general recurrence relation of (5.27) is

$$\begin{array}{l} [(k+r)^2+\lambda^2]c_k + [(k+r-1)(k+r-2)+3\lambda^2 - l(l+1)]c_{k-1} + \\ 3\lambda^2 c_{k-2} + \lambda^2 c_{k-3} = 0, \end{array} \quad (5.40)$$

with a radius of convergence, $R$, given by

$$\frac{1}{R} = \lim_{k\to\infty} \left| \frac{c_k}{c_{k-1}} \right| = |-1| = 1 . \quad (5.41)$$

From the ratio test [34], a series converges if

$$L = \lim_{k\to\infty} \left| \frac{c_k}{c_{k-1}} \right| |x| < 1 . \quad (5.42)$$

Thus, the interval of convergence is $-1 < x \leq 1$, which does not lie in the desired range. Also, because of the negative sign in (5.40), one can conclude it is an alternating series. Therefore, such a solution cannot be used even at the ends of the dipole (at which points $x$ may exceed 1), leave alone at the far point. (It may be added that the indicial roots are not real.)

## 5.2.2.2. THE SOLUTIONS FOR THE TRANSFORMED EQUATION BY USING THE FROBENIUS METHOD

In order to check the singularity of (5.30), one has to write it in terms of $x$ explicitly, i.e.

$$Z'' + \left( \frac{4\lambda^2 (x+1)^4 - 4l(l+1)x^2 - 4[l(l+1)-1]x + 1}{4x^2 (x+1)^2} \right) Z = 0, \tag{5.43}$$

which also has a regular singular point at $x = 0$. Thus, substituting (5.33) into (5.42) gives

$$x^r \left\{ \begin{array}{l} 4(x^2 + 2x^3 + x^4) \sum\limits_{n=1} (n+r)(n+r-1) c_n x^{n-2} + \\ \left\{ 4\lambda^2 (x^4 + 4x^3 + 6x^2 + 4x + 1) - 4l(l+1)x^2 - 4[l(l+1)-1]x + 1 \right\} \sum\limits_{n=0} c_n x^n \end{array} \right\} = 0. \tag{5.44}$$

Hence, the indicial equation in this case is

$$4r(r-1) + 4\lambda^2 + 1 = 0, \tag{5.45}$$

with the two indicial roots

$$r_{\pm} = \frac{1}{2} \pm i\lambda . \tag{5.46}$$

Thus, (5.43) can be written as

$$x^r \left\{ \begin{array}{l} \sum\limits_{n=0} 4(n+r)(n+r-1) c_n x^n + \sum\limits_{n=0} 6(n+r)(n+r-1) c_n x^{n+1} + \\ \sum\limits_{n=0} 4(n+r)(n+r-1) c_n x^{n+2} + 4\lambda^2 [\sum\limits_{n=0} c_n x^{n+4} + 4\sum\limits_{n=0} c_n x^{n+3} + \\ 6\sum\limits_{n=0} c_n x^{n+2} + 4\sum\limits_{n=0} c_n x^{n+1} + \sum\limits_{n=0} c_n x^n ] - 4l(l+1) \sum\limits_{n=0} c_n x^{n+2} - \\ 4[l(l+1)-1] \sum\limits_{n=0} c_n x^{n+1} + \sum\limits_{n=0} c_n x^n \end{array} \right\} = 0. \tag{5.47}$$

Hence, the radius of convergence, $R$, is

$$\frac{1}{R} = \lim_{n \to \infty} \left| \frac{c_n}{c_{n-1}} \right| = \left| -\frac{6}{4} - \frac{c_{n-2}}{c_{n-1}} \right| . \tag{5.48}$$

which depends on the values of the $c$'s, i.e. it needs to be evaluated numerically. Also, from the ratio test, the interval of convergence is even less than the previous one.

Therefore, we have the same problem as with the previous method, which does not allow a series solution of the equation at the far point. (Again, the indicial roots of the singularity are not real In this case.)

## 5.2.3. THE ASYMPTOTIC SOLUTION OF THE EQUATION

The solution of the equation at the position of the observer, which is the far point, was not determined by the foregoing methods. All of them have problems of divergence as $x \to \infty$. We will try to find ways to avoid this problem. One way is to change variables, which makes the power series convergent as $x \to \infty$ [34, 35]. The other is to use an asymptotic analysis [36].

### 5.2.3.1. USING THE CHANGE OF VARIABLES TO FIND THE ASYMPTOTIC SOLUTION

As far as we know, the equation (5.16) has no exact analytical solution. Thus, we tried to approximate the solution at the far point by changing variables. One of the possible changes is

$$x = \frac{1}{y}. \tag{5.49}$$

If the solution is convergent at y = 0, it is convergent at $x = \infty$ too. Thus from (5.48), $dz/dx = -y^2 dz/dy$, $d^2z/dx^2 = 2y^3\, dz/dy + y^4 dz^2/dy^2$, and $x^2\,(x + 1) = (1 + y)/y^3$. Substituting all these expressions in (5.16), we have

$$y^4(y+1)z'' + y^3(y+2)z' + [\lambda^2(y+1)^3 - y^2 l(l+1)]z = 0, \tag{5.50}$$

where $z' = dz/dy$. In canonical form, (5.50) can be written as

$$z'' + \frac{(y+2)}{y(y+1)} z' + \frac{[\lambda^2 (y+1)^3 - y^2 l(l+1)]}{y^4 (y+1)} z = 0 . \tag{5.51}$$

To find a power series solution to (5.50), including $y = 0$ in the interval of convergence, one has to use the Frobenius' method with $y_0 = 0$. To do so, we have to make sure that the singularity at $y = 0$, is regular. Using the same method introduced in §5.2.2, gives

$$\lim_{y\to 0} y p(y) = \lim_{y\to 0} y \frac{(y+2)}{y(y+1)} \to 1, \qquad \text{(finite)}, \tag{5.52}$$

$$\lim_{y\to 0} y^2 q(y) = \lim_{y\to 0} y^2 \frac{[\lambda^2 (y+1)^3 - y^2 l(l+1)]}{y^4 (y+1)} = \lim_{y\to 0} \frac{\lambda^2 (y+1)^2}{y^2} - \frac{l(l+1)}{(y+1)} \to \infty . \tag{5.53}$$

Therefore, the Frobenius method cannot be used to solve (5.50) at $y = 0$.

### 5.2.3.2. USING THE ASYMPTOTIC ANALYSIS TO FIND THE ASYMPTOTIC SOLUTION

Instead of solving (5.7) as it is, one can solve its asymptotic form. In this subsection, we will apply the same procedures introduced in [36] to find the solution as $x \to \infty$. Although this is an approximate analytical method, it gives a simple damped oscillating solution. In this method, we will use the transformed equation (5.29). It can be written as

$$Z'' + f(x) Z = 0 , \tag{5.54}$$

where $f(x) = q(x) - (1/2)p'(x)$-$(1/4)p^2(x)$. It can be explicitly written items of $x$ as

$$f(x) = \frac{4\lambda^2 (x+1)^4 - 4l(l+1)x^2 - 4[l(l+1) - 1]x + 1}{4x^2 (x+1)^2} . \tag{5.55}$$

To apply the asymptotic method, $x^2 f(x)$ must be a non-analytic function as $x \to \infty$, i.e. it has an irregular singularity at infinity [36]. Hence,

$$\lim_{x\to\infty} x^2 f(x) \to \infty , \tag{5.56}$$

which satisfies the requirement of applicability.

The next step is to find the limit of $f(x)$ as $x \to \infty$, which is

$$\lim_{x\to\infty} f(x) \to \lambda^2 . \tag{5.57}$$

Since the limit is only of $O(1)$, we can write the general solution to (5.54) as

$$Z(x) = Ae^{i\lambda x} + Be^{-i\lambda x}, \tag{5.58}$$

where, $A$ and $B$ are constants. Clearly, (5.58) is the wave equation. We have only an incoming wave, hence we must take $A = 0$. Thus,

$$Z(x) = Be^{-i\lambda x}. \tag{5.59}$$

Hence, the general solution for $z(x)$, as $x \to \infty$, can be obtained by using (5.30-31) as

$$z(x) = \sqrt{\frac{x+1}{x}} Ae^{-i\lambda x}, \tag{5.60}$$

whose real part is an oscillating function, see Fig. 5.2.

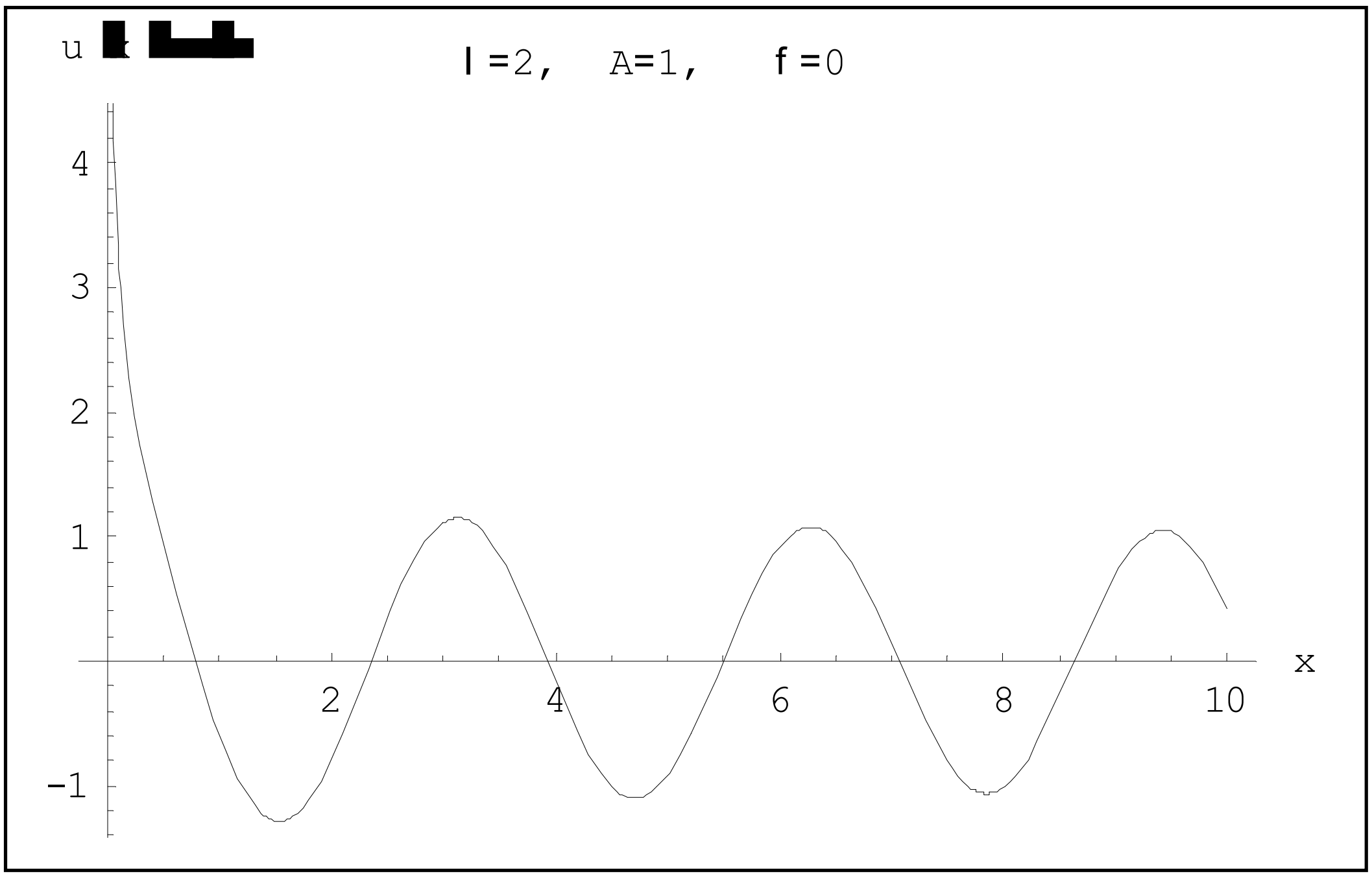

**Fig.5.2.** The plot of real part of $z(x)$ versus $x$, at $\lambda = 2$, and $A = 1$.

Therefore, we can use (5.60) as a solution for (5.16) at the far point, which is the position of the observer. This solution represents $u(x)$ in the Green function (5.9). Also, we can use the power series (5.20) with the recurrence relations (5.24-25) as a solution for (5.16) on the surface of the NS, which is the position of the radiation source. This solution represents $\upsilon(\mathrm{x})$ in the Green function (5.9). At the end of this sub-section, one can construct the Green's function, which is the cornerstone to obtain the energy spectra of the fast, oblique, rotating magnetic (electric) dipole incorporating the effect of mass, in the low angular momentum *l*. In the case of high *l,* we need to rewrite (5.16) and approximate it to simplify the problem. This will be discussed in the next sub-section.

## 5.2.4. THE SOLUTION OF THE EQUATION AT HIGH ANGULAR MOMENTUM

At very high angular momentum $l$ and low azimuthal number $m$, the first term in the square bracket of (5.17) will become negligible relative to the second. Then the equation will become

$$x(x+1)z''(x) + z'(x) - l(l+1)z(x) = 0\,, \tag{5.61}$$

which has a hypergeometric series solution [33] of the form $F(\alpha, \beta; \gamma; x)$, where $\alpha, \beta = \dfrac{-1 \pm \sqrt{1+4l(l+1)}}{2}$, $\gamma = 1$. This is a solution in the interval I = (-1, 1) about $x = 0$. To find the solution outside this interval in the desaird range we can use this relation

$$F(\alpha, \beta; \gamma; -x) = (1+x)^{-\alpha} F(\alpha, \gamma-\beta; \gamma; \frac{x}{x+1})\,. \tag{5.62}$$

Substituting $\alpha$, $\beta$; $\gamma$, the hypergeometric function on the right hand side of (5.62) will become $F(\alpha,\alpha; \gamma, x)$ which is the spherical Bessel function. Thus, in this limit, we can use the massless solution [32].

Also, the large $l$ and $m$, components of the radiation will become negligible, because both $R^e{}_{lm}$ and $R^o{}_{lm}$ (5.8) become negligible too. The only case we have to consider at high $l$ is $R^o{}_{lm}$ for relatively low $m$. The mass appears only in $\lambda$ . Therefore, the effect of the mass is negligible at high $l$ and we can also use the massless solution.

For these reasons, we will use our solution at low $l$, and the solution for the massless case at high $l$. Comparing the values of the first and the second term in the square bracket of (5.17) and the values of $R^e{}_{lm}$ and $R^o{}_{lm}$ at certain $l$ and for different $m$, gives a very small effect of mass above certain $l$ for each inclination angle $\chi$, and velocity, $\beta$. Generally speaking, above $l = 200$, the effect of mass can be neglected for all cases.

# 5.3. THE EXPECTED ENERGY SPECTRA OF SUB-MILLISECOND PULSARS

This section can be considered as the core of this thesis. Since, we have derived the Green's function incorporating the mass effect (§5.2), and without incorporating it, (§4.3.2), we can construct the energy spectra of the electromagnetic radiation from the fast, oblique, rotating magnetic dipole for both cases. The second case, in fact, was obtained in the previous chapter. In this section we will first write down a Mathematica program, which plots the energy radiated, $E_l$ (5.1), against the angular momentum number $l$, by the dipole, first at low $l$ incorporating the mass effect (5.7). Then it will be matched to the massless solution (5.2). That program will be operated at different rotational velocities, $\beta$'s, and inclination angles, $\chi$'s.

## 5.3.1. THE MATHEMATICA PROGRAM

There are several reasons for choosing the Mathematica package to do the numerical calculations and then to plot the energy spectra. First, it is a complete mathematical software that allows you to perform any mathematical operation. Second, it is easy to program and get results. Third, it can do the calculation and plot the results at the same time although we will use different software (Microsoft Excel XP) to do so. The main problem with it is its limited speed of calculation. Thus, we first used Mathematica 2.2, and then to reduce the time of calculation we used the upgraded version, Mathematica 4.1. In this subsection, we will introduce the program we wrote and explain the purpose of each command.

In fact the program, which can perform all of the numerical calculation and plots the energy spectra at relatively low *l,* is just one page, Box. 5.1. We can divide the program into three main parts. The first is the numerical inputs. The second is the evaluation of the Green's function. The third is the plotting of the energy spectra, depending on the numerical values of the first parts. The other part of the program (for high *l*, assumed to be the same as the massless case, § 4.3) is introduced in Box. 5.2. Since, the layout of Mathematica 4.1 is similar to our notation and easier to read and understand, we display the program in this version. The same program, written in Mathematica 2.2, is displayed in Appendix I. The expressions in the square brackets, displayed on the right margin, are not included in the program.

**Box.5.1. Mathematica 4.1 program for the case of massive dipole.**

**Part I**

bmax = 0.99 — [$\beta_{max} = \upsilon/c$ = 0.8, 0.85, 0.90, 0.95, 0.99]

c = p / 2 — [Inclination angle = π/2, π/4, π/6, π/8]

b = bmax Sin [illegible] — [$\beta$ at certain $\chi$]

Mns = 0.5 — [Mass of NS = 0.5, 1.0, 1.5, 2.0, 2.5, 3.0 $M_{\odot}$]

a = 8 — [Radius of NS in gravitational unit ≈ 12 km]

w = bmax / a — [Angular frequency (gravitational unit)]

rs = 2 Mns — [Schwarzschild radius $r_s$ )

l = m w rs — [ λ, defined in (5.16)]

a = [illegible] Sin [illegible] - 1 [illegible] rs — [$x$($a$, $M_{ns}$), (5.3)]

x0 = a — [$x_0$, (5.8)]

R = $10^{15}$ — [$R \to \infty$, (5.9) = $10^{15}$]

Mg = 1 — [Magnetic moment =1, it is common factor in (5.1)]

**Part II**

c [illegible]

c [illegible]

c [illegible] + 1 [illegible] + 1 [illegible] / 2 a² [illegible] + 1 [illegible] — [$c_2$, (5.23)]

v1 = c [illegible] — [$\upsilon(\alpha)$, (5.19)]

dv1 = c [illegible] — [$\upsilon'(\alpha)$, to use in $G$ (5.9)]

d2v1 = 2 c [illegible] — [$\upsilon''(\alpha)$, to use in $G'$(5.8)]

u [illegible] 1 / x [illegible] x [illegible] — [$u(x)$, (5.60), $A$ = 1]

du [illegible] — [$u'(x)$, used in $G$ (5.9)]

G1 = [illegible] u [illegible] / u R [illegible] v1 du R [illegible] — [$G_l(x)$, (5.9)]

**Part III**

dG1 = u R [illegible] R [illegible] v [illegible] - dv1 du R [illegible] / [illegible] R [illegible] v1 du [illegible] — [$G'_l(x)$, used in (5.8)]

D = [illegible] L [illegible] + 1 [illegible]

L = [illegible] L [illegible] + 1 [illegible] + 1 [illegible] — [Δ, Λ, and $K_{lm}$, used in (5.8)]

K [illegible] SphericalHarmonicY [illegible] 0 [illegible] SphericalHarmonicY [illegible] c , p [illegible]

Rev1 [illegible] p Mg K [illegible] / a rs D [illegible] 0 + [illegible]

Rod1 [illegible] p Mg w L K [illegible] / a D — [$R^e$ and $R^o$, defined by (5.8)]

e1 [illegible] 2 [illegible] w [illegible] bs [illegible] bs Rod1 [illegible] / 4 p, m=1 — [$E_l$, radiated energy, (5.1)]

S11 = N [illegible], 201, 10 [illegible] — [Evaluate $E_l$, at $l$ = 1-201, at a step of 10, to avoid even $l$]

**Box.5.2. Mathematica 4.1 program for the case of mass-less dipole.**

| | |
|---|---|
| `a2 = a m w Sin` | [$x_0$, (4.73)] |
| `x02 = a2` | |
| `G2 ... 2 x ... BesselJ L + 1/2, x` | [$G_l(x)$, (5.18)] |
| `dG2 ... G2` | [$G'_l(x)$, used in (5.2)] |
| `D = L + 1` | |
| `L = L + 1 ... + 1` | [Δ, Λ, and $K_{lm}$, used in (5.2)] |
| `K ... SphericalHarmonicY ... 0 ... SphericalHarmonicY ... c, p` | |
| `Rev2 ... p Mg ... D ... G2 x02` | [$R^e$ and $R^o$, defined by (5.2)] |
| `Rod2 ... p MgwSin c ... G2 x02 / x02 D` | |
| `e2 ... 2 w ... bs ... bs Rod2 ... / 4 p`, `m=1` | [$E_l$, radiated energy, (5.1)] |
| `S1 = N ... , 301, 10` | [Evaluate $E_l$, at $l$ = 1-301, at a step of 10, to avoid even $l$] |
| `S12 = 10^10 S1` | [To rescale the values of $E_l$ to mach that for massive case] |
| `Sr1 = Join` | [Join the two series at similar point to get a complete spectrum] |
| `Sp1 = ListPlot ... Joined® True` | [Plot the continuous energy spectrum] |

## 5.3.1.1. THE VALIDITY OF THE PROGRAM

In order to examine the validity of the first program Box.5.1, we have to check two points. One is to determine the radius of convergence in the power series solution, $\upsilon$(x), and how it changes with angular momentum, velocity, and mass of the dipole. The second is to reproduce the results of §4.3.2, if possible. We should obtain the same result as $M_{NS} \to 0$.

The interval of convergence, $I$, is defined [34] as the set of all values of $x$, for which the series converges (exists). In other words, it has a finite value if it converges. The radius of convergence, $R$, is the maximum distance between the center of the series, α, and the point $x$, up to which it converges, i.e. $I = (\alpha - R,\ \alpha + R)$. Where the series is alternating, the end point is also to be included.

As mentioned in §5.2.1.1, $R$ cannot be evaluated directly because its value depends on some other constants, $c_n$, which are functions of other parameters. Those parameters are the angular momentum, $l$, the gravitational mass of the dipole, and the dimensionless variable, λ. Therefore, we can only use numerical methods to evaluate the radius of convergence and the corresponding number of summations, $n$. For this purpose, one has to specify these parameters and determine the $n$ after which the series gives the same finite value in the interval $I$. The plots of $\upsilon(x)$ as defined in (5.19), show that the radius of convergence decreases as $l$ and $M$ increase, see Figs.5.3-4. On the other hand, the plot of $\upsilon(x)$ does not show any change as $\beta$ changes, see Fig.5.5.

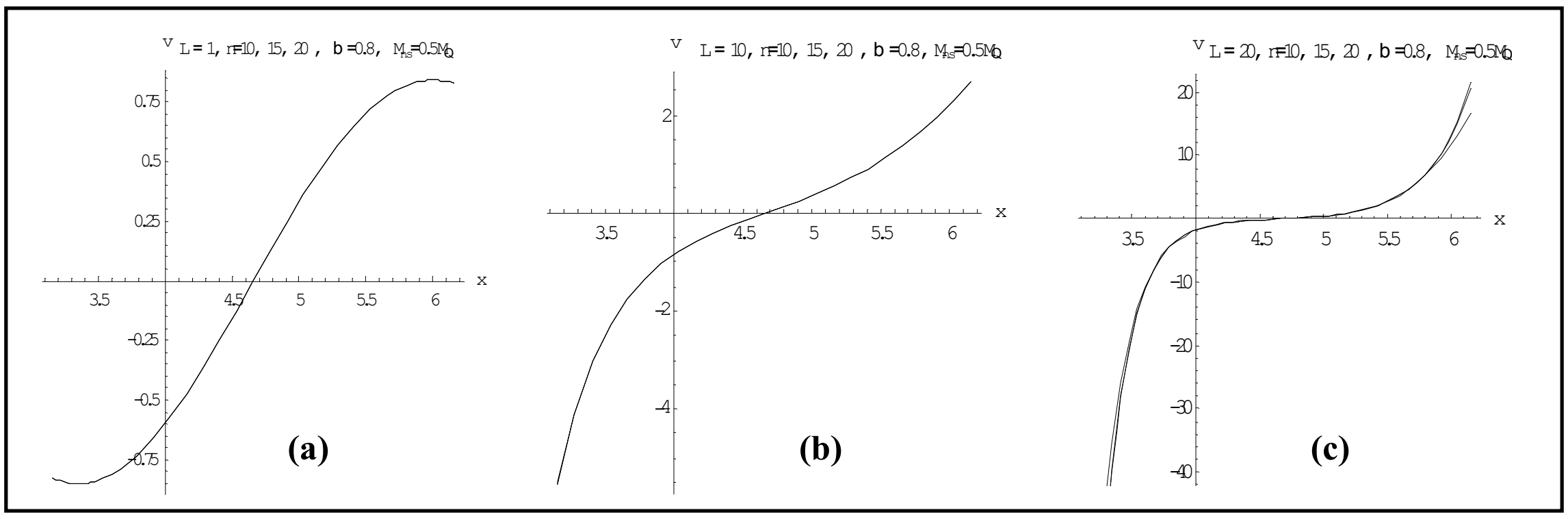

**Fig.5.3.** The plot of υ($x$) versus $x$ at different angular momentum $l$, where $a$ = 8, $\chi = \pi/4$. At (a) $l$= 1, the three power series are almost the same. They deviate from each other as $l$ increase, (b) and (c).

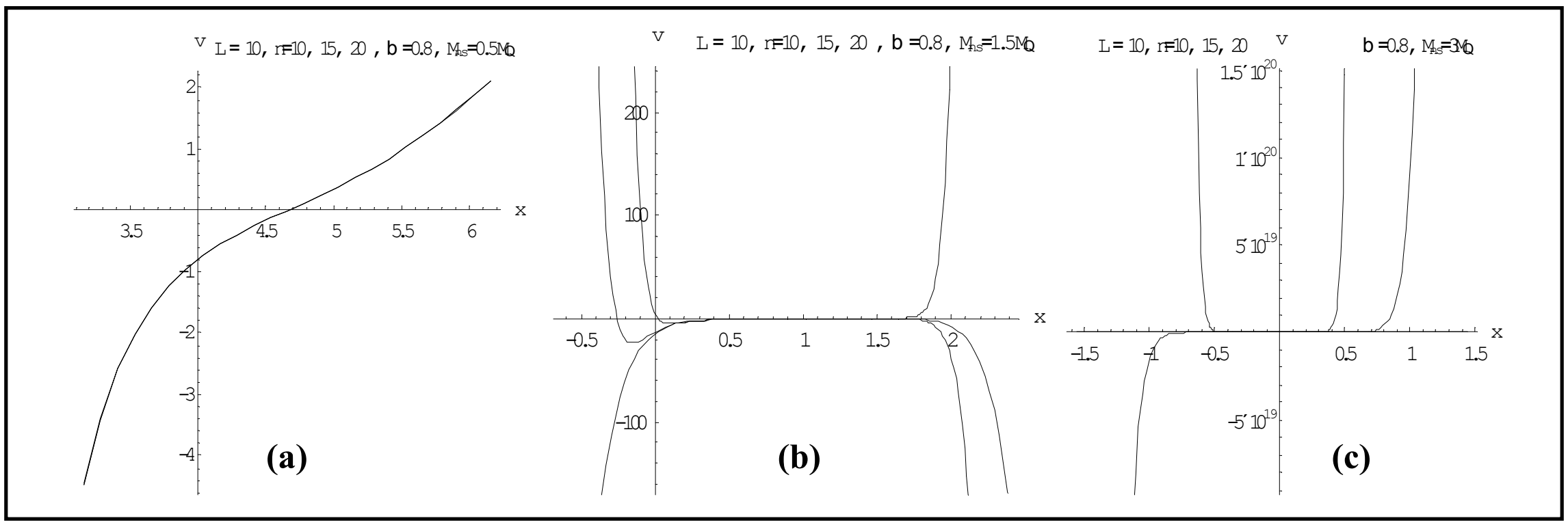

**Fig.5.4.** The plot of υ($x$) versus $x$ at different masses $M$, where $a$ = 8, $\chi = \pi/4$. The deviation from the center of the series is $\alpha$. Obviously, the differences between the three power series become higher as $M$ increases.

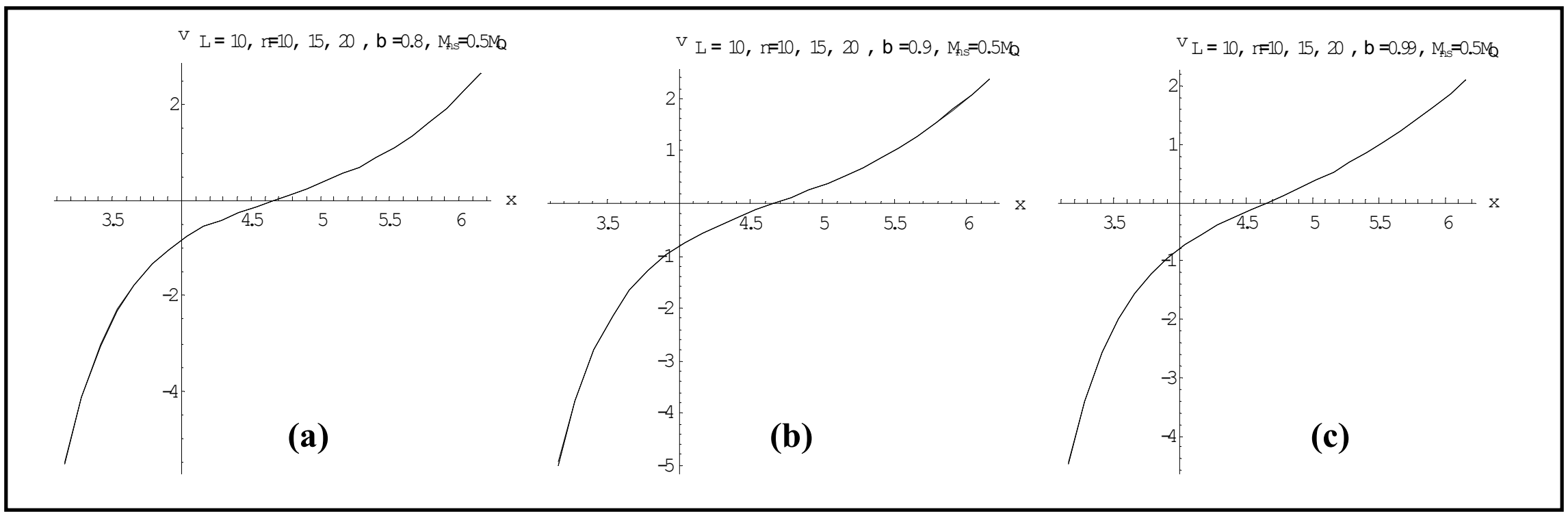

**Fig.5.5.** The plot of υ($x$) versus $x$ at different velocities $\beta$, where $a$ = 8, $\chi = \pi/4$. The three power series are almost the same in all cases.

Because of the problem of convergence, we can only draw part of the energy spectra, for $1 \leq l \leq 11$, (even $l$ are excluded as they give zero), by using the power series with $n = 20$, to ensure that we have a valid solution of the equation (5.16). Also, notice that the values of these series are almost the same at their centers, which is the value we will use. Thus, we can go for higher $l$. Moreover, we cannot make a direct comparison of our results with the mass-less case because (5.4a) becomes singular as $M$ tends to zero. There is an additional problem of comparison if we try to use the entire power series, because we have to truncate our calculations at $l = 11$ instead of more than 600. To go to the required $l$ values, above the surface of the NS, we have to increase $n$ enormously, or to take it at the center of the series, i.e. ($x = \alpha,$ at exactly the surface of the NS). The first choice is not possible with Mathematica 4.1. To obtain the required solutions, different (faster) software or some other language (such as FORTRAN) would have to be used. For our purposes, however, we can estimate the effect of the mass on the radiation with only the small $l$ part of the spectrum. In fact, at higher $l$, the effect of mass is negligible as described in § 5. 2. 4.

## 5.3.2. THE ENERGY SPECTRA OF THE MASSIVE DIPOLE

In this subsection, we will display complete energy spectra of the obliquely, fast rotating dipole, which include the effect of the mass for low $l$ and are similar to the massless case for high $l$. As a result of our calculations, we can obtain these spectra by running the above program, after inserting desired inputs and limits. In fact we obtain the values from the program and we plot them and match between the massive and massless case by using Microsoft Excel XP software. However, the area under each spectrum, which is total energy per frequency, $\varepsilon$, is evaluated by using Origin 6.1 software.

In order to construct an approximate relation between the changes in energy due to the change of gravitational mass, we have to consider the angular frequency, $\omega$, and in turn the rotational velocity of the dipole, which affects the relativistic factor, $\gamma$ (4.22, 42). For this purpose, we will draw the spectra with different masses, $M$ (0.5, 1.0, 1.5, 2.0, 2.5, 3.0 $M_{\odot}$) and maximum (equatorial) velocities, $\beta_{max}$ (0.8, 0.85, 0.90, 0.95, 0.99), at certain inclination angle, $\chi$, dipole radius, $a$, and large distance $R$ (5.5). We will choose, $\chi = \pi/2$, $\pi/4$, $\pi/6$, and $\pi/8$, but we will not take $\chi \to 0$, because the results will not be stable at that angle. We consider the dipole as if it is embedded in the NS. The problem is that the rescaled and shifted distance parameter, $x_0$, becomes negative and close to $-1$, where convergence becomes problematical, especially at higher values of $M$. The large distance we use is $R = 10^{15}$ and the dipole radius is $a = 8$ (~12 km). In this subsection, the plots of $E_l$ versus $l$, for each angle, $\chi$ will be displayed.

## 5.3.2.1. THE ENERGY SPECTRA FOR $\chi=\pi/2$ (Orthogonal Rotator)

The results for $\chi = \pi/2$ can be seen in the Figs.5.6-10 below.

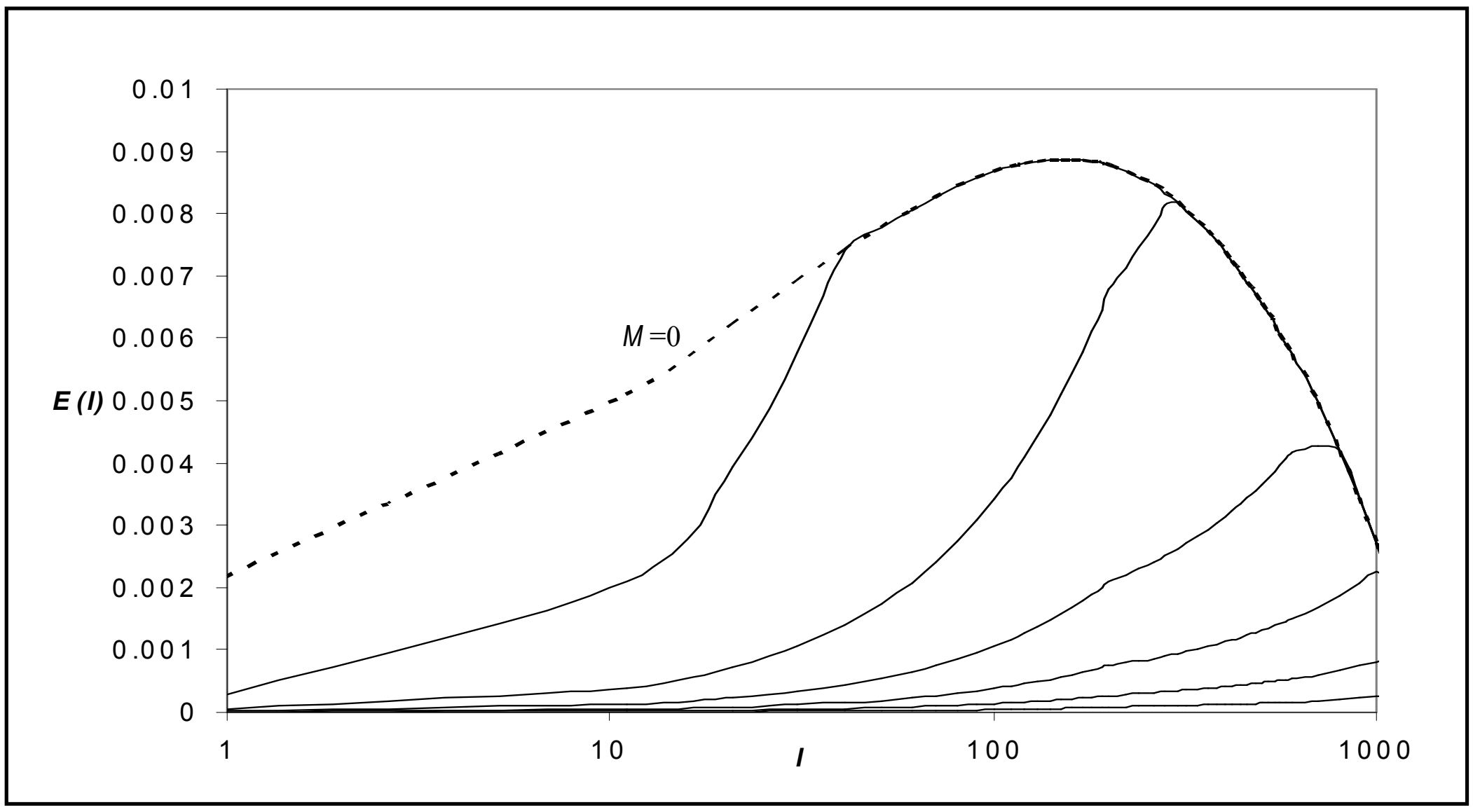

**Fig.5.6.** Energy spectra, at $\beta = 0.99$, $\chi = \pi/2$, for dipole of masses (0.5, 1, 1.5, 2, 2.5, 3) $M_\odot$, from the upper line to the lower one. The dashed curve is for the massless case.

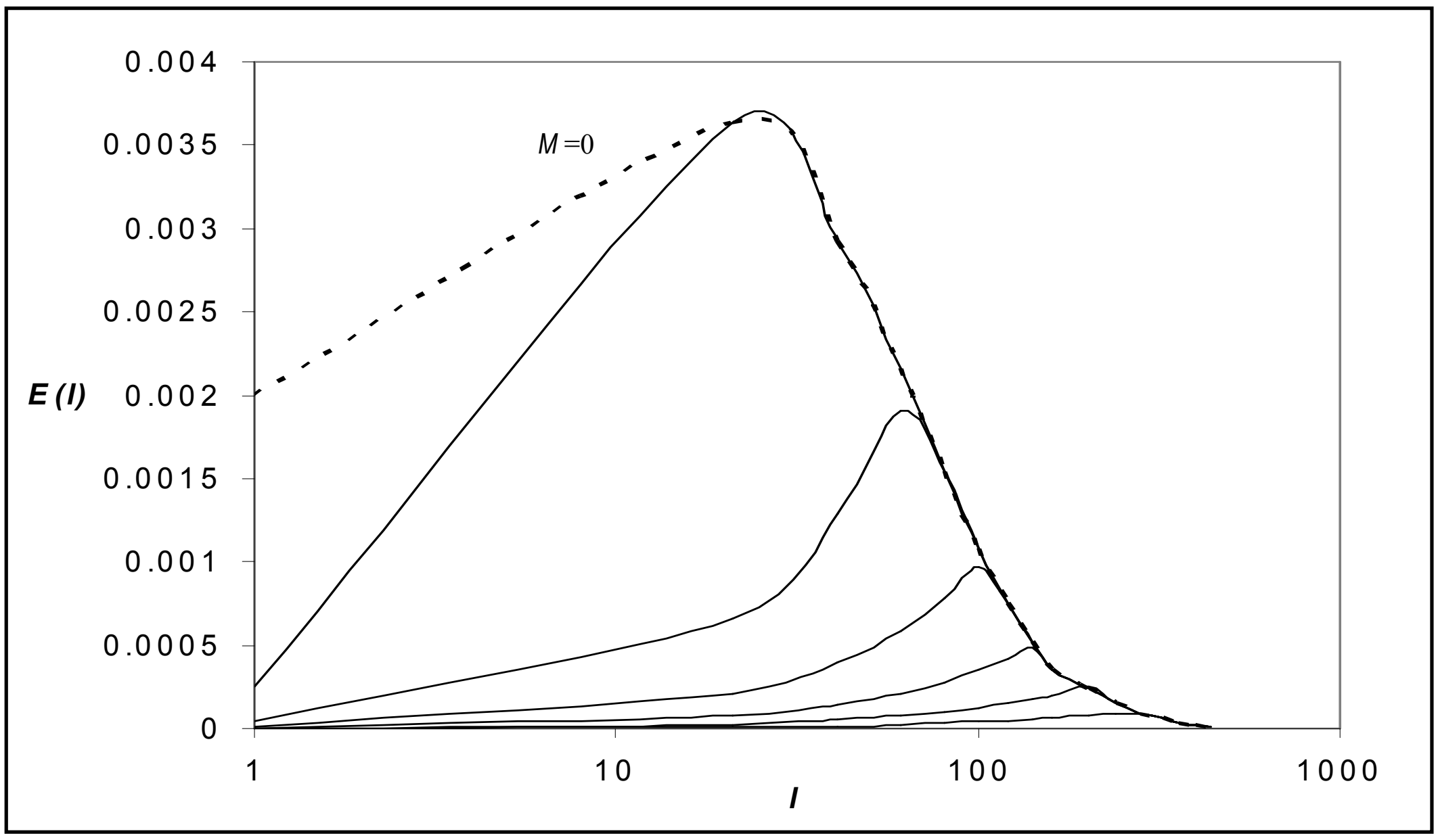

**Fig.5.7.** Energy spectra, at $\beta = 0.95$, $\chi = \pi/2$, for dipole of masses (0.5, 1, 1.5, 2, 2.5, 3) $M_\odot$, from upper line to the lower one. The dashed curve is for the massless case.

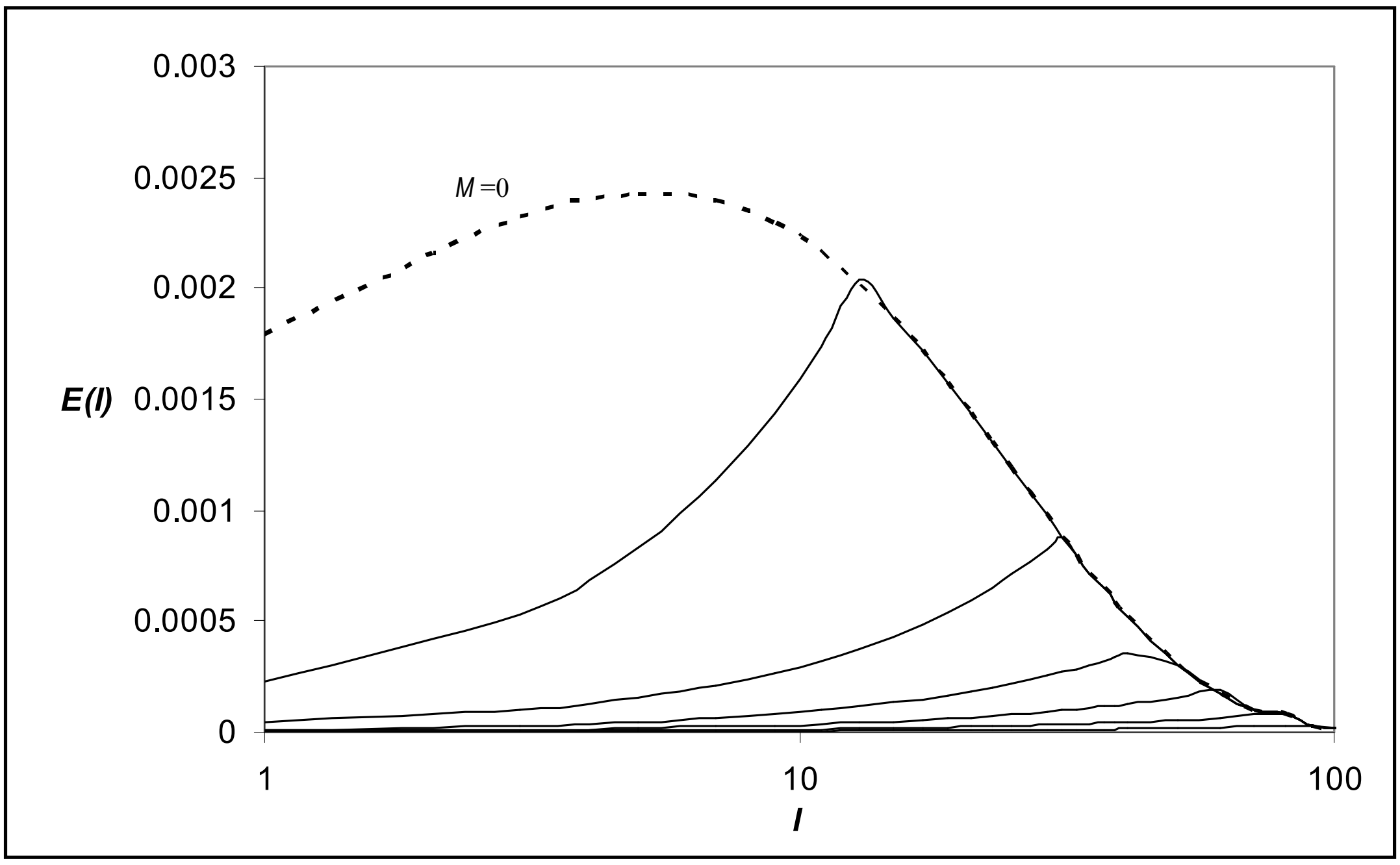

**Fig.5.8.** Energy spectra, at $\beta$ = 0.90, $\chi$ =π/2, for dipole of masses (0.5, 1, 1.5, 2, 2.5, 3) $M_{\odot}$, from upper line to the lower one.. The dashed curve is for the massless case.

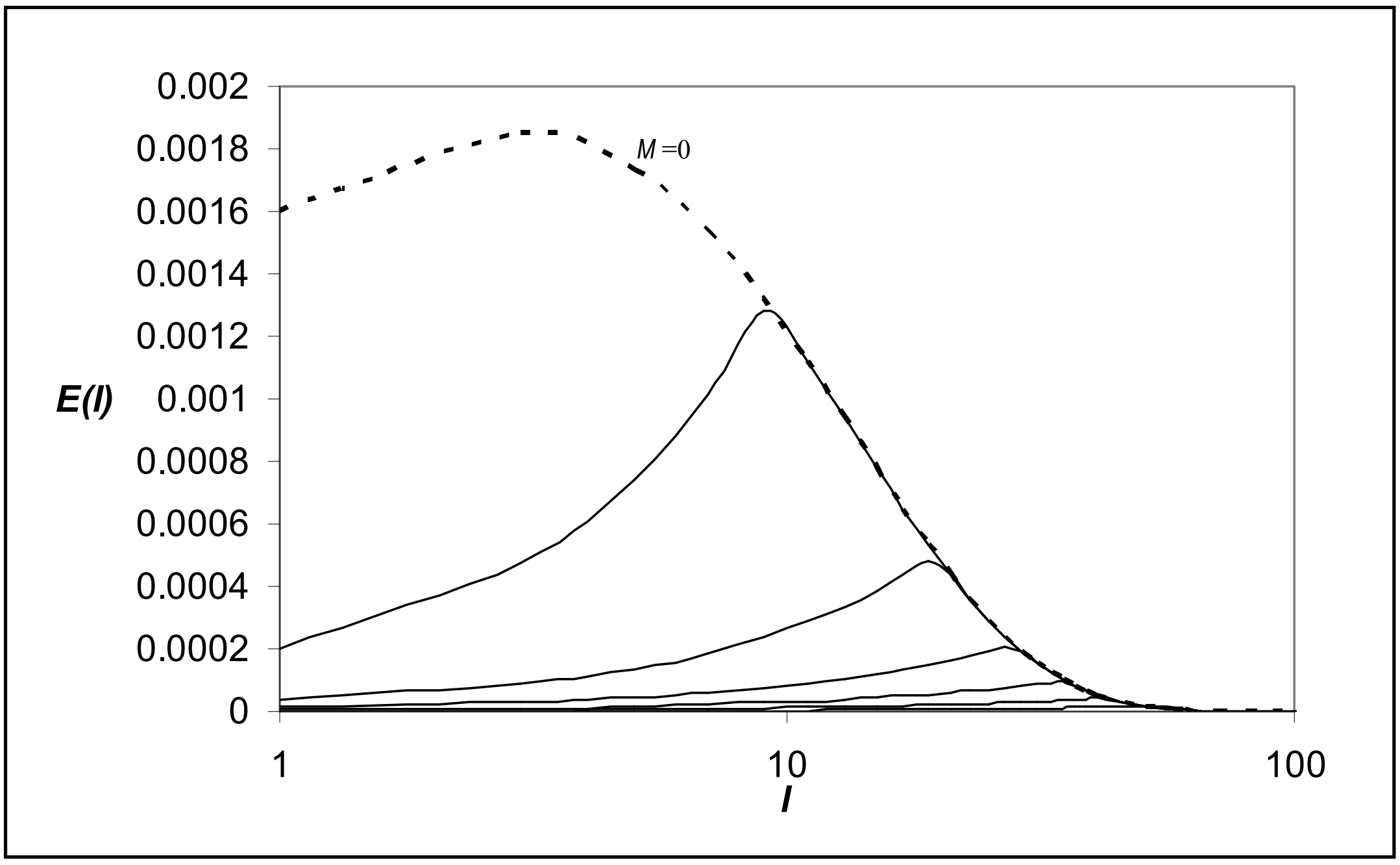

**Fig.5.9.** Energy spectra, at $\beta$ = 0.85, $\chi$ =π/2, for dipole of masses (0.5, 1, 1.5, 2, 2.5, 3) $M_{\odot}$, from upper line to the lower one. The dashed curve is for the massless case.

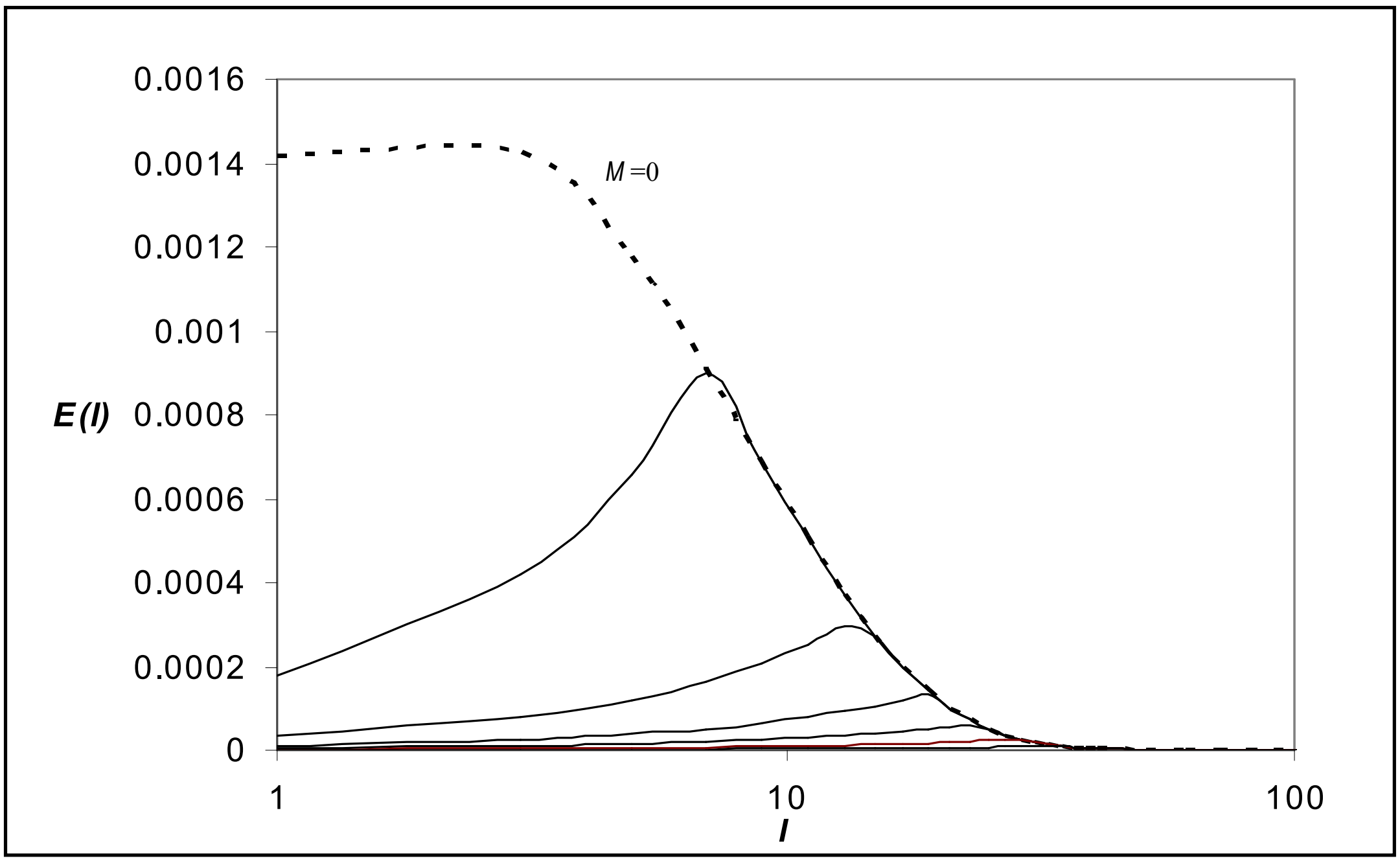

**Fig.5.10.** Energy spectra, at $\beta$ = 0.80, $\chi$ =$\pi$/2, for dipole of masses (0.5, 1, 1.5, 2, 2.5, 3) $M_{\odot}$, from upper line to the lower one. The dashed curve is for the massless case.

Since we are interested only on the cases at which mass is not neglected, some of the data, obtained from Fig.5.6-10 in the relatively low $l$ ($\leq$ 101), can be tabulated in tables 5.1-2.

**Table.5.1.** The values of energies radiated by different massive rotating dipoles, with $\chi = \pi/2$, $\beta$ =0.99, 0.95.

| $\beta_{max}$ | $\beta$ | $\omega$(GU) | $l$ | $E$-0.5 $M_{\odot}$ | $E$-1 $M_{\odot}$ | $E$-1.5 $M_{\odot}$ | $E$-2 $M_{\odot}$ | $E$-2.5 $M_{\odot}$ | $E$-3 $M_{\odot}$ |
|---|---|---|---|---|---|---|---|---|---|
| 0.99 | 0.99 | 0.12375 | 1 | 0.000275 | 5.05E-05 | 1.56E-05 | 5.61E-06 | 2.02E-06 | 6.23E-06 |
| 0.99 | 0.99 | 0.12375 | 21 | 0.003939 | 0.000723 | 0.000223 | 8.04E-05 | 2.89E-05 | 8.93E-06 |
| 0.99 | 0.99 | 0.12375 | 41 | 0.007603 | 0.001396 | 0.000431 | 0.000155 | 5.59E-05 | 1.72E-05 |
| 0.99 | 0.99 | 0.12375 | 61 | 0.011267 | 0.002069 | 0.000639 | 0.00023 | 8.28E-05 | 2.55E-05 |
| 0.99 | 0.99 | 0.12375 | 81 | 0.014931 | 0.002742 | 0.000846 | 0.000305 | 0.00011 | 3.39E-05 |
| 0.99 | 0.99 | 0.12375 | 101 | 0.018595 | 0.003415 | 0.001054 | 0.000379 | 0.000137 | 4.22E-05 |
| 0.95 | 0.95 | 0.11875 | 1 | 0.000253 | 4.65E-05 | 1.43E-05 | 5.16E-06 | 1.86E-06 | 5.74E-07 |
| 0.95 | 0.95 | 0.11875 | 21 | 0.003627 | 0.000666 | 0.000206 | 7.4E-05 | 2.66E-05 | 8.22E-06 |
| 0.95 | 0.95 | 0.11875 | 41 | 0.007001 | 0.001286 | 0.000397 | 0.000143 | 5.14E-05 | 1.59E-05 |
| 0.95 | 0.95 | 0.11875 | 61 | 0.010375 | 0.001906 | 0.000588 | 0.000212 | 7.62E-05 | 2.35E-05 |
| 0.95 | 0.95 | 0.11875 | 81 | 0.013749 | 0.002525 | 0.000779 | 0.000281 | 0.000101 | 3.12E-05 |
| 0.95 | 0.95 | 0.11875 | 101 | 0.017123 | 0.003145 | 0.000971 | 0.000349 | 0.000126 | 3.88E-05 |

**Table.5.2.** The values of energies radiated by different massive rotating dipoles, for $\chi = \pi/2$, $\beta = 0.90$, 0.85,0.80.

| βmax | β | ω(GU) | l | E-0.5 M⊙ | E-1 M⊙ | E-1.5 M⊙ | E-2 M⊙ | E-2.5 M⊙ | E-3 M⊙ |
|---|---|---|---|---|---|---|---|---|---|
| 0.90 | 0.90 | 0.1125 | 1 | 0.000227 | 4.17E-05 | 1.29E-05 | 4.63E-06 | 1.67E-06 | 5.15E-07 |
| 0.90 | 0.90 | 0.1125 | 21 | 0.003255 | 0.000598 | 0.000185 | 6.64E-05 | 2.39E-05 | 7.38E-06 |
| 0.90 | 0.90 | 0.1125 | 41 | 0.006283 | 0.001154 | 0.000356 | 0.000128 | 4.62E-05 | 1.42E-05 |
| 0.90 | 0.90 | 0.1125 | 61 | 0.009311 | 0.00171 | 0.000528 | 0.00019 | 6.84E-05 | 2.11E-05 |
| 0.90 | 0.90 | 0.1125 | 81 | 0.01234 | 0.002266 | 0.0007 | 0.000252 | 9.07E-05 | 2.8E-05 |
| 0.90 | 0.90 | 0.1125 | 101 | 0.015368 | 0.002823 | 0.000871 | 0.000314 | 0.000113 | 3.48E-05 |
| 0.85 | 0.85 | 0.10625 | 1 | 0.000203 | 3.72E-05 | 1.15E-05 | 4.13E-06 | 1.49E-06 | 4.59E-07 |
| 0.85 | 0.85 | 0.10625 | 21 | 0.002904 | 0.000533 | 0.000165 | 5.93E-05 | 2.13E-05 | 6.58E-06 |
| 0.85 | 0.85 | 0.10625 | 41 | 0.005605 | 0.001029 | 0.000318 | 0.000114 | 4.12E-05 | 1.27E-05 |
| 0.85 | 0.85 | 0.10625 | 61 | 0.008306 | 0.001526 | 0.000471 | 0.00017 | 6.1E-05 | 1.88E-05 |
| 0.85 | 0.85 | 0.10625 | 81 | 0.011007 | 0.002022 | 0.000624 | 0.000225 | 8.09E-05 | 2.5E-05 |
| 0.85 | 0.85 | 0.10625 | 101 | 0.013708 | 0.002518 | 0.000777 | 0.00028 | 0.000101 | 3.11E-05 |
| 0.80 | 0.80 | 0.1 | 1 | 0.000179 | 3.3E-05 | 1.02E-05 | 3.66E-06 | 1.32E-06 | 4.07E-07 |
| 0.80 | 0.80 | 0.1 | 21 | 0.002572 | 0.000472 | 0.000146 | 5.25E-05 | 1.89E-05 | 5.83E-06 |
| 0.80 | 0.80 | 0.1 | 41 | 0.004965 | 0.000912 | 0.000281 | 0.000101 | 3.65E-05 | 1.13E-05 |
| 0.80 | 0.80 | 0.1 | 61 | 0.007357 | 0.001351 | 0.000417 | 0.00015 | 5.41E-05 | 1.67E-05 |
| 0.80 | 0.80 | 0.1 | 81 | 0.00975 | 0.001791 | 0.000553 | 0.000199 | 7.16E-05 | 2.21E-05 |
| 0.80 | 0.80 | 0.1 | 101 | 0.012142 | 0.00223 | 0.000688 | 0.000248 | 8.92E-05 | 2.75E-05 |

As mentioned earlier, to find the total energy per frequency emitted by a dipole, $\varepsilon$, we have to evaluate areas under each curve. It is easy to do that by using the Origin 6.1 program. The results are shown in Table.5.3.

**Table.5.3.** The total energies emitted by different massive rotating dipoles, with $\chi = \pi/2$.

| $\beta_{max}$ | $\beta$ | $\omega$(GU) | $M$ =0 | $M$ =0.5$M$ | $M$ = $M$ | $M$ =1.5$M$ | $M$ =2$M$ | $M$ =2.5$M$ | $M$ =3$M$ |
|---|---|---|---|---|---|---|---|---|---|
| 0.99 | 0.99 | 0.12375 | 6.81547 | 6.74091 | 5.80237 | 3.58306 | 1.83644 | 0.80341 | 0.25799 |
| 0.95 | 0.95 | 0.11875 | 0.30874 | 0.2904 | 0.1913 | 0.12004 | 0.07499 | 0.04551 | 0.02202 |
| 0.90 | 0.90 | 0.1125 | 0.06984 | 0.05572 | 0.03055 | 0.01644 | 0.00928 | 0.00473 | 0.00198 |
| 0.85 | 0.85 | 0.10625 | 0.02799 | 0.02004 | 0.00987 | 0.00515 | 0.00273 | 0.00133 | 5.96E-04 |
| 0.80 | 0.80 | 0.1 | 0.0142 | 0.00927 | 0.00421 | 0.00212 | 0.00109 | 5.36E-04 | 2.28E-04 |

Notice that, we are mainly interested in values around $\chi = \pi/2$ as this is where the NS can experience relativistic behavior ($\gamma$ becomes very high), since we assume that the dipole is embedded in the NS. Also, the very fast rotating pulsars are expected to be near-orthogonal rotators [32].

## 5.3.2.2. THE ENERGY SPECTRA FOR $\chi=\pi/4$

Although the $\chi = \pi/4$ case does not have much significance, we display its spectra to study the effect of the inclination angle. The spectra for $\chi = \pi/4$, are shown in Figs.5.11-15, for different values of $\beta$. Notice that $x + 1 = r\sin\chi/r_s$. Hence, $\lambda^2(x + 1)^3$ will become much smaller than $l(l+1)x$ in the square bracket of (5.17). Thus, we expect to match the massless case for smaller values of $l$.

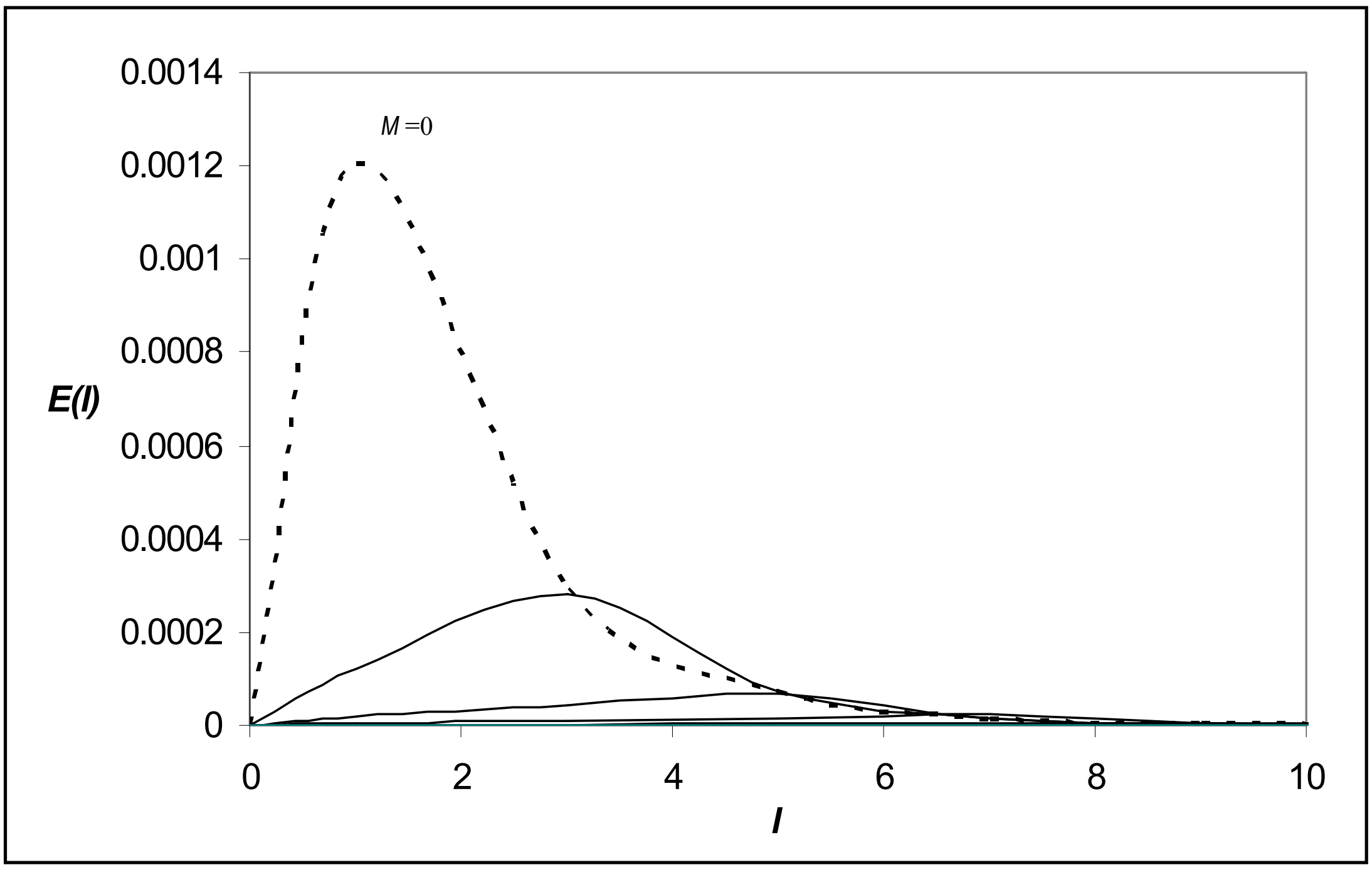

**Fig.5.11.** Energy spectra, at $\beta_{max} = 0.99$, $\chi = \pi/4$ for dipole of masses (0.5, 1, 1.5, 2, 2.5) $M_{\odot}$, from upper line to the lower one. The dashed curve is for the massless case.

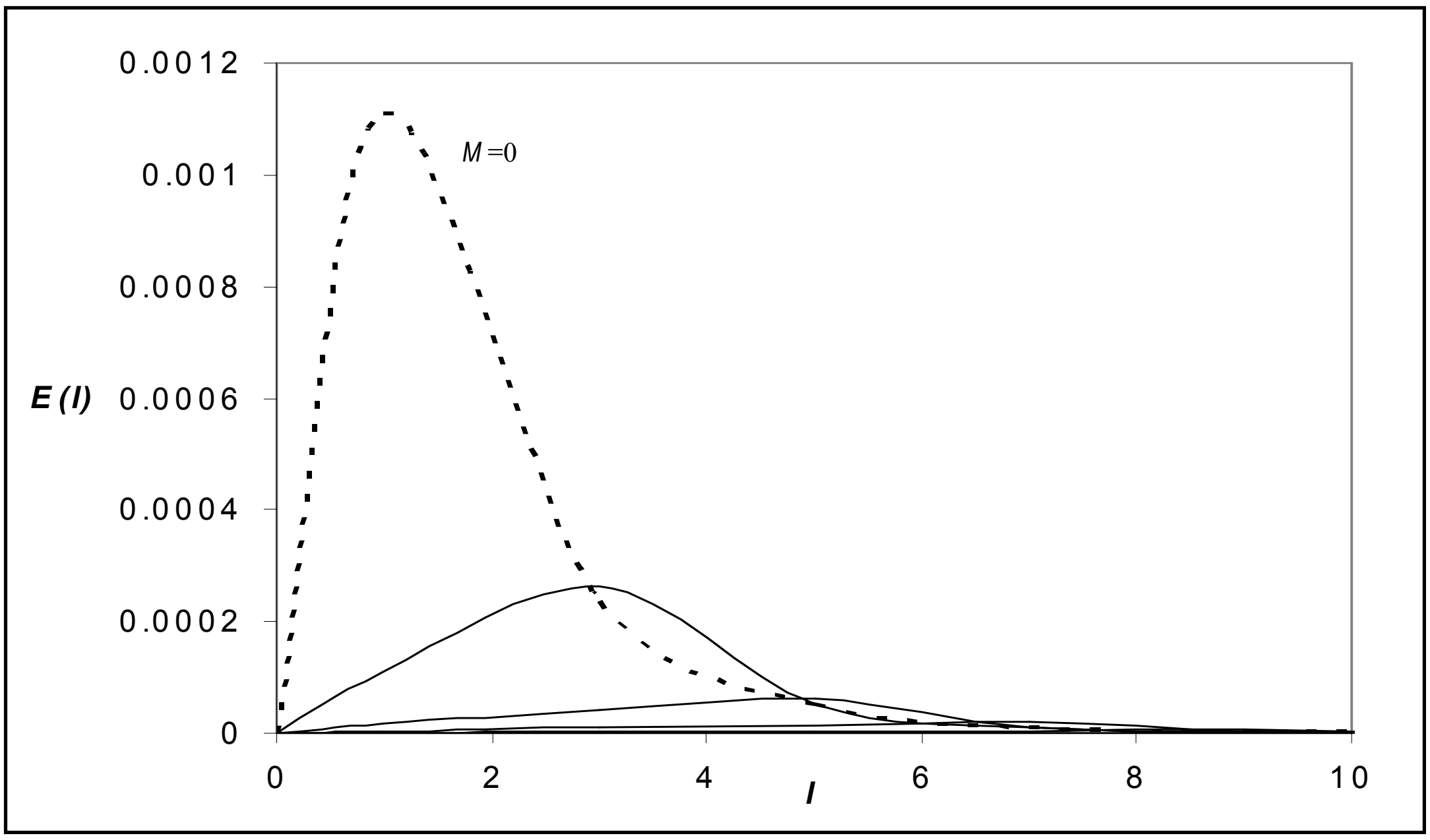

**Fig.5.12.** Energy spectra, at $\beta_{max}$ = 0.95, $\chi$ =π/4 for dipole of masses (0.5, 1, 1.5, 2, 2.5) $M_{\odot}$, from upper line to the lower one. The dashed curve is for the massless case.

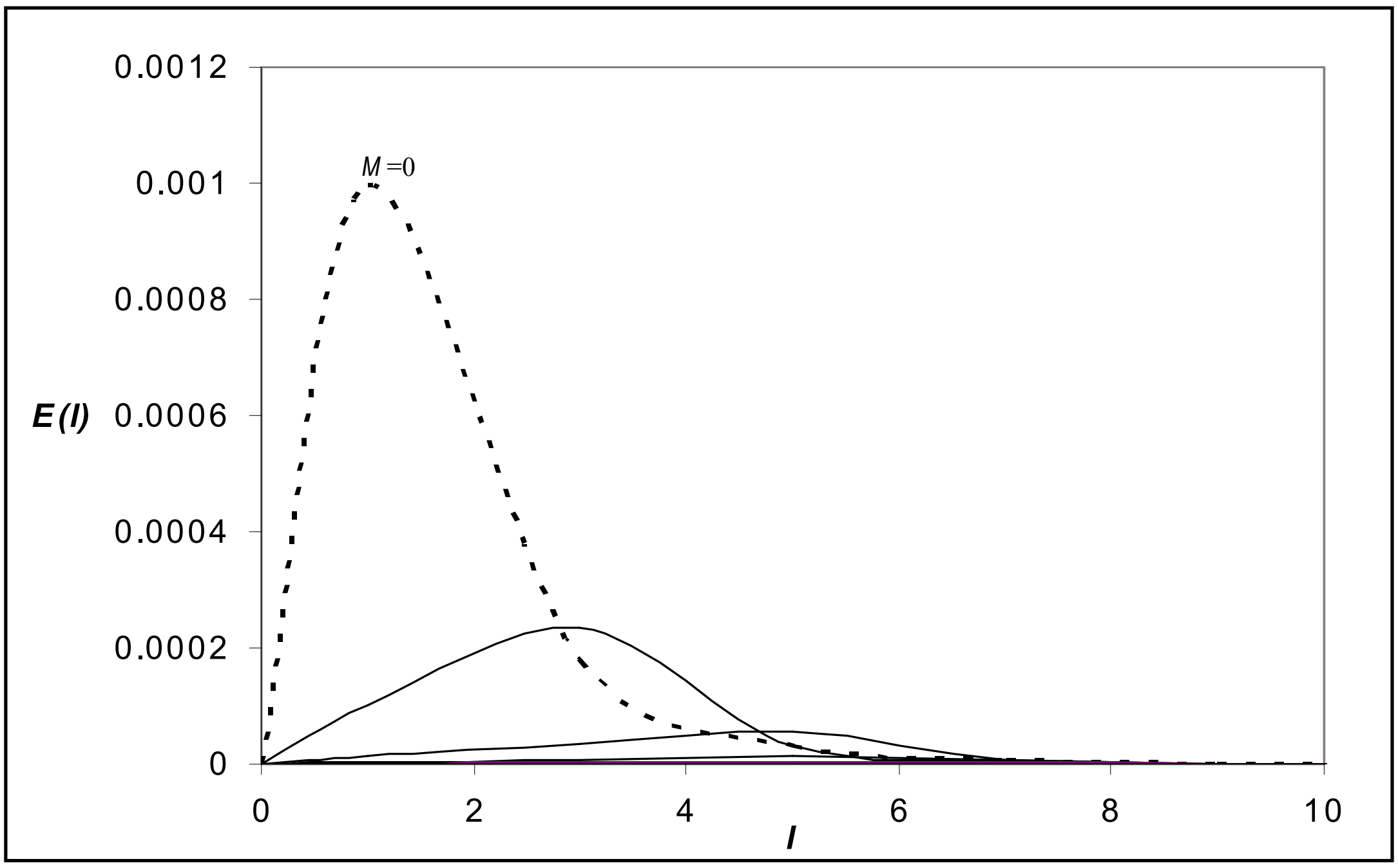

**Fig.5.13.** Energy spectra, at $\beta_{max}$= 0.90, $\chi$ =π/4, for dipole of masses (0.5, 1, 1.5, 2, 2.5) $M_{\odot}$, from upper line to the lower one. The dashed curve is for the massless case.

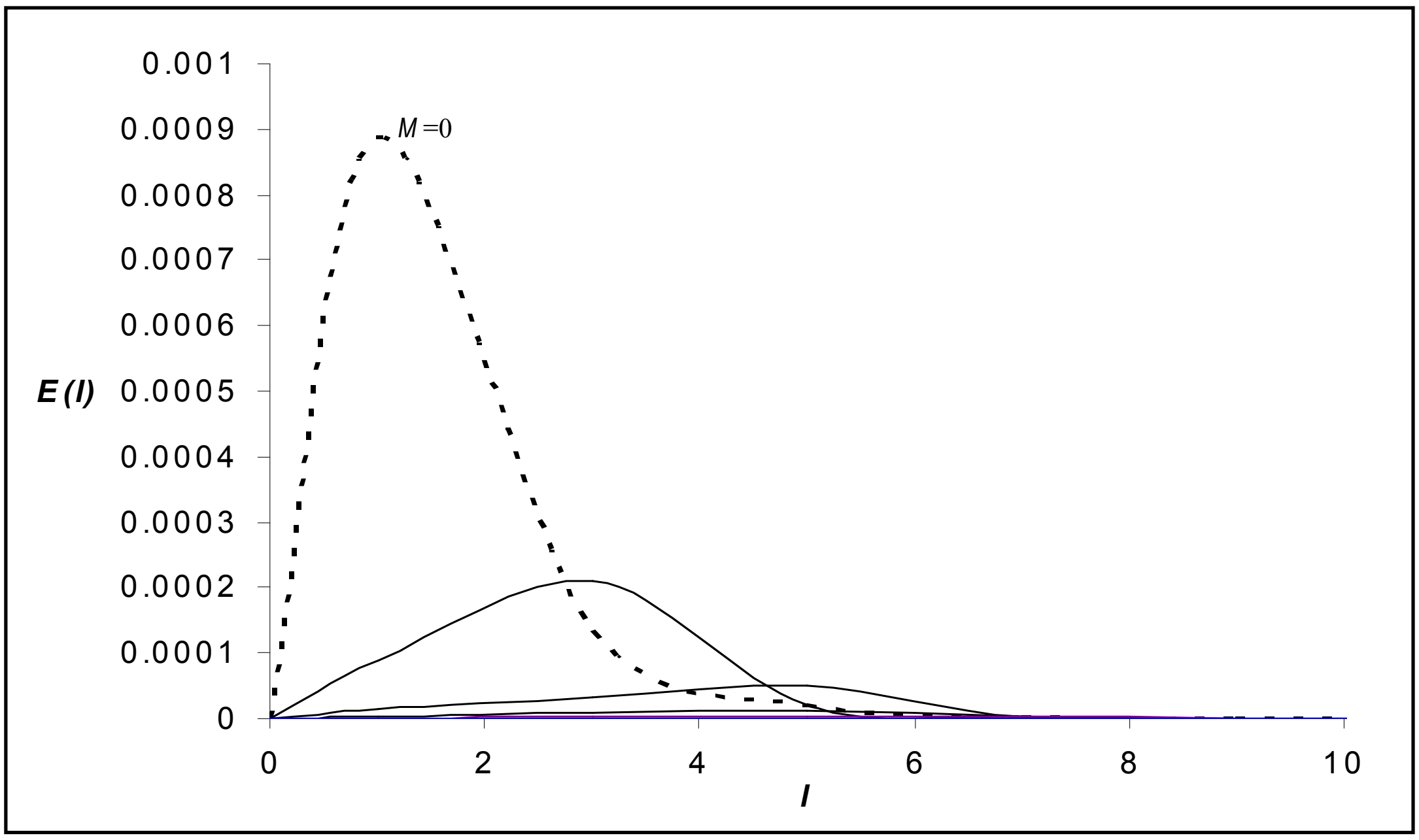

**Fig.5.14.** Energy spectra, at $\beta_{max}$ = 0.85, $\chi$ =$\pi$/4, for dipole of masses (0.5, 1, 1.5, 2, 2.5) $M_{\odot}$, from upper line to the lower one. The dashed curve is for the massless case.

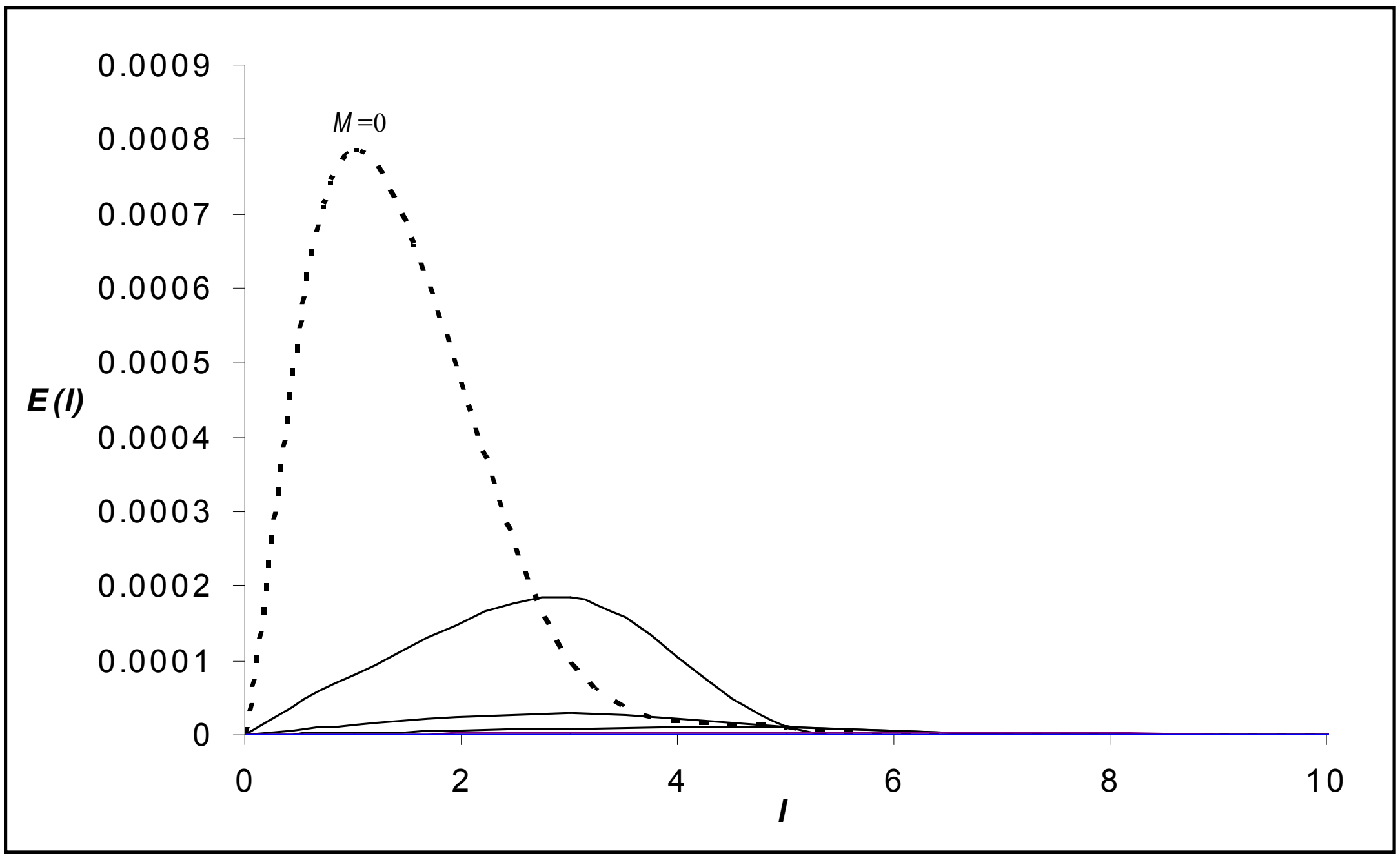

**Fig.5.15.** Energy spectra, at $\beta_{max}$ = 0.80, $\chi$ =$\pi$/4, for dipole of masses (0.5, 1, 1.5, 2, 2.5) $M_{\odot}$, from upper line to the lower one. The dashed curve is for the massless case.

Again some of the data, obtained from Figs.5.11-15, is given in Tables 5.3-4. Notice that we extend the value of $l$ up to 101, just to compare the spectra for $\chi = \pi/2$ even though the effect of mass may be neglected for the higher part of this range.

**Table.5.4.** The values of energies radiated by different massive rotating dipoles, for $\chi = \pi/4$.

| $\beta_{max}$ | $\beta$ | $\omega$(GU) | $l$ | $E$-0.5 $M_\odot$ | $E$-1 $M_\odot$ | $E$-1.5 $M_\odot$ | $E$-2 $M_\odot$ | $E$-2.5 $M_\odot$ | $E$-3 $M_\odot$ |
|---|---|---|---|---|---|---|---|---|---|
| 0.99 | 0.70 | 0.12375 | 1 | 0.000122 | 1.87E-05 | 4.40E-06 | 9.62E-07 | 9.68E-08 | 1.83E-08 |
| 0.99 | 0.70 | 0.12375 | 21 | 0.001743 | 0.000269 | 6.3E-05 | 1.38E-05 | 1.39E-06 | 2.63E-07 |
| 0.99 | 0.70 | 0.12375 | 41 | 0.003365 | 0.000519 | 0.000122 | 2.66E-05 | 2.68E-06 | 5.07E-07 |
| 0.99 | 0.70 | 0.12375 | 61 | 0.004986 | 0.000769 | 0.00018 | 3.95E-05 | 3.97E-06 | 7.51E-07 |
| 0.99 | 0.70 | 0.12375 | 81 | 0.006608 | 0.001019 | 0.000239 | 5.23E-05 | 5.26E-06 | 9.95E-07 |
| 0.99 | 0.70 | 0.12375 | 101 | 0.00823 | 0.001269 | 0.000298 | 6.51E-05 | 6.55E-06 | 1.24E-06 |
| 0.95 | 0.672 | 0.11875 | 1 | 0.000112 | 1.73E-05 | 4.05E-06 | 8.86E-07 | 8.91E-08 | 1.69E-08 |
| 0.95 | 0.672 | 0.11875 | 21 | 0.001605 | 0.000247 | 5.81E-05 | 1.27E-05 | 1.28E-06 | 2.42E-07 |
| 0.95 | 0.672 | 0.11875 | 41 | 0.003098 | 0.000478 | 0.000112 | 2.45E-05 | 2.47E-06 | 4.67E-07 |
| 0.95 | 0.672 | 0.11875 | 61 | 0.004592 | 0.000708 | 0.000166 | 3.63E-05 | 3.65E-06 | 6.92E-07 |
| 0.95 | 0.672 | 0.11875 | 81 | 0.006085 | 0.000938 | 0.00022 | 4.81E-05 | 4.85E-06 | 9.17E-07 |
| 0.95 | 0.672 | 0.11875 | 101 | 0.007578 | 0.001168 | 0.000274 | 6E-05 | 6.04E-06 | 1.14E-06 |
| 0.9 | 0.636 | 0.1125 | 1 | 0.000101 | 1.55E-05 | 3.64E-06 | 7.95E-07 | 8.00E-08 | 1.52E-08 |
| 0.9 | 0.636 | 0.1125 | 21 | 0.001441 | 0.000222 | 5.21E-05 | 1.14E-05 | 1.15E-06 | 2.17E-07 |
| 0.9 | 0.636 | 0.1125 | 41 | 0.002781 | 0.000429 | 0.000101 | 2.2E-05 | 2.21E-06 | 4.19E-07 |
| 0.9 | 0.636 | 0.1125 | 61 | 0.004121 | 0.000635 | 0.000149 | 3.26E-05 | 3.28E-06 | 6.21E-07 |
| 0.9 | 0.636 | 0.1125 | 81 | 0.005461 | 0.000842 | 0.000198 | 4.32E-05 | 4.34E-06 | 8.23E-07 |
| 0.9 | 0.636 | 0.1125 | 101 | 0.006801 | 0.001048 | 0.000246 | 5.38E-05 | 5.40E-06 | 1.03E-06 |
| 0.85 | 0.601 | 0.10625 | 1 | 8.97E-05 | 1.38E-05 | 3.24E-06 | 7.09E-07 | 7.13E-08 | 1.35E-08 |
| 0.85 | 0.601 | 0.10625 | 21 | 0.001285 | 0.000198 | 4.65E-05 | 1.02E-05 | 1.02E-06 | 1.94E-07 |
| 0.85 | 0.601 | 0.10625 | 41 | 0.00248 | 0.000382 | 8.97E-05 | 1.96E-05 | 1.97E-06 | 3.74E-07 |
| 0.85 | 0.601 | 0.10625 | 61 | 0.003676 | 0.000567 | 0.000133 | 2.91E-05 | 2.93E-06 | 5.54E-07 |
| 0.85 | 0.601 | 0.10625 | 81 | 0.004871 | 0.000751 | 0.000176 | 3.85E-05 | 3.87E-06 | 7.34E-07 |
| 0.85 | 0.601 | 0.10625 | 101 | 0.006067 | 0.000935 | 0.000219 | 4.8E-05 | 4.82E-06 | 9.14E-07 |
| 0.8 | 0.566 | 0.1 | 1 | 7.94E-05 | 1.22E-05 | 2.87E-06 | 6.28E-07 | 6.32E-08 | 1.20E-08 |
| 0.8 | 0.566 | 0.1 | 21 | 0.001138 | 0.000175 | 4.12E-05 | 9.01E-06 | 9.06E-07 | 1.72E-07 |
| 0.8 | 0.566 | 0.1 | 41 | 0.002197 | 0.000339 | 7.95E-05 | 1.74E-05 | 1.75E-06 | 3.30E-07 |
| 0.8 | 0.566 | 0.1 | 61 | 0.003256 | 0.000502 | 0.000118 | 2.58E-05 | 2.59E-06 | 4.90E-07 |
| 0.8 | 0.566 | 0.1 | 81 | 0.004315 | 0.000665 | 0.000156 | 3.41E-05 | 3.44E-06 | 6.46E-07 |
| 0.8 | 0.566 | 0.1 | 101 | 0.005374 | 0.000828 | 0.000194 | 4.25E-05 | 4.28E-06 | 8.04E-07 |

Notice that for $M$ more than about 2.82 $M$ the value of $\alpha = x_0 = (a \sin\chi / 2M_{NS} - 1)$ becomes zero at $\chi$ = π/4. Thus, all $c_n$, defined in (5.24-25) become infinite, and hence the Green's function, (5.9), is indeterminate. This makes it impossible to use the procedure beyond that mass. We have given the table for $M = 3M_{\odot}$ for completeness, but the value of the energy radiated is not reliable there. This problem did not arise for $\chi$ = π/2 as the radius, in solar mass units, was greater than 3. However, as seen later, it arises for even smaller masses for smaller $\chi$. For larger dipole size the problem would appear for still smaller values of $\chi$.

Again the total energy per frequency, $\varepsilon$, emitted by a dipole, can be obtained by using Origin 6.1. The results are shown in Table.5.5.

**Table.5.5.** The total energies emitted by different massive rotating dipoles, with $\chi$ = π/4.

| $\beta_{max}$ | $\beta$ | $\omega$(GU) | $M$ =0 | $M$ =0.5$M$ | $M$ = $M$ | $M$ =1.5$M$ | $M$ =2$M$ | $M$ =2.5$M$ |
|---|---|---|---|---|---|---|---|---|
| 0.99 | 0.70 | 0.12375 | 0.00257 | 9.34E-04 | 2.95E-04 | 1.13E-04 | 3.70E-05 | 5.45E-06 |
| 0.95 | 0.672 | 0.11875 | 0.00226 | 8.16E-04 | 2.58E-04 | 1.00E-04 | 3.30E-05 | 4.18E-06 |
| 0.9 | 0.636 | 0.1125 | 0.00193 | 6.96E-04 | 2.22E-04 | 6.19E-05 | 2.08E-05 | 3.24E-06 |
| 0.85 | 0.601 | 0.10625 | 0.00164 | 5.98E-04 | 1.93E-04 | 5.00E-05 | 1.75E-05 | 2.69E-06 |
| 0.8 | 0.566 | 0.1 | 0.0014 | 5.15E-04 | 1.01E-04 | 4.16E-05 | 1.51E-05 | 1.80E-06 |

.

## 5.3.2.3. THE ENERGY SPECTRA FOR $\chi=\pi/6$

This case is even less significant than the previous one, but is included for completeness. The spectra for $\chi$ = π/6, at different $\beta$s, are shown in Figs.5.16-20, and some data are tabulated in Table.5.6. Because of the reason mentioned before, we expect to match the massless case in even smaller values of $l$ and we extend its tabulated values to 101 to compare it with the other cases. Here, the problem, mentioned above, arises at $M = 2M_\odot$, and so the values of radiated energy are not meaningful in the last three columns. Also, the total radiated energies for the first three cases are shown in Table 5.7.

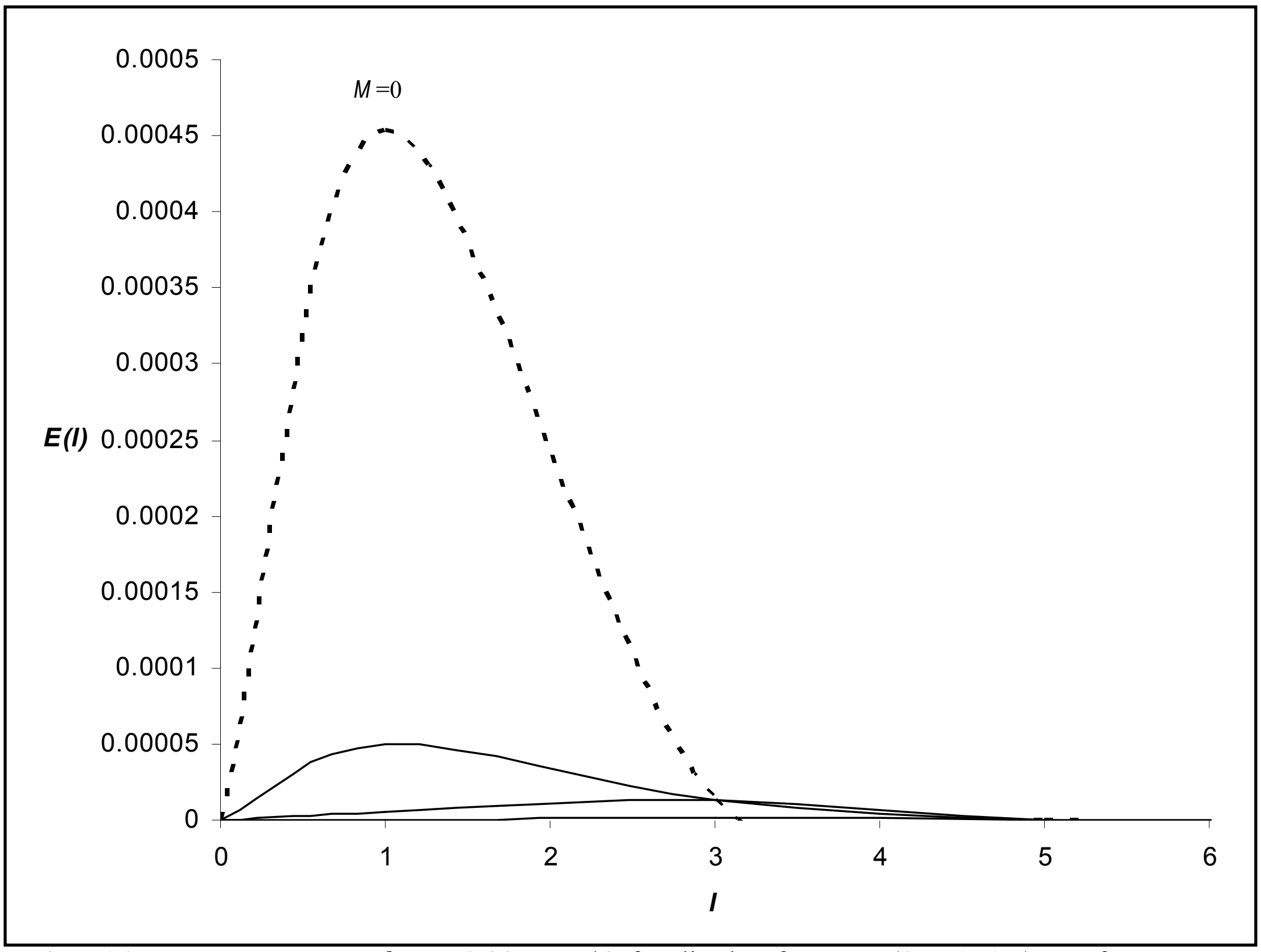

**Fig.5.16.** Energy spectra, at $\beta_{max}$ = 0.99, $\chi$ =π/6, for dipole of masses (0.5, 1, 1.5) $M_\odot$, from upper line to the lower one. The dashed curve is for the massless case.

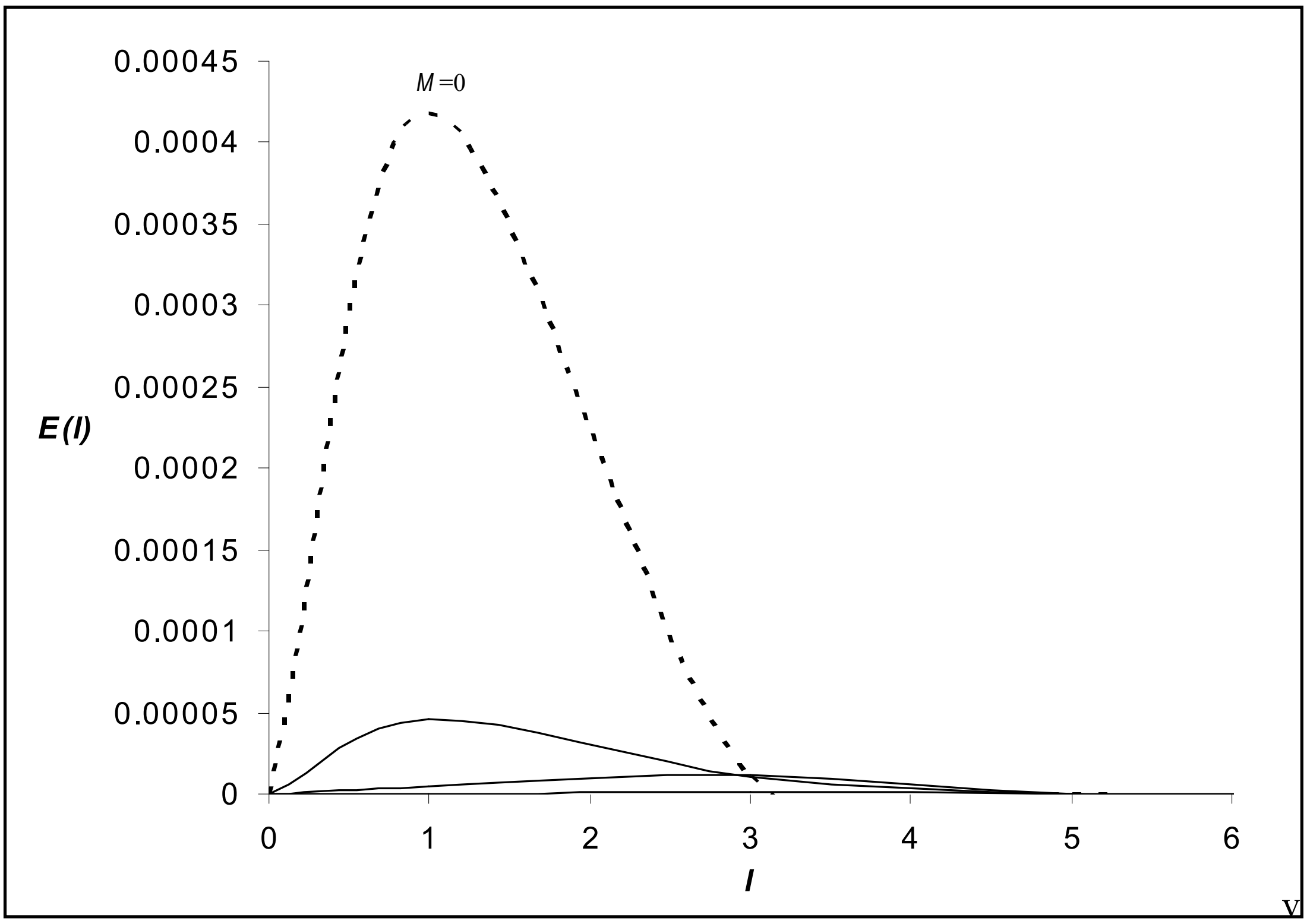

**Fig.5.17.** Energy spectra, at $\beta_{max} = 0.95$, $\chi = \pi/6$, for dipole of masses (0.5, 1, 1.5) $M_{\odot}$, from upper line to the lower one. The dashed curve is for the massless case.

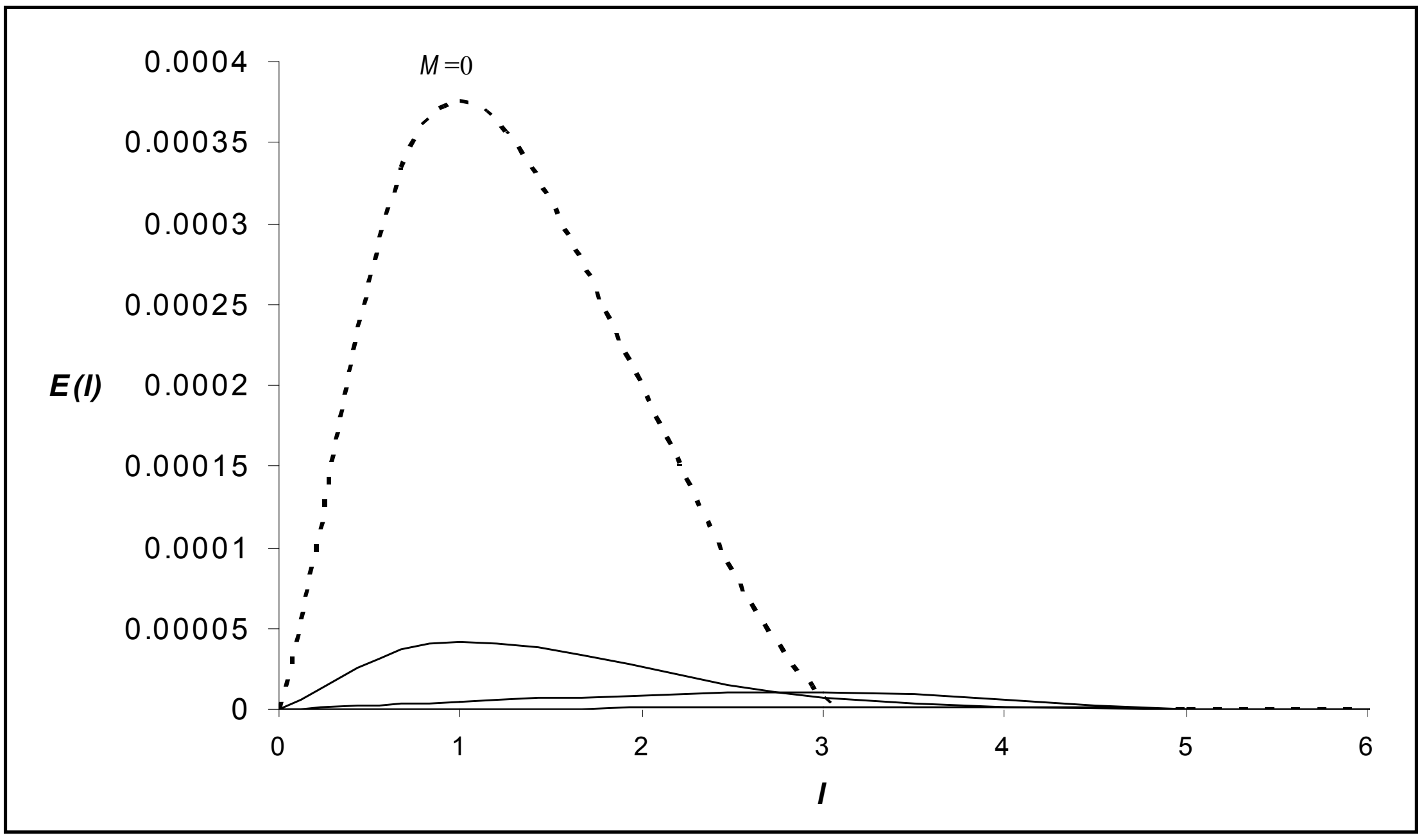

**Fig.5.18.** Energy spectra, at $\beta_{max} = 0.90$, $\chi = \pi/6$, for dipole of masses (0.5, 1, 1.5) $M_{\odot}$, from upper line to the lower one. The dashed curve is for the massless case.

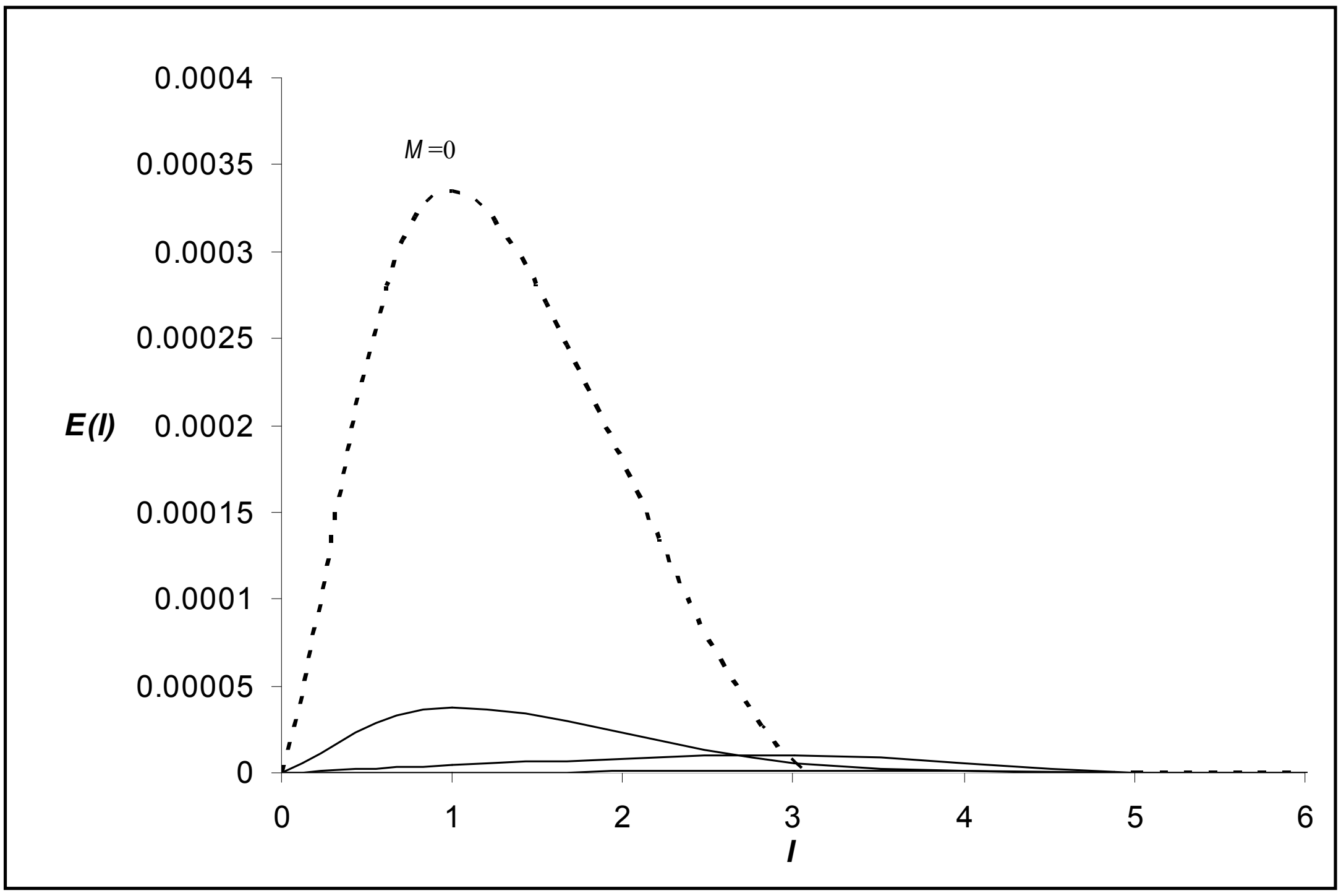

**Fig.5.19.** Energy spectra, at $\beta_{max} = 0.85$, $\chi = \pi/6$, for dipole of masses (0.5, 1, 1.5) $M_{\odot}$, from upper line to the lower one. The dashed curve is for the massless case.

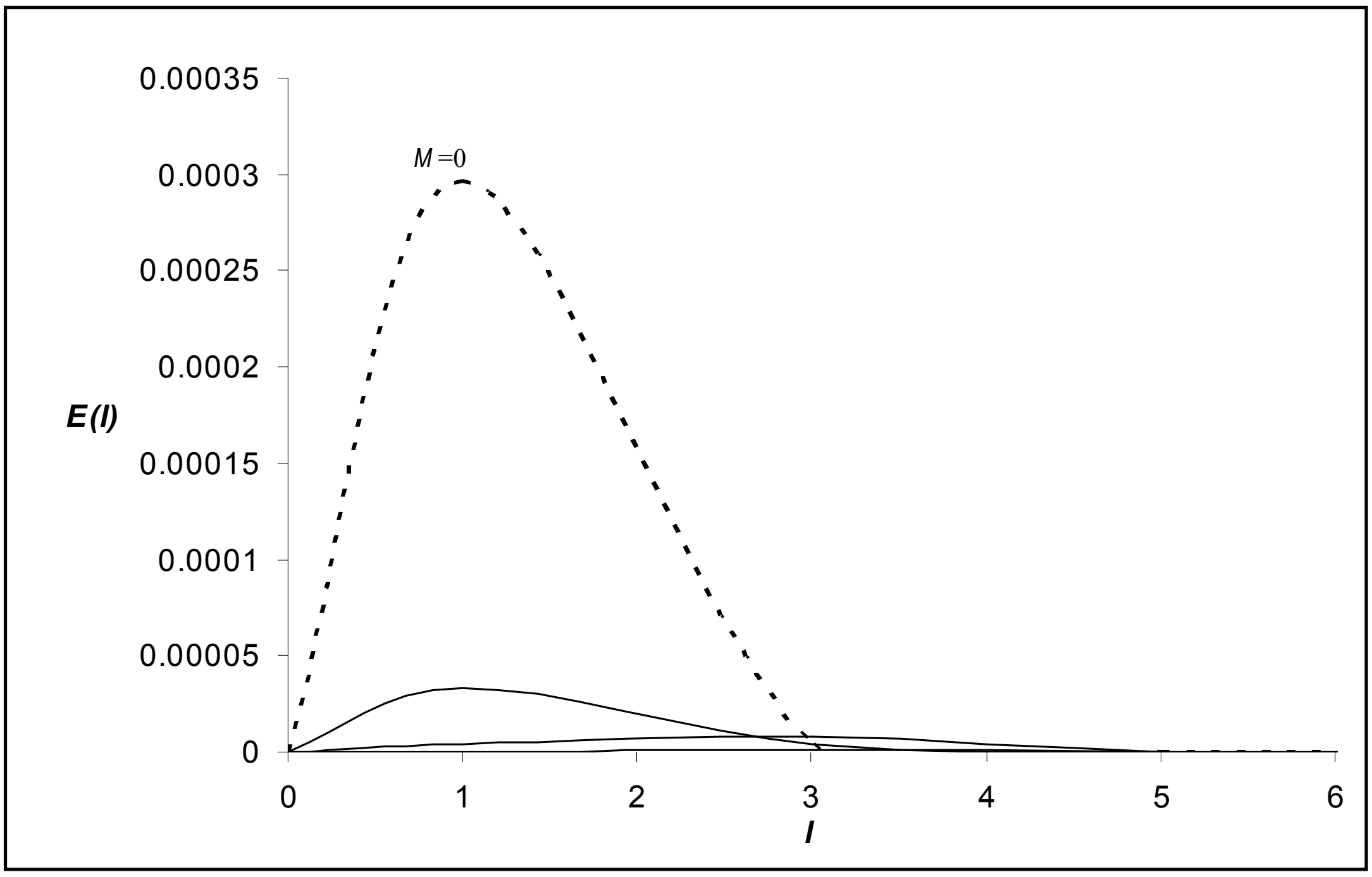

**Fig.5.20.** Energy spectra, at $\beta_{max} = 0.80$, $\chi = \pi/6$, for dipole of masses (0.5, 1, 1.5) $M_{\odot}$, from upper line to the lower one. The dashed curve is for the massless case.

**Table.5.6.** The values of energies radiated by different massive rotating dipoles, with $\chi = \pi/6$.

| $\beta_{max}$ | $\beta$ | $\omega$(GU) | $l$ | $E$-0.5 $M_\odot$ | $E$-1 $M_\odot$ | $E$-1.5 $M_\odot$ | $E$-2 $M_\odot$ | $E$-2.5 $M_\odot$ | $E$-3 $M_\odot$ |
|---|---|---|---|---|---|---|---|---|---|
| 0.99 | 0.495 | 0.12375 | 1 | 5.05E-05 | 5.61E-06 | 6.23E-07 | Indeterminate | 2.24E-07 | 6.23E-07 |
| 0.99 | 0.495 | 0.12375 | 21 | 0.000723 | 8.04E-05 | 8.93E-06 | Indeterminate | 3.22E-06 | 8.93E-06 |
| 0.99 | 0.495 | 0.12375 | 41 | 0.001396 | 0.000155 | 1.724E-05 | Indeterminate | 6.20E-06 | 1.72E-05 |
| 0.99 | 0.495 | 0.12375 | 61 | 0.002069 | 0.00023 | 2.56E-05 | Indeterminate | 9.18E-06 | 2.56E-05 |
| 0.99 | 0.495 | 0.12375 | 81 | 0.002742 | 0.000305 | 3.39E-05 | Indeterminate | 1.22E-05 | 3.39E-05 |
| 0.99 | 0.495 | 0.12375 | 101 | 0.003415 | 0.000379 | 4.217E-05 | Indeterminate | 1.51E-05 | 4.22E-05 |
| 0.95 | 0.475 | 0.11875 | 1 | 4.65E-05 | 5.16E-06 | 5.74E-07 | Indeterminate | 2.07E-07 | 5.74E-07 |
| 0.95 | 0.475 | 0.11875 | 21 | 0.000666 | 7.4E-05 | 8.22E-06 | Indeterminate | 2.96E-06 | 8.22E-06 |
| 0.95 | 0.475 | 0.11875 | 41 | 0.001286 | 0.000143 | 1.59E-05 | Indeterminate | 5.70E-06 | 1.59E-05 |
| 0.95 | 0.475 | 0.11875 | 61 | 0.001906 | 0.000212 | 2.35E-05 | Indeterminate | 8.47E-06 | 2.35E-05 |
| 0.95 | 0.475 | 0.11875 | 81 | 0.002525 | 0.000281 | 3.12E-05 | Indeterminate | 1.12E-05 | 3.12E-05 |
| 0.95 | 0.475 | 0.11875 | 101 | 0.003145 | 0.000349 | 3.88E-05 | Indeterminate | 1.39E-05 | 3.88E-05 |
| 0.9 | 0.450 | 0.1125 | 1 | 4.17E-05 | 4.63E-06 | 5.15E-07 | Indeterminate | 1.85E-07 | 5.15E-07 |
| 0.9 | 0.450 | 0.1125 | 21 | 0.000598 | 6.64E-05 | 7.38E-06 | Indeterminate | 2.66E-06 | 7.38E-06 |
| 0.9 | 0.450 | 0.1125 | 41 | 0.001154 | 0.000128 | 1.42E-05 | Indeterminate | 5.13E-06 | 1.42E-05 |
| 0.9 | 0.450 | 0.1125 | 61 | 0.00171 | 0.00019 | 2.11E-05 | Indeterminate | 7.60E-06 | 2.11E-05 |
| 0.9 | 0.450 | 0.1125 | 81 | 0.002266 | 0.000252 | 2.80E-05 | Indeterminate | 1.01E-05 | 2.80E-05 |
| 0.9 | 0.450 | 0.1125 | 101 | 0.002823 | 0.000314 | 3.49E-05 | Indeterminate | 1.25E-05 | 3.49E-05 |
| 0.85 | 0.425 | 0.10625 | 1 | 3.72E-05 | 4.13E-06 | 4.60E-07 | Indeterminate | 1.65E-07 | 4.60E-07 |
| 0.85 | 0.425 | 0.10625 | 21 | 0.000533 | 5.93E-05 | 6.58E-06 | Indeterminate | 2.37E-06 | 6.58E-06 |
| 0.85 | 0.425 | 0.10625 | 41 | 0.001029 | 0.000114 | 1.27E-05 | Indeterminate | 4.58E-06 | 1.27E-05 |
| 0.85 | 0.425 | 0.10625 | 61 | 0.001526 | 0.00017 | 1.88E-05 | Indeterminate | 6.78E-06 | 1.88E-05 |
| 0.85 | 0.425 | 0.10625 | 81 | 0.002022 | 0.000225 | 2.50E-05 | Indeterminate | 9.00E-06 | 2.50E-05 |
| 0.85 | 0.425 | 0.10625 | 101 | 0.002518 | 0.00028 | 3.11E-05 | Indeterminate | 1.12E-05 | 3.11E-05 |
| 0.8 | 0.40 | 0.1 | 1 | 3.3E-05 | 3.66E-06 | 4.07E-07 | Indeterminate | 1.46E-07 | 4.07E-07 |
| 0.8 | 0.40 | 0.1 | 21 | 0.000472 | 5.25E-05 | 5.83E-06 | Indeterminate | 2.10E-06 | 5.83E-06 |
| 0.8 | 0.40 | 0.1 | 41 | 0.000912 | 0.000101 | 1.13E-05 | Indeterminate | 4.05E-06 | 1.13E-05 |
| 0.8 | 0.40 | 0.1 | 61 | 0.001351 | 0.00015 | 1.67E-05 | Indeterminate | 6.01E-06 | 1.67E-05 |
| 0.8 | 0.40 | 0.1 | 81 | 0.001791 | 0.000199 | 2.21E-05 | Indeterminate | 7.95E-06 | 2.21E-05 |
| 0.8 | 0.40 | 0.1 | 101 | 0.00223 | 0.000248 | 2.75E-05 | Indeterminate | 9.90E-06 | 2.75E-05 |

**Table.5.7.** The total energies emitted by different massive rotating dipoles, for $\chi = \pi/6$.

| $\beta_{max}$ | $\beta$ | $\omega$(GU) | $M$ =0 | $M$ =0.5$M$ | $M$ = $M$ | $M$ =1.5$M$ |
|---|---|---|---|---|---|---|
| 0.99 | 0.495 | 0.12375 | 7.07E-04 | 1.02E-04 | 3.54E-05 | 4.61E-06 |
| 0.95 | 0.475 | 0.11875 | 6.48E-04 | 9.04E-05 | 3.24E-05 | 4.07E-06 |
| 0.9 | 0.450 | 0.1125 | 5.78E-04 | 7.76E-05 | 2.88E-05 | 3.49E-06 |
| 0.85 | 0.425 | 0.10625 | 5.13E-04 | 6.67E-05 | 2.57E-05 | 3.01E-06 |
| 0.8 | 0.40 | 0.1 | 4.53E-04 | 5.71E-05 | 2.27E-05 | 2.61E-06 |

## 5.3.2.4. THE ENERGY SPECTRA FOR $\chi=\pi/8$

Finally, the spectra for $\chi$ = π/8, at different $\beta$s, are shown in Figs.5.21-25, and some data are tabulated in Table.5.8. Here, the problem arises just after $M = 1.5M_\odot$. Thus, we only need for the first three cases. The total radiated energies per frequencies for these cases are shown in Table 5.9. We do not consider the energy of the dipoles of masses greater than $1.5M_\odot$ for the same reason mentioned before.

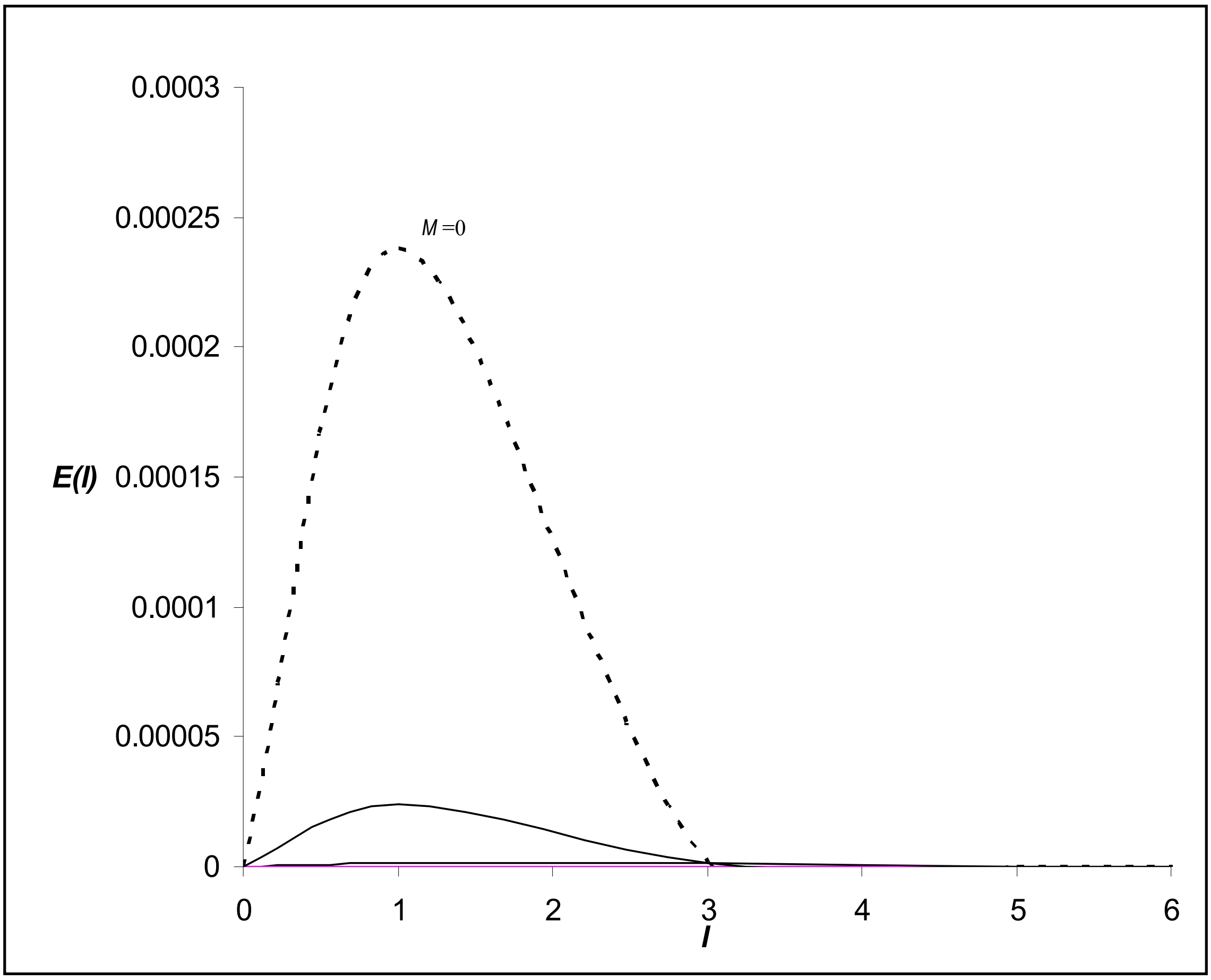

**Fig.5.21.** Energy spectra, at $\beta_{max}$ = 0.99, $\chi$ =π/8, for dipole of masses (0.5, 1, 1.5) $M_\odot$. The dashed curve is for the massless case.

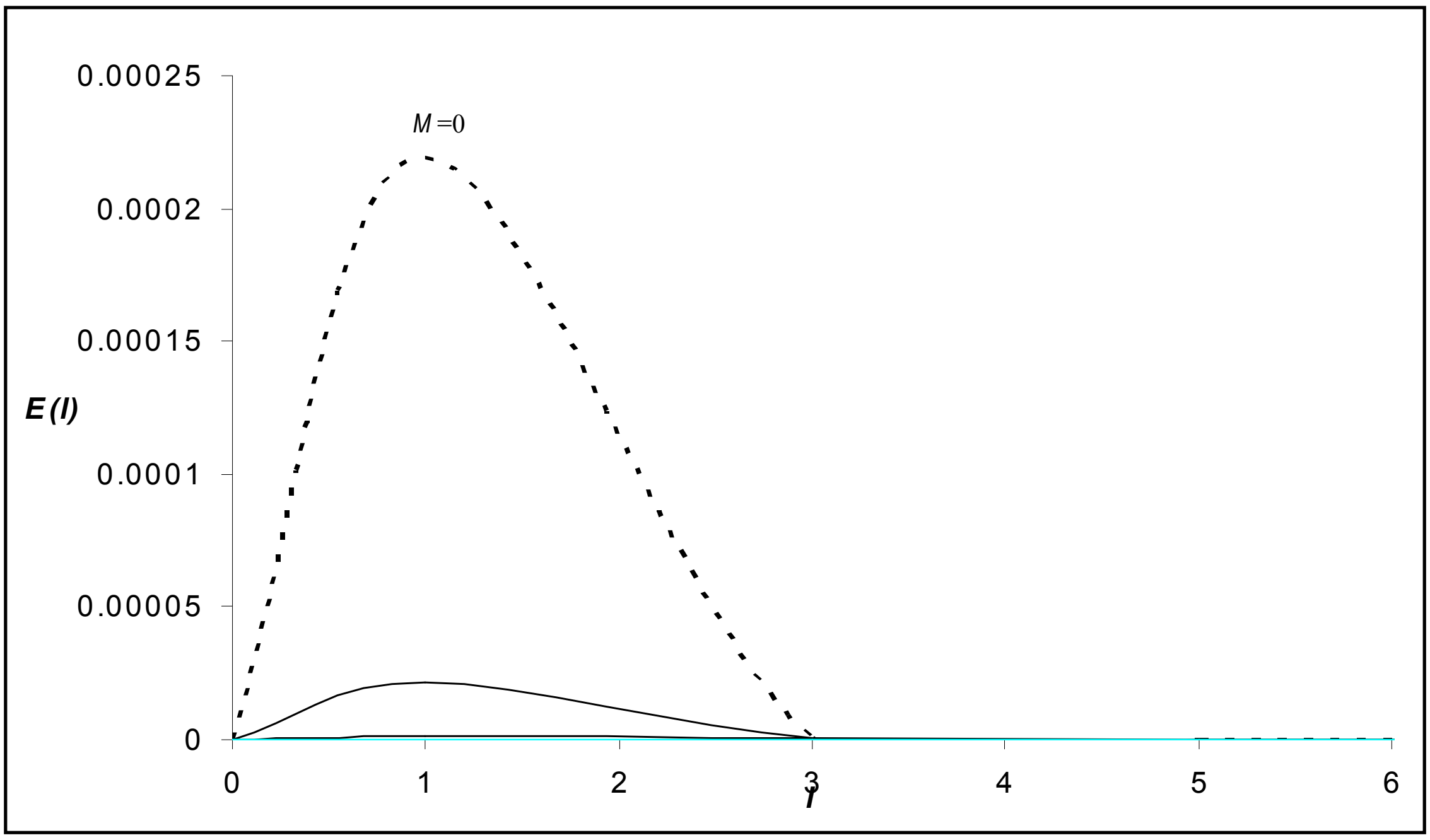

**Fig.5.22.** Energy spectra, at $\beta_{max}$ = 0.95, $\chi$ =$\pi$/8, for dipole of masses (0.5, 1, 1.5) $M_{\odot}$. The dashed curve is for the massless case.

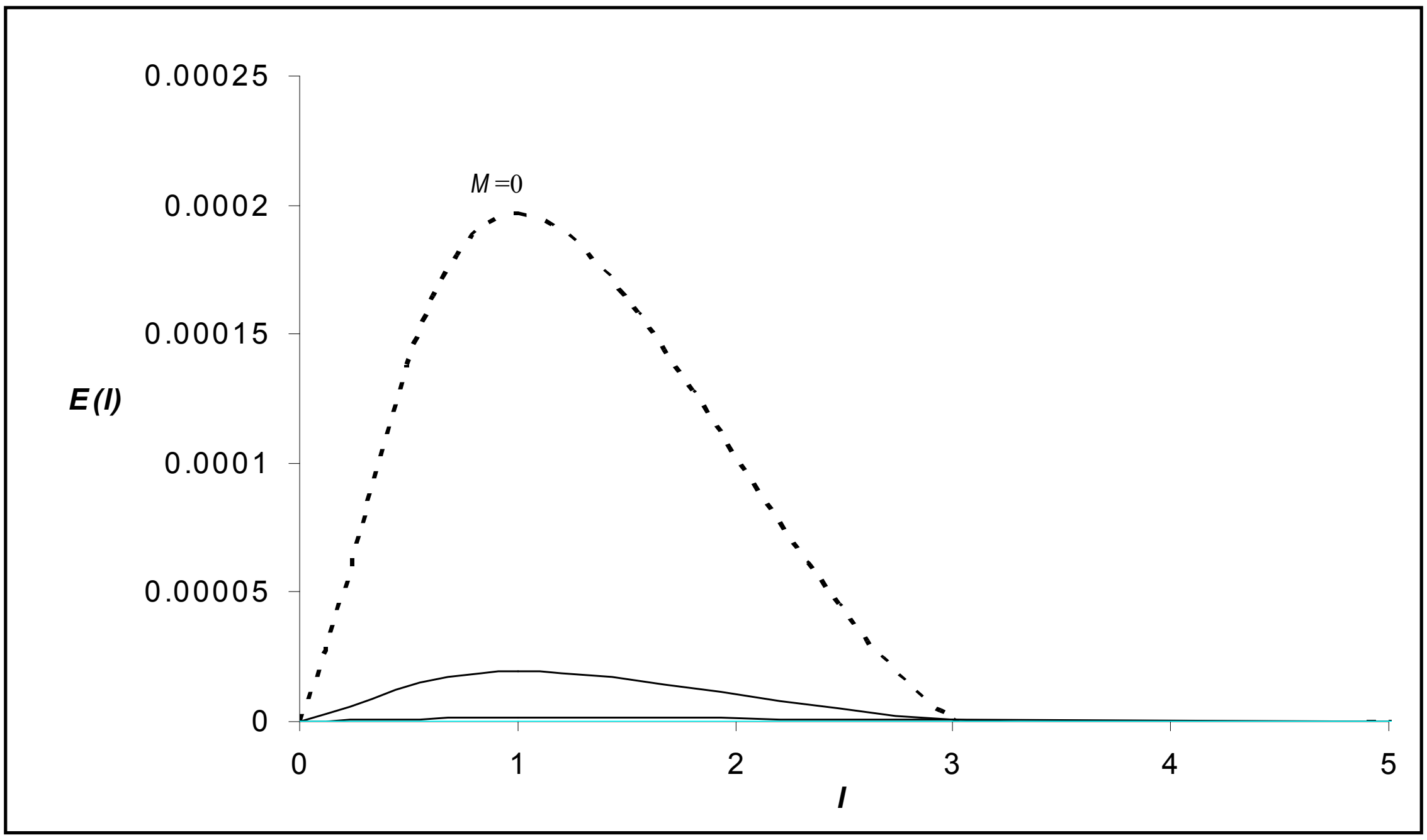

**Fig.5.23.** Energy spectra, at $\beta_{max}$ = 0.90, $\chi$ =$\pi$/8, for dipole of masses (0.5, 1, 1.5) $M_{\odot}$. The dashed curve is for the massless case.

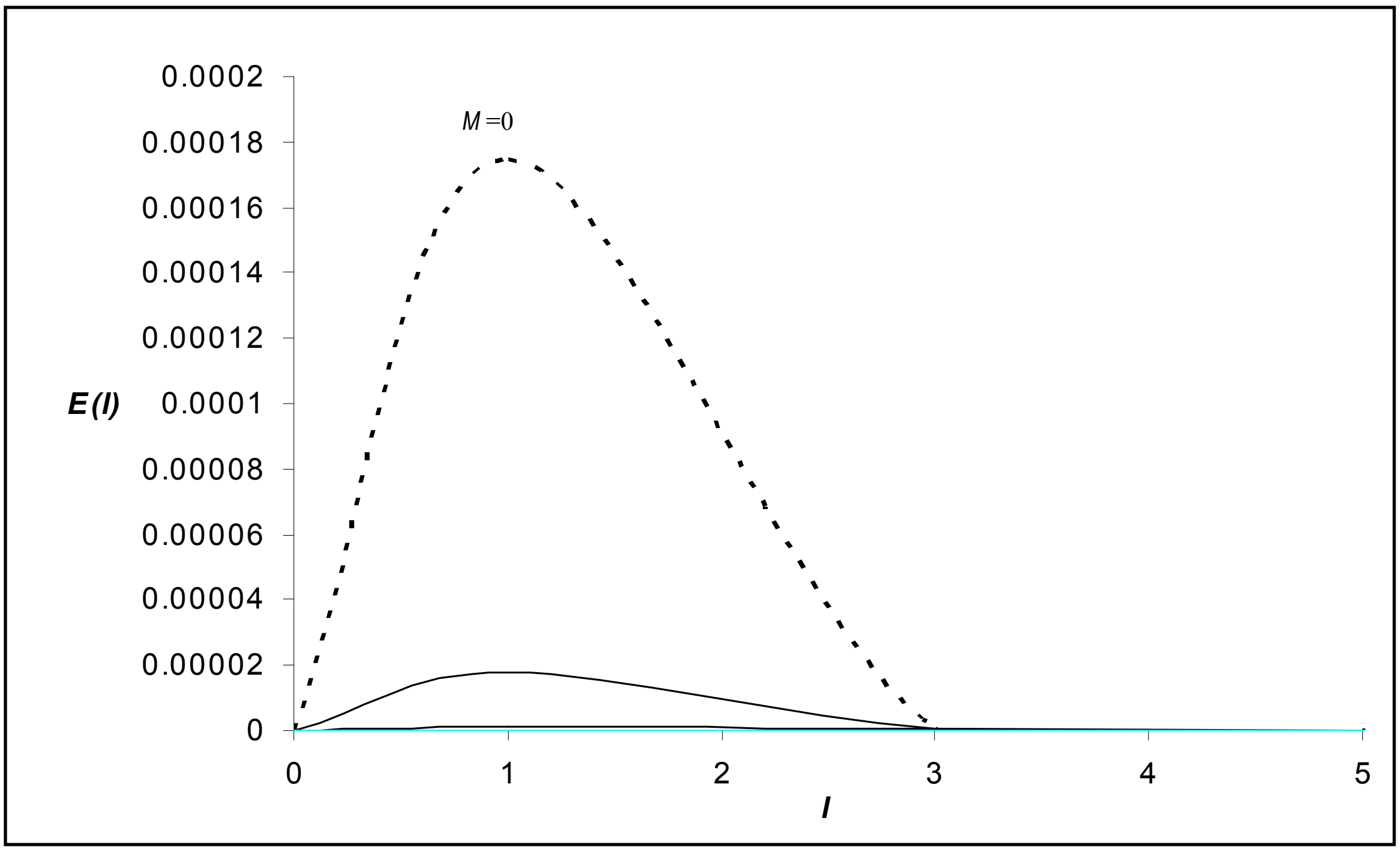

**Fig.5.24.** Energy spectra, at $\beta_{max}$ = 0.85, $\chi$ =$\pi$/8, for dipole of masses (0.5, 1, 1.5) $M_{\odot}$. The dashed curve is for the massless case.

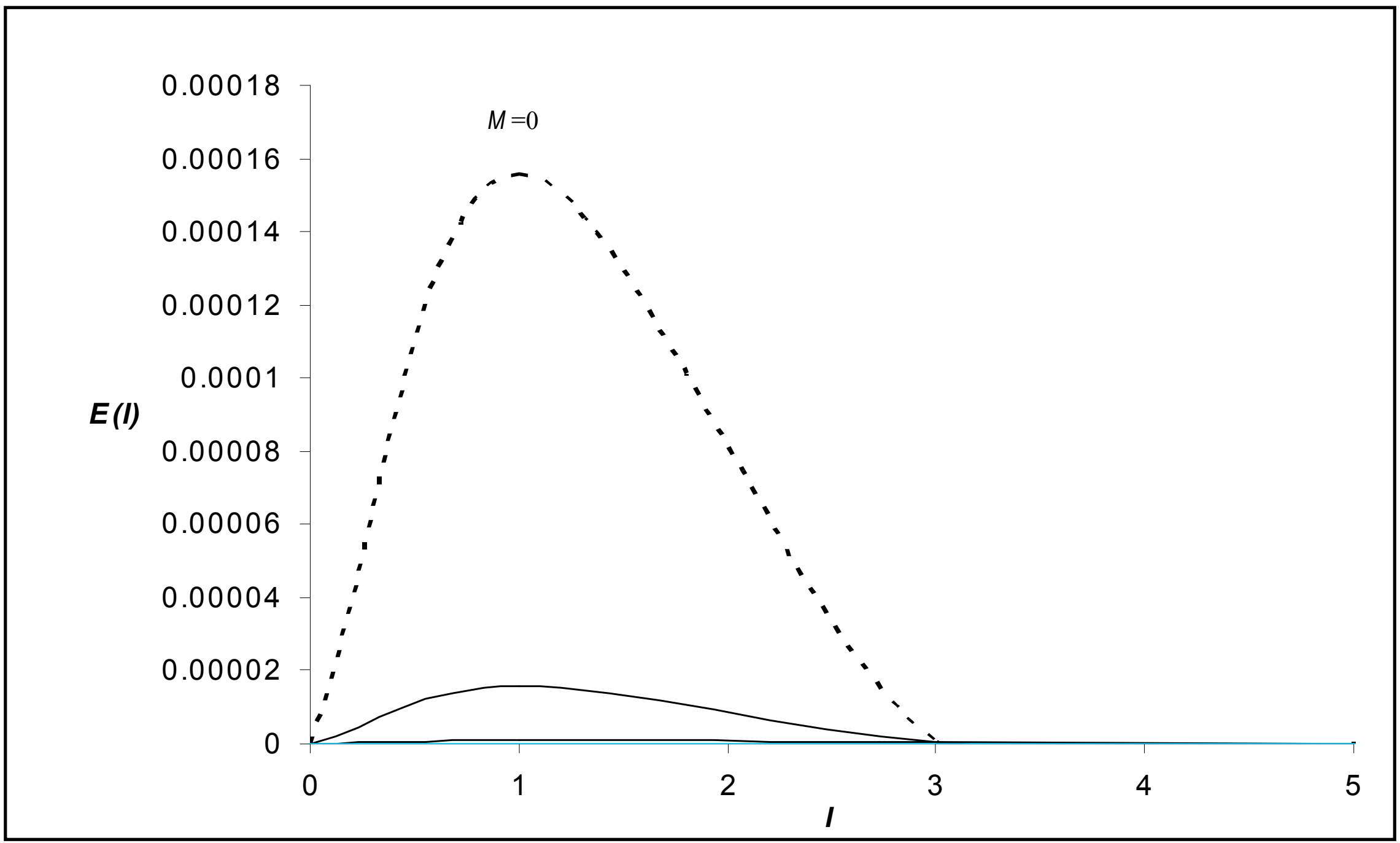

**Fig.5.25.** Energy spectra, at $\beta_{max}$ = 0.80, $\chi$ =$\pi$/8, for dipole of masses (0.5, 1, 1.5) $M_{\odot}$. The dashed curve is for the massless case.

**Table.5.8.** The values of energies radiated by different massive rotating dipoles, for $\chi = \pi/8$.

| $\beta_{max}$ | $\beta$ | $\omega$(GU) | $l$ | $E$-0.5 $M_{\odot}$ | $E$-1 $M_{\odot}$ | $E$-1.5 $M_{\odot}$ |
|---|---|---|---|---|---|---|
| 0.99 | 0.379 | 0.12375 | 1 | 2.38E-05 | 1.58E-06 | 2.35E-09 |
| 0.99 | 0.379 | 0.12375 | 21 | 0.000342 | 2.26E-05 | 3.37E-08 |
| 0.99 | 0.379 | 0.12375 | 41 | 0.000659 | 4.37E-05 | 6.51E-08 |
| 0.99 | 0.379 | 0.12375 | 61 | 0.000977 | 6.48E-05 | 9.64E-08 |
| 0.99 | 0.379 | 0.12375 | 81 | 0.001295 | 8.58E-05 | 1.28E-07 |
| 0.99 | 0.379 | 0.12375 | 101 | 0.001613 | 1.07E-04 | 1.59E-07 |
| 0.95 | 0.364 | 0.11875 | 1 | 2.19E-05 | 1.45E-06 | 2.17E-09 |
| 0.95 | 0.364 | 0.11875 | 21 | 0.000315 | 2.08497E-05 | 3.11E-08 |
| 0.95 | 0.364 | 0.11875 | 41 | 0.000607 | 4.02E-05 | 6.00E-08 |
| 0.95 | 0.364 | 0.11875 | 61 | 0.0009 | 5.96E-05 | 8.90E-08 |
| 0.95 | 0.364 | 0.11875 | 81 | 0.001192 | 7.90488E-05 | 1.18E-07 |
| 0.95 | 0.364 | 0.11875 | 101 | 0.001485 | 9.84E-05 | 1.47E-07 |
| 0.9 | 0.344 | 0.1125 | 1 | 1.97E-05 | 1.30E-06 | 1.95E-09 |
| 0.9 | 0.344 | 0.1125 | 21 | 0.000282 | 1.87E-05 | 2.79E-08 |
| 0.9 | 0.344 | 0.1125 | 41 | 0.000545 | 3.61E-05 | 5.39E-08 |
| 0.9 | 0.344 | 0.1125 | 61 | 0.000808 | 5.35E-05 | 7.98E-08 |
| 0.9 | 0.344 | 0.1125 | 81 | 0.00107 | 7.1E-05 | 1.06E-07 |
| 0.9 | 0.344 | 0.1125 | 101 | 0.001333 | 8.84E-05 | 1.32E-07 |
| 0.85 | 0.325 | 0.10625 | 1 | 1.76E-05 | 1.16E-06 | 1.74E-09 |
| 0.85 | 0.325 | 0.10625 | 21 | 0.000252 | 1.67E-05 | 2.49E-08 |
| 0.85 | 0.325 | 0.10625 | 41 | 0.000486 | 3.22E-05 | 4.81E-08 |
| 0.85 | 0.325 | 0.10625 | 61 | 0.00072 | 4.78E-05 | 7.12E-08 |
| 0.85 | 0.325 | 0.10625 | 81 | 0.000955 | 6.33E-05 | 9.44E-08 |
| 0.85 | 0.325 | 0.10625 | 101 | 0.001189 | 7.88E-05 | 1.18E-07 |
| 0.8 | 0.306 | 0.1 | 1 | 1.56E-05 | 1.03E-06 | 1.54E-09 |
| 0.8 | 0.306 | 0.1 | 21 | 0.000223 | 1.48E-05 | 2.20E-08 |
| 0.8 | 0.306 | 0.1 | 41 | 0.000431 | 2.85E-05 | 4.25E-08 |
| 0.8 | 0.306 | 0.1 | 61 | 0.000638 | 4.23E-05 | 6.29E-08 |
| 0.8 | 0.306 | 0.1 | 81 | 0.000846 | 5.61E-05 | 8.34E-08 |
| 0.8 | 0.306 | 0.1 | 101 | 0.001053 | 6.98E-05 | 1.04E-07 |

**Table.5.9.** The total energies emitted by different massive rotating dipoles, for $\chi = \pi/8$.

| $\beta_{max}$ | $\beta$ | $\omega$(GU) | $M$ =0 | $M$ =0.5$M$ | $M$ = $M$ | $M$ =1.5$M$ |
|---|---|---|---|---|---|---|
| 0.99 | 0.379 | 0.12375 | 3.60E-04 | 3.82E-05 | 4.81E-06 | 2.80E-08 |
| 0.95 | 0.364 | 0.11875 | 3.31E-04 | 3.48E-05 | 4.09E-06 | 2.24E-08 |
| 0.9 | 0.344 | 0.1125 | 2.97E-04 | 3.09E-05 | 3.35E-06 | 1.74E-08 |
| 0.85 | 0.325 | 0.10625 | 2.63E-04 | 2.73E-05 | 2.73E-06 | 1.38E-08 |
| 0.8 | 0.306 | 0.1 | 2.35E-04 | 2.40E-05 | 2.24E-06 | 1.11E-08 |

# 5.4. THE DISCUSSION OF THE RESULTS AND THE CONCLUSION

Referring to the main objective of this thesis, which is *to study the effect of the gravitational mass on the electromagnetic radiation from an oblique, relativistically rotating dipole*, we can derive various conclusions. Since we have three physical parameters ($M$, $\chi$, $\omega$), which affect the energy $E_l$ and $\varepsilon$, we will discuss each effect in this section. We will state the effect of the first parameter ($M$), which is the crux of our work, and then go on to give the others for completeness.

## 5.4.1. THE EFFECT OF THE GRAVITATIONAL MASS ON THE ENERGY

From the foregoing spectra, Figs. 5.6-25, and their corresponding tables, Tables.5.1-9, one sees a dramatic reduction of the radiation due to the increase of mass, in the physical regions. In order to evaluate this reduction, we determine the ratios between the "energies", i.e. $E_l$ and $\varepsilon$, at different masses. Thus, we come up with Tables.5.10-17 for $\chi = \pi/2, \pi/4, \pi/6$, and $\pi/8$ .

**Table.510.** The ratios between radiated energies, $E_l$, from different massive dipoles, for $\chi = \pi/2$.

| $\beta_{max}$ | $\omega$(GU) | $l$ | E(0.5)/E(1) | E(1)/E(1.5) | E(1.5)/E(2) | E(2)/E(2.5) | E(2.5)/E(3) |
|---|---|---|---|---|---|---|---|
| 0.99 | 0.12375 | 1 | 5.44445 | 3.239994 | 2.776881 | 2.778603 | 3.24077 |
| 0.99 | 0.12375 | 21 | 5.444444 | 3.239994 | 2.777783 | 2.777778 | 3.24056 |
| 0.99 | 0.12375 | 41 | 5.44443 | 3.240008 | 2.777771 | 2.777788 | 3.239988 |
| 0.99 | 0.12375 | 61 | 5.444424 | 3.240004 | 2.777781 | 2.777774 | 3.239995 |
| 0.99 | 0.12375 | 81 | 5.444423 | 3.240015 | 2.777776 | 2.777782 | 3.239687 |
| 0.99 | 0.12375 | 101 | 5.444407 | 3.240016 | 2.777788 | 2.77777 | 3.239625 |
| 0.95 | 0.11875 | 1 | 5.444443 | 3.239988 | 2.777866 | 2.777838 | 3.239805 |
| 0.95 | 0.11875 | 21 | 5.44444 | 3.239993 | 2.777785 | 2.777776 | 3.240151 |
| 0.95 | 0.11875 | 41 | 5.444454 | 3.239996 | 2.777771 | 2.777783 | 3.239994 |
| 0.95 | 0.11875 | 61 | 5.44446 | 3.239999 | 2.777778 | 2.777778 | 3.239994 |
| 0.95 | 0.11875 | 81 | 5.444448 | 3.240001 | 2.777782 | 2.777767 | 3.240014 |
| 0.95 | 0.11875 | 101 | 5.44444 | 3.240006 | 2.777774 | 2.777772 | 3.240013 |
| 0.90 | 0.1125 | 1 | 5.444446 | 3.24 | 2.778291 | 2.77485 | 3.242718 |
| 0.90 | 0.1125 | 21 | 5.444433 | 3.240002 | 2.777776 | 2.777775 | 3.240637 |
| 0.90 | 0.1125 | 41 | 5.444441 | 3.240005 | 2.777776 | 2.777775 | 3.24 |
| 0.90 | 0.1125 | 61 | 5.444453 | 3.239994 | 2.777781 | 2.777775 | 3.240008 |
| 0.90 | 0.1125 | 81 | 5.444435 | 3.240012 | 2.777776 | 2.777775 | 3.24 |
| 0.90 | 0.1125 | 101 | 5.444463 | 3.239996 | 2.777776 | 2.777775 | 3.240014 |
| 0.85 | 0.10625 | 1 | 5.444452 | 3.240008 | 2.777891 | 2.778226 | 3.239713 |
| 0.85 | 0.10625 | 21 | 5.444442 | 3.239991 | 2.777783 | 2.777775 | 3.242006 |
| 0.85 | 0.10625 | 41 | 5.444449 | 3.240001 | 2.777774 | 2.77778 | 3.239991 |
| 0.85 | 0.10625 | 61 | 5.444448 | 3.239996 | 2.777783 | 2.777771 | 3.239998 |
| 0.85 | 0.10625 | 81 | 5.444467 | 3.240007 | 2.777769 | 2.777784 | 3.239993 |
| 0.85 | 0.10625 | 101 | 5.444465 | 3.240008 | 2.777768 | 2.777784 | 3.239994 |
| 0.80 | 0.1 | 1 | 5.444431 | 3.24001 | 2.777854 | 2.778452 | 3.239125 |
| 0.80 | 0.1 | 21 | 5.444443 | 3.240004 | 2.777776 | 2.777774 | 3.240141 |
| 0.80 | 0.1 | 41 | 5.444446 | 3.239998 | 2.777789 | 2.777769 | 3.239998 |
| 0.80 | 0.1 | 61 | 5.44444 | 3.240001 | 2.77779 | 2.777771 | 3.240006 |
| 0.80 | 0.1 | 81 | 5.444451 | 3.239999 | 2.777791 | 2.777762 | 3.24 |
| 0.80 | 0.1 | 101 | 5.44444 | 3.239993 | 2.777793 | 2.777765 | 3.240013 |
| Average Ratios | | | 5.44442 | 3.24001 | 2.777695 | 2.777856 | 3.239881 |

**Table.5.11.** The ratios between total energies $\varepsilon$ emitted by the different massive rotating dipoles, for $\chi = \pi/2$.

| $\chi$ | $\beta$ | E(0)/E(0.5) | E(0.5)/E(1) | E(1)/E(1.5) | E(1.5)/E(2) | E(2)/E(2.5) | E(2.5)/E(3) |
|---|---|---|---|---|---|---|---|
| $\pi/2$ | 0.80 | 1.5318231 | 2.2019002 | 1.9858491 | 1.94495413 | 2.03465636 | 2.3464867 |
| $\pi/2$ | 0.85 | 1.3967066 | 2.0303951 | 1.9165049 | 1.88644689 | 2.05263158 | 2.2315212 |
| $\pi/2$ | 0.90 | 1.2534099 | 1.8238953 | 1.8582725 | 1.77155172 | 1.96194503 | 2.3888889 |
| $\pi/2$ | 0.95 | 1.0631543 | 1.5180345 | 1.5936355 | 1.60074677 | 1.64776972 | 2.0667575 |
| $\pi/2$ | 0.99 | 1.0110608 | 1.1617511 | 1.6193896 | 1.95109015 | 2.28580675 | 3.114113 |
| Average Ratio | | 1.2512309 | 1.7471953 | 1.7947303 | 1.83095793 | 1.99656189 | 2.4295534 |

**Table.5.12.** The ratios between radiated energies, $E_l$, from different massive dipoles, for $\chi = \pi/4$.

| $\beta_{max}$ | $\omega$(GU) | $l$ | E(0.5)/E(1) | E(1)/E(1.5) | E(1.5)/E(2) | E(2)/E(2.5) |
|---|---|---|---|---|---|---|
| 0.99 | 0.12375 | 1 | 6.486799 | 4.261136 | 4.573805 | 9.938017 |
| 0.99 | 0.12375 | 21 | 6.486799 | 4.262482 | 4.571344 | 9.922086 |
| 0.99 | 0.12375 | 41 | 6.486801 | 4.262468 | 4.571357 | 9.933284 |
| 0.99 | 0.12375 | 61 | 6.486802 | 4.262463 | 4.571361 | 9.937204 |
| 0.99 | 0.12375 | 81 | 6.486802 | 4.26246 | 4.571364 | 9.939202 |
| 0.99 | 0.12375 | 101 | 6.486802 | 4.262458 | 4.571365 | 9.940412 |
| 0.95 | 0.11875 | 1 | 6.486779 | 4.26284 | 4.571106 | 9.943883 |
| 0.95 | 0.11875 | 21 | 6.486798 | 4.262483 | 4.57136 | 9.921641 |
| 0.95 | 0.11875 | 41 | 6.486786 | 4.262496 | 4.571336 | 9.924453 |
| 0.95 | 0.11875 | 61 | 6.486797 | 4.262477 | 4.571353 | 9.95263 |
| 0.95 | 0.11875 | 81 | 6.48678 | 4.262503 | 4.571324 | 9.925938 |
| 0.95 | 0.11875 | 101 | 6.486779 | 4.262504 | 4.571322 | 9.926242 |
| 0.9 | 0.1125 | 1 | 6.486802 | 4.262724 | 4.572327 | 9.939985 |
| 0.9 | 0.1125 | 21 | 6.486774 | 4.262475 | 4.57135 | 9.945986 |
| 0.9 | 0.1125 | 41 | 6.486803 | 4.262483 | 4.571338 | 9.955204 |
| 0.9 | 0.1125 | 61 | 6.486797 | 4.262466 | 4.571369 | 9.940183 |
| 0.9 | 0.1125 | 81 | 6.486818 | 4.262488 | 4.571331 | 9.960074 |
| 0.9 | 0.1125 | 101 | 6.486822 | 4.262488 | 4.57133 | 9.96107 |
| 0.85 | 0.10625 | 1 | 6.486774 | 4.265802 | 4.569817 | 9.943899 |
| 0.85 | 0.10625 | 21 | 6.486777 | 4.26247 | 4.57136 | 9.967451 |
| 0.85 | 0.10625 | 41 | 6.486787 | 4.262476 | 4.571353 | 9.961574 |
| 0.85 | 0.10625 | 61 | 6.486779 | 4.262488 | 4.571347 | 9.925529 |
| 0.85 | 0.10625 | 81 | 6.486792 | 4.262479 | 4.571349 | 9.958475 |
| 0.85 | 0.10625 | 101 | 6.486793 | 4.262479 | 4.571349 | 9.957842 |
| 0.8 | 0.1 | 1 | 6.486776 | 4.265854 | 4.570064 | 9.936709 |
| 0.8 | 0.1 | 21 | 6.486782 | 4.262493 | 4.569267 | 9.944812 |
| 0.8 | 0.1 | 41 | 6.486794 | 4.262477 | 4.571346 | 9.933429 |
| 0.8 | 0.1 | 61 | 6.486786 | 4.262476 | 4.57135 | 9.946371 |
| 0.8 | 0.1 | 81 | 6.486801 | 4.262469 | 4.572444 | 9.927429 |
| 0.8 | 0.1 | 101 | 6.486802 | 4.262467 | 4.572668 | 9.926203 |
| Average Ratios | | | 6.486783 | 4.262532 | 4.571307 | 9.930131 |

**Table.5.13.** The ratios between total energies $\varepsilon$ emitted by the different massive rotating dipoles, for $\chi = \pi/8$.

| $\chi$ | $\beta_{max}$ | E(0)/E(0.5) | E(0.5)/E(1) | E(1)/E(1.5) | E(1.5)/E(2) | E(2)/E(2.5) |
|---|---|---|---|---|---|---|
| $\pi/4$ | 0.80 | 2.7169958 | 5.1001673 | 2.4263201 | 2.75662185 | 8.40453352 |
| $\pi/4$ | 0.85 | 2.7437642 | 3.1000576 | 3.8560412 | 2.849284 | 6.53483228 |
| $\pi/4$ | 0.90 | 2.7730921 | 3.1348768 | 3.5862668 | 2.97177311 | 6.43357248 |
| $\pi/4$ | 0.95 | 2.768261 | 3.1605931 | 2.579413 | 3.0302691 | 7.90115576 |
| $\pi/4$ | 0.99 | 2.7529618 | 3.1609528 | 2.6172668 | 3.05191784 | 6.78964011 |
| Average Ratio | | 2.751015 | 3.5313295 | 3.0130616 | 2.93197318 | 7.21274683 |

**Table.5.14.** The ratios between radiated energies, $E_l$, from different massive dipoles, for $\chi = \pi/6$.

| $\beta_{max}$ | $\omega$(GU) | $l$ | E(0.5)/E(1) | E(1)/E(1.5) |
|---|---|---|---|---|
| 0.99 | 0.12375 | 1 | 9.00E+00 | 9.00E+00 |
| 0.99 | 0.12375 | 21 | 9.00E+00 | 9.00E+00 |
| 0.99 | 0.12375 | 41 | 9.00E+00 | 9.00E+00 |
| 0.99 | 0.12375 | 61 | 9.00E+00 | 9.00E+00 |
| 0.99 | 0.12375 | 81 | 9.00E+00 | 9.00E+00 |
| 0.99 | 0.12375 | 101 | 9.00E+00 | 9.00E+00 |
| 0.95 | 0.11875 | 1 | 9.01E+00 | 8.99E+00 |
| 0.95 | 0.11875 | 21 | 9.00E+00 | 9.00E+00 |
| 0.95 | 0.11875 | 41 | 9.00E+00 | 9.00E+00 |
| 0.95 | 0.11875 | 61 | 9.00E+00 | 9.00E+00 |
| 0.95 | 0.11875 | 81 | 9.00E+00 | 9.00E+00 |
| 0.95 | 0.11875 | 101 | 9.00E+00 | 9.00E+00 |
| 0.9 | 0.1125 | 1 | 9.01E+00 | 8.99E+00 |
| 0.9 | 0.1125 | 21 | 9.00E+00 | 9.00E+00 |
| 0.9 | 0.1125 | 41 | 9.00E+00 | 9.00E+00 |
| 0.9 | 0.1125 | 61 | 9.00E+00 | 9.00E+00 |
| 0.9 | 0.1125 | 81 | 9.00E+00 | 9.00E+00 |
| 0.9 | 0.1125 | 101 | 9.00E+00 | 9.00E+00 |
| 0.85 | 0.10625 | 1 | 9.01E+00 | 8.98E+00 |
| 0.85 | 0.10625 | 21 | 9.00E+00 | 9.01E+00 |
| 0.85 | 0.10625 | 41 | 9.00E+00 | 9.00E+00 |
| 0.85 | 0.10625 | 61 | 9.00E+00 | 9.00E+00 |
| 0.85 | 0.10625 | 81 | 9.00E+00 | 9.00E+00 |
| 0.85 | 0.10625 | 101 | 9.00E+00 | 9.00E+00 |
| 0.80 | 0.1 | 1 | 9.01E+00 | 8.99E+00 |
| 0.80 | 0.1 | 21 | 9.00E+00 | 9.00E+00 |
| 0.80 | 0.1 | 41 | 9.00E+00 | 9.00E+00 |
| 0.80 | 0.1 | 61 | 9.00E+00 | 9.00E+00 |
| 0.80 | 0.1 | 81 | 9.00E+00 | 9.00E+00 |
| 0.80 | 0.1 | 101 | 9.00E+00 | 9.00E+00 |
| | | Average Ratios | 9.00E+00 | 9.00E+00 |

**Table.5.15.** The ratios between total energies $\varepsilon$ emitted by the different massive rotating dipoles, for $\chi = \pi/6$.

| $\chi$ | $\beta_{max}$ | E(0)/E(0.5) | E(0.5)/E(1) | E(1)/E(1.5) |
|---|---|---|---|---|
| $\pi/4$ | 0.80 | 7.9320504 | 2.5180908 | 8.6856431 |
| $\pi/4$ | 0.85 | 7.7003218 | 2.5957767 | 8.5293698 |
| $\pi/4$ | 0.90 | 7.4414588 | 2.6920199 | 8.2735307 |
| $\pi/4$ | 0.95 | 7.1678212 | 2.7922742 | 7.9599742 |
| $\pi/4$ | 0.99 | 6.9442497 | 2.8795677 | 7.6697479 |
| Average Ratio | | 7.4371804 | 2.6955459 | 8.2236532 |

**Table.5.16.** The ratios between radiated energies, $E_l$, from different massive dipoles, with $\chi = \pi/8$.

| $\beta_{max}$ | $\omega$(GU) | $l$ | E(0.5)/E(1) | E(1)/E(1.5) |
|---|---|---|---|---|
| 0.99 | 0.12375 | 1 | 1.51E+01 | 6.72E+02 |
| 0.99 | 0.12375 | 21 | 1.51E+01 | 6.72E+02 |
| 0.99 | 0.12375 | 41 | 1.51E+01 | 6.72E+02 |
| 0.99 | 0.12375 | 61 | 1.51E+01 | 6.72E+02 |
| 0.99 | 0.12375 | 81 | 1.51E+01 | 6.72E+02 |
| 0.99 | 0.12375 | 101 | 1.51E+01 | 6.72E+02 |
| 0.95 | 0.11875 | 1 | 1.51E+01 | 6.68E+02 |
| 0.95 | 0.11875 | 21 | 1.51E+01 | 6.70E+02 |
| 0.95 | 0.11875 | 41 | 1.51E+01 | 6.71E+02 |
| 0.95 | 0.11875 | 61 | 1.51E+01 | 6.71E+02 |
| 0.95 | 0.11875 | 81 | 1.51E+01 | 6.71E+02 |
| 0.95 | 0.11875 | 101 | 1.51E+01 | 6.71E+02 |
| 0.90 | 0.1125 | 1 | 1.52E+01 | 6.68E+02 |
| 0.90 | 0.1125 | 21 | 1.51E+01 | 6.71E+02 |
| 0.90 | 0.1125 | 41 | 1.51E+01 | 6.71E+02 |
| 0.90 | 0.1125 | 61 | 1.51E+01 | 6.71E+02 |
| 0.90 | 0.1125 | 81 | 1.51E+01 | 6.71E+02 |
| 0.90 | 0.1125 | 101 | 1.51E+01 | 6.71E+02 |
| 0.90 | 0.10625 | 1 | 1.52E+01 | 6.68E+02 |
| 0.90 | 0.10625 | 21 | 1.51E+01 | 6.71E+02 |
| 0.90 | 0.10625 | 41 | 1.51E+01 | 6.71E+02 |
| 0.90 | 0.10625 | 61 | 1.51E+01 | 6.71E+02 |
| 0.90 | 0.10625 | 81 | 1.51E+01 | 6.71E+02 |
| 0.90 | 0.10625 | 101 | 1.51E+01 | 6.71E+02 |
| 0.85 | 0.1 | 1 | 1.51E+01 | 6.67E+02 |
| 0.85 | 0.1 | 21 | 1.51E+01 | 6.70E+02 |
| 0.85 | 0.1 | 41 | 1.51E+01 | 6.70E+02 |
| 0.85 | 0.1 | 61 | 1.51E+01 | 6.71E+02 |
| 0.85 | 0.1 | 81 | 1.51E+01 | 6.71E+02 |
| 0.85 | 0.1 | 101 | 1.51E+01 | 6.71E+02 |
| 0.80 | 0.12375 | 1 | 1.51E+01 | 6.69E+02 |
| 0.80 | 0.12375 | 21 | 1.51E+01 | 6.72E+02 |
| 0.80 | 0.12375 | 41 | 1.51E+01 | 6.72E+02 |
| 0.80 | 0.12375 | 61 | 1.51E+01 | 6.72E+02 |
| 0.80 | 0.12375 | 81 | 1.51E+01 | 6.72E+02 |
| 0.80 | 0.12375 | 101 | 1.51E+01 | 6.72E+02 |
| | | Average Ratios | 1.51E+01 | 6.715E+02 |

**Table.5.17.** The ratios between total energies $\varepsilon$ emitted by the different massive rotating dipoles, for $\chi = \pi/8$.

| $\chi$ | $\beta_{max}$ | E(0)/E(0.5) | E(0.5)/E(1) | E(1)/E(1.5) |
|---|---|---|---|---|
| $\pi/4$ | 0.80 | 9.7583376 | 10.71969 | 201.72605 |
| $\pi/4$ | 0.85 | 9.6360837 | 10.01234 | 198.49334 |
| $\pi/4$ | 0.90 | 9.5981787 | 9.2389201 | 192.94874 |
| $\pi/4$ | 0.95 | 9.5083003 | 8.5232976 | 182.23644 |
| $\pi/4$ | 0.99 | 9.4221353 | 7.9374904 | 171.57049 |
| Average Ratio | | 9.5846071 | 9.2863476 | 189.39501 |

Notice that the ratios $E_l$, remain more or less the same for each $\chi$. However, the ratios $\varepsilon$, change for different velocities, $\beta$. The ratios decrease for higher values of $\beta$ in the most relativistic case, i.e. for $\chi = \pi/2$, as seen from Table 5.11 and Fig.5.26. However, these ratios are more or less the same for the other cases. As such, we can expect the same sort of behavior for higher values of $l$.

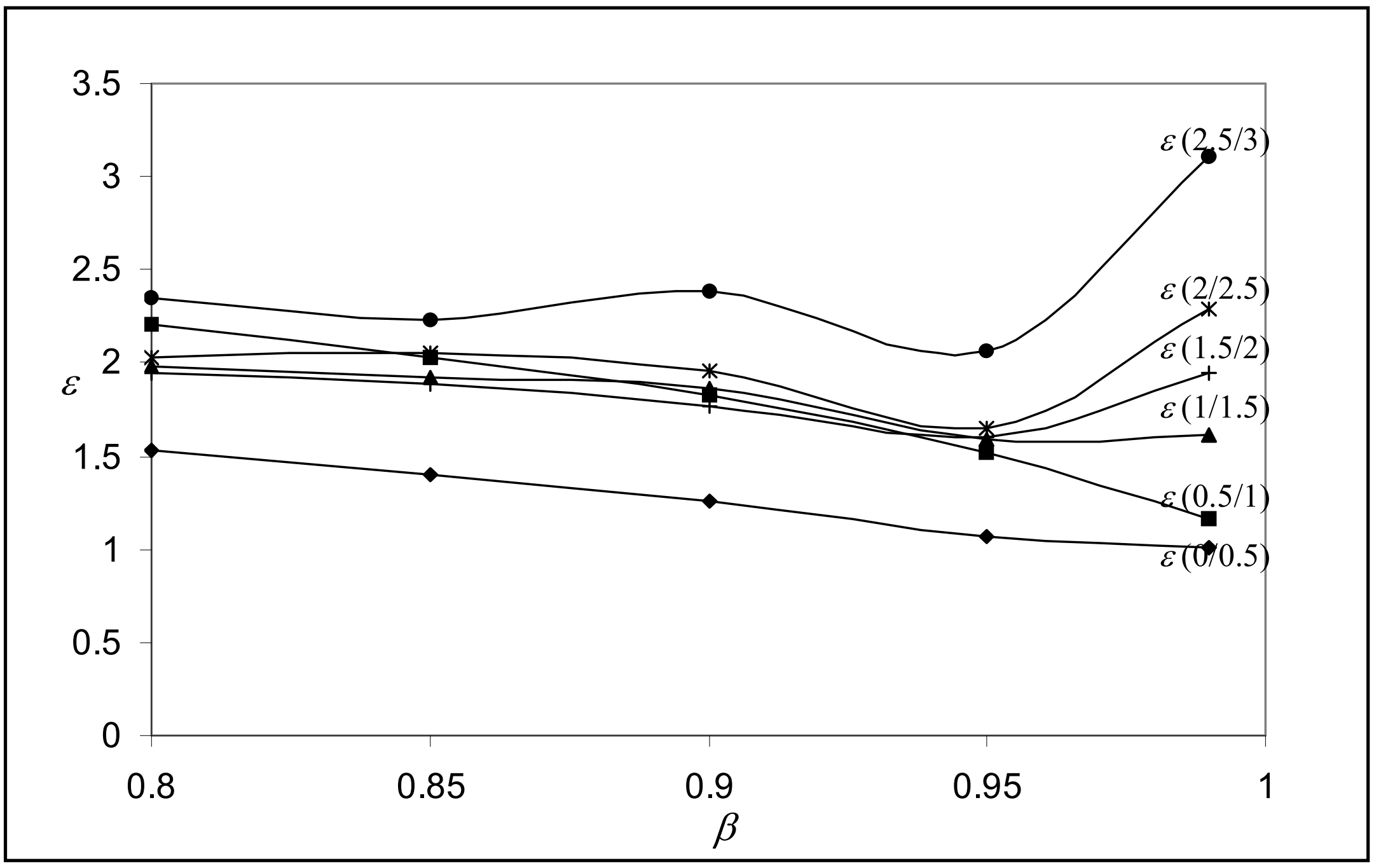

**Fig.5.26**. The plot of the reduction ratios of $\varepsilon$ versus the dipole's velocities $\beta$ for different masses of the dipole. The numbers in brackets are the factors of solar mass $M$ .

Thus, from the above spectra and tables, we arrive at the following conclusions:

1) ***The increase of gravitational mass causes an enormous reduction in the energy of the dipole.***

The reduction of the radiation, $\varepsilon$, due to the gravitational mass is in a range where it cannot be neglected. For example, the amount of radiation $\varepsilon$, decreases by a factor of more than 1.7, 3.5, and 9 when the mass is changed from 0.5 $M_\odot$ to 1$M_\odot$ for $\chi = \pi/2$, $\pi/4$, and $\pi/8$ respectively. The total reduction from 0.5 $M_\odot$ to 3$M_\odot$ is over 34 for the first case. As such, the tremendous relativistic enhancement of radiation found earlier [27, 28, 32], is strongly suppressed by the gravitational mass, see Fig.5.27. Presumably, most of the extra energy goes into increasing the relativistic angular momentum of the star and does not get radiated.

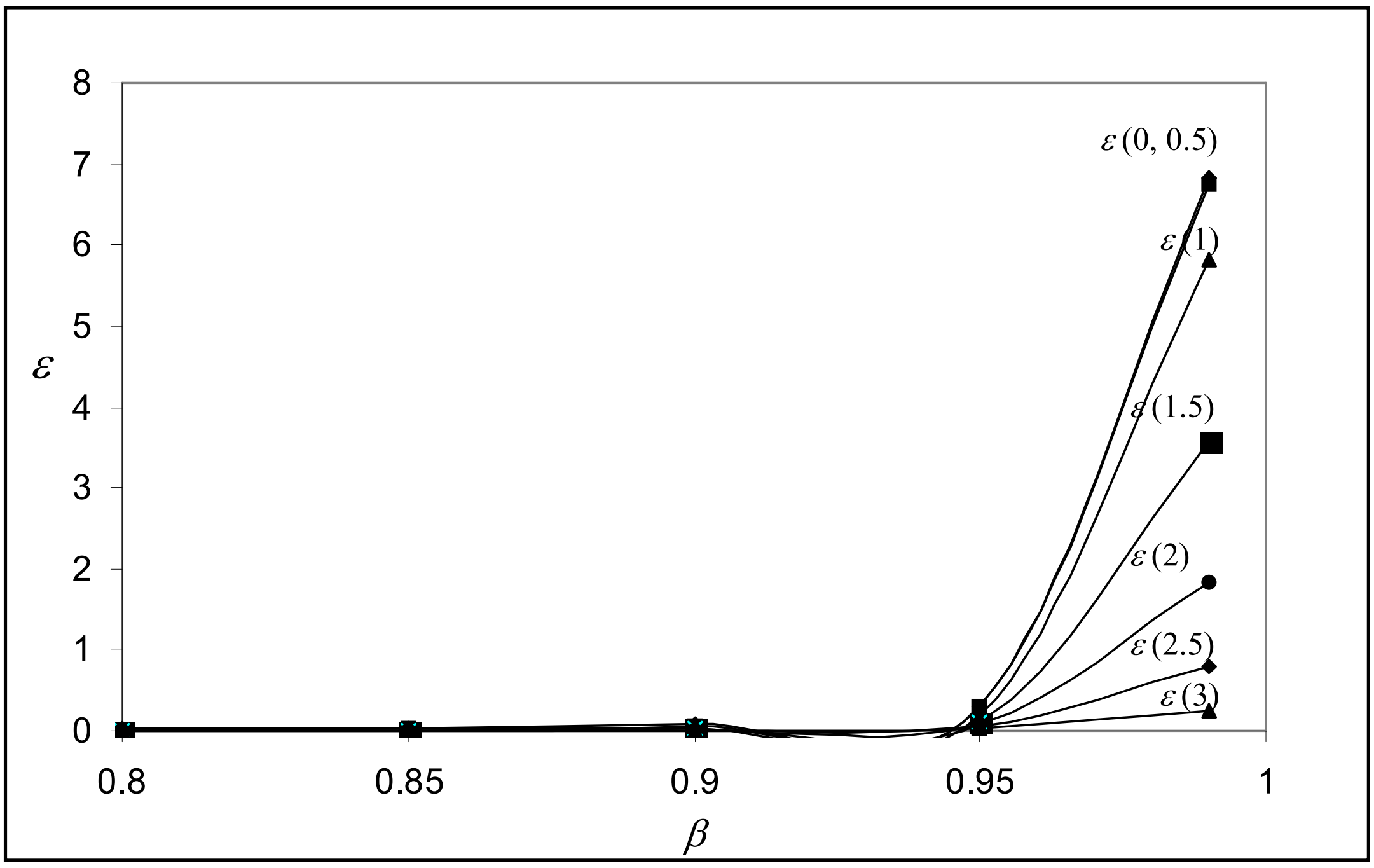

**Fig.5.27**. The plot of $\varepsilon$ versus the dipole's velocities $\beta$, for different masses of the dipole. The numbers in brackets are the factors of solar mass $M$ .

It is to be noted that the suppression of $E_l$ remains more or less constant for all high rotational velocities and all small angular momenta. However, it varies as inclination angles change. It is likely that the result would hold true for relatively high angular momentum but not for lower rotational speeds. Hence

2) ***In all cases the rate of reduction of $E_l$ is more or less independent of the angular momentum, l, and the high rotational velocity, β, but it varies with the angle of inclination, χ.***

As we change masses from 1 to 1.5 to 2 to 2.5 to 3 $M_{\odot}$, we find a change in the magnitude of the suppression of $E_l$, which varies from about 15 to 2.78. The interesting feature is that the suppression is least at a certain mass, *M* see Fig.5.27. However, the rate of suppression of $\varepsilon$ increases as the mass increases and becomes more or less constant around certain mass see Fig.5.28. The reason for this odd behavior is that the energy depends on the rotational speed and the mass in a highly non-linear manner. Thus

3) ***The rate of reduction of $E_l$ varies at higher and smaller masses, and becomes almost fixed around a minimum value. However, the rate of reduction of ε increases as the mass increases and become constant around certain value.***

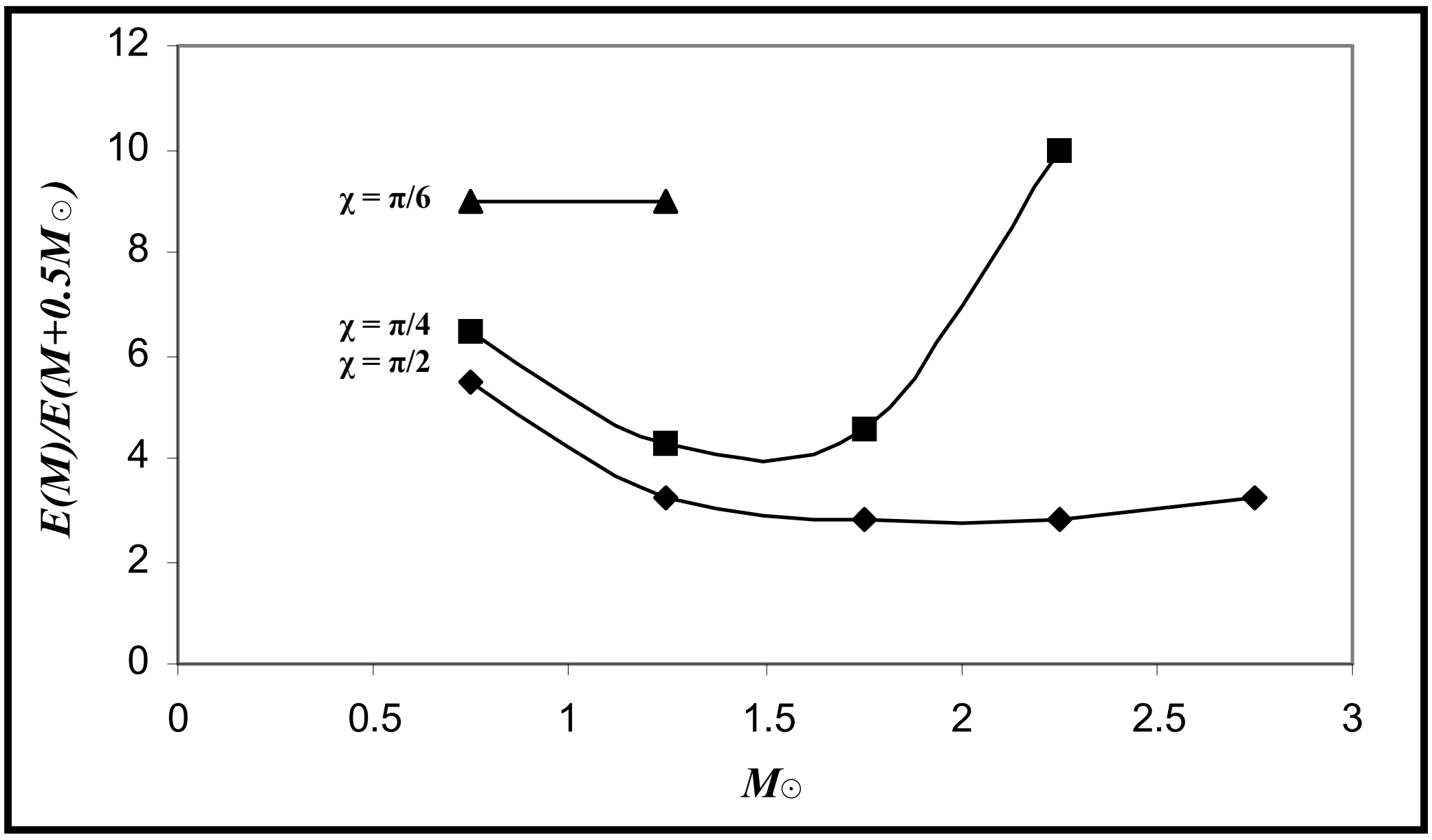

**Fig.5.28**. The plot of the reduction ratios of $E_l$ versus the dipole's mass, for $\chi = \pi/2$, $\pi/4$ and $\pi/6$.

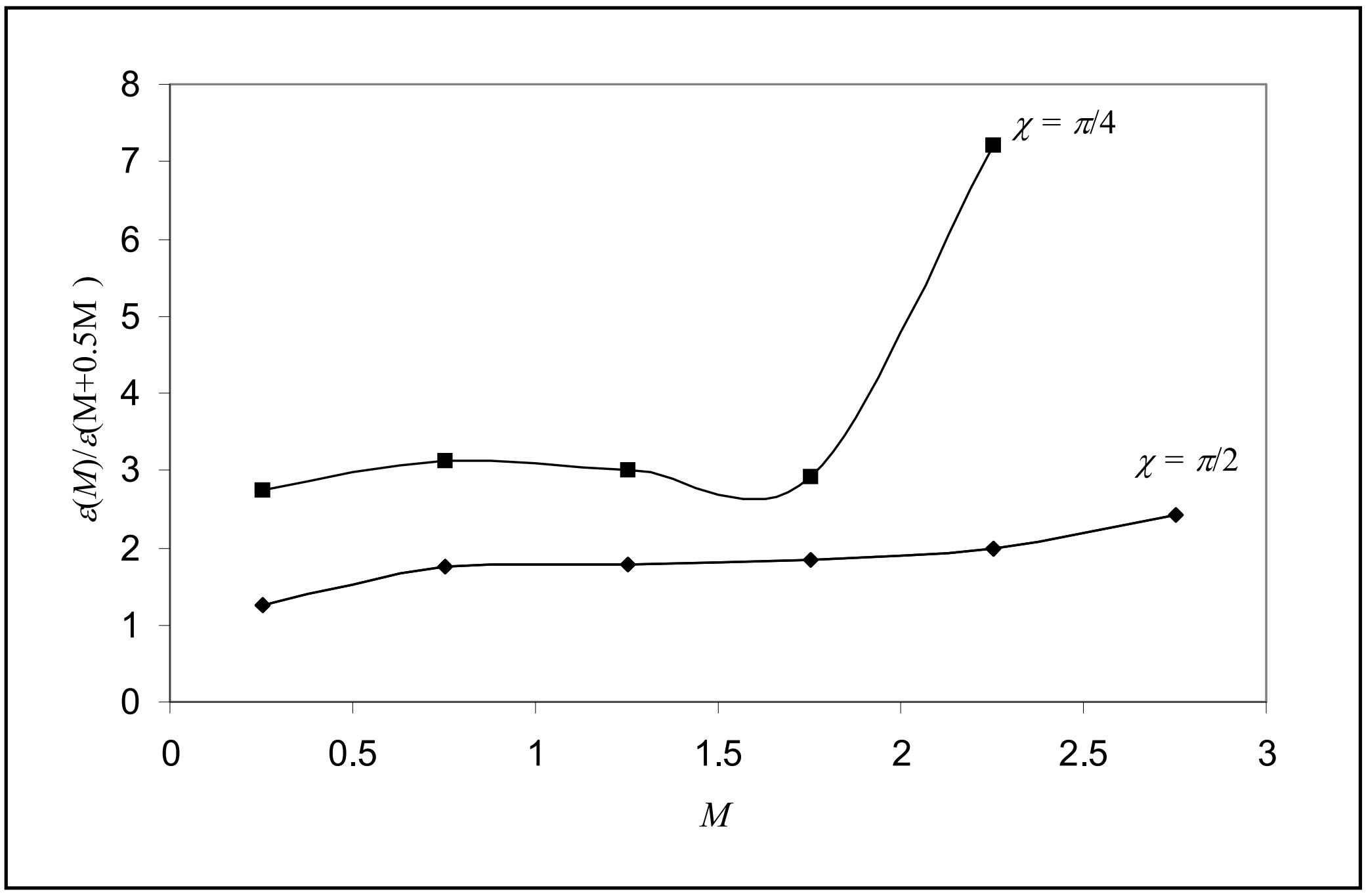

**Fig.5.29**. The plot of the reduction ratios of $\varepsilon$ versus the dipole's mass, for $\chi = \pi/2$, and $\pi/4$. The unit of $M$ axis is solar mass ($M$ ).

We are not interested in angles < π/4, because the relativistic enhancement is negligible for these angles.

Notice that the effect of the gravitational mass on the radiation appears at relatively low angular momenta $l$. As seen from Fig.5.6-25, the radiation of the massive dipole becomes similar to that of the massless dipole at relatively low angular momenta $l$ as the mass decreases, the rotational velocity, $\beta$, increases, and the inclination angle, $\chi$, approaches π/2. The closest behavior to the massless case is for $\chi = \pi/2$ and $\beta = 0.99$, see Fig. 5.6. However, for very high $l$, this effect becomes negligible, and the spectra are like the massless case, as described in § 5.2.4.

## 5.4.2. THE EFFECT OF INCLINATION ANGLE ON THE ENERGY

The relation, defined by (5.1), between $E_l$ and $\chi$ at fixed $\omega$, can be seen from (5.8). Note that $K_{lm}^2(\chi)$ is similar to $\sin^2\chi$. Therefore, we expect that the energy will increase as $\chi$ increases towards $\pi/2$. It is easy to see this effect in Tables.5.1-8 (see also Fig.5.27). It was shown by *Belinsky et al* [28] (our equation (4.22)), that the total radiation field depends on obliquity as $\sin^2\chi$.

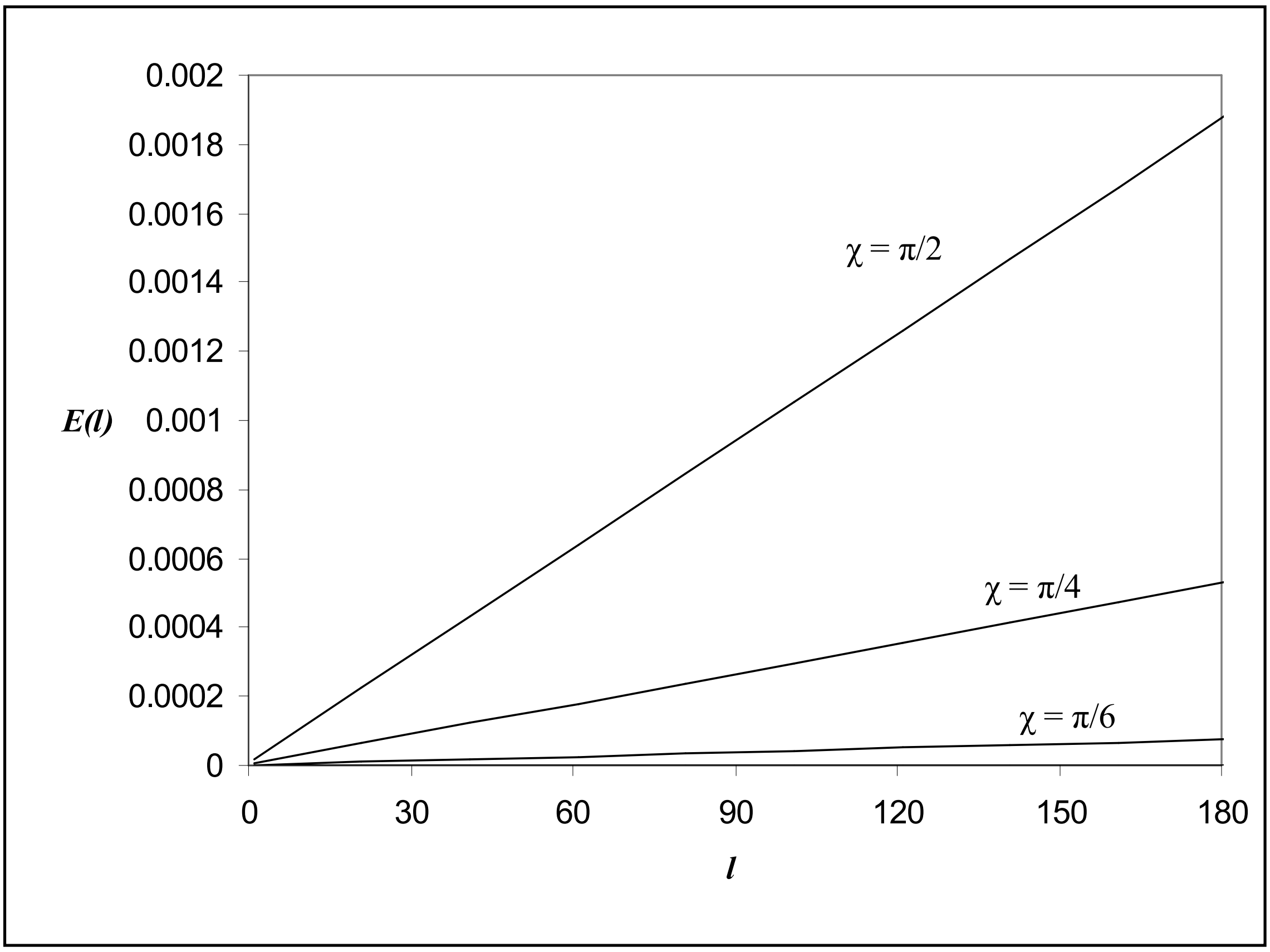

**Fig.5.30.** The energy spectra for a dipole of mass = 1.5 $M_{\odot}$, $\beta_{max}$= 0.99, for $\chi$ = π/2, π/4, and π/6 from upper line to the lower one respectively.

Taking the radiated energy, $E_l$ for $M = 1.5\ M_{\odot}$, at $l = 101$, and $\beta_{max} = 0.99$, we can construct Fig.2.28. Bear in mind that $E_l$ calculated is not reliable for $\chi = \pi/8$.

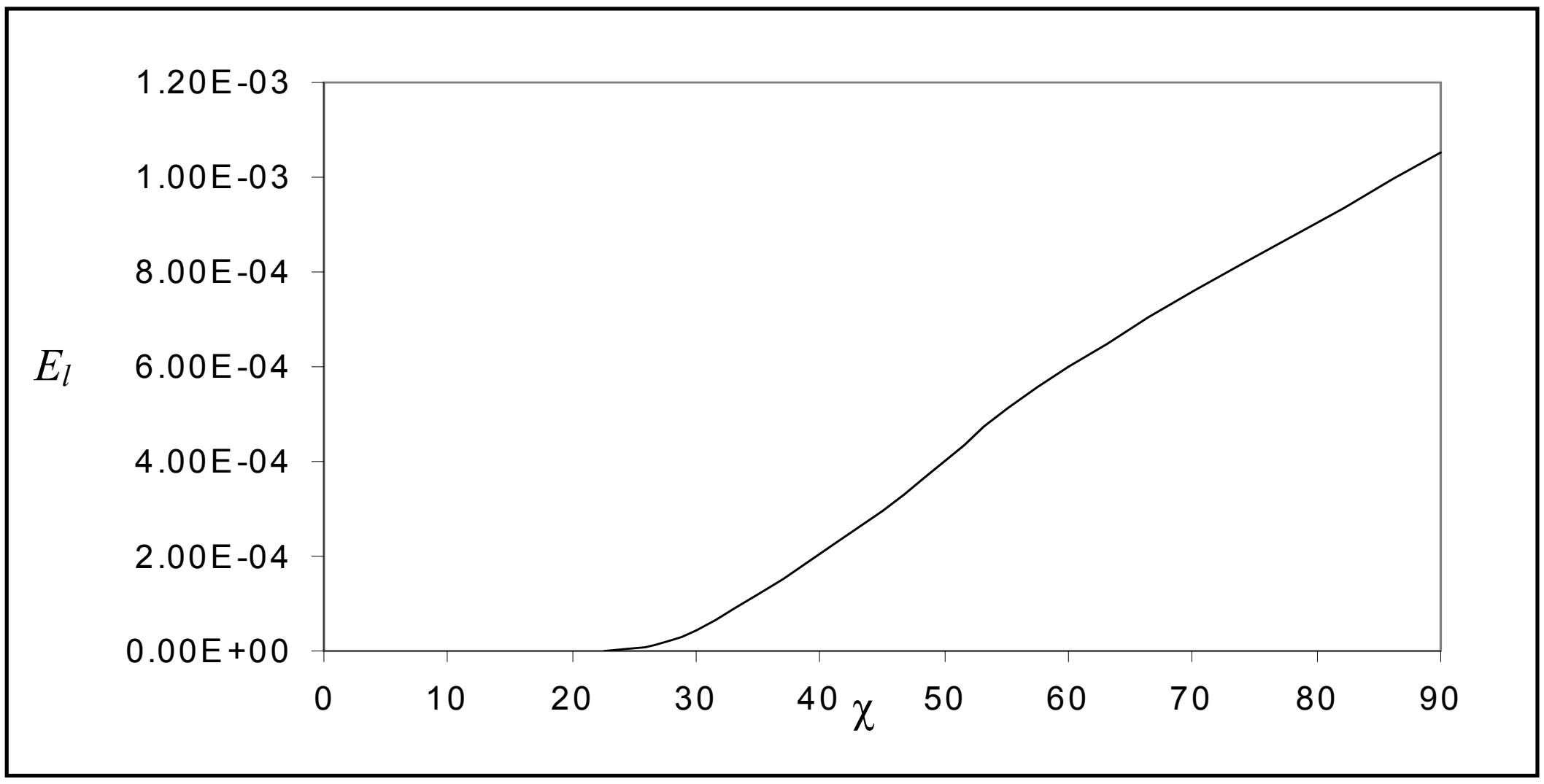

**Fig.5.31.** The plot of $E_{l,}$ versus $\chi$, for a dipole of mass = 1.5 $M_{\odot}$, for $l = 101$, $\beta_{max} = 0.99$. It is similar to the behavior of $\sin^2\chi$.

Plotting $\varepsilon$ versus $\chi$, for a dipole of $M = 1.5\ M_{\odot}$, and $\beta_{max} = 0.99$, see Fig.5.30, there is a huge increase in $\varepsilon$ as the $\chi \to \pi/2$.

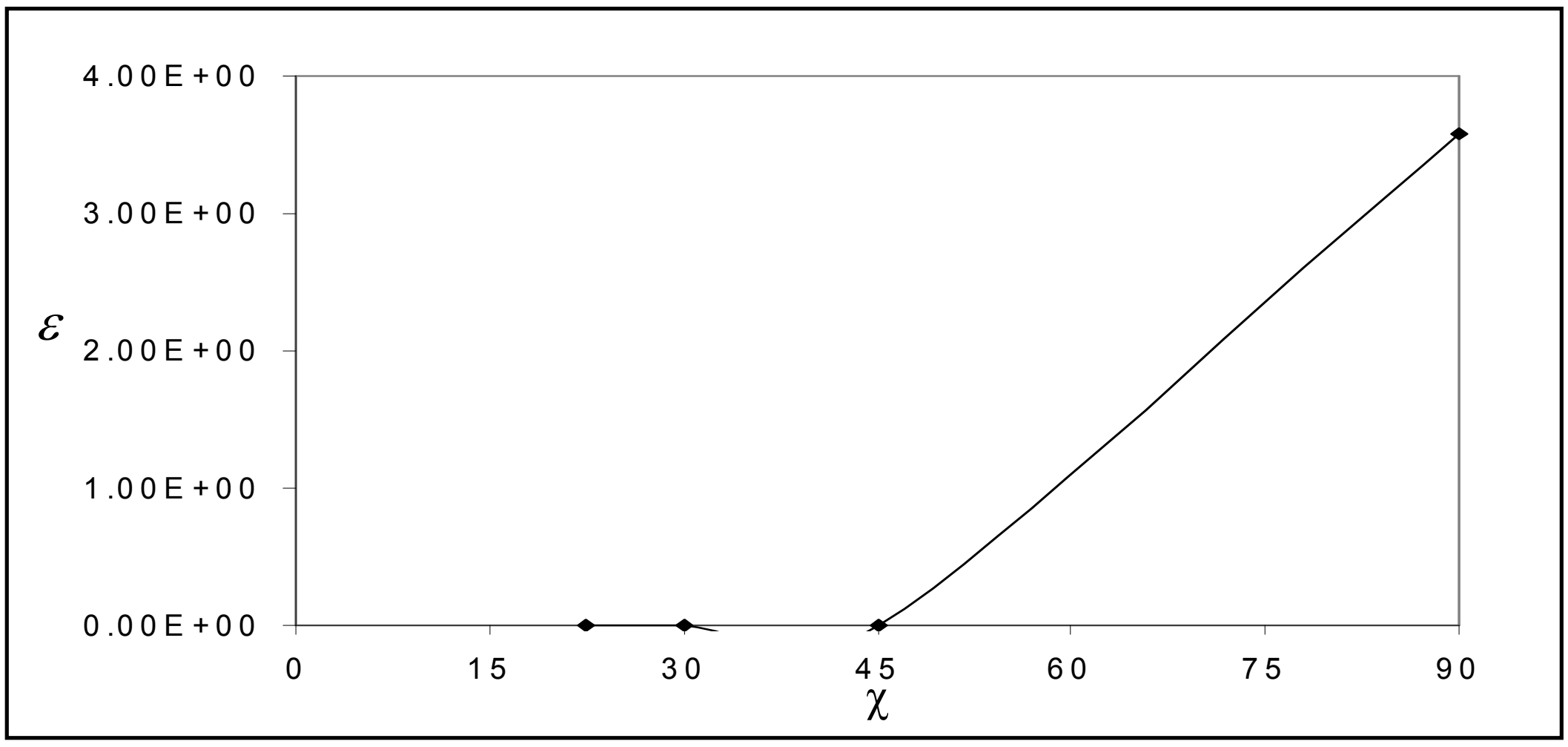

**Fig.5.32.** The plot of $\varepsilon$ versus $\chi$, for a dipole of mass = 1.5 $M_{\odot}$, and $\beta_{max} = 0.99$.

## 5.4.3. THE EFFECT OF ANGULAR FREQUENCY ON THE ENERGY

It is easy to notice from tables 5.1-9 that the radiated energy increases as the angular frequency, ω, and velocity, $\beta$, of the dipole increase. At a certain angular frequency, the angular velocity varies as the inclination angle, $\chi$, changes. Therefore, we expect to have different plots of $E_l$ versus $\omega$, for each $\chi$. To do so, we can construct Table. 5.18. The plot of $E_l$ versus $\omega$, at given angular momentum and mass can be seen in Fig.5.29.

**Table.5.18.** The amount of radiation energy, $E_l$ at different angular frequency, ω, at $l$ = 101, $M$ = 1.5 $M_{\odot}$

| ω | $E$ : π/2 | ω | $E$ - π/4 | ω | $E$ - π/6 |
|---|---|---|---|---|---|
| 0.1 | 0.000688 | 0.1 | 0.000194 | 0.1 | 2.75E-05 |
| 0.10625 | 0.000777 | 0.10625 | 0.000219 | 0.10625 | 3.11E-05 |
| 0.1125 | 0.000871 | 0.1125 | 0.000246 | 0.1125 | 3.49E-05 |
| 0.11875 | 0.000971 | 0.11875 | 0.000274 | 0.11875 | 3.88E-05 |
| 0.12375 | 0.001054 | 0.12375 | 0.000298 | 0.12375 | 4.22E-05 |

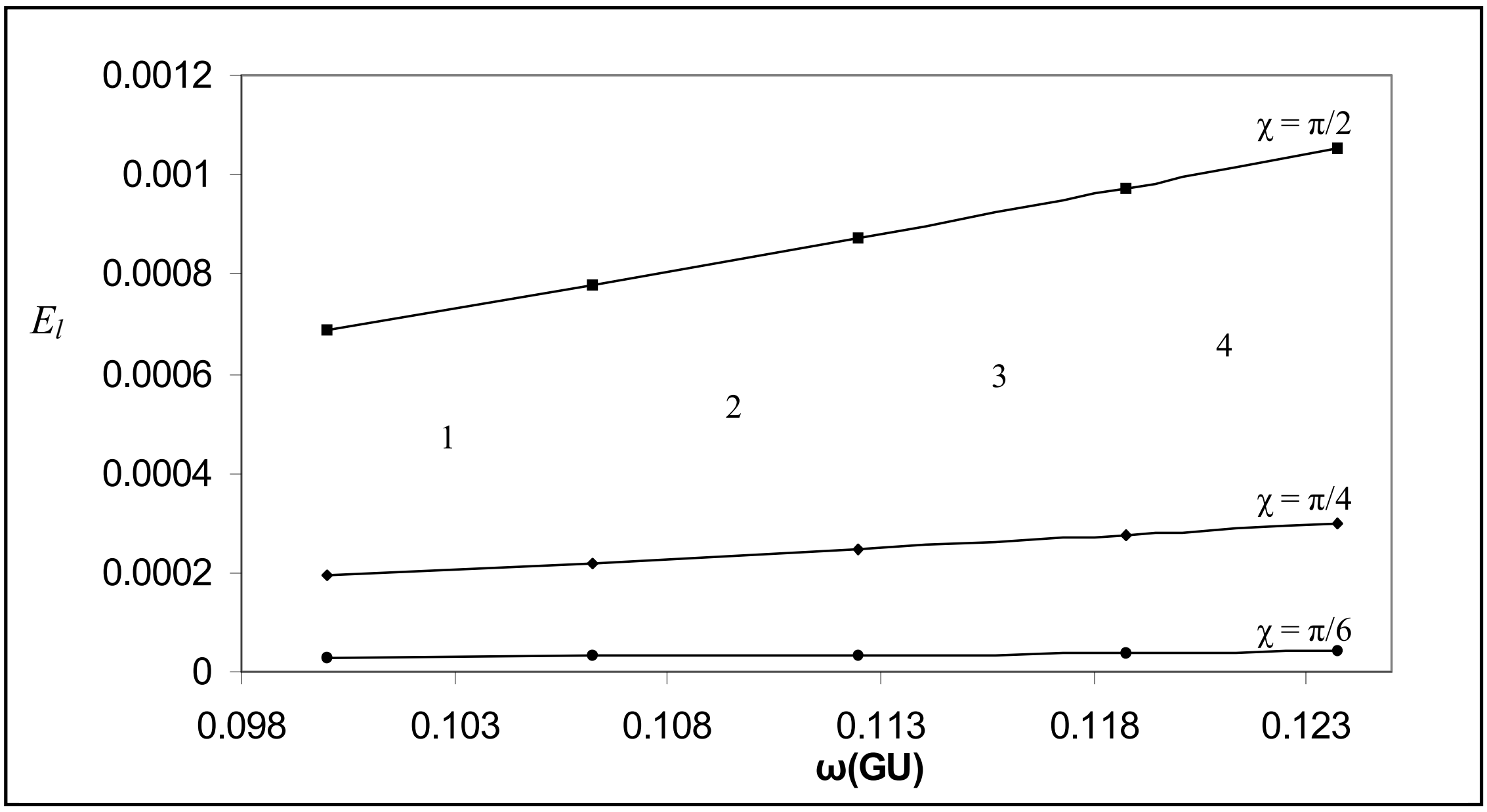

**Fig.5.33.** The relation between $E_l$ and ω, at $l$ = 101, $M$ = 1.5 $M_{\odot}$. The angular frequency, ω, is given in gravitational unit (GU). The numbers refer to lines segments.

The plot of $\varepsilon$ versus $\omega$, for a dipole of $M = 1.5\ M_\odot$, and $\chi=\pi/2$, shows that $\varepsilon$ goes faster than $\omega^2$, probably close to $\omega^3$, see Fig.5.32. Since $\varepsilon$ is the total energy per frequency, the total energy goes as $\omega^4$, which is similar to the classical result (3.63).

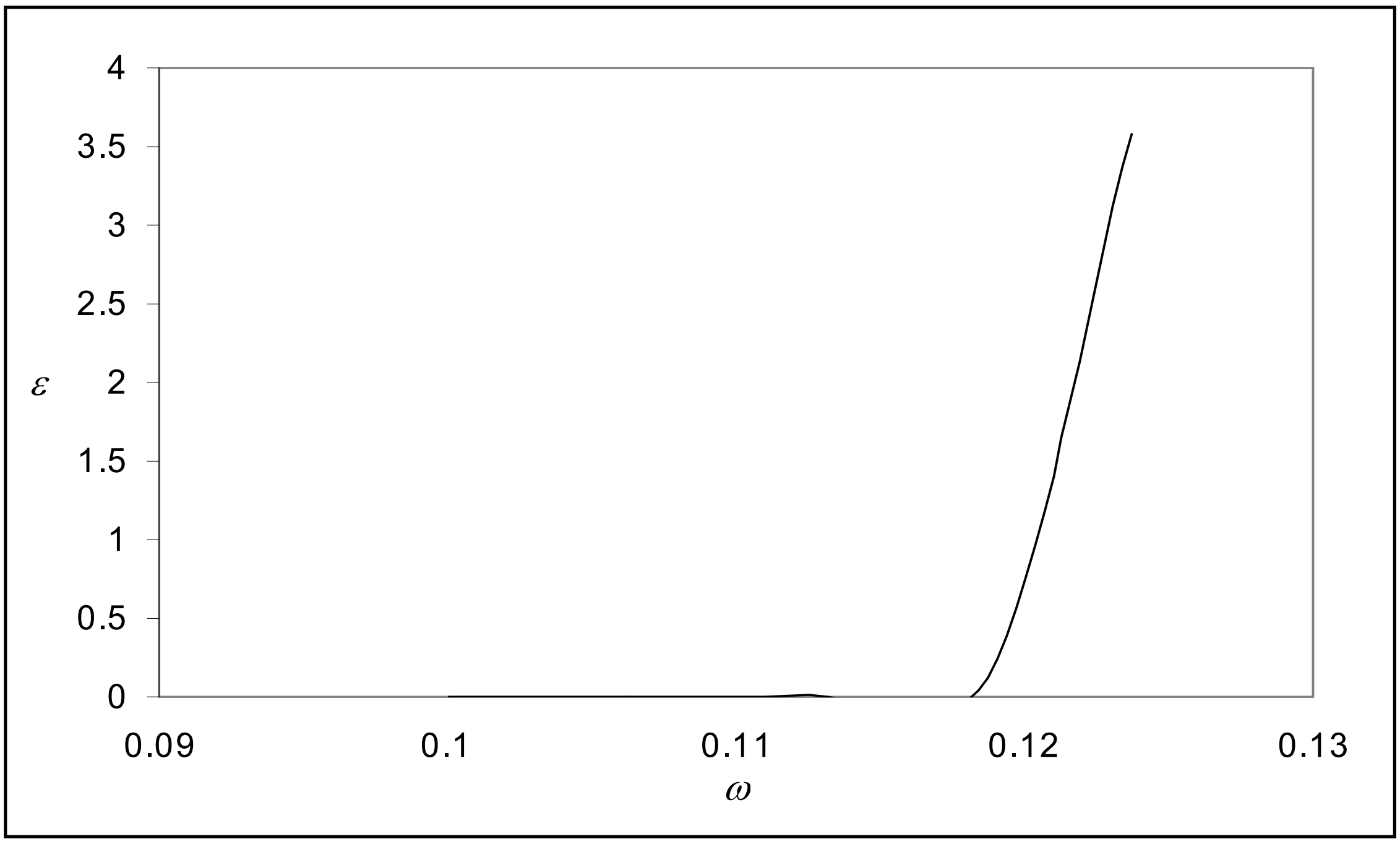

**Fig.5.34.** The relation between $\varepsilon$ and $\omega$, for $M = 1.5\ M_\odot$. The angular frequency, $\omega$, is given in gravitational units (GU).

Also, we can construct Table. 5.19. Plotting $E_l$ against $\beta$ gives different lines for each $\chi$, see Fig. 5.33.

**Table.5.19.** The amount of radiation energy, $E_l$, at different angular velocities, $\beta$, at $l = 101$, $M = 1.5\ M_\odot$

| $\beta$ | E-π/2 | $\beta$ | E-π/4 | $\beta$ | E-π/6 |
|---|---|---|---|---|---|
| 0.8 | 0.000688 | 0.566 | 0.000194 | 0.4 | 2.75E-05 |
| 0.85 | 0.000777 | 0.601 | 0.000219 | 0.425 | 3.11E-05 |
| 0.9 | 0.000871 | 0.636 | 0.000246 | 0.45 | 3.49E-05 |
| 0.95 | 0.000971 | 0.672 | 0.000274 | 0.475 | 3.88E-05 |
| 0.99 | 0.001054 | 0.7 | 0.000298 | 0.495 | 4.22E-05 |

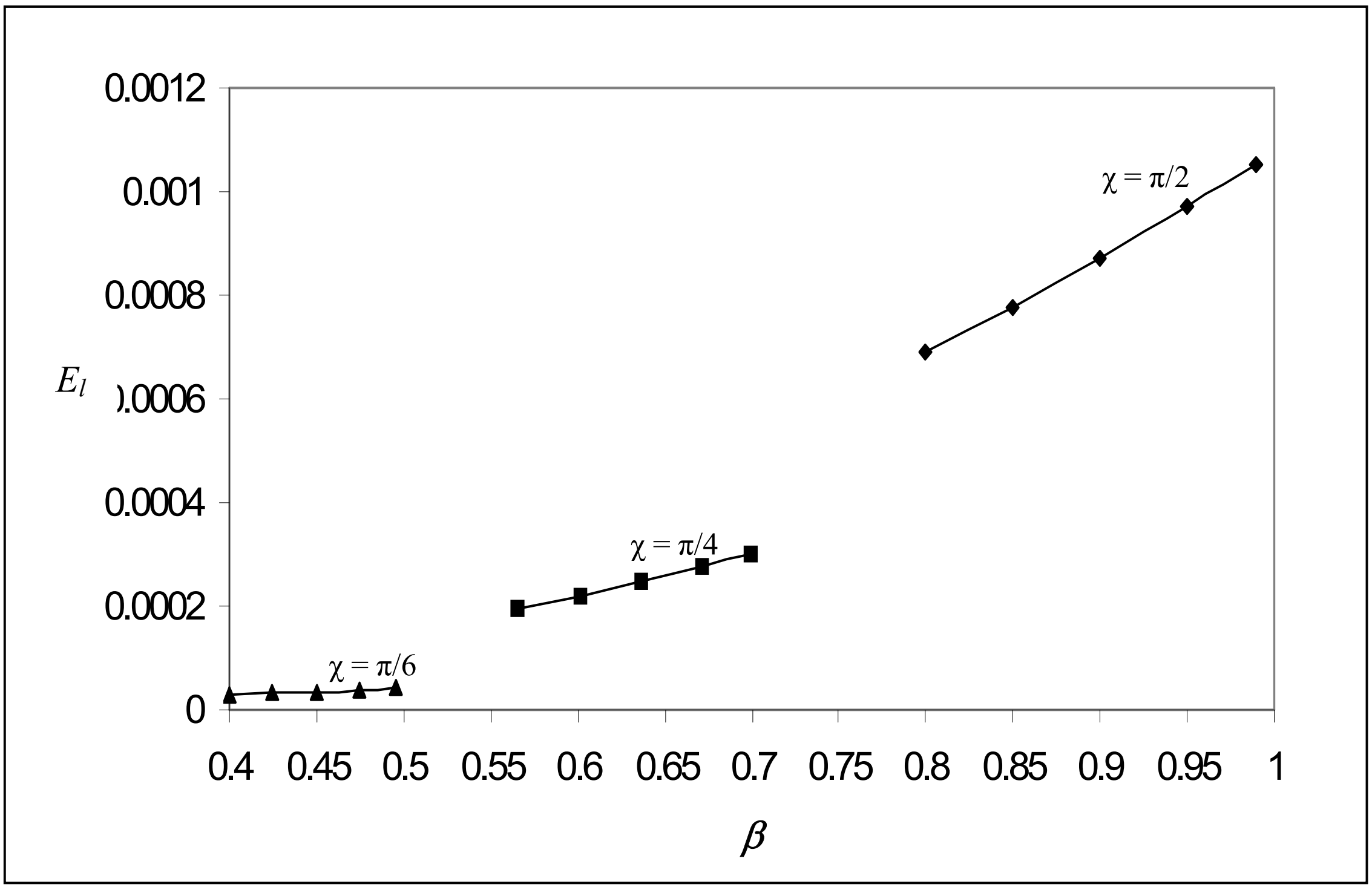

**Fig.5.35**. The relation between $E_l$ and$\beta$, at $l$ = 101, $M$ = 1.5 $M_{\odot}$. The angular velocity, $\beta$, is given in gravitational unit ($c$ = 1).

From Tables.5.17 we can obtain the relation between $E$ and $\omega$ by calculating the instantaneous slopes of different line segments shown in Fig.5.29. To do so, we construct Table.5.20 for the dipole of $M$ = 1.5 $M_{\odot}$, and $\chi = \pi/2$, $\pi/4$, and $\pi/6$.

**Table.5.20.** The slopes of each line segments shown in fig. 5.29, for a dipole of $M$ = 1.5 $M_{\odot}$, for $\chi = \pi/2, \pi/4, \pi/6$.

| Segments | ω | Slope: $E$ - π/2 | Slope: $E$ - π/4 | Slope: $E$ - π/6 |
|---|---|---|---|---|
| 1 | (0.1- 0.10625) | 0.01419664 | 0.0040085 | 0.0005688 |
| 2 | (0.10625 - 0.1125) | 0.0150576 | 0.0042515 | 0.000601 |
| 3 | (0.1125 - 0.11875) | 0.01591808 | 0.0044943 | 0.0006381 |
| 4 | (0.11875 - 0.12375) | 0.0166922 | 0.0047138 | 0.000666 |

From Table.5.19, we conclude that the slope increases as $\omega$ increases, but at a small rate in this relativistic range of $\omega$ for these values of $l$. For example, it changes from 0.0142 to 0.0167 as $\omega$ changes from 0.1 to 0.12375 (from $\beta$ = 0.80 to 0.99), for $\chi = \pi/2$, and $l$ = 101. Hence, the relation is not perfectly linear. However, in the limited range of values used, to a good approximation the relation between the radiated energy, $E$, and the angular frequency $\omega$, of the dipole, at certain $M$, $l$, and $\chi$, can be regarded as linear.

In fact, it seems that $E_l \sim \omega^2\sin^2\chi$. For the full range of values, it would presumably follow the classical behavior, $\varepsilon \sim \omega^3\sin^2\chi$, see (3.63). Again, this slope can be taken as only approximately reliable. Thus, once again, we see that the (special) relativistic enhancement expected is lost to the (general) relativistic increase of angular momentum.

For completeness and to compare with that of massless dipole spectra, $\chi = \pi/2$, shown in Fig.4.6, we can draw Fig.5.34 for a dipole of $M = 1.5\ M_\odot$.

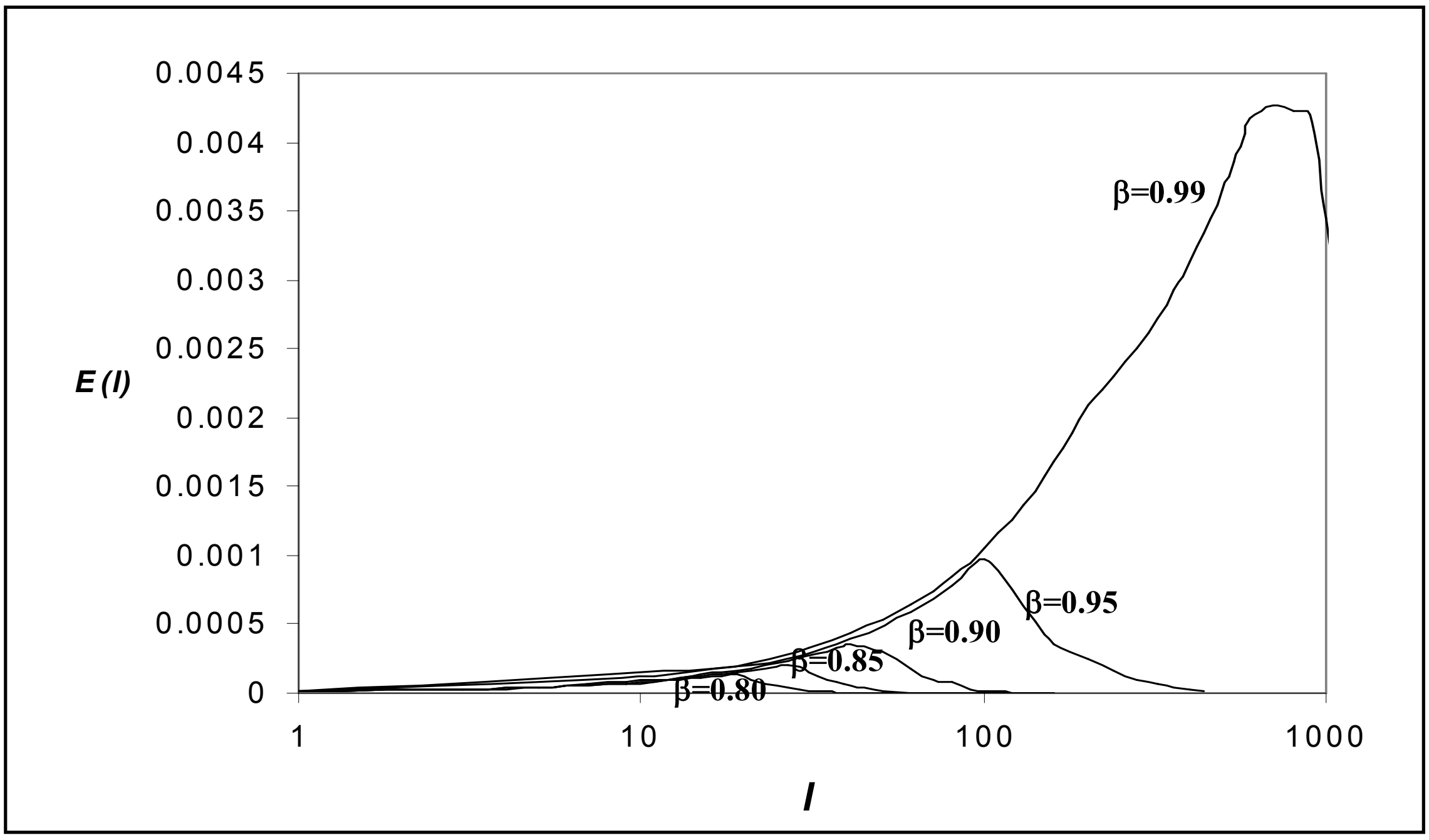

**Fig.5.36.** The radiation spectra for a massive magnetic (electric) dipole, $M = 1.5\ M_\odot$, for $\chi = \pi/2$ with different velocities.

# 5.5. SUMMARY AND CONCLUSIONS

Finally, we summarize some highlights of this work:

1) The main factor for radiation from neutron stars (NSs) is their magnetic field [40];

2) For the purpose of calculating the radiation from the NS, its magnetic field can be approximated by a dipole field [27] (the first calculation of the electromagnetic fields of the slow magnetized star was obtained by *Deutsch* [24]);

3) There are several reasons to consider very fast rotating NSs [32];

4) An exact solution for the radiation field from an arbitrarily fast rotating magnetic dipole, obtained by *Belinsky et al* [27], coincides with that of Deutsch in the non-relativistic limit;

5) A relativistic analysis of the Deutsch model, which corresponds with the result of Belinsky et al, was also obtained by *De Paolis* and *Qadir* [28];

6) The fast rotation introduces a relativistic scaling factor ($\sim \gamma^4$) which enhances the radiation;

7) Another approach to the problem, from a different direction, was adopted by extending the analysis of *Haxton* and *Ruffini* [31], to dipoles, by *De Paolis*, *Ingrosso* and *Qadir* [32] without incorporating the gravitational mass effect;

8) Using the same approach we incorporate the effect of the gravitational mass;

9) It is found that the effect of the gravitational mass is very significant in suppressing the relativistic enhancement factor by more than two orders of magnitude;

10) The effect of the gravitational mass is negligible at very high angular momentum, i.e. the spectra of massive and massless dipole [32] matched.

11) This is an indication that most of the angular momentum of the NS is retained as rotational kinetic energy instead of being radiated as electromagnetic energy, which indicates that the (special) relativistic enhancement expected is lost to the (general) relativistic increase of angular momentum;

12) The relation between the radiated energy, $E_l$, and the angular frequency is quadratic, and the relation between the total energy per frequency, $\varepsilon$, appears to be cubic as would be expected classically;

13) The next step (not done here) is to extend the analysis of Haxton and Ruffini to a *Kerr metric* [22], where the general relativistic effect of rotating masses is incorporated as well;

14) For actual application to pulsars, many other factors must be incorporated, such as the surrounding magnetosphere [41, 42], and so on.

# APPENDIX.A. MATHEMATICA 2.2 PROGRAM

Mathematica 4.1 program displayed in § 5.3.1 (Box.5.1), was first written by using Mathematica 2.2 as follow:

```
In[1]: Bmax=0.99
In[2]: vi=0
In[3]: xi=Pi/4
In[4]: B=Bmax*Sin[xi]
In[5]: Mns=0.5
In[6]: r=8
In[7]: w=Bmax/r
In[8]: y=2*Mns*w*m
In[9]: a=r*Sin[xi]/2Mns-1
In[10]: x0=a
In[11]: d=10^15
In[12]: Mg=1
In[13]: c[0]=0
In[14]: c[1]=1
In[15]: c[2]=-1/(2a^2(a+1))*(a*c[1]+(y^2(a+1)^3-a*L(L+1))c[0])
In[16]: v = c[0]
In[17]: dv = c[1]
In[18]: d2v = 2*c[2]
In[19]: u[x_]= Sqrt[(1+x)/x]*Exp [-Iy*x]
In[20]: du[x_]=D[u[x],x]
In[21]: G[L_,x_]=v*u[d]/(v]*du[d]-u[d]*dv)
In[22]: dG[L_,x_]=(u[d]*dv*(u[d]dv-vdu[d])-u[d]*v*(u[d]*d2v-dv*du[d])/(u[d]*dv-v*du[d])^2
In[23]: K[L_,m_]:=SphericalHarmonicY[L,m,xi,0]-SphericalHarmonicY[L,m,Pi-xi,Pi]
In[24]: Rev[L_,m_]:=Pi*Mg/(r*Mns*(L(L+1))^0.5)*K[L,m]*(x0/(x0+1)*dG[L,x0])
In[28]: Rod[L_,m_]:=2*Pi*Mg*w* (L*(L+1)-m(m+1))^0.5/(r*(L(L+1))^0.5)*K[L,m+1]*G[L,x0]
In[29]: e[L_]:=Sum[(m*w)^2/(2Pi)*(Abs[Rev[L,m]]^2+Abs[Rod[L,m]]^2),{m,1,L}]
In[30]: S=Table[e[L],{L,1,101,2}]//N
In[31]: ListPlot[S,PlotJoined->True]
```

# THE VITA

The author, Anwar Saleh Al-Muhammad, was born in 1973 at Al-Qattef where he finished his elementary and high school. In 1991 he joined King Fahd University of Petroleum and Minerals (KFUPM) and in 1996 he got a B.S degree in physics. In 1997, he started the M.S program of physics in the same university and he accomplished the M.S thesis in 2002.